\documentclass[fleqn,usenatbib]{mnras}

\usepackage{newtxtext,newtxmath}

\usepackage[T1]{fontenc}
\usepackage{ae,aecompl}

\usepackage{array}
\usepackage{eqnarray}
\usepackage{graphicx}	
\usepackage{multicol}        
\usepackage{bm}		
\usepackage{pdflscape}	
\usepackage{threeparttable}
\usepackage[dvipsnames]{xcolor}

\def\kms{\mbox{$\rm km\,s^{-1}$}}
\def\vlos{\mbox{$v_{\rm los}$}}

\def\logg{\mbox{log~{\it g}}}

\defcitealias{Mackey:2013aa}{Ma13}
\defcitealias{Kamann:2018ab}{K18}
\defcitealias{Song:2019aa}{S19}

\title[Masses and $M/L$ ratios of star clusters -- II]{Dynamical masses and mass-to-light ratios of resolved massive star clusters -- II. Results for 26 star clusters in the Magellanic Clouds\thanks{This paper is dedicated to the memory of Prof. Paul Hodge (1934-2019), a mentor to many of the authors of this paper and a pioneering giant of the study of Magellanic Cloud star clusters.}}

\author[Y.-Y. Song et al.]{Ying-Yi Song,$^{1}$\thanks{E-mail: yysong@umich.edu, songyingyi@gmail.com}\href{https://orcid.org/0000-0002-6270-8851}{\includegraphics[scale=0.6]{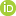}}
Mario Mateo,$^{1}$
John I. Bailey, III,$^{2}$\href{https://orcid.org/0000-0002-4272-263X}{\includegraphics[scale=0.6]{orcid.png}}
Matthew G. Walker,$^{3}$\href{https://orcid.org/0000-0003-2496-1925}{\includegraphics[scale=0.6]{orcid.png}}
\newauthor Ian U. Roederer,$^{1,4}$\href{https://orcid.org/0000-0001-5107-8930}{\includegraphics[scale=0.6]{orcid.png}}
Edward W. Olszewski,$^{5}$\href{https://orcid.org/0000-0002-7157-500X}{\includegraphics[scale=0.6]{orcid.png}}
Megan Reiter,$^{6,1}$\href{https://orcid.org/0000-0002-3887-6185}{\includegraphics[scale=0.6]{orcid.png}}
and
Anthony Kremin$^{7,8}$
\\
$^{1}$Department of Astronomy, University of Michigan, 1085 S. University Avenue, Ann Arbor, MI 48109, USA\\
$^{2}$Department of Physics, University of California Santa Barbara, Santa Barbara, CA 93106, USA \\
$^{3}$McWilliams Center for Cosmology, Department of Physics, Carnegie Mellon University, 5000 Forbes Avenue, Pittsburgh, PA 15213, USA\\
$^{4}$Joint Institute for Nuclear Astrophysics---Center for the Evolution of the Elements (JINA-CEE), USA\\
$^{5}$Steward Observatory, The University of Arizona, 933 N Cherry Avenue, Tucson, AZ 85721, USA \\
$^{6}$UK Astronomy Technology Centre, Blackford Hill, Edinburgh, EH9 3HJ, UK \\
$^{7}$Department of Physics, University of Michigan, Ann Arbor, MI 48109, USA \\
$^{8}$Lawrence Berkeley National Laboratory, 1 Cyclotron Road, Berkeley, CA 94720, USA \\
}

\date{Accepted XXX. Received YYY; in original form ZZZ}

\pubyear{2021}

\begin{document}
\label{firstpage}
\pagerange{\pageref{firstpage}--\pageref{lastpage}}
\maketitle

\begin{abstract}
We present spectroscopy of individual stars in 26 Magellanic Cloud (MC) star clusters with the aim of estimating dynamical masses and $V$-band mass-to-light ($M/L_V$) ratios over a wide range in age and metallicity. We obtained 3137 high-resolution stellar spectra with M2FS on the \textit{Magellan}/Clay Telescope. Combined with 239 published spectroscopic results of comparable quality, we produced a final sample of 2787 stars with good quality spectra for kinematic analysis in the target clusters. Line-of-sight velocities measured from these spectra and stellar positions within each cluster were used in a customized expectation-maximization (EM) technique to estimate cluster membership probabilities. Using appropriate cluster structural parameters and corresponding single-mass dynamical models, this technique ultimately provides self-consistent total mass and $M/L_V$ estimates for each cluster. Mean metallicities for the clusters were also obtained and tied to a scale based on calcium IR triplet metallicites. We present trends of the cluster $M/L_V$ values with cluster age, mass and metallicity, and find that our results run about 40~per~cent on average lower than the predictions of a set of simple stellar population (SSP) models. Modified SSP models that account for internal and external dynamical effects greatly improve agreement with our results, as can models that adopt a strongly bottom-light IMF.  To the extent that dynamical evolution must occur, a modified IMF is not required to match data and models. In contrast, a bottom-heavy IMF is ruled out for our cluster sample as this would lead to higher predicted $M/L_V$ values, significantly increasing the discrepancy with our observations.
\end{abstract}

\begin{keywords}
techniques: spectroscopic -- stars: kinematics and dynamics -- stars: abundances -- galaxies: star clusters -- Magellanic Clouds
\end{keywords}


\section{Introduction}
\label{sec:intro}

The baryonic mass-to-light ($M/L$) ratios of galaxies rely crucially on the detailed star-formation and chemical-enrichment histories of the multiple stellar populations that can exist in such stellar systems.  As such, $M/L$ ratios offer a convenient way to estimate the baryonic masses of galaxies from photometric observations, and numerous models to do so have been developed over the years \citep[e.g.][]{Bruzual:2003aa, Maraston:2005aa, Vazdekis:2010aa, Conroy:2009aa, Conroy:2010aa}.  The application of such models to data from large-scale photometric and spectroscopic surveys more-or-less directly determines much of what we know about how galaxies evolve with redshift \citep[e.g.][]{Kauffmann:2003aa, Bell:2003aa, Blanton:2007aa, Tojeiro:2009aa, Chen:2012aa, Maraston:2013aa}, highlighting the need to test these models wherever possible.  To be useful, such tests should involve independent estimates of the masses of stellar systems from kinematic observations along with high-quality luminosity measurements.  Moreover, an ideal test should involve systems with {\it simple} evolutionary histories so that $M/L$ ratio can be tied to a specific age and metallicity (or, a small range in both).  Although masses and luminosities can certainly be measured for galaxies, their complex evolutionary histories make them ill-suited to test model predictions of $M/L$ ratios.

In contrast, star clusters provide a better alternative to test $M/L$ predictions given their comparatively simple dynamical states and generally small internal variations of age and metallicity in a given system.  Such tests boil down to using kinematic data to derive cluster masses and photometric measurements to estimate total luminosities.
From both measurements, $M/L$ ratios can be estimated in as purely empirical a manner as possible.  Note that the questions such tests address go beyond determining which set of models is `best', but can also explore to what extent---and, if so, why---$M/L$ ratios vary among stellar systems of similar age and metallicity.

From an observational standpoint, two spectroscopic methods can be used to determine the dynamical masses of star clusters: (1) measuring the central radial velocity dispersions from integrated-light spectroscopy \citep[e.g.][]{Illingworth:1976ab, Mandushev:1991aa, Zaritsky:2012aa, Zaritsky:2013aa, Zaritsky:2014aa}, and (2) collecting radial velocities of individual member stars \citep[e.g.][]{Gunn:1979aa, Meylan:1986aa, Lupton:1987aa, Lupton:1989aa, Mateo:1991aa, Fischer:1992aa, Fischer:1992ab, Fischer:1993aa, Suntzeff:1992aa}. In recent years, the second approach has become significantly more practical with the development of wide-field multi-object spectrographs \citep[MOSs, e.g.][]{ Lane:2010ab, Mackey:2013aa, Kimmig:2015aa, Song:2019aa} and comparatively wide-field integral field units \citep[IFUs, e.g.][]{Kamann:2016aa, Kamann:2018aa, Kamann:2018ab}. In addition to spectroscopy, proper motion data are also available for cluster mass determination---though mostly for the globular clusters in the Milky Way---via the \textit{Hubble Space Telescope (HST)} observations \citep[e.g.][]{Bellini:2014aa, Watkins:2015aa} and the \textit{Gaia} Data Release 2 \citep[][hereafter \textit{Gaia} DR2]{Gaia-Collaboration:2018aa} astrometry \citep[e.g.][]{Baumgardt:2019aa}. 
In contrast to the mass estimates, luminosities of star clusters are determined almost exclusively using calibrated surface brightness/density profiles \citep[e.g.][]{McLaughlin:2005aa, Song:2019aa}.
Together, masses and luminosities measured in these ways can be combined to produce empirical $M/L$ estimates \citep[e.g.][]{McLaughlin:2005aa, Strader:2009aa, Strader:2011aa, Kimmig:2015aa, Baumgardt:2018aa}.

Simple stellar population (SSP) models can predict the $M/L$ ratios of star clusters as a function of age and metallicity.  Numerous studies have compared predictions from such models with empirical $M/L$ ratios, mostly for old globular clusters either in the Milky Way \citep[e.g.][]{Pryor:1993aa, McLaughlin:2005aa, Kimmig:2015aa, Baumgardt:2017aa, Baumgardt:2018aa,Dalgleish:2020aa} or local group (LG) galaxies \citep[e.g.][]{Larsen:2002aa, Strader:2009aa, Strader:2011aa}. 
Some of these studies have revealed variations of $M/L$ ratios with respect to metallicity and cluster masses that do not appear to conform to model expectations \citep[e.g.][]{Strader:2009aa, Strader:2011aa, Kimmig:2015aa, Baumgardt:2017aa, Dalgleish:2020aa}.

Much less common are high-precision $M/L$ measurements for young and intermediate-age clusters.  Such systems can, in principle, greatly expand the parameter space in age and metallicity over which SSP models can be tested.  
At a distance of 50-60 kpc, the populous clusters of the Magellanic Clouds (MCs) provide excellent laboratories to broadly test SSP models in just this manner.  These clusters are compact enough in the sky for both integrated-light spectroscopy and photometry, but also close enough to allow spectroscopy of individual stellar members in sufficient numbers to produce good-quality statistical samples.  Further, the fact that both integrated-light and individual-star methods can be applied effectively to many MC clusters makes them particularly useful test cases to understand the relative systematics that may arise from each technique.  Indeed, both approaches have been employed in past studies of MC clusters.   \citet{Zaritsky:2012aa, Zaritsky:2013aa, Zaritsky:2014aa} obtained integrated-light spectroscopic observations to measure $M/L$ ratios for a sample of 17 clusters younger than 7~Gyr in the Large and Small Magellanic Clouds (LMC and SMC, respectively). To date, high-precision individual-star spectroscopic studies have been carried out for six additional MC clusters  \citep{Fischer:1992aa, Fischer:1992ab, Fischer:1993aa, Mackey:2013aa, Kamann:2018ab, Song:2019aa, Patrick:2020aa}. 

In an earlier paper \citep[][hereafter \citetalias{Song:2019aa}]{Song:2019aa},  we described the basic methodology of our survey and the analysis of the data as applied to the SMC cluster NGC~419 and the LMC cluster NGC~1846.  Both clusters had been subjects of recent high-quality multi-object spectroscopic studies \citep{Kamann:2018ab, Mackey:2013aa} and so represented apt test cases to compare with independent empirical dynamical analyses.  In general, we found acceptable agreement with past work once differences in kinematic precision were taken into account.

\begin{figure}
   \centering
   \includegraphics[width=0.47\textwidth]{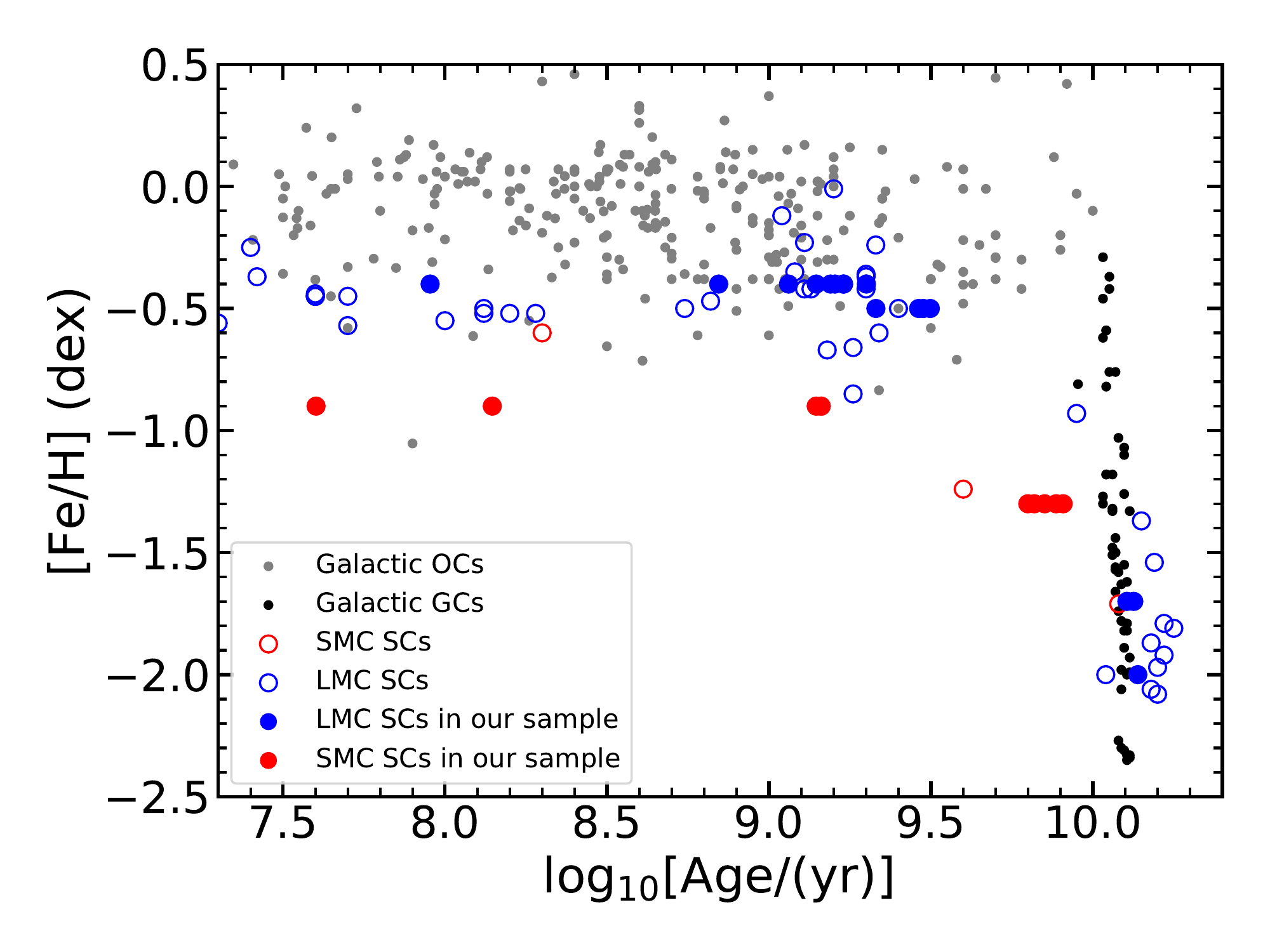}
   \caption{A representative plot of metallicity versus $\log{\rm (Age)}$ for star clusters in the Milky Way and the Magellanic Clouds. Filled circles denote the 16 LMC (blue) and 10 SMC (red) clusters studied in this paper, respectively. The metallicities and ages of these clusters correspond to the values listed in \autoref{tab:basic}. Open circles correspond to other clusters in the LMC (blue) and SMC (red) from the collected catalog of \citet{Pessev:2006aa, Pessev:2008aa}. For the Galactic clusters, we present the globular clusters listed in \citet{VandenBerg:2013aa} (black dots) and the Galactic open clusters listed in \citet{Dias:2002aa} (gray dots).}
   \label{fig:amr}
\end{figure}

This present paper updates the methodology from \citetalias{Song:2019aa} as applied to our full kinematic sample consisting of 26 MC clusters with high-quality kinematic data (10 in the SMC, 16 in the LMC).  These clusters were chosen to span the range from $\sim$100 Myr to $\sim$13 Gyr in age, and from $-2.0$ to $-0.4$ in $\rm [Fe/H]$ (see \autoref{fig:amr} and \autoref{tab:basic}) in order to provide the most leverage on our tests of $V$-band $M/L$ ($M/L_V$) predictions from populations models. As in \citetalias{Song:2019aa}, the present study employs spectroscopic observations obtained using the Michigan/\textit{Magellan} Fiber System (M2FS) from which we measure kinematics and metallicities of samples of individual stars associated with the clusters in our sample. We derive dynamical masses and $M/L_V$ ratios of all 26 clusters, along with independent spectroscopic estimates of the mean metallicities of the clusters.

This paper is organized as follows. The technical details of our study are described in Sections~\ref{sec:data}--\ref{sec:mem}.  Specifically, in \autoref{sec:data} we describe our target selection, cluster center determination, observational and data reduction procedures adopted for all clusters in our survey. \autoref{sec:analysis} describes improvements of the Bayesian spectral fitting method we introduced in \citetalias{Song:2019aa} to derive velocity and physical parameters from M2FS spectra. In \autoref{sec:mem}, we obtain the final stellar sample for cluster kinematic analysis and assign a cluster membership probability to each star. In  \autoref{sec:results} we report the dynamical and chemical results for all clusters in our sample.  We compare these results critically with those from previous studies. The key scientific results of our analysis are provided \autoref{sec:discuss}, where we discuss the trends of our $M/L$ results with respect to various physical cluster parameters.  We also compare these results with predictions from standard SSP models and with models that attempt to account for internal and external dynamical processes that can alter cluster $M/L$ ratios.  \autoref{sec:summary} summarizes our methodology and results, and provides a brief outline of our conclusions.

\section{Data Overview}
\label{sec:data}

\begin{table*}
\caption{General properties of star clusters in our sample.}
\label{tab:basic}
\begin{threeparttable}
\begin{tabular}{llrrcrrrcc}
\hline
    \multicolumn{1}{c}{Galaxy} & 
    \multicolumn{1}{c}{Cluster} & 
    \multicolumn{1}{c}{$V_{\rm ap}$} & 
    \multicolumn{1}{c}{Aper.$^{\rm a}$} & 
    \multicolumn{1}{c}{Ref.$^{\rm b}$} & 
    \multicolumn{1}{c}{Age} & 
    \multicolumn{1}{c}{$(m-M)_0$} & 
    \multicolumn{1}{c}{$A_V$} & 
    \multicolumn{1}{c}{Ref.$^{\rm b}$} &
    \multicolumn{1}{c}{$\rm [Fe/H]^{\rm c}$} \\
    \multicolumn{1}{c}{} & 
    \multicolumn{1}{c}{} & 
    \multicolumn{1}{c}{(mag)} & 
    \multicolumn{1}{c}{(arcsec)} & 
    \multicolumn{1}{c}{} & 
    \multicolumn{1}{c}{(Gyr)} & 
    \multicolumn{1}{c}{(mag)} & 
    \multicolumn{1}{c}{(mag)} & 
    \multicolumn{1}{c}{} &
    \multicolumn{1}{c}{(dex)} \\ 
    \multicolumn{1}{c}{(1)} & 
    \multicolumn{1}{c}{(2)} & 
    \multicolumn{1}{c}{(3)} & 
    \multicolumn{1}{c}{(4)} & 
    \multicolumn{1}{c}{(5)} & 
    \multicolumn{1}{c}{(6)} & 
    \multicolumn{1}{c}{(7)} & 
    \multicolumn{1}{c}{(8)} & 
    \multicolumn{1}{c}{(9)} & 
    \multicolumn{1}{c}{(10)}\\
\hline
SMC & Kron 3 & $11.41\pm0.09$ & 50.0 & 1 & $7.10\pm0.70$ & $18.80\pm0.04$ & $0.08\pm0.03$ & 7 & -1.3 \\ 
SMC & Lindsay 1 & $13.32\pm0.05$ & 31.0 & 2 & $7.70\pm0.70$ & $18.69\pm0.04$ & $0.15\pm0.03$ & 7 & -1.3 \\ 
SMC & NGC 152 & $12.33\pm0.05$ & 50.0 & 1 & $1.40\pm0.20$ & $18.93\pm0.04$ & $0.16\pm0.04$ & 8 & -0.9 \\ 
SMC & NGC 330 & $9.60\pm0.01$ & 31.0 & 2 & $0.04\pm0.00$ & $18.80\pm0.04$ & $0.36\pm0.04$ & 9 & -0.9 \\ 
SMC & NGC 339 & $12.12\pm0.04$ & 50.0 & 1 & $6.30\pm0.50$ & $18.75\pm0.08$ & $0.19\pm0.03$ & 7  & -1.3\\ 
SMC & NGC 361 & $12.24\pm0.01$ & 31.0 & 2 & $8.10\pm1.20$ & $18.49\pm0.04$ & $0.40\pm0.04$ & 8  & -1.3\\ 
SMC & NGC 411 & $11.81\pm0.07$ & 50.0 & 1 & $1.45\pm0.05$ & $18.82\pm0.03$ & $0.16\pm0.02$ & 10 & -0.9\\ 
SMC & NGC 416 & $11.42\pm0.00$ & 31.0 & 2 & $6.60\pm0.80$ & $18.76\pm0.07$ & $0.25\pm0.03$ & 7  & -1.3\\ 
SMC & NGC 419 & $10.30\pm0.16$ & 50.0 & 1 & $1.45\pm0.05$ & $18.85\pm0.03$ & $0.15\pm0.02$ & 10 & -0.9\\ 
SMC & NGC 458 & $11.73\pm0.03$ & 31.0 & 2 & $0.14\pm0.03$ & $19.11\pm0.20$ & $0.12\pm0.06$ & 11 & -0.9\\ 
LMC & Hodge 4 & $13.33\pm0.02$ & 19.0 & 3 & $2.14\pm0.00$ & $18.37\pm0.03$ & $0.12\pm0.04$ & 12 & -0.5\\ 
LMC & NGC 1466 & $11.59\pm0.03$ & 30.0 & 4 & $13.38\pm2.00$ & $18.66\pm0.03$ & $0.16\pm0.04$ & 13 & -1.7 \\ 
LMC & NGC 1751 & $11.67\pm0.13$ & 50.0 & 1 & $1.40\pm0.05$  & $18.50\pm0.03$ & $0.38\pm0.02$ & 10 & -0.4 \\ 
LMC & NGC 1783 & $10.39\pm0.03$ & 50.0 & 1 & $1.70\pm0.05$  & $18.49\pm0.03$ & $0.00\pm0.02$ & 10 & -0.4 \\ 
LMC & NGC 1806 & $11.00\pm0.05$ & 50.0 & 1 & $1.60\pm0.05$  & $18.50\pm0.03$ & $0.05\pm0.03$ & 10 & -0.4 \\ 
LMC & NGC 1831 & $11.18\pm0.02$ & 30.0 & 4 & $0.70\pm0.10$  & $18.23\pm0.09$ & $0.03\pm0.06$ & 14 & -0.4 \\ 
LMC & NGC 1841 & $11.43\pm0.02$ & 93.5 & 3 & $13.77\pm1.70$ & $18.34\pm0.04$ & $0.35\pm0.04$ & 13 & -2.0 \\ 
LMC & NGC 1846 & $10.68\pm0.20$ & 50.0 & 1 & $1.70\pm0.05$  & $18.42\pm0.03$ & $0.07\pm0.02$ & 10 & -0.4 \\ 
LMC & NGC 1850 & $9.57\pm0.20$ & 25.0 & 3 & $0.09\pm0.05$   & $18.45\pm0.03$ & $0.37\pm0.02$ & 15 & -0.4 \\ 
LMC & NGC 1978 & $10.20\pm0.02$ & 50.0 & 1 & $2.00\pm0.00$  & $18.55\pm0.04$ & $0.16\pm0.04$ & 16 & -0.4 \\ 
LMC & NGC 2121 & $12.37\pm0.01$ & 31.0 & 5 & $2.90\pm0.50$  & $18.24\pm0.04$ & $0.22\pm0.06$ & 14 & -0.5 \\ 
LMC & NGC 2155 & $12.59\pm0.48$ & 50.0 & 1 & $3.00\pm0.25$  & $18.32\pm0.04$ & $0.06\pm0.03$ & 14 & -0.5 \\ 
LMC & NGC 2203 & $11.29\pm0.15$ & 75.0 & 3 & $1.55\pm0.05$  & $18.37\pm0.03$ & $0.16\pm0.02$ & 10 & -0.4 \\ 
LMC & NGC 2209 & $13.15\pm0.01$ & 34.0 & 3 & $1.15\pm0.05$  & $18.37\pm0.03$ & $0.23\pm0.02$ & 17 & -0.4 \\ 
LMC & NGC 2257 & $12.62\pm0.02$ & 30.5 & 3 & $12.74\pm2.00$ & $18.25\pm0.04$ & $0.12\pm0.04$ & 13 & -1.7 \\ 
LMC & SL 663 & $22.13\pm0.24^{\rm d}$ & 0.0 & 6 & $3.15\pm0.40$    & $18.32\pm0.07$ & $0.22\pm0.06$ & 14 & -0.5 \\  
\hline
\end{tabular}
\begin{tablenotes}
   \item $^{\rm a}$ Aperture radius used for measuring the $V_{\rm ap}$ magnitude in column 3. 
   \item $^{\rm b}$ References: (1) \citet{Goudfrooij:2006aa}; (2) \citet{Alcaino:1978aa}; (3) \citet{Bica:1996aa}; (4) \citet{van-den-Bergh:1981aa}; (5) \citet{Bernard:1975aa}; (6) \citet{McLaughlin:2005aa}; (7) \citet{Glatt:2008ab}; (8) \citet{Crowl:2001aa}; (9) \citet{Milone:2018aa}; (10) \citet{Goudfrooij:2014aa}; (11) \citet{Alcaino:2003aa}; (12) \citet{Grocholski:2007aa}; (13) \citet{Wagner-Kaiser:2017aa}; (14) \citet{Kerber:2007aa}; (15) \citet{Correnti:2017aa}; (16) \citet{Martocchia:2018aa}; (17) \citet{Correnti:2014aa}.
   \item $^{\rm c}$ The adopted [Fe/H] values were estimated from the age-metallicity relations for the LMC and SMC clusters, respectively. For the LMC clusters, we assumed that $\rm [Fe/H]=-0.4$ if 0-2 Gyr, $\rm [Fe/H]=-0.5$ if 2-4 Gyr, $\rm [Fe/H]=-1.7$ for NGC~1466 and NGC~2257, and $\rm [Fe/H]=-2.0$ for NGC~1841; while for the SMC clusters, we assumed that $\rm [Fe/H]=-0.9$ if 0-4 Gyr, and $\rm [Fe/H]=-1.3$ if 6-9 Gyr. These age-metallicity relations were averaged from multiple papers that fitted cluster CMDs with the Padova isochrones (\citealt{Kerber:2007aa, Grocholski:2007aa, Milone:2009aa, Goudfrooij:2014aa, Milone:2018aa} for LMC, and \citealt{Crowl:2001aa, Glatt:2008ab, Goudfrooij:2014aa, Milone:2018aa} for SMC). The final cluster metallicities from this study are listed in \autoref{tab:FeH}.
   \item $^{\rm d}$ SL~663 has no $V$-band aperture photometry in the literature. We have adopted the best-fit $V$-band extinction-corrected central surface brightness from \citet{McLaughlin:2005aa}. The value listed here is in unit of $\rm mag\,arcsec^{-2}$.
\end{tablenotes}
\end{threeparttable}
\end{table*}

\subsection{Cluster Candidates}
\label{sec:cluster_select}
We selected cluster candidates from catalogs of MC clusters with good-quality age and metallicity estimates and for which we could expect to obtain samples of a few dozen stellar members.  \autoref{tab:basic} lists the general properties of the star clusters in our final sample and specifies all the literature sources used to select our spectroscopic targets.  Columns 1 and 2 in the table specify the host galaxy and the most common names of each cluster. The V-band aperture magnitude is listed in column 3, with the corresponding aperture radius listed in column 4.  Column 5 lists the sources for these photometric results. Columns 6 through 8 list age, distance modulus and extinction values taken from the sources listed in column 9; these three parameters were estimated from the cited sources from comparisons of a given cluster's color-magnitude diagram (CMD) with modern synthetic isochrones (see below).  We gave preference to clusters with deep \textit{HST} photometry, but if no \textit{HST} data were available good-quality ground-based photometry was also used . Most of the studies listed in \autoref{tab:basic} used the isochrones of the Padova group \citep{Girardi:2000aa, Girardi:2002aa, Bressan:2012aa}. In a few cases, the sources listed in column 9 employed multiple sets of isochrones from different synthetic groups;  for consistency, we adopted only their best-fit results using Padova/PARSEC isochrones. 
Column 10 lists the adopted metallicities that estimated from the age-metallicity relations for the LMC and SMC clusters, respectively (see the corresponding table note for details).
\autoref{fig:amr} plots the ages and metallicities from \autoref{tab:basic} of our sample, illustrating the wide range of these parameters sampled by the clusters in our study. 

\autoref{tab:structure} lists the positions and structural parameters of the clusters in our sample. These parameters are essential for the background/sky subtraction processes (see \autoref{sec:sky_sub}) and to establish the central velocity dispersions of the clusters (see \autoref{sec:mem}).  The cluster positions listed in \autoref{tab:structure} (columns 3 and 4) were derived from the \textit{Gaia} DR2. Details of how we determined these centers can be found in Appendix~\ref{sec:cluster_center}\footnote{We have not tested the impact of centering errors on the cluster structural parameters. Hence the robustness of the central velocity dispersion measurements (see \autoref{sec:vd}) relative to small shifts of the cluster centers is based on the assumption that the structural parameters are unchanged to such shifts.}. For the cluster structural parameters, columns 5 and 6 give the concentration parameter and the King radius of an empirical number density profile by \citet[][hereafter the K62 profile]{King:1962aa}.  Column 7 lists the references for these structural parameters.  Columns 8 and 9 give the same parameters but for a dynamical model developed by \citet[][hereafter the K66 model]{King:1966aa}. 
In cases where only the K62 profile or K66 model is available from the literature, we transformed the structural parameters from one to the other, as described in \citetalias{Song:2019aa}.  This approach ensured that the central surface brightness, core radius (defined as the radius at the half of the central surface brightness) and total luminosity agree for both profiles. 

\begin{table*}
\caption{Positions and structural parameters of star clusters in our sample.}
\label{tab:structure}
\begin{threeparttable}
\begin{tabular}{llrrrrcrrc}
\hline
    \multicolumn{1}{c}{Galaxy} &
    \multicolumn{1}{c}{Cluster} & 
    \multicolumn{1}{c}{$\alpha_{\rm J2000}$} & 
    \multicolumn{1}{c}{$\delta_{\rm J2000}$} & 
    \multicolumn{1}{c}{$c_{\rm K62}$$^{\rm a}$} & 
    \multicolumn{1}{c}{$r_{0,\rm K62}$}  & 
    \multicolumn{1}{c}{Ref.$^{\rm b}$} & 
    \multicolumn{1}{c}{$c_{\rm K66}$$^{\rm a}$} & 
    \multicolumn{1}{c}{$r_{0,\rm K66}$} & 
    \multicolumn{1}{c}{Ref.$^{\rm b}$} \\
    \multicolumn{1}{c}{} & \multicolumn{1}{c}{} & \multicolumn{1}{c}{(hh mm ss)} & \multicolumn{1}{c}{(dd mm ss)} & \multicolumn{1}{c}{} & \multicolumn{1}{c}{(arcsec)} & \multicolumn{1}{c}{} & \multicolumn{1}{c}{} & \multicolumn{1}{c}{(arcsec)} & \multicolumn{1}{c}{} \\
    \multicolumn{1}{c}{(1)} & 
    \multicolumn{1}{c}{(2)} & 
    \multicolumn{1}{c}{(3)} & 
    \multicolumn{1}{c}{(4)} & 
    \multicolumn{1}{c}{(5)} & 
    \multicolumn{1}{c}{(6)} & 
    \multicolumn{1}{c}{(7)} & 
    \multicolumn{1}{c}{(8)} & 
    \multicolumn{1}{c}{(9)} & 
    \multicolumn{1}{c}{(10)}\\
\hline
SMC & Kron 3 & 00 24 45.98 & $-$72 47 37.9 & $0.79^{+0.01}_{-0.01}$ & $29.62^{+0.69}_{-0.69}$ & 1 & $1.14^{+0.05}_{-0.04}$ & $23.24^{+1.26}_{-1.31}$ & 7 \\ 
SMC & Lindsay 1 & 00 03 53.50 & $-$73 28 15.0 & $0.55^{+0.02}_{-0.02}$ & $66.76^{+2.50}_{-2.50}$ & 1 & $0.60^{+0.02}_{-0.02}$ & $75.40^{+2.82}_{-2.82}$ & ... \\ 
SMC & NGC 152 & 00 32 56.76 & $-$73 06 56.4 & $0.98^{+0.08}_{-0.09}$ & $26.70^{+2.65}_{-1.77}$ & ... & $1.09^{+0.09}_{-0.10}$ & $27.39^{+2.72}_{-1.82}$ & 7 \\ 
SMC & NGC 330 & 00 56 18.55 & $-$72 27 45.1 & $1.35^{+0.09}_{-0.08}$ & $8.34^{+0.74}_{-0.69}$ & ... & $1.41^{+0.09}_{-0.08}$ & $8.55^{+0.76}_{-0.71}$ & 7 \\ 
SMC & NGC 339 & 00 57 47.62 & $-$74 28 14.6 & $0.79^{+0.01}_{-0.01}$ & $32.91^{+0.59}_{-0.59}$ & 1 & $0.82^{+0.21}_{-0.17}$ & $33.18^{+6.00}_{-4.34}$ & 7 \\ 
SMC & NGC 361 & 01 02 11.11 & $-$71 36 25.3 & $0.88^{+0.24}_{-0.18}$ & $26.55^{+4.90}_{-3.63}$ & ... & $0.99^{+0.27}_{-0.20}$ & $27.27^{+5.03}_{-3.73}$ & 7 \\ 
SMC & NGC 411 & 01 07 55.64 & $-$71 46 03.1 & $0.87^{+0.03}_{-0.03}$ & $14.50^{+0.89}_{-0.89}$ & 2 & $1.38^{+0.11}_{-0.11}$ & $9.30^{+1.09}_{-0.99}$ & 7 \\ 
SMC & NGC 416 & 01 07 59.03 & $-$72 21 20.5 & $0.81^{+0.02}_{-0.02}$ & $12.56^{+0.48}_{-0.48}$ & 1 & $0.89^{+0.08}_{-0.07}$ & $12.05^{+0.86}_{-0.79}$ & 7 \\ 
SMC & NGC 419 & 01 08 17.26 & $-$72 53 01.8 & $1.11^{+0.03}_{-0.03}$ & $14.51^{+0.93}_{-0.93}$ & 1, 2 & $1.20^{+0.03}_{-0.03}$ & $14.90^{+0.95}_{-0.95}$ & ... \\ 
SMC & NGC 458 & 01 14 52.94 & $-$71 32 60.0 & $0.98^{+0.10}_{-0.11}$ & $12.01^{+1.84}_{-1.36}$ & ... & $1.09^{+0.11}_{-0.12}$ & $12.32^{+1.89}_{-1.40}$ & 7 \\ 
LMC & Hodge 4 & 05 32 25.64 & $-$64 44 07.7 & $6.54^{+5.10}_{-1.25}$ & $14.27^{+1.12}_{-0.61}$ & ... & $2.45^{+1.91}_{-0.47}$ & $14.41^{+1.13}_{-0.62}$ & 7 \\ 
LMC & NGC 1466 & 03 44 32.71 & $-$71 40 18.0 & $1.00^{+0.06}_{-0.05}$ & $10.81^{+0.68}_{-0.65}$ & ... & $1.11^{+0.07}_{-0.06}$ & $11.06^{+0.70}_{-0.66}$ & 7 \\ 
LMC & NGC 1751 & 04 54 12.91 & $-$69 48 26.8 & $0.84^{+0.03}_{-0.03}$ & $22.00^{+1.57}_{-1.57}$ & 2, 3 & $0.95^{+0.03}_{-0.03}$ & $22.64^{+1.61}_{-1.61}$ & ... \\ 
LMC & NGC 1783 & 04 59 08.75 & $-$65 59 15.6 & $0.96^{+0.02}_{-0.02}$ & $37.70^{+1.76}_{-1.76}$ & 2, 3 & $1.08^{+0.02}_{-0.02}$ & $38.58^{+1.80}_{-1.80}$ & ... \\ 
LMC & NGC 1806 & 05 02 11.86 & $-$67 59 08.5 & $0.90^{+0.02}_{-0.02}$ & $24.50^{+1.12}_{-1.12}$ & 2, 3 & $1.01^{+0.02}_{-0.02}$ & $25.15^{+1.15}_{-1.15}$ & ... \\ 
LMC & NGC 1831 & 05 06 16.12 & $-$64 55 06.9 & $1.08^{+0.05}_{-0.05}$ & $18.24^{+0.90}_{-1.06}$ & ... & $1.18^{+0.05}_{-0.05}$ & $18.65^{+0.92}_{-1.08}$ & 7 \\ 
LMC & NGC 1841 & 04 45 24.38 & $-$83 59 53.3 & $0.53^{+0.26}_{-0.53}$ & $42.76^{+77.76}_{-9.61}$ & ... & $0.57^{+0.28}_{-0.57}$ & $49.14^{+89.36}_{-11.04}$ & 7 \\ 
LMC & NGC 1846 & 05 07 34.47 & $-$67 27 37.8 & $0.79^{+0.03}_{-0.03}$ & $26.00^{+1.59}_{-1.59}$ & 2, 4 & $0.90^{+0.03}_{-0.03}$ & $26.92^{+1.64}_{-1.64}$ & ... \\ 
LMC & NGC 1850 & 05 08 46.32 & $-$68 45 38.7 & $0.99^{+0.02}_{-0.02}$ & $12.40^{+0.49}_{-0.49}$ & 5 & $0.98^{+0.02}_{-0.02}$ & $12.72^{+0.51}_{-0.51}$ & ... \\ 
LMC & NGC 1978 & 05 28 44.73 & $-$66 14 09.3 & $1.16^{+0.05}_{-0.05}$ & $17.30^{+0.78}_{-0.78}$ & ... & $1.26^{+0.05}_{-0.05}$ & $17.70^{+0.80}_{-0.80}$ & 8 \\ 
LMC & NGC 2121 & 05 48 12.74 & $-$71 28 45.2 & $0.60^{+0.13}_{-0.13}$ & $43.21^{+9.28}_{-5.90}$ & ... & $0.67^{+0.15}_{-0.15}$ & $47.30^{+10.16}_{-6.46}$ & 7 \\ 
LMC & NGC 2155 & 05 58 31.98 & $-$65 28 41.0 & $0.84^{+0.17}_{-0.14}$ & $19.20^{+2.78}_{-1.98}$ & ... & $0.95^{+0.19}_{-0.16}$ & $19.76^{+2.86}_{-2.04}$ & 7 \\ 
LMC & NGC 2203 & 06 04 42.17 & $-$75 26 15.8 & $0.69^{+0.02}_{-0.02}$ & $32.90^{+1.61}_{-1.61}$ & 2 & $0.79^{+0.02}_{-0.02}$ & $34.61^{+1.69}_{-1.69}$ & ... \\ 
LMC & NGC 2209 & 06 08 36.06 & $-$73 50 07.9 & $0.85^{+0.02}_{-0.02}$ & $27.30^{+1.40}_{-1.40}$ & 6 & $0.83^{+0.02}_{-0.02}$ & $28.09^{+1.44}_{-1.44}$ & ... \\ 
LMC & NGC 2257 & 06 30 12.62 & $-$64 19 40.0 & $0.81^{+0.30}_{-0.21}$ & $30.99^{+6.33}_{-3.88}$ & ... & $0.91^{+0.34}_{-0.24}$ & $32.02^{+6.54}_{-4.01}$ & 7 \\ 
LMC & SL 663 & 05 42 28.20 & $-$65 21 50.2 & $2.19^{+2.93}_{-1.14}$ & $27.85^{+7.10}_{-1.57}$ & ... & $1.86^{+2.49}_{-0.97}$ & $28.58^{+7.29}_{-1.61}$ & 7 \\ 
\hline
\end{tabular}
\begin{tablenotes}
   \item $^{\rm a}$ Concentration parameter $c\equiv \log_{10}{r_{\rm t}/r_0}$, where $r_0$ and $r_{\rm t}$ are the King radius and truncation radius, respectively.
   \item $^{\rm b}$ References: (1) \citet{Glatt:2009aa}; (2) \citet{Goudfrooij:2014aa}; (3) \citet{Goudfrooij:2011aa}; (4) \citet{Goudfrooij:2009aa}; (5) \citet{Correnti:2017aa}; (6) \citet{Correnti:2014aa}; (7) \citet{McLaughlin:2005aa}; (8) \citet{Fischer:1992ab}.
\end{tablenotes}
\end{threeparttable}
\end{table*}

\subsection{Target Selection Within Cluster Fields}
\label{sec:star_select}

For any given cluster, we selected a variety of specific types of targets for spectroscopic analysis.  The primary science targets were typically drawn from the red giant branch (RGB) of a cluster's CMD.  These targets were prioritized according to their proximity to their respective cluster center.  Additional science targets beyond the formal tidal radii of clusters were also included to allow us to determine the kinematic and chemical distribution of the local field populations. For both science target selections, we identified apparently isolated stars as potential spectroscopic targets. We regarded a star to be isolated when the integrated flux of all other stars in the corresponding photometric catalog within 1 arcsec of the star adds up to $\leq$ 20\%\ of the candidate star's flux.  

The scientific targets were selected in the following manner. Before the \textit{Gaia} DR2 became available on April 2018, our target selection generally relied on archival \textit{Hubble Space Telescope (HST)} images (see below), the Magellanic Clouds Photometric Survey (MCPS) $UBVI$ catalog \citep{Zaritsky:2002aa, Zaritsky:2004aa}, and the Johnson $BV$ photometry by \citet{Jeon:2014aa}.  Some of these sources exhibited rather large astrometric offsets and/or scale errors, which made the observational data based on these sources less reliable.  
For all adopted \textit{HST} images, short exposures with the F555W and F814W (or F450W when F814W was not available) filters were reduced using the WFPC2 or ACS modules from the \texttt{DOLPHOT} package \citep{Dolphin:2000aa}, and the resulting CMDs were mainly used to select science targets within cluster tidal radii. For the selection of science targets beyond the cluster tidal radii, the MCPS catalog, however, was exclusively used. For NGC~1466, NGC~1841 and NGC~2257 (which are not covered by MCPS), we used \citet{Jeon:2014aa}'s catalog obtained by the Cerro Tololo Inter-American Observatory (CTIO) 0.9 m SMARTS telescope. 
Once the \textit{Gaia} DR2 catalog was available, we exclusively selected candidates based on the \textit{Gaia} CMDs and requiring that the parallax of any candidate was $\leq 0.15$ mas. The catalogues used for each cluster's target selection are listed in column 7 of \autoref{tab:obs}.
 
We adopted the positions of targets from different catalogues in the following way. The coordinates for targets selected and observed before \textit{Gaia} DR2 were tied to the NOMAD astrometric system \citep{Zacharias:2004aa}.
We cross-matched all stars brighter than 17.5 mag in $V$-band from \textit{HST} images or MCPS catalog with stars in the NOMAD catalog, and transformed the coordinates onto the NOMAD frame.
For some \textit{HST}-selected stars, rather large astrometric corrections of up to 1--2 arcsec were necessary.  For reference, the M2FS fiber apertures are 1.2 arcsec in diameter, and systematic precision of 0.25 arcsec is typically required. After \textit{Gaia} DR2 and to ensure the accuracy of positional-dependent kinematic analysis, we cross-matched all targets with \textit{Gaia} DR2 and adopted their J2000 positions.

Fiber positions to sample the backgrounds in and around the clusters were also identified for many, but not all, of the cluster fields of our sample.
Before the release of \textit{Gaia} DR2, background/sky locations were selected by eye within the tidal radii of clusters were identified using F555W \textit{HST} images when available.  For regions outside the clusters' tidal radii, background/sky positions were randomly chosen from the DSS red-band images out to at least 2 arcmin from the clusters' centers. After the release of \textit{Gaia} DR2, background/sky positions within the clusters were selected by identifying random positions that had no stars within a radius of 3 arcsec in \textit{Gaia} DR2.

\subsection{Observations and Data Reduction}
\label{sec:obs}

\begin{table*}
 \caption{Observations.}
 \label{tab:obs}
 \begin{threeparttable}
 \begin{tabular}{llcllrcrc}
\hline
\multicolumn{1}{c}{Galaxy} & 
\multicolumn{1}{c}{Cluster} & 
\multicolumn{1}{c}{Obs. Date} & 
\multicolumn{1}{c}{Exp. Time (On Target)} & 
\multicolumn{1}{c}{Exp. Time (Offset)} & 
\multicolumn{1}{c}{$N_{\rm star}$} & 
\multicolumn{1}{c}{Sources$^{\rm a}$} & 
\multicolumn{1}{c}{$N_{\rm sky}$} & 
\multicolumn{1}{c}{Sky Sub. Method} \\
 \multicolumn{1}{c}{} & 
 \multicolumn{1}{c}{} & 
 \multicolumn{1}{c}{(UT)} & 
 \multicolumn{1}{c}{(s)} & 
 \multicolumn{1}{c}{(s)} & 
 \multicolumn{1}{c}{} & 
 \multicolumn{1}{c}{} & 
 \multicolumn{1}{c}{} & 
 \multicolumn{1}{c}{} \\
    \multicolumn{1}{c}{(1)} & 
    \multicolumn{1}{c}{(2)} & 
    \multicolumn{1}{c}{(3)} & 
    \multicolumn{1}{c}{(4)} & 
    \multicolumn{1}{c}{(5)} & 
    \multicolumn{1}{c}{(6)} & 
    \multicolumn{1}{c}{(7)} & 
    \multicolumn{1}{c}{(8)} & 
    \multicolumn{1}{c}{(9)} \\
\hline
SMC & Kron 3 & 2018-08-22 & 3$\times$1200 & 2$\times$300 & 120 & 1 & 12 & A \\ 
SMC & Lindsay 1 & 2018-08-21 & 1$\times$1200$+$1$\times$1500$+$1$\times$1200 & 1$\times$600 & 120 & 1 & 12 & A \\ 
SMC & NGC 152 & 2018-08-20 & 3$\times$1200 & 2$\times$300 & 120 & 1 & 12 & A \\ 
SMC & NGC 330 & 2018-08-18 & 1$\times$1200$+$2$\times$1500 & 2$\times$480 & 39 & 1 & 15 & A \\ 
SMC & NGC 339 & 2016-09-12 & 5$\times$2400 & 2$\times$600 & 127 & 2, 3 & 3 & A \\ 
SMC & NGC 361 & 2018-11-26 & 3$\times$1200 & 3$\times$600 & 120 & 1 & 12 & A$^{*}$ \\ 
SMC & NGC 411 & 2018-08-15 & 3$\times$1500 & 2$\times$600 & 120 & 1 & 12 & A \\ 
SMC & NGC 416 & 2018-08-21 & 3$\times$1200 & 2$\times$300 & 120 & 1 & 12 & A \\ 
SMC & NGC 419 & 2017-09-21 & 4$\times$2100 & 2$\times$480 & 123 & 2, 3 & 5 & A \\ 
SMC & NGC 458 & 2018-08-15 & 3$\times$1200 & 2$\times$300 & 82 & 1 & 16 & A \\ 
LMC & Hodge 4 & 2018-12-02 & 2$\times$1080$+$1$\times$840 & 2$\times$240 & 119 & 1 & 12 & A \\ 
LMC & NGC 1466 & 2016-12-11 & 3$\times$2400 & 0 & 46 & 4, 5 & 42 & B \\ 
LMC & NGC 1751 & 2019-02-24 & 3$\times$1730 & 2$\times$450 & 118 & 1 & 12 & A \\ 
LMC & NGC 1783 & 2018-02-16 & 1$\times$1800$+$2$\times$2000 & 0 & 114 & 2, 6 & 9 & B \\ 
LMC & NGC 1806 & 2018-11-30 & 1$\times$1200$+$2$\times$960 & 2$\times$240 & 120 & 1 & 12 & A \\ 
LMC & NGC 1831 & 2018-12-01 & 1$\times$1500$+$1$\times$1200$+$1$\times$1500 & 2$\times$300 & 119 & 1 & 12 & A \\ 
LMC & NGC 1841 & 2018-02-25 & 2$\times$1200 & 0 & 112 & 4 & 12 & B \\ 
LMC & NGC 1846 & 2018-02-21 & 4$\times$1200 & 0 & 111 & 2, 6 & 13 & B \\ 
LMC & NGC 1850 & 2018-02-19 & 3$\times$1200 & 0 & 113 & 2, 7 & 13 & B \\ 
LMC & NGC 1978 & 2017-03-03 & 3$\times$1800 & 0 & 130 & 2, 8 & 1 & B \\ 
LMC & NGC 1978 & 2017-11-10 & 3$\times$2000 & 0 & 129 & 2, 8 & 0 & -- \\ 
LMC & NGC 2121 & 2018-02-23 & 3$\times$1200 & 0 & 117 & 2, 7 & 8 & B \\ 
LMC & NGC 2155 & 2018-12-08 & 3$\times$1800 & 2$\times$600 & 119 & 1 & 12 & A \\ 
LMC & NGC 2203 & 2018-12-04 & 1$\times$1200$+$1$\times$1000$+$2$\times$1800 & 4$\times$600  & 119 & 1 & 12 & A \\ 
LMC & NGC 2203 & 2019-03-06 & 1$\times$1600$+$2$\times$1200 & 2$\times$600 & 118 & 1 & 12 & A \\ 
LMC & NGC 2209 & 2018-12-05 & 1$\times$1200$+$1$\times$1800 & 1$\times$480$+$1$\times$600 & 119 & 1 & 11 & A \\ 
LMC & NGC 2209 & 2019-03-01 & 3$\times$1850 & 0 & 118 & 1 & 11 & B \\ 
LMC & NGC 2257 & 2018-05-18 & 3$\times$1800 & 0 & 116 & 4 & 8 & B \\ 
LMC & SL 663 & 2019-02-28 & 3$\times$1750 & 0 & 118 & 1 & 12 & B \\ 
\hline
\end{tabular}
\begin{tablenotes}
   \item $^{\rm a}$ Sources for target selection: (1) GDR2 \citep{Gaia-Collaboration:2018aa}; (2) MCPS \citep{Zaritsky:2002aa, Zaritsky:2004aa}; (3) \textit{HST} GO-10396 (PI: Gallagher); (4) \citet{Jeon:2014aa}; (5) \textit{HST} GO-5897 (PI: Bolte); (6) \textit{HST} GO-10595 (PI: Goudfrooij); (7) \textit{HST} SNAP-5475 (PI: Shara); (8) \textit{HST} GO-9891 (PI: Gilmore).
   \item $^{*}$ The adopted $n_k$ applied to all other observations (see \autoref{sec:sky_sub}).

\end{tablenotes}
\end{threeparttable}
\end{table*}

The spectral data used in this study were obtained with the M2FS \citep{Mateo:2012aa} on the \textit{Magellan}/Clay Telescope over 26 nights during a campaign lasting from September 2016 to March 2019. The detailed spectral configuration parameters are the same as described in \citetalias{Song:2019aa}. In summary, the single-order spectra ranged from 5130 to 5192 \AA\ in wavelength with a mean resolution of 18,000.  In parallel to these single-order observations, we usually also obtained data using a broad order-isolating filter that allowed us to obtain spectra covering 23 orders from 4058 to 5524 \AA\ for up to five targets (typically four stars and one background position). For this study, we use only the same order employed in the single-order spectra from these multi-order data. \autoref{tab:obs} lists the full set of observations, including the observing date, the on-target exposures, the offset exposures and the number of background/sky positions ($N_{\rm sky}$) assigned together with scientific targets. Though not detailed in the table, additional calibration data (e.g. flats, aperture reference spectra, ThArNe arc spectra, twilights, darks and biases) were obtained throughout the relevant M2FS runs.   These calibration data were required by the standard data reduction steps described in \citetalias{Song:2019aa}.

For most clusters we also obtained exposures while the telescope was deliberately offset from the nominal target positions in order to provide another way to determine background contamination. These offset exposures were taken in order to sample the local background for every target throughout the cluster and in the corresponding field.  Such offset exposures are not available for all cluster fields (see \autoref{tab:obs}).  The total exposure times on these offset positions ranged from 10\% to 50\% of the on-target exposure times (see \autoref{tab:obs} for the actual on- and off-target exposure times for all clusters).  We obtained offset background measurements of this sort for 18 of the 29 visits for the 26 clusters in our sample (we visited three clusters---NGC~1978, NGC~2203 and NGC~2209---on two separate occasions each; see \autoref{tab:obs}). 
  
All data were processed using an M2FS pipeline based on IRAF\footnote{IRAF is a collection of astronomical data reduction software originally written at the National Optical Astronomy Observatory (NOAO).}. The principal end products of this pipeline are the sky-subtracted spectra and their associated variances. In the pipeline, the reduction processes were largely the same as those described in \citetalias{Song:2019aa}, except for the last step---an improved background/sky subtraction---as described in the following subsection.

\subsection{Background Correction}
\label{sec:bg_cor}

Background/sky contamination is significant in our dataset and challenging to measure, in part due to the presence of unresolved light from the clusters and also due to telluric components arising from sunlight scattered within the atmosphere and reflected off the moon.  Together, these sources cause the backgrounds to vary in intensity and spectral character as a function of location relative to cluster centers. Background uncertainties generally have minor impact on the quality of kinematic measurements in the Mgb spectral region used in this study (see \autoref{sec:obs}).  However, good background measurements are required to obtain reliable stellar parameters such as surface gravity and metallicity.  Since both of these parameters are used to help determine cluster membership (see \autoref{sec:mem}), the precision of background subtraction has indirect impact on the derived dynamical properties of many of the clusters in our sample.  In this section, we describe how we sampled the backgrounds in our cluster fields and the methods we developed and tested to apply background corrections to our spectral data.

\subsubsection{Background sampling}
\label{sec:sky_obs}

For most of the clusters in our sample we assigned from 8 to 16 dedicated sky fibers within the tidal limits of the clusters (see \autoref{sec:star_select} regarding how background locations were identified). There were some exceptions.  NGC~1466 had no predetermined background positions (see \autoref{sec:star_select}); in this case we plugged 42 unassigned fibers to random open holes in the cluster's plug plate over as much of the full radial extent of the cluster as possible.  For NGC~339 and NGC~419, only 3 and 5 sky fibers, respectively, were assigned within their tidal radii.  In the case of NGC~1978, our first visit (of two) had only one sky-fiber assigned, while the second visit had none (see \autoref{sec:sky_sub} for more on this system).  In all cases, the fairly limited number of sky fibers reflects the relatively large (14 arcsec) minimum spacing between fibers which made it difficult to pack a large number of background fibers in the central regions of the clusters.

In order to sample the background in more locations, we began to obtain observations at offset positions (see \autoref{sec:obs}).  The advantages of this approach are that the local background can be sampled close to each target star (typically about 5 arcsec), and we obtain denser background sampling near the clusters' centers.  The disadvantages are that the offset positions were often contaminated by relatively bright stars, the off-target exposure times were typically only 10--50\% the total time on the science targets (see \autoref{sec:obs} and \autoref{tab:obs}), and the offset exposures were not clearly contemporaneous with target observations which can compromise their utility due to changes in observing conditions (e.g. moonrise/set, airmass changes, onset of twilight).

\subsubsection{Background subtraction}
\label{sec:sky_sub}

In \citetalias{Song:2019aa} we developed two techniques, dubbed `Method A' and `Method B', to estimate background contributions to M2FS data in and around MC star clusters. `Method A' is suitable for clusters that have dedicated sky fibers {\it and} offset-sky observations.  Method B can be used when offset-sky observations are not available, only sky fibers which may or (usually) may not sample the inner parts of a cluster adequately. This method requires having a set of Method-A clusters available in the sample. Details regarding how these methods were applied in this study can be found in Appendix~\ref{sec:sky_sub_app}.

For the sake of consistency, the final adopted background profiles for our sample were obtained by applying Method B to all clusters regardless of whether a given cluster had both dedicated sky fibers and offset-sky data.  We then adopted the `median minus 1-$\sigma$' profile as the best estimate for a clusters background-light profile.  In practice, this  profile corresponds to that of NGC~361, the cluster whose background profile is at the 16 percentile (the fifth of 17 profiles; that is, `median minus 1-$\sigma$') of the rank-ordered distribution of the background profiles.\footnote{ The use of the `median minus 1-$\sigma$' background profile was found to produce consistently flat metallicity profiles within the clusters (see \autoref{sec:rm_FeH} below).  In contrast, the median background profiles produce strong inward-rising radial metallicity gradients in nearly all the clusters of our sample.  This is precisely what one would expect if the background is being oversubtracted in the central cluster regions. We take this as corroborating evidence that the `median minus 1-$\sigma$' backgrounds are to be preferred.}
The background spectrum for every target star within a given cluster was calculated as the normalized median background spectrum from sky/offset fibers located beyond the tidal radius scaled by the adopted background profile for that cluster.

One cluster in our sample---NGC~1978---has inadequate data for reliable background subtraction.  Specifically, our first visit deployed only one sky-fiber position, and no offset observations were obtained.  Our second observation of this cluster lacked both a sky-fiber position and offset observations.  As a result, neither Method A nor B be applied in this case.  This limits our ability to obtain useful surface gravities and metallicities for the stars in NGC~1978.  The kinematic data, however, remain useful, though as we shall see, somewhat enigmatic compared to any other cluster in our sample.  We will return to this special case in \autoref{sec:MLv}.

\subsection{Total Working Sample}
\label{sec:working_sample}

\begin{table*}
 \caption{Previoius Kinematic Data of Clusters in Our Sample$^{\rm a}$}
 \label{tab:archive}
 \begin{center}
 \begin{threeparttable}
 \begin{tabular}{clrrrrll}
\hline
    \multicolumn{1}{c}{Galaxy}  &  
    \multicolumn{1}{c}{Cluster}  &
    \multicolumn{1}{c}{$N_{\rm sample}$} &  
    \multicolumn{1}{c}{$N_{\rm common}$} &  
    \multicolumn{1}{c}{ $V_{\rm sys, prev}$} &  
    \multicolumn{1}{c}{ $\Delta V_{\rm sys}$} &
    \multicolumn{1}{c}{Source} &
    \multicolumn{1}{c}{Code} \\
    \multicolumn{1}{c}{} &  
    \multicolumn{1}{c}{} &  
    \multicolumn{1}{c}{} &  
    \multicolumn{1}{c}{} & 
    \multicolumn{1}{c}{$\rm (km\,s^{-1})$}  &  
    \multicolumn{1}{c}{$\rm (km\,s^{-1})$}  &
    \multicolumn{1}{c}{} & 
    \multicolumn{1}{c}{} \\ 
    \multicolumn{1}{c}{(1)} & 
    \multicolumn{1}{c}{(2)} & 
    \multicolumn{1}{c}{(3)} & 
    \multicolumn{1}{c}{(4)} & 
    \multicolumn{1}{c}{(5)} &
    \multicolumn{1}{c}{(6)} &
    \multicolumn{1}{c}{(7)} &
    \multicolumn{1}{c}{(8)} \\
\hline 
SMC & NGC~330 & 16 & 5 & $153.7\pm1.0$ & $-0.87\pm1.27$ & \citet{Patrick:2020aa} & Pa20\\ 
LMC & NGC~1783 & 6 & 0 & $277.6\pm1.0$ & $1.98\pm1.02$ & \citet{Mucciarelli:2008aa} & Mu08\\ 
LMC & NGC~1806 & 8 & 2 & $228.6\pm0.5$ & $0.93\pm0.87$ & \citet{Mucciarelli:2014aa} & Mu14\\ 
LMC & NGC~1846 & 105 & 17 & $239.1\pm1.0$ & $0.21\pm0.46$ & \citet{Mackey:2013aa} & Ma13\\ 
LMC & NGC~1850 & 52 & 2 & $251.4\pm2.0$ & $-2.43\pm2.16$ & \citet{Fischer:1993aa} & Fi93\\ 
LMC & NGC~1978 & 35 & 8 & $293.3\pm1.0$  & $-0.19\pm1.11$ & \citet{Fischer:1992ab} & Fi92\\ 
LMC & NGC~1978 & 11 & 4 & $293.1\pm0.8$  & $0.01\pm1.03$ & \citet{Ferraro:2006aa} & Fe06\\ 
LMC & NGC~2257 & 6 & 1 & $299.4\pm1.5$ & $2.42\pm1.53$ & \citet{Mucciarelli:2010aa} & Mu10\\  
\hline
\end{tabular}
\begin{tablenotes}
 \item $^{\rm a}$ {Columns 3 and 4 list the number of stars in the studies listed in column 7 with data of sufficient kinematic precision for inclusion in our current sample ($N_{\rm sample}$) and the number of stars in common with our current sample ($N_{\rm common}$), respectively.  Columns 5 and 6 list the systemic cluster velocities for each cluster from the cited sources and the difference in the sense $\Delta V_{\rm sys} =  V_{\rm sys, M2FS}-V_{\rm sys,prev}$. Column 8 lists for each source cited in Column 7 a short code used in \autoref{tab:sample_all}. }
\end{tablenotes}
\end{threeparttable}
\end{center}
\end{table*}
\begin{table*}
\caption{Sample of all stars. (The full table is available online as supplementary material.)}
\label{tab:sample_all}
\begin{threeparttable}
\begin{tabular}{cllrrccccc}
\hline
    \multicolumn{1}{c}{Galaxy} & 
    \multicolumn{1}{c}{Cluster} & 
    \multicolumn{1}{c}{ID} & 
    \multicolumn{1}{c}{RAJ2000} & 
    \multicolumn{1}{c}{DEJ2000} & 
    \multicolumn{1}{c}{$G$} & 
    \multicolumn{1}{c}{$G_{\rm BP}-G_{\rm RP}$} & 
    \multicolumn{1}{c}{$T_{\rm eff}$} & 
    \multicolumn{1}{c}{\textit{Gaia} DR2 ID} &
    \multicolumn{1}{c}{...} \\
    \multicolumn{1}{c}{} & 
    \multicolumn{1}{c}{} & 
    \multicolumn{1}{c}{} &
    \multicolumn{1}{c}{(deg)} & 
    \multicolumn{1}{c}{(deg)} & 
    \multicolumn{1}{c}{(mag)} & 
    \multicolumn{1}{c}{(mag)} & 
    \multicolumn{1}{c}{(K)} & 
    \multicolumn{1}{c}{} & 
    \multicolumn{1}{c}{} \\  
    \multicolumn{1}{c}{(1)} & 
    \multicolumn{1}{c}{(2)} & 
    \multicolumn{1}{c}{(3)} & 
    \multicolumn{1}{c}{(4)} & 
    \multicolumn{1}{c}{(5)} &
    \multicolumn{1}{c}{(6)} & 
    \multicolumn{1}{c}{(7)} & 
    \multicolumn{1}{c}{(8)} & 
    \multicolumn{1}{c}{(9)} & 
    \multicolumn{1}{c}{...} \\ 
\hline 
SMC & Kron 3 & K3-1-b002 & 6.420511 & -72.862261 & 18.35 & 1.20 & 4869 & 4688782811694646272 & \multicolumn{1}{c}{...} \\ 
SMC & Kron 3 & K3-1-b004 & 6.366614 & -72.850370 & 18.29 & 1.26 & 4854 & 4688782983493342336 & \multicolumn{1}{c}{...} \\ 
SMC & Kron 3 & K3-1-b005 & 6.307460 & -72.842177 & 17.28 & 1.45 & 4577 & 4688783022152125440 & \multicolumn{1}{c}{...} \\ 
 \multicolumn{1}{c}{} & \multicolumn{1}{c}{} & \multicolumn{1}{c}{} & \multicolumn{1}{c}{} & \multicolumn{1}{c}{$\vdots$} & \multicolumn{1}{c}{} & \multicolumn{1}{c}{} & \multicolumn{1}{c}{} & \multicolumn{1}{c}{} \\
LMC & NGC 1978 & N1978-3-r049 & 82.190452 & -66.235682 & 16.14 & 1.04 & 3840 & 4660340443101129856 & \multicolumn{1}{c}{...} \\ 
LMC & NGC 1978 & N1978-3-b034 & 82.209310 & -66.256802 & 15.94 & 1.90 & 3755 & 4660340271299657216 & \multicolumn{1}{c}{...} \\ 
LMC & NGC 1978 & N1978-3-b037 & 82.211433 & -66.245587 & 15.76 & 1.95 & 3755 & 4660340370035622144 & \multicolumn{1}{c}{...} \\ 
 \multicolumn{1}{c}{} & \multicolumn{1}{c}{} & \multicolumn{1}{c}{} & \multicolumn{1}{c}{} & \multicolumn{1}{c}{$\vdots$} & \multicolumn{1}{c}{} & \multicolumn{1}{c}{} & \multicolumn{1}{c}{} & \multicolumn{1}{c}{} \\
LMC & NGC 1978 & 06 & 82.193775 & -66.236033 & \multicolumn{1}{c}{--} & \multicolumn{1}{c}{--} & \multicolumn{1}{c}{--} & \multicolumn{1}{c}{--} & \multicolumn{1}{c}{...} \\ 
LMC & NGC 1978 & 07 & 82.192219 & -66.238693 & \multicolumn{1}{c}{--} & \multicolumn{1}{c}{--} & \multicolumn{1}{c}{--} & \multicolumn{1}{c}{--} & \multicolumn{1}{c}{...} \\ 
 \multicolumn{1}{c}{} & \multicolumn{1}{c}{} & \multicolumn{1}{c}{} & \multicolumn{1}{c}{} & \multicolumn{1}{c}{$\vdots$} & \multicolumn{1}{c}{} & \multicolumn{1}{c}{} & \multicolumn{1}{c}{} & \multicolumn{1}{c}{} \\
\hline
\end{tabular}
\end{threeparttable}
\end{table*}

\begin{table*}
\contcaption{Sample of all stars. (The full table is available online as supplementary material.)}
\begin{threeparttable}
\begin{tabular}{crcccrrcc}
\hline
    \multicolumn{1}{c}{...} & 
    \multicolumn{1}{c}{S/N$^{\rm a}$} & 
    \multicolumn{1}{c}{$v_{\rm los}$} & 
    \multicolumn{1}{c}{$\log{g}$} & 
    \multicolumn{1}{c}{${\rm [Fe/H]}_{\rm raw}$} & 
    \multicolumn{1}{c}{$P_{M}$} &
    \multicolumn{1}{c}{$P_{M}^{\prime}$} &
    \multicolumn{1}{c}{Flag$^{\rm b}$} &
    \multicolumn{1}{c}{Source$^{\rm c}$} \\
    \multicolumn{1}{c}{} &
    \multicolumn{1}{c}{} & 
    \multicolumn{1}{c}{$\rm (km\,s^{-1})$} & 
    \multicolumn{1}{c}{$\rm (dex)$} & 
    \multicolumn{1}{c}{$\rm (dex)$} &
    \multicolumn{1}{c}{} & 
    \multicolumn{1}{c}{} & 
    \multicolumn{1}{c}{} & 
    \multicolumn{1}{c}{} \\ 
    \multicolumn{1}{c}{...} &
    \multicolumn{1}{c}{(10)} &
    \multicolumn{1}{c}{(11)} & 
    \multicolumn{1}{c}{(12)} & 
    \multicolumn{1}{c}{(13)} &
    \multicolumn{1}{c}{(14)} &
    \multicolumn{1}{c}{(15)} & 
    \multicolumn{1}{c}{(16)} & 
    \multicolumn{1}{c}{(17)} \\
\hline 
 \multicolumn{1}{c}{...} & 4.1 & $111.20\pm0.78$ & $1.63\pm0.21$ & $-1.03\pm0.08$ & 0.00 & 0.00 & 0000000 & M2FS \\ 
 \multicolumn{1}{c}{...} & 4.3 & $130.36\pm0.49$ & $1.35\pm0.20$ & $-0.88\pm0.08$ & 0.00 & 0.00 & 0000000 & M2FS \\ 
 \multicolumn{1}{c}{...} & 7.3 & $160.47\pm0.31$ & $1.21\pm0.09$ & $-0.75\pm0.04$ & 0.00 & 0.00 & 0000000 & M2FS \\
 \multicolumn{1}{c}{} & \multicolumn{1}{c}{} & \multicolumn{1}{c}{} & \multicolumn{1}{c}{} & \multicolumn{1}{c}{$\vdots$} & \multicolumn{1}{c}{} & \multicolumn{1}{c}{} & \multicolumn{1}{c}{} & \multicolumn{1}{c}{} \\
 \multicolumn{1}{c}{...} & 4.4 & $292.97\pm0.52$ & $0.51\pm0.10$ & $-0.89\pm0.07$ & 1.00 & 1.00 & 0000010 & M2FS \\ 
 \multicolumn{1}{c}{...} & 13.5 & $292.38\pm0.16$ & $0.84\pm0.04$ & $-0.65\pm0.04$ & 0.99 & 0.99 & 0000000 & M2FS+Fi92+Mu08 \\ 
 \multicolumn{1}{c}{...} & 14.7 & $295.43\pm0.26$ & $0.84\pm0.05$ & $-0.66\pm0.04$ & 1.00 & 1.00 & 0000000 & M2FS+Fi92 \\ 
 \multicolumn{1}{c}{} & \multicolumn{1}{c}{} & \multicolumn{1}{c}{} & \multicolumn{1}{c}{} & \multicolumn{1}{c}{$\vdots$} & \multicolumn{1}{c}{} & \multicolumn{1}{c}{} & \multicolumn{1}{c}{} & \multicolumn{1}{c}{} \\
 \multicolumn{1}{c}{...} & \multicolumn{1}{c}{--} & $291.21\pm1.30$ & \multicolumn{1}{c}{--} & \multicolumn{1}{c}{--} & 1.00 & 1.00 & 0000000 & Fi92 \\ 
 \multicolumn{1}{c}{...} & \multicolumn{1}{c}{--} & $288.76\pm0.20$ & \multicolumn{1}{c}{--} & \multicolumn{1}{c}{--} & 1.00 & 1.00 & 0000000 & Fi92+Mu08 \\
 \multicolumn{1}{c}{} & \multicolumn{1}{c}{} & \multicolumn{1}{c}{} & \multicolumn{1}{c}{} & \multicolumn{1}{c}{$\vdots$} & \multicolumn{1}{c}{} & \multicolumn{1}{c}{} & \multicolumn{1}{c}{} & \multicolumn{1}{c}{} \\
\hline
\end{tabular}
\begin{tablenotes}
 \item $^{\rm a}$ {Median S/N per pixel of M2FS spectrum.}
 \item $^{\rm b}$ {These seven-digit flags denote with a `1' the following: Rejection due to poor skew/kurtosis values in the Bayesian spectra fits (Digit 1);  Excessive velocity error (Digit 2); Carbon star (Digit 3); Foreground dwarf (Digit 4); Member of a non-cluster/non-MC population (Digit 5); Has a large color offset for $T_{\rm eff}$ determination (Digit 6); Likely metallicity non-member (Digit 7).  Details about how these flags are set can be found in \autoref{sec:rm_PSK}, \autoref{sec:rm_odd}, \autoref{sec:rm_dwarf}, \autoref{sec:rm_vlos} and \autoref{sec:rm_FeH}.}
 \item $^{\rm c}$ {Sources for the LOS velocities.  M2FS denotes this work.  Other codes are listed in column 4 of  \autoref{tab:archive}.}
\end{tablenotes}
\end{threeparttable}
\end{table*}

\begin{figure}
   \centering
   \includegraphics[width=0.47\textwidth]{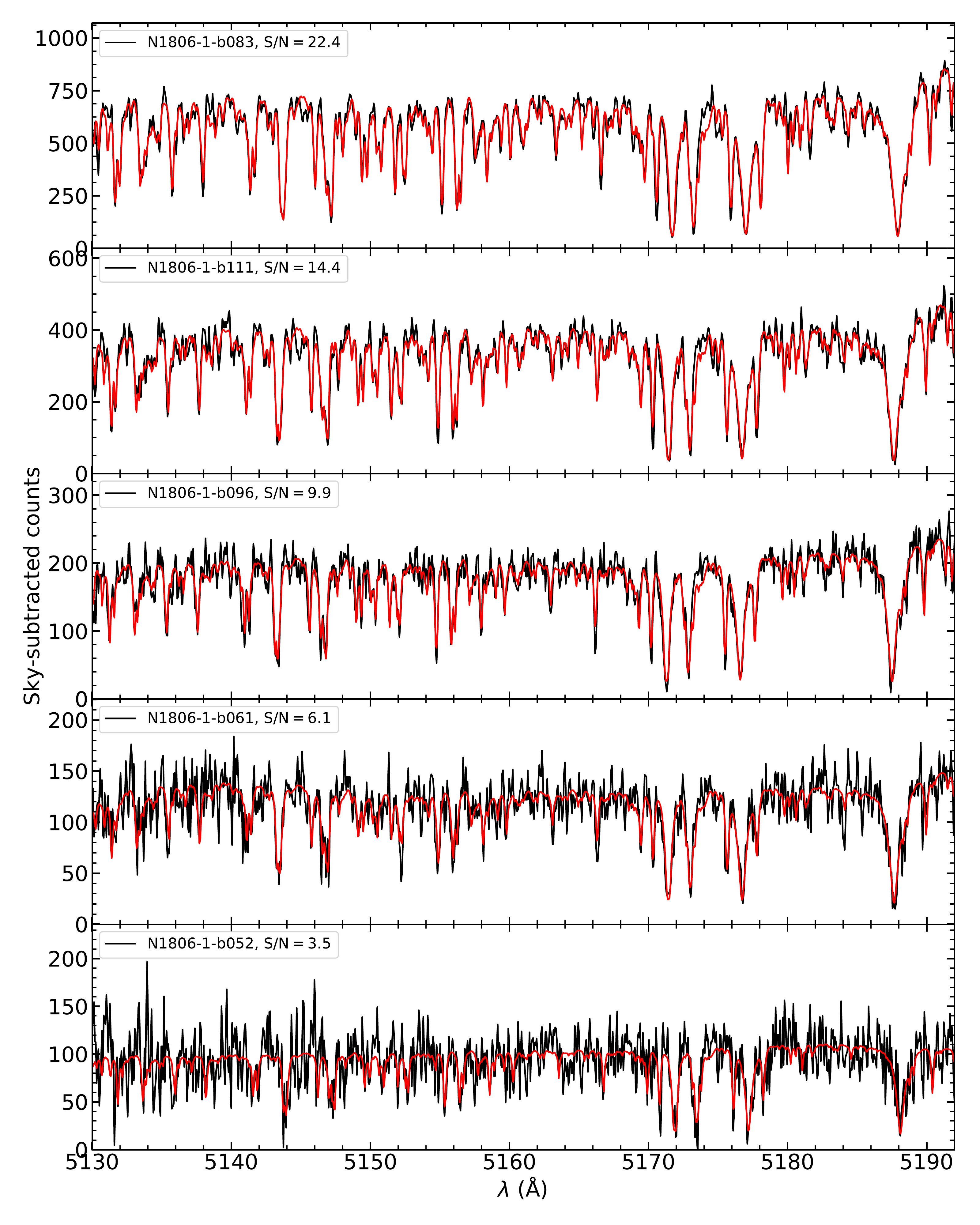}
   \caption{Representative spectra (black) for five stars observed with M2FS in NGC~1806 which conveniently span almost the full range of S/N of usable spectra within our full dataset. The spectra shown here have been corrected for backgrounds as described in \autoref{sec:sky_sub}. The red lines are the best-fitting spectral models determined with the Bayesian spectral fitting analysis described in \autoref{sec:spec_fit}. The legends within each panel list the target ID (following the naming system of \autoref{tab:sample_all} and the median S/N per pixel of each spectrum.}
   \label{fig:spec_sample}
\end{figure}

Based on the reduction steps described in \autoref{sec:obs} and \autoref{sec:bg_cor}, our full M2FS dataset consists of 3137 background-subtracted target spectra of 2901 distinct targets in the fields of the 26 MC star clusters we observed for this study (see \autoref{tab:basic}). The final total includes the effect of two clusters, NGC~2203 and NGC~2209, where we have combined background-subtracted spectra results, suitably weighted, from independent visits to each cluster (see \autoref{tab:obs}). Some representative examples of background-subtracted spectra that span nearly the full range of S/N in our sample are shown in  \autoref{fig:spec_sample}. A full listing of the results from the 2901 spectra is provided in \autoref{tab:sample_all}.

We have expanded this dataset with previously published-kinematic data of comparable quality for targets in and near the clusters of our sample (\autoref{tab:basic}). We restricted our sources to those with typical velocity precisions of less than about 3~\kms\ \citep{Fischer:1992ab, Fischer:1993aa, Ferraro:2006aa, Mucciarelli:2008aa, Mucciarelli:2010aa, Mucciarelli:2014aa, Mackey:2013aa, Patrick:2020aa}. \autoref{tab:archive} provides some details of the data obtained from these sources, including the total number of targets previously studied and the number of stars in common with our M2FS sample.   The table also lists the cluster systemic velocities from the previous studies and the difference relative to the corresponding systemic velocities obtained from the new M2FS data in the sense M2FS results minus previously published results. Details regarding how we combined earlier and M2FS data are described in \autoref{sec:combine_samples}, and final consolidated results are incorporated in \autoref{tab:sample_all}.  We critically compare repeat measurements of common stars in \autoref{sec:comp}. With the addition of these earlier datasets, our final comprehensive sample consists of 3376 spectroscopic results for 3095 distinct sources in our target clusters.

\section{Spectral Analysis}
\label{sec:analysis}

\subsection{Bayesian Fitting of M2FS Spectra}
\label{sec:spec_fit}

Our analysis of the background-corrected spectra employed the same Bayesian formalism described in \citet{Walker:2015aa, Song:2017aa} and adopted in \citetalias{Song:2019aa}. The outputs of this analysis include estimates of line-of-sight (LOS) velocity, surface gravity and metallicity for every star along with associated uncertainties. Some examples of the best-fit spectra obtained with this method are shown in \autoref{fig:spec_sample}.

For this study we altered one important aspect of the analysis compared to \citetalias{Song:2019aa}.  Namely, the effective temperatures of all stars were treated as priors based on the $G_{\rm BP}-G_{\rm RP}$ colors provided in the \textit{Gaia} DR2 rather than from $V-I$ colors from, for example, \textit{HST} images (the method we used for NGC~419 and NGC~1846 in \citetalias{Song:2019aa}).  The resulting temperatures were then forced to remain fixed throughout the spectral-fitting process. The details of this procedure are described in the following section.

\subsection{Effective Temperature Priors}
\label{sec:Teff}

\begin{figure*}
   \centering
    \includegraphics[width=0.33\textwidth]{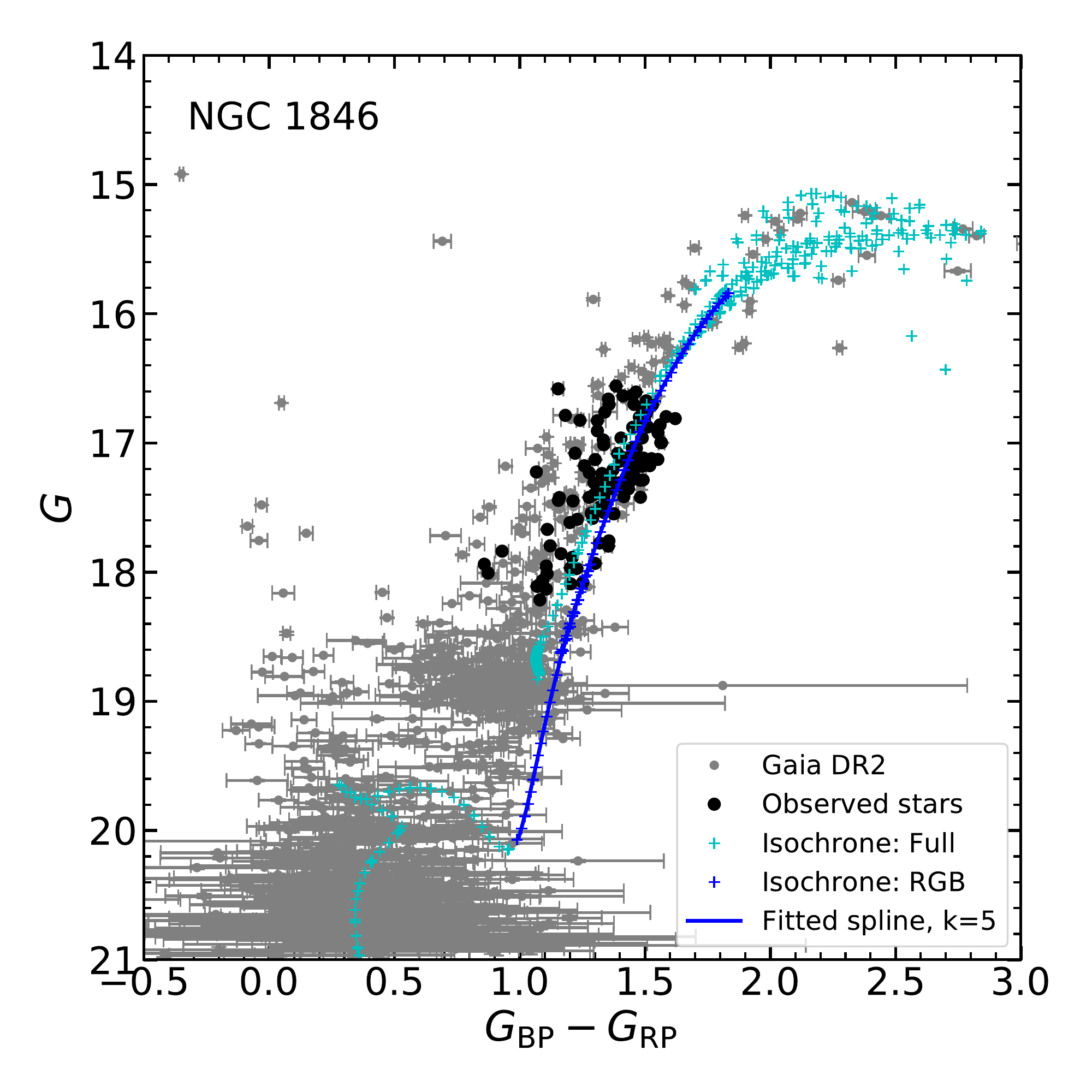} 
    \includegraphics[width=0.33\textwidth]{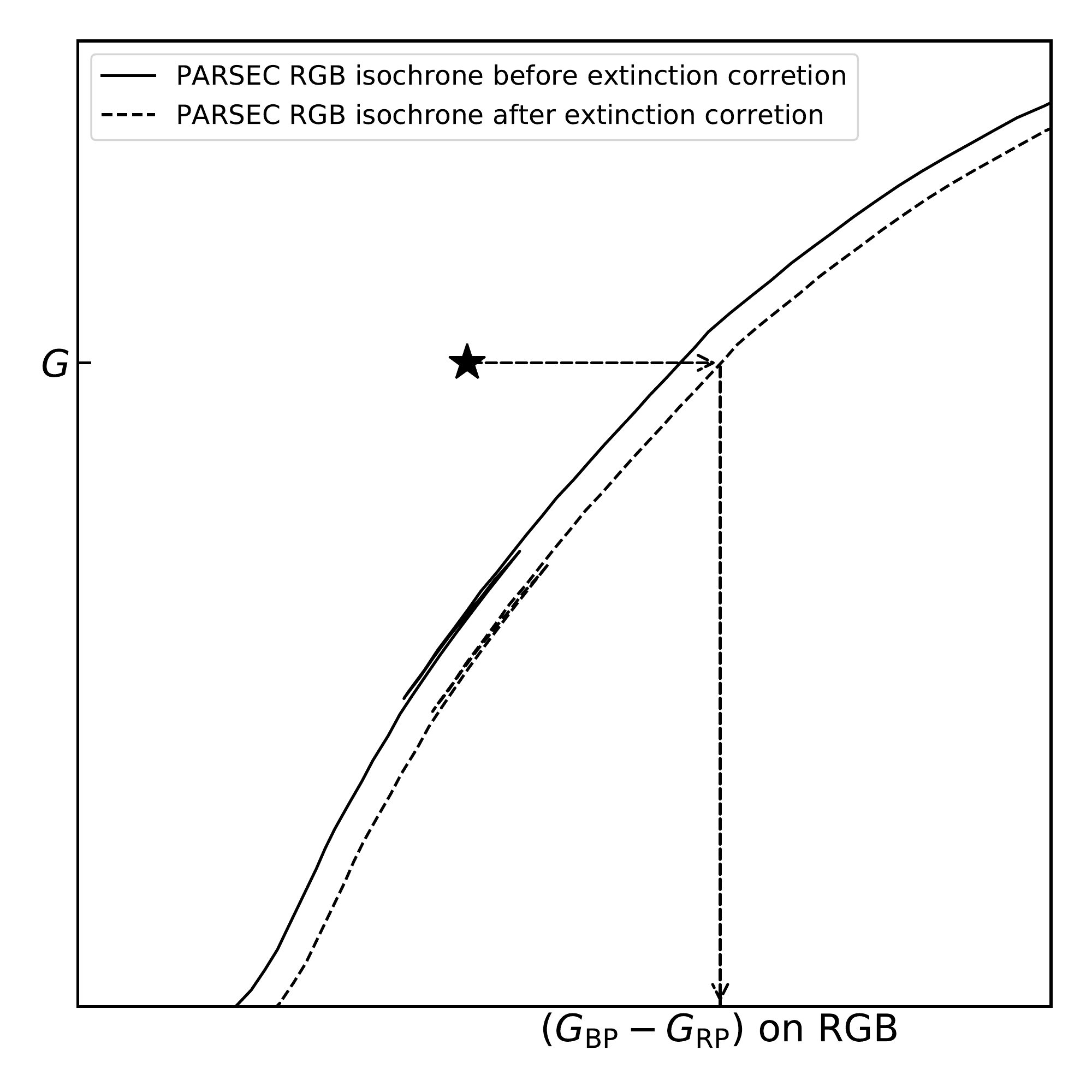}
    \includegraphics[width=0.33\textwidth]{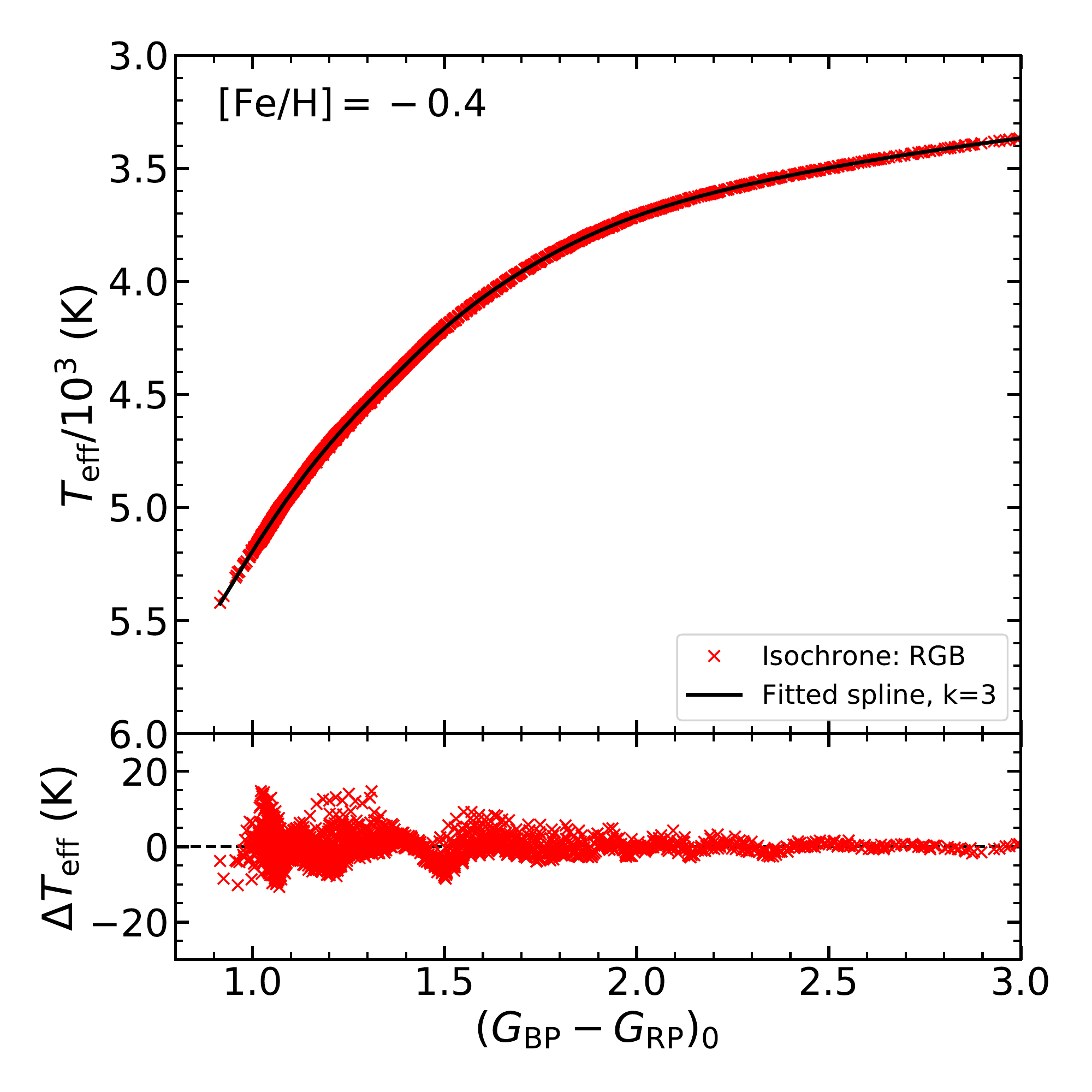}
   \caption{(Left panel) The color-magnitude diagram for NGC~1846 based on \textit{Gaia} DR2 and using the DR2 photometric system.  A PARSEC model isochrone \citep{Bressan:2012aa} using cluster's age and metallicity estimates (see \autoref{tab:basic}) is overplotted (light blue crosses). Gray dots correspond to all \textit{Gaia} DR2 stars located 2 arcmin from the cluster center, while black dots represent the stars observed in our sample. The blue crosses highlight the RGB sequence of the isochrone, and the blue solid curve is a spline fit to this RGB segment. 
   (Middle panel) Following the discussion in \autoref{sec:Teff}, this plot illustrates schematically how a given star in the CMD is projected to the adopted RGB sequence to determine the star's model-based $G_{\rm BP}-G_{\rm RP}$ color. 
   (Right panel) The red symbols show an example of the \textit{Gaia} DR2 color/temperature relations obtained from the PARSEC models. The relation shown here is suitable for the clusters, such as NGC~1846, with a mean metallicity of $\rm [Fe/H] \sim -0.4$ (see \autoref{tab:basic}). These individual color-temperature points were fit to a third-order spline (black line; fit residuals are shown in the lower subpanel).  We used this spline fit to convert the adjusted colors (see middle panel) to effective temperature for each star.  These temperatures were fixed during the Bayesian spectral fitting procedure (see \autoref{sec:spec_fit}).}
   \label{fig:Teff}
\end{figure*}

\begin{figure}
   \centering
    \includegraphics[width=0.47\textwidth]{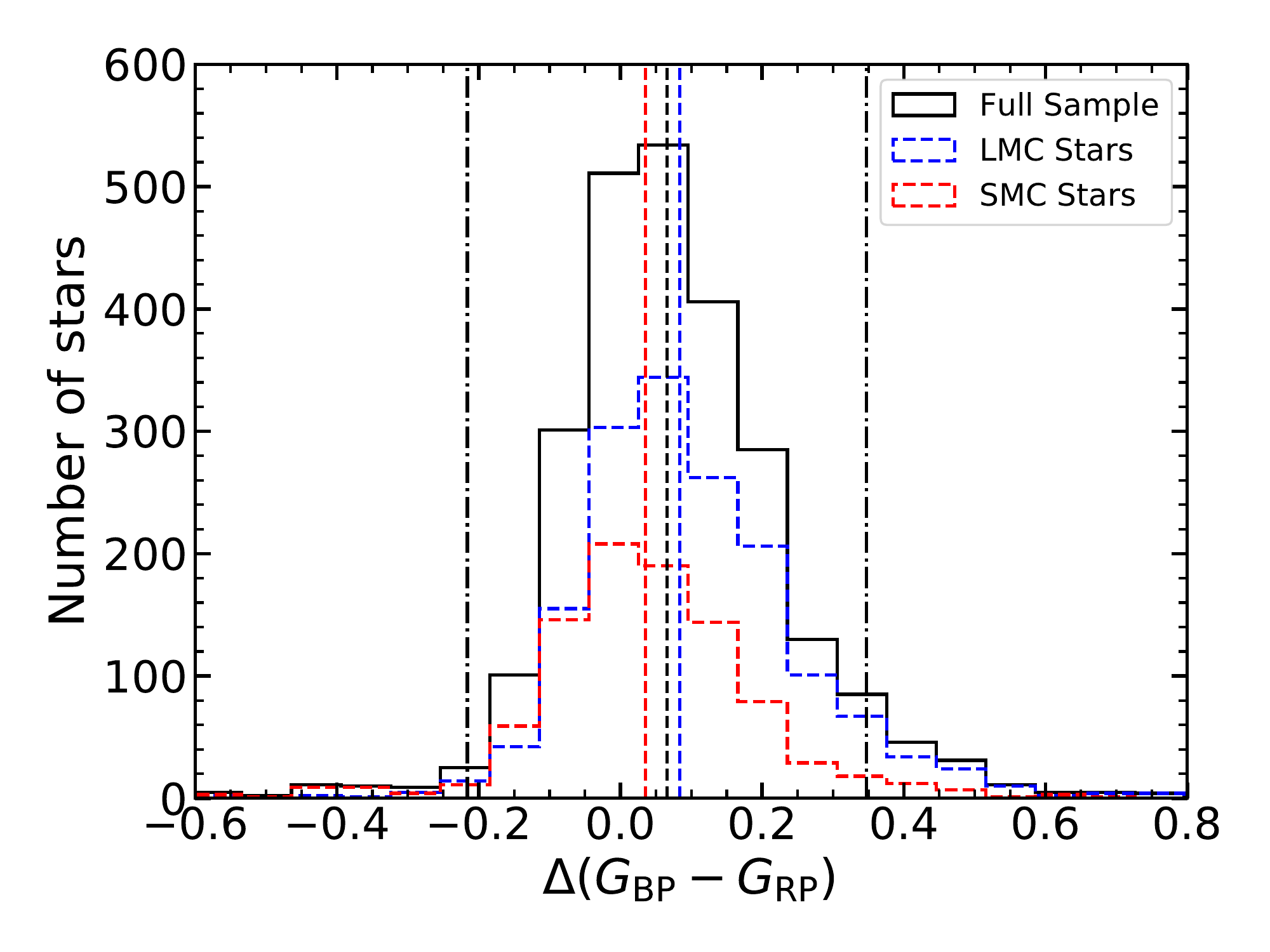}
   \caption{A histogram of the color shifts, $\Delta{(G_{\rm BP}-G_{\rm RP})}$, between the \textit{Gaia} DR2 color of a given star in our dataset and our adjusted color obtained as described in \autoref{sec:Teff} and \autoref{fig:Teff}.
   For 2543 stars in the full sample, the biweight mean offset $C_{\rm BI,dcolor}$ in $\Delta{(G_{\rm BP}-G_{\rm RP})}$ (denoted by the dashed line) is 0.065 with a $2S_{\rm BI,dcolor}$ range of  $\pm 0.281$ (denoted by the dot-dash lines). A total of 203 stars are located outside the dashed lines, though the distribution is clearly skewed toward positive values (68 stars on the negative side and 135 on the positive side). This likely represents the fact that some stars located blueward of the adopted RGB sequences in individual clusters (see the middle panel of \autoref{fig:Teff}) are hotter than red giants. For the purpose of metallicity determination described in \autoref{sec:rm_FeH}, we use the stars within two biweight dispersion scales of the biweight mean, i.e. $|\Delta{(G_{\rm BP}-G_{\rm RP})}-C_{\rm BI, dcolor}|\leq2S_{\rm BI, dcolor}$. This ensures that only stars with relatively small color shifts---likely true RGB stars---are used to determine the cluster metallicity. The blue and red dashed histograms show the distributions in $\Delta{(G_{\rm BP}-G_{\rm RP})}$ for the LMC (1595 stars) and the SMC (948 stars).   The mean offsets of these histograms are $C_{\rm BI,dcolor}\pm 2S_{\rm BI,dcolor}=0.084\pm0.283$ for the LMC and $0.036\pm0.269$ for the SMC.}
   \label{fig:dcolor}
\end{figure}

As argued in \citet{Song:2017aa}, an effective temperature prior ($T_{\rm eff}$) helps to break the temperature-metallicity degeneracy in the posterior probability density functions (PDFs) calculated using our Bayesian analysis (see \autoref{sec:spec_fit}). This degeneracy affects our data strongly due to the limited wavelength range (5130 to 5192 \AA) of the single-order spectra we obtained (see \autoref{sec:obs}).  To help mitigate this problem, our application of the Bayesian spectral fitting procedure fixes the effective temperature throughout the optimization process for a given star. 

One complication with this approach is that if we were to use the \textit{Gaia} DR2 colors to estimate $T_{\rm eff}$ directly, our temperature estimates would likely exhibit significant systematic and random uncertainties.  There are two reasons for this.  First, many of the stars in our sample are comparatively faint ($G>17$ mag) for the \textit{Gaia} sample, and so their formal photometric errors are moderate \citep{Evans:2018aa}.  Second, given the high source densities in many of our fields, background variations and contamination due to crowding can lead to significant additional uncertainty in the \textit{Gaia} $G_{\rm BP}-G_{\rm RP}$ colors \citep{Weiler:2018aa, Maiz-Apellaniz:2018aa}.  Moreover, in clusters, blending of red giants with typically hotter, fainter cluster members tends to drive colors systematically to the blue.  

These issues are particularly problematic for $T_{\rm eff}$ determinations in RGB stars since even a small shift in color can lead to a significant temperature change which, in turn, degrades the precision of the metallicity estimate of a star. To mitigate these problems, we have devised a method to estimate $T_{\rm eff}$ that depends primarily on the observed $G$ magnitude of a star and that relies on the fact that
the stars in our sample were almost exclusively selected to be located on the RGBs of their respective clusters (see \autoref{sec:star_select}). This methodology could not be applied to NGC~330, NGC~458 and NGC~1850 as they lack an extended RGB owing to their young ages.

The method starts by identifying an isochrone that matches the adopted age and metallicity of a given cluster (see \autoref{tab:basic}).  For this study we have chosen to use the PARSEC isochrones \citep[version 1.2S,][]{Bressan:2012aa} converted to the \textit{Gaia} DR2 photometric system following the prescription of \citet{Maiz-Apellaniz:2018aa}\footnote{In practice, we computed tailor-made isochrones for each cluster in the appropriate photometric system using the PARSEC website (\url{http://stev.oapd.inaf.it/cgi-bin/cmd}).}.
We then plotted the synthetic isochrones onto the observational $G$ vs $G_{\rm BP}-G_{\rm RP}$ plane (an example is shown in the left panel of \autoref{fig:Teff}).  In all cases, we shifted the isochrones to the observational plane with the appropriate reddening values and distance moduli for each cluster (\autoref{tab:basic}).  As suggested by the PARSEC model website, these corrections used the extinction parameters $A_G=0.86A_V$, $E(G_{\rm BP}-G_{\rm RP})=0.42A_V$ and $R_V=3.1$ within the extinction relations from \citet{Cardelli:1989aa} and \citet{ODonnell:1994aa}.   For computational convenience, we then fit each isochrone in the observational plane to a third-order spline to smooth out round-off noise and some of the non-monotonic evolutionary behavior in the isochrones themselves.  An example of such a fit is illustrated in the left panel of \autoref{fig:Teff}. 

The \textit{Gaia} DR2 photometry for every cluster was corrected to the `true' $G$-band system as suggested by \citet{Anders:2019aa} before plotting the data in the observational CMD with their respective isochrones.
Each star in our spectroscopic sample was then projected at constant $G$ magnitude onto the spline fits to the RGBs on a cluster-by-cluster basis.  The corresponding $G_{\rm BP}-G_{\rm RP}$ color at that point in the isochrone was adopted as `the' color for that star. The middle panel of \autoref{fig:Teff} provides a schematic illustration of this process, and was adopted for all clusters older than about 0.5 Gyr since their RGB sequences could be reasonably defined from the models. 

For the three clusters in our sample younger than 0.5 Gyr (NGC~330, NGC~458 and NGC~1850), we did not correct the colors in this manner but simply adopted the published \textit{Gaia} DR2 colors corrected as prescribed by \citet{Anders:2019aa}.  For these systems, the red (super)giants tend to be brighter than 17 mag in $G$-band and hence exhibit relatively small color errors, and these stars are less susceptible to crowding/background-related photometric errors.  Moreover, these younger stars tend to cluster in a region in the CMD rather than on a distinct giant branch \citep[see, e.g.][]{Mermilliod:1981aa, Alcaino:2003aa, Correnti:2017aa,  Milone:2018aa}; hence they are ill-suited for the temperature-determination procedure we used for the rest of our sample which assumes a well-defined red giant locus with little color spread at a given luminosity.

\autoref{fig:dcolor} shows a histogram of the color shifts in $\Delta(G_{\rm BP}-G_{\rm RP})$ that we determined for every star with a tabulated \textit{Gaia} DR2 color using the method described above.  The mode of this distribution is offset from, but close to zero, confirming that most of these objects are consistent with being RGB/AGB stars.  This supports the efficacy of our approach.  Since the slope of the giant branch is approximately $\delta(G_{\rm BP}-G_{\rm RP})/\delta G \sim 0.2$, the color error is about 20\% the $G$-magnitude error for a given star {\it if it is on the RGB as assumed}.  Consequently, the resulting error distribution of the adopted colors is considerably narrower than the distribution of the `raw' $\Delta(G_{\rm BP}-G_{\rm RP})$ implied by \autoref{fig:dcolor}. It is also evident in \autoref{fig:dcolor} that the $\Delta(G_{\rm BP}-G_{\rm RP})$ distribution exhibits a moderate tail to positive values.  This implies that some comparatively hot stars fail our assumption that they are on the RGB and hence they have been assigned colors that are systematically too red. 

The fact that the mean offset in $\Delta(G_{\rm BP}-G_{\rm RP})$ differs between the LMC and SMC (\autoref{fig:dcolor}) suggests that our color corrections may depend on metallicity (larger effect for larger metallicities in the LMC). We will return to this problem in \autoref{sec:rm_FeH} and \autoref{sec:FeH} where we use \autoref{fig:dcolor} to help us determine mean cluster metallicities.

The final step in estimating $T_{\rm eff}$ values for our target stars was to convert their $G_{\rm BP}-G_{\rm RP}$ colors---either the `raw' values for stars in the three young clusters or the model-based values determined using the procedure above---to temperatures.  We again used the PARSEC RGB sequence for each cluster (appropriately reddened and adopting the ages and metallicities in \autoref{tab:basic}) to define a color-temperature relation for each case (see the right panel of \autoref{fig:Teff} for an example). The final $T_{\rm eff}$ values were then tabulated for every star and these values are provided in \autoref{tab:sample_all}.

\subsection{Velocity Uncertainty Correction}
\label{sec:e_vlos}

To reliably measure small internal velocity dispersion of systems such as the star clusters in our sample,  high-quality kinematic data and reliable error estimates are essential (see, e.g. \citealp{Kamann:2016aa}; \citetalias{Song:2019aa}).  In \citetalias{Song:2019aa}, we analyzed repeat measurements of individual stars to empirically assess the quality of the uncertainties computed by the Bayesian spectral analysis for the cases of NGC~419 and NGC~1846.  We found that individual velocity uncertainties returned by the Bayesian analysis underestimated the true uncertainties by approximately 23\% and 12\%, respectively.   

In this study we explore this question again for the two clusters in our sample with the lowest derived central velocity dispersions, NGC~2155 and SL~663 (see \autoref{sec:mem}). 
As in \citetalias{Song:2019aa}, we compared the velocities from individual exposures, $s_i$,  and then fit a Gaussian to the error distribution expressed in units of the formal error, $\sigma_i$, returned by the Bayesian analysis for the $i$-th spectrum  (see \citealp{Kamann:2016aa} for analogous analysis). 
One complication with this test is that there are small systematic velocity shifts in M2FS data that correlates with the temperature of the instrument near the fiber pseudo slit (see \citealp{Walker:2015ab}).  The spectra from the individual exposures for both clusters were first corrected for these shifts and referenced to a common velocity scale before generating the $s/\sigma$ distribution.  The results of this analysis show that the velocity uncertainties for these clusters were similarly underestimated as found in \citetalias{Song:2019aa}, i.e. approximately 16\% and 18\% for NGC~2155 and SL~663, respectively (see Section 3.1.1 and Figure 4 of \citetalias{Song:2019aa} for details).

Together with our earlier results from \citetalias{Song:2019aa}, we conclude that our application of the Bayesian spectral analysis underestimates the true velocity errors by $17\% \pm 3\%$. Consequently, we have increased the velocity uncertainties from all Bayesian spectral analysis results by 17\% prior to carrying out any dynamical analyses of these systems (see \autoref{sec:mem} and \autoref{sec:results})\footnote{The origin of the 17\% correction likely lies in the fact that our data-reduction pipeline rebins the data during the extraction and wavelength-calibration steps.  This is confirmed by independent reductions of similar data by M.~G.~Walker (private communication) who finds that unbinned analyses lead to a much smaller correction consistent with unity. His results also independently confirm the velocity-uncertainty correction factor of 17\%\ that we obtained.  We consider the correction to be sufficiently small  and sufficiently well-determined to use it to adjust our formal velocity error estimates for the purposes of this study.}.

\subsection{Toward a Final M2FS Spectral Sample}
\label{sec:rm_all}

\subsubsection{Rejecting spectra with poorly determined parameters}
\label{sec:rm_PSK}

Our Bayesian analysis as described in \autoref{sec:spec_fit} generates posterior parameter distribution functions (PDFs) for 14 free parameters (the 15th, $T_{\rm eff}$, is fixed as described in \autoref{sec:Teff}) used to characterize the model spectra (see \citealt{Song:2017aa} for details).  The analysis measures the skew, $S$, and kurtosis, $K$, of all of these PDFs.  We use the $S$-$K$ values for the PDF of the LOS velocity ($v_{\rm los}$) parameter for every spectrum to carry out an initial quality cut on the data.   After inspection of the distribution of the skew/kurtosis values for the LOS velocities, we found that the same cuts in $S$ and $K$ from \citet{Walker:2015aa} were suitable for our data, namely $|S|>1$, or $|K-3|>1$. This `SK cut' led to the rejection of 71 spectra from our sample.  These cases are flagged in column 17 of \autoref{tab:sample_all}; see Appendix~\ref{sec:spec_PSK} for further details. We will not use these stars in any subsequent analyses in this paper.

\subsubsection{Stars with anomalous velocity uncertainties}
\label{sec:rm_odd}

\begin{figure}
   \centering
    \includegraphics[width=0.47\textwidth]{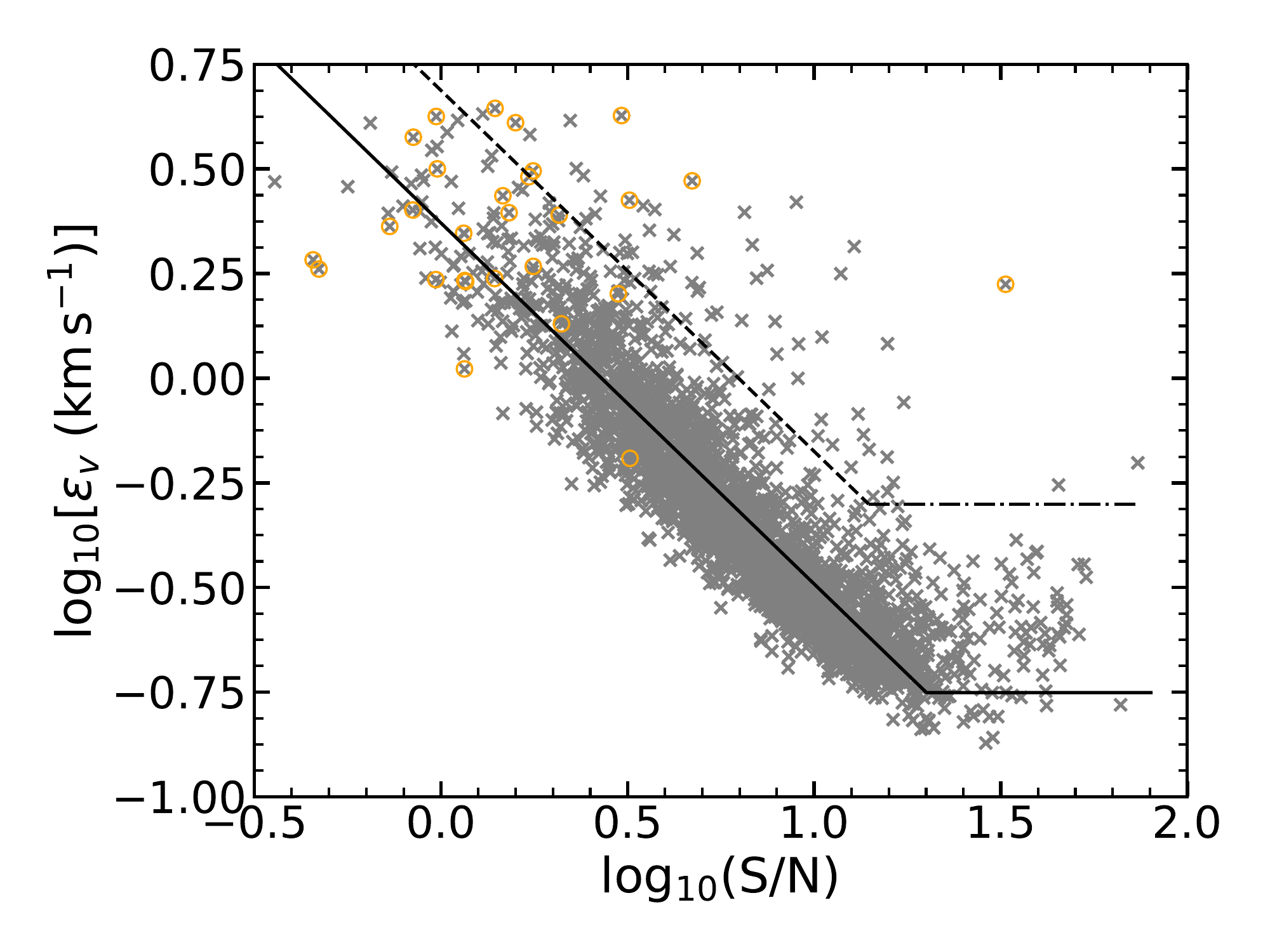}     
   \caption{Corrected LOS velocity uncertainties versus median S/N per pixel for all stars from the 26 clusters in log-log plane. The solid lines correspond to a linear relation in the log-log plane fit to all points with $\rm S/N \leq 20$ (excluding the yellow circles rejected by the SK discussion in \autoref{sec:rm_PSK}). For $\rm S/N>20$, we assume $\varepsilon_v$ is constant at 0.18~\kms.  The dashed and dot-dashed lines together denote the upper acceptance limits in these plots as described in \autoref{sec:rm_odd}.}
   \label{fig:e_vlos}
\end{figure}

\autoref{fig:e_vlos} shows the distribution of corrected velocity uncertainties, $\varepsilon_v$, as a function of median S/N ratio for every pixel in every target spectrum obtained for this study in log-log space.
The well-populated ridge lines reveal a robust global trend of velocity error with S/N for our M2FS dataset.  The approximately linear trend visible in the log-log space indicates that a good fit to these data has the form of a power law.  

With the aim of using the relation between $\varepsilon_v$ and S/N to remove outliers from our kinematic sample, we fit a linear relation in the log-log plane using an unweighted least-squares fit to all points corresponding to spectra with median $\rm S/N \leq 20$.  Targets removed from the SK-cut described in \autoref{sec:rm_PSK} were not used to compute this fit.  The fitted relation is shown as a solid straight line in the log-log plot (the coefficients for the fit equations are provided in \autoref{fig:e_vlos}). We then computed logarithmic residuals ($\Delta=\log_{10}{\varepsilon_{v}}-\log_{10}{\varepsilon_{v, \rm fit}}$) to estimate the Tukey's biweight location ($C_{{\rm BI}, \Delta}$) and biweight scale ($S_{{\rm BI}, \Delta}$) of the data about the fitted line in the log-log plot \citep{Beers:1990aa}.  The dashed straight line in the log-log plot in \autoref{fig:e_vlos} indicates the line offset upward by $C_{\rm BI}+3S_{\rm BI}$ from the fitted relation; the dashed line in the linear plot (left panel) shows the power law relation corresponding to this offset line.  We also chose to keep all stars with high-S/N spectra that resulted in  $\varepsilon_v \leq 0.5$~\kms; this condition is shown as a dot-dash line in both panels.  

We define the dashed/dot-dashed lines in \autoref{fig:e_vlos} as a `rejection boundary.' Excluding the 71 stars already rejected as SK outliers (\autoref{sec:rm_PSK}), 62 additional stars lie above this boundary.  We inspected each of these 62 spectra and found that we could classify them into distinct categories, each of which are flagged in column 17 of \autoref{tab:sample_all}. Details of the classifications of these spectra are provided in Appendix~\ref{sec:spec_odd}.

In \citetalias{Song:2019aa} we identified eight stars in NGC 419 and NGC 1846 (four in each) that we rejected using similar criteria as described here but based only on the small samples of these two clusters.  The four stars in our NGC~419 sample (three C stars and one blended source)  rejected in that paper have also been rejected in the present sample.   However, the four stars in our NGC~1846 sample that were rejected in \citetalias{Song:2019aa} consist of three weak C stars and one mildly blended source.  All four of these stars were {\it not} rejected in the present analysis using our full spectral sample.  This very slightly---and statistically indistinguishably---alters our final dynamical results below (see \autoref{sec:results}) for NGC~1846 compared to the results reported in \citetalias{Song:2019aa}.

\subsubsection{Foreground dwarfs}
\begin{figure}
   \centering
   \includegraphics[width=0.47\textwidth]{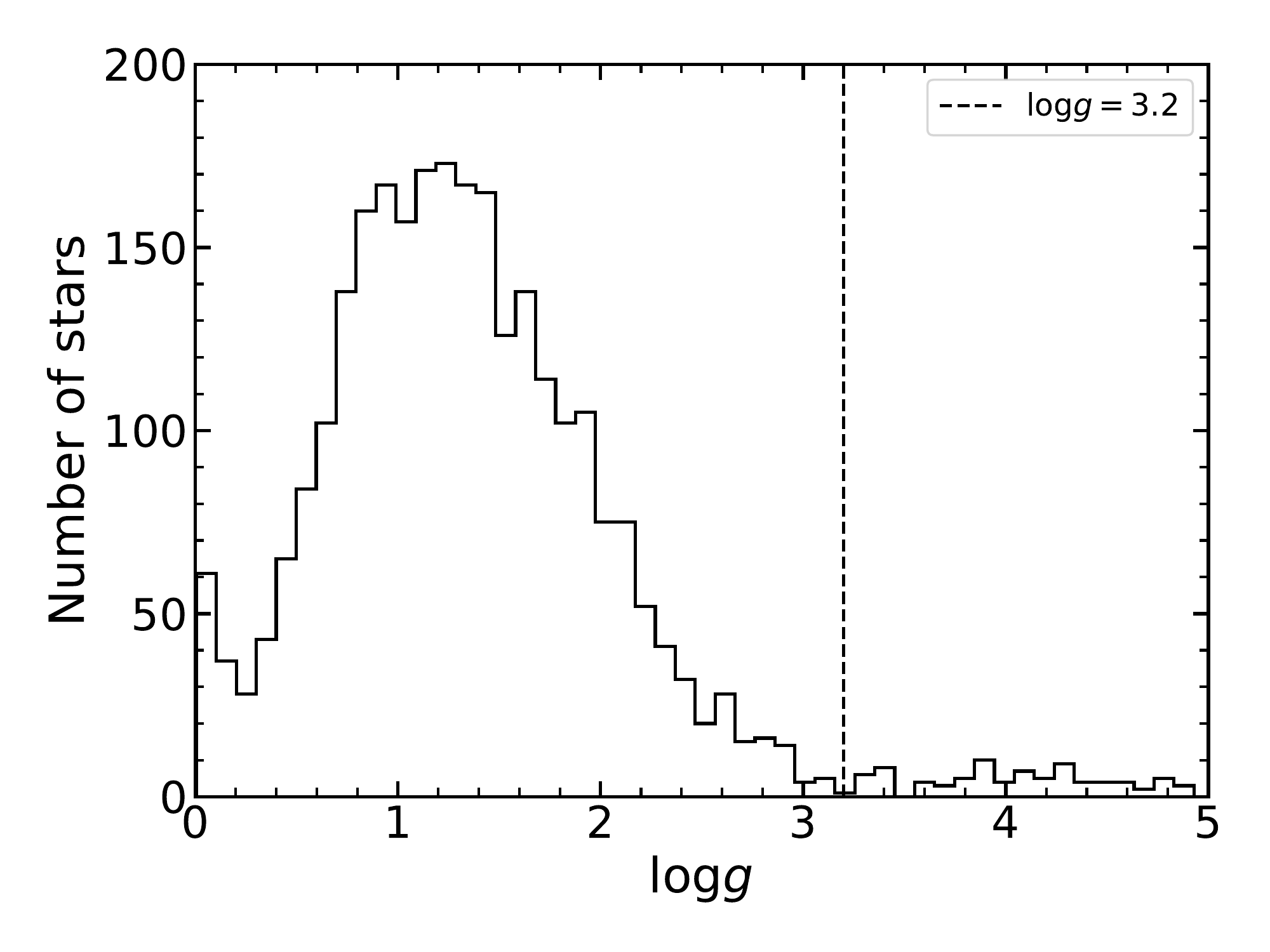}
   \includegraphics[width=0.47\textwidth]{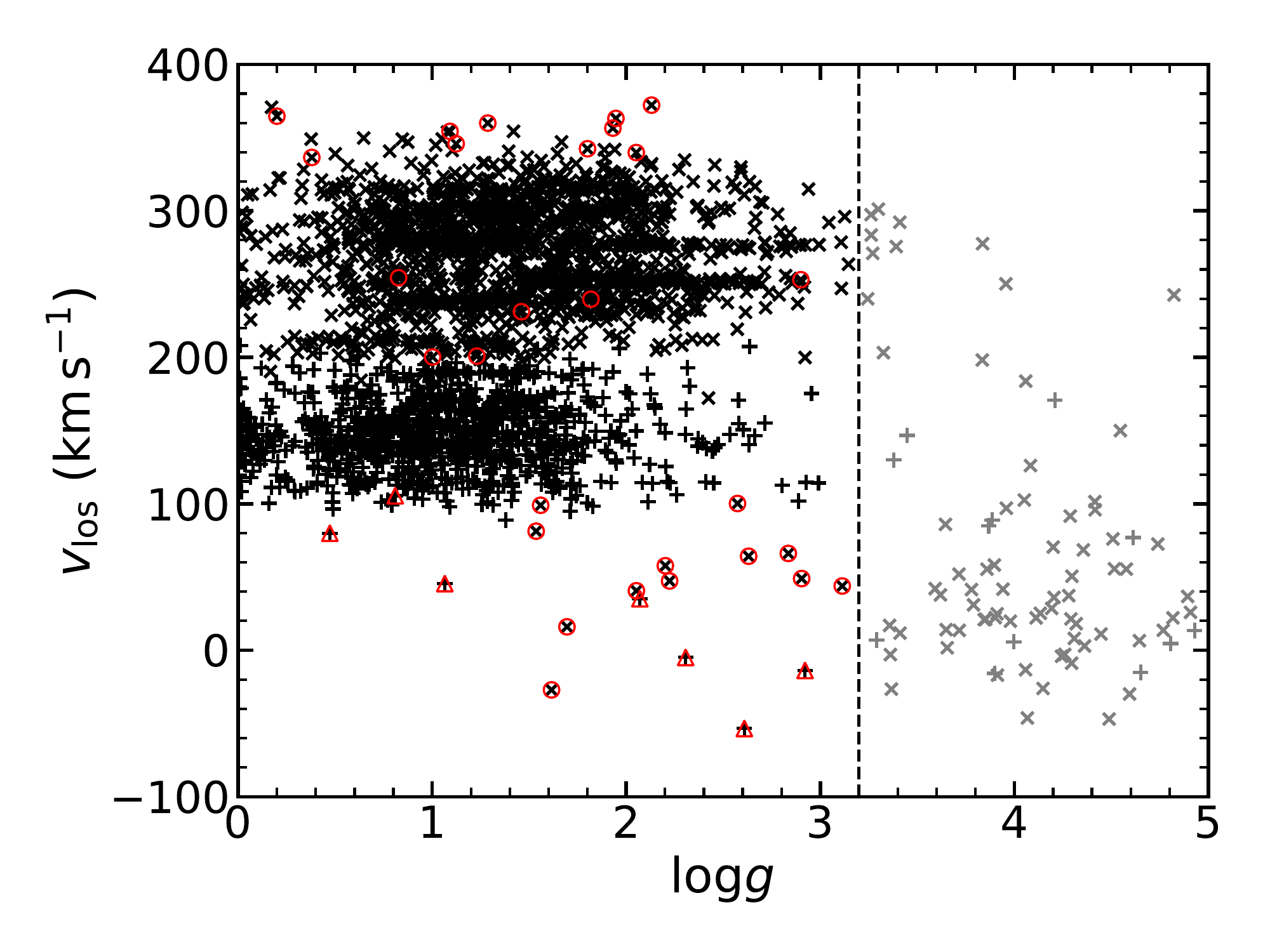}
   \caption{(Top panel) A histogram of the surface gravity for 2764 stars in our M2FS sample after removing the stars with either poorly determined parameters (see \autoref{sec:rm_PSK}) or anomalous velocity uncertainties (see \autoref{sec:rm_odd}). The vertical dashed line corresponds to an adopted empirical criterion, $\logg=3.2$, to divide giants and (foreground) dwarfs, as described in \autoref{sec:rm_dwarf}. The 84 stars to the right of dashed line were removed from our sample as likely foreground dwarfs.
   (Bottom panel) A plot on the \vlos-\logg\ plane of 2764 stars in our M2FS sample after removing the stars with either poorly determined parameters (see \autoref{sec:rm_PSK}) or anomalous velocity uncertainties (see \autoref{sec:rm_odd}). Crosses denote stars observed in LMC fields, while plus signs refer to stars observed in SMC fields. The dashed vertical line correspond to the criterion, $\log{g}=3.2$, used in \autoref{sec:rm_dwarf} to separate a total of 84 (foreground) dwarfs (gray crosses and plus signs) from giants (black crosses and plus signs) in our sample. Red circles (LMC) and red triangles (SMC) denote the 35 rejected stars that we concluded are associated with a third population as discussed in \autoref{sec:rm_vlos}.}
   \label{fig:hist_logg}
\end{figure}

Surface gravity is one of the parameters returned by the Bayesian analysis (\autoref{sec:spec_fit}) and this can be used to identity foreground dwarf stars in our sample.  As seen in \autoref{fig:hist_logg}, there is a clear tail in the distribution of $\log{g}$ extending to larger values. We have adopted $\log{g} \geq 3.2$ as the dividing value between giants (lower $\log{g}$) and dwarfs.   A total of 84 stars in our sample were removed using this criterion and are flagged in column 17 of \autoref{tab:sample_all}. Representative spectra of these stars are illustrated in Appendix~\ref{sec:spec_dwarf}.

At this stage, our Bayesian spectral-fitting results have allowed us to produce  a well-defined M2FS sample consisting of 2680 spectra of (mostly) RGB stars in our 26 target clusters (see \autoref{tab:sample_all}). \autoref{fig:spec_sample} presents some representative spectra of normal stars in our sample along with their Bayesian spectral fits.

\label{sec:rm_dwarf}

\section{Cluster Membership Analysis}
\label{sec:mem}

We closely follow the expectation-maximization (EM) analysis used in \citetalias{Song:2019aa} to determine simultaneously the cluster dynamics and the cluster membership probabilities of all stars in our samples. 
For most of the clusters, we assume in the EM analysis that there are two stellar populations in each kinematic cluster sample: one corresponding to the cluster, and one corresponding to the LMC/SMC field stars. 
Following \citetalias{Song:2019aa}, we still adopt single-mass K66 models\footnote{The K66 models are realized through \texttt{LIMEPY} \citep{Gieles:2015aa} as described in \citetalias{Song:2019aa}.} (as listed in \autoref{tab:structure}) to generate the projected velocity dispersion profile of each cluster, while the field population is represented by a Gaussian distribution.
In our EM analysis for two old LMC clusters, NGC~1466 and NGC~1841, however, we only assume a single population (also in the form of the K66 model) because both clusters are so distant from their parent galaxy (the LMC) that they lack significant numbers of LMC field stars. 

 An issue that arises at this stage is that most clusters contain stars that are either associated with a third population not included in our model (see equation 6 of \citetalias{Song:2019aa}), or stars that have metallicities that deviate significantly enough from their respective cluster mean metallicity to call in question their membership (keeping in mind that metallicity is not used directly in assigning membership probabilities in our EM analysis; see \citetalias{Song:2019aa}).  We describe here how we have used preliminary EM analysis results and metallicity estimates from the Bayesian spectral fitting (\autoref{sec:spec_fit}) to develop a final sample for kinematic analysis.  We also describe how we have supplemented this final sample with results from previous studies.

\subsection{Removing Stars From a Third Stellar Population}
\label{sec:rm_vlos}

Stars from a stellar population not explicitly included in our EM model can distort the estimation of the field population and hence the membership probability of other stars in the sample. We flagged such stars on a cluster-by-cluster basis using the following criterion:
\begin{equation}
\label{eq:rm_vlos}
\frac{\Delta\vlos}{ {\sigma}} \equiv \frac{ \left|v_{{\rm los},i}-\overline{\vlos}\right| }{ \sqrt{\varepsilon_{\vlos, i}^2+\sigma_{\vlos}^2}}\geq 3,
\end{equation}
where $v_{{\rm los},i}$ and $\varepsilon_{\vlos, i}$ are the LOS velocity and its associated uncertainty of the $i$-th star in a cluster sample, respectively, while $\overline{\vlos}$ and $\sigma_{\vlos}^2$ are weighted average velocity and the corresponding weighted standard deviation of the corresponding cluster sample.  Using this process, a total of 35 stars in 17 cluster samples were identified as third-population contaminants (bottom panel in \autoref{fig:hist_logg}); all are flagged in column 17 of \autoref{tab:sample_all}.  The generally low surface gravities of these rejected stars suggests they are likely Galactic halo giants or giants associated with an extended halo population of the MCs, neither of which were explicitly accounted for in our EM model (see \autoref{sec:mem} and \citetalias{Song:2019aa}).

Using the remaining sample of 2645 stars, we reran the EM analysis for every cluster to assign a preliminary membership probabilities, $P_{M,i}$, to each star.

\subsection{Removing Stars With Anomalous Metallicities}
\label{sec:rm_FeH}

\begin{table}
 \caption{Mean Cluster Metallicity on the CG97 Scale.}
 \label{tab:FeH}
 \begin{center}
 \begin{threeparttable}
 \begin{tabular}{clcccc}
\hline
    \multicolumn{1}{c}{Galaxy}  &  
    \multicolumn{1}{c}{Cluster}  &  
    \multicolumn{1}{c}{$N_{\rm rem}$}  &
    \multicolumn{1}{c}{$N_{\rm [Fe/H]}$}  &  
    \multicolumn{1}{c}{${\rm [Fe/H]}$}  &  
    \multicolumn{1}{c}{$\sigma_{{\rm [Fe/H]}}$}  \\
    \multicolumn{1}{c}{} &  
    \multicolumn{1}{c}{} &  
    \multicolumn{1}{c}{} &  
    \multicolumn{1}{c}{} &
    \multicolumn{1}{c}{$\rm (dex)$}  &  
    \multicolumn{1}{c}{$\rm (dex)$}  \\
    \multicolumn{1}{c}{(1)} & 
    \multicolumn{1}{c}{(2)} & 
    \multicolumn{1}{c}{(3)} & 
    \multicolumn{1}{c}{(4)} & 
    \multicolumn{1}{c}{(5)} &
    \multicolumn{1}{c}{(6)} \\
\hline 
SMC & Kron~3 & 1 & 41 & -0.96 & 0.15\\ 
SMC & Lindsay~1 & 6 & 80 & -0.98 & 0.13\\ 
SMC & NGC~152 & 2 & 22 & -0.73 & 0.11\\ 
SMC & NGC~330 & 1 & 4 & -0.65 & 0.10\\ 
SMC & NGC~339 & 8 & 35 & -1.01 & 0.17\\ 
SMC & NGC~361 & 0 & 20 & -0.75 & 0.17\\ 
SMC & NGC~411 & 2 & 19 & -0.66 & 0.09\\ 
SMC & NGC~416 & 0 & 16 & -0.80 & 0.17\\ 
SMC & NGC~419 & 2 & 35 & -0.66 & 0.15\\ 
SMC & NGC~458 & 0 & 14 & -0.70 & 0.20\\ 
LMC & Hodge~4 & 2 & 30 & -0.49 & 0.12\\ 
LMC & NGC~1466 & 1 & 22 & -1.40 & 0.16\\ 
LMC & NGC~1751 & 1 & 20 & -0.46 & 0.14\\ 
LMC & NGC~1783 & 2 & 52 & -0.54 & 0.10\\ 
LMC & NGC~1806 & 1 & 27 & -0.53 & 0.12\\ 
LMC & NGC~1831 & 3 & 49 & -0.41 & 0.15\\ 
LMC & NGC~1841 & 3 & 64 & -1.96 & 0.12\\ 
LMC & NGC~1846 & 2 & 36 & -0.49 & 0.08\\ 
LMC & NGC~1850 & 0 & 26 & -0.31 & 0.20\\ 
LMC & NGC~1978 & 3 & 37 & -0.49 & 0.10\\ 
LMC & NGC~2121 & 3 & 38 & -0.54 & 0.11\\ 
LMC & NGC~2155 & 3 & 36 & -0.59 & 0.12\\ 
LMC & NGC~2203 & 2 & 64 & -0.45 & 0.12\\ 
LMC & NGC~2209 & 1 & 50 & -0.52 & 0.15\\ 
LMC & NGC~2257 & 1 & 57 & -1.64 & 0.11\\ 
LMC & SL~663 & 1 & 19 & -0.51 & 0.11\\ 
\hline
\end{tabular}
\begin{tablenotes}
\item 
\end{tablenotes}
\end{threeparttable}
\end{center}
\end{table}

We can also use metallicities obtained from the Bayesian analysis (see \autoref{sec:spec_fit}) to identify stars with chemical abundances that differ significantly from the mean of their respective cluster's $\rm [Fe/H]$ distributions. Such deviant metallicities may indicate non-membership, but they could also result from issues such as poor background subtraction or poor temperature assignments. 

We applied two criteria to identify potential metallicity non-members.  First, the preliminary membership probabilities had to satisfy the condition $P_{M,i} \geq 0.5$. Second, the color shifts, $\Delta{(G_{\rm BP}-G_{\rm RP})}$ to assign effective temperature to each star (see \autoref{sec:Teff}) had to satisfy the condition $|\Delta{(G_{\rm BP}-G_{\rm RP})}-C_{\rm BI, dcolor}|\leq2S_{\rm BI, dcolor}$, where $C_{\rm BI, dcolor}$ and $S_{\rm BI, dcolor}$ are the biweight mean offset and dispersion scale determined in \autoref{sec:Teff}, respectively. This criterion was designed to avoid metallicity offsets due to systematically invalid temperature estimates (see \autoref{sec:Teff} and \autoref{fig:dcolor} for details.  Stars identified in this manner as likely metallicity outliers are flagged in column 17 of \autoref{tab:sample_all}. 

Using all stars satisfying these two criteria, we then calculated for each cluster an initial pair of metallicity parameters---the weighted mean metallicity ($\rm \overline{[Fe/H]}_{SC}$) and weighted standard deviation ($\sigma_{\rm [Fe/H],SC}$). Following this process, probable metallicity non-members were identified using the criterion
\begin{equation}
\label{eq:rm_FeH}
\frac{\Delta{\rm [Fe/H]}}{ {\sigma}} \equiv \frac{ \left|{\rm [Fe/H]}_{i}-{\rm \overline{[Fe/H]}_{SC}}\right| }{ \sqrt{\varepsilon_{{\rm [Fe/H]}, i}^2+\sigma_{\rm [Fe/H],SC}^2}}\geq 2,
\end{equation}
where ${\rm [Fe/H]}_{i}$ and $\varepsilon_{{\rm [Fe/H]},i}$ are the metallicity and its uncertainty of the $i$-th star in the EM sample, respectively. 
This procedure resulted in the removal of 51 stars from 22 clusters, and each is flagged in column 17 of \autoref{tab:sample_all}. The number of stars that satisfy \autoref{eq:rm_FeH} (i.e., metallicity non-members) are listed for each cluster in column 3 of \autoref{tab:FeH}, while the total number that do not satisfy this criterion are listed in column 4. The metallicities listed in columns 5--6 of \autoref{tab:FeH} are determined using methods describe below in \autoref{sec:FeH}. After applying this metallicity criterion, a total of 2594 stars in 26 clusters remain in our M2FS kinematic sample.

\subsection{Combining Previous Samples}
\label{sec:combine_samples}

We have identified all previously published individual stellar velocity measurements with a typical single-star precision better than 3~\kms\ (see also \autoref{tab:archive}) associated with the clusters in our sample. 
In this section, we identify and compare stars common to our sample and samples from previous studies with the aim of expanding our final kinematic sample.

We first compared published coordinates of individual stars with those of stars associated with our cluster sample.  Stars with coordinates agreeing to within a separation of 1 arcsec or less were considered to be the same star. 
There were no ambiguous cases in which more than one star from previously published catalogue satisfied this positional matching tolerance.
For NGC~1978, we also cross-matched the \citet{Fischer:1992ab} and \citet{Ferraro:2006aa} samples to identify any stars in common. These two samples were found to have nine stars in common, three of which are also found in our M2FS sample. 

Before combining results from different samples, we measured the systemic velocity offsets for clusters in common among the various datasets. These offsets were computed by subtracting the published systemic velocities (see column 6 in \autoref{tab:archive}) from our preliminary EM results (see \autoref{sec:rm_vlos}), and the uncertainties in the offsets were estimated by using the sum of errors in quadrature. When combining results from different samples, we then applied the offsets to the previously published velocities while keeping the published velocity uncertainties unchanged.  In all cases, we adopted the M2FS velocity zero points for the velocity scale (\autoref{tab:EM}). For each star with multiple measurements, the weighted mean velocities and their weighted errors were calculated after applying these offsets. This process allowed us to increase the sample sizes for seven clusters (see \autoref{tab:archive}).

Most of the stars matched by position as described above agree well in velocity.
However, two stars have velocity differences 3 times greater than their combined total velocity errors. The star in NGC~1846, whose detail had been discussed in Section 3.1.2 in \citetalias{Song:2019aa}, is a likely binary.  The other star, in NGC~330, is a well-known binary system identified in earlier studies \citep[see e.g.][and the references therein]{Patrick:2020aa}. Both stars were removed from our final kinematic sample.  For the NGC~1846 binary, there are too few measurements to determine a reliable systemic velocity, while for the NGC~330 binary, its mean velocity and metallicity, though both somewhat poorly determined, make it a likely non-member of the cluster.  Indeed, for this reason, this star had already been flagged for removal in \autoref{sec:rm_FeH}.

\subsection{Final Kinematic Sample}
\label{sec:final_sample}

After the addition of 193 more stars from the literature, our final working sample consists of 2787 stars in 26 clusters. We will refer to this as the `Kinematic' sample, and use it exclusively for the final EM analysis following the procedures described in \citetalias{Song:2019aa} and \autoref{sec:mem}.  

A summary of the initial analysis of the full Kinematic sample is provided in \autoref{tab:EM}: Column 3 lists the total number of stars in our Kinematic sample for each cluster, while column 4 lists the number of stars within each cluster's tidal radius (see \autoref{tab:structure}); columns 5 and 6 list the estimates of the cluster systemic velocity and projected central velocity dispersion of each cluster; columns 7 and 8 list the mean velocity and velocity dispersion of the field population when applicable.

\section{Dynamical and Chemical Results}
\label{sec:results}

\subsection{Cluster Systemic Velocity and Velocity Dispersion}
\label{sec:vd}

\begin{figure*}
   \centering
   \includegraphics[width=0.47\textwidth]{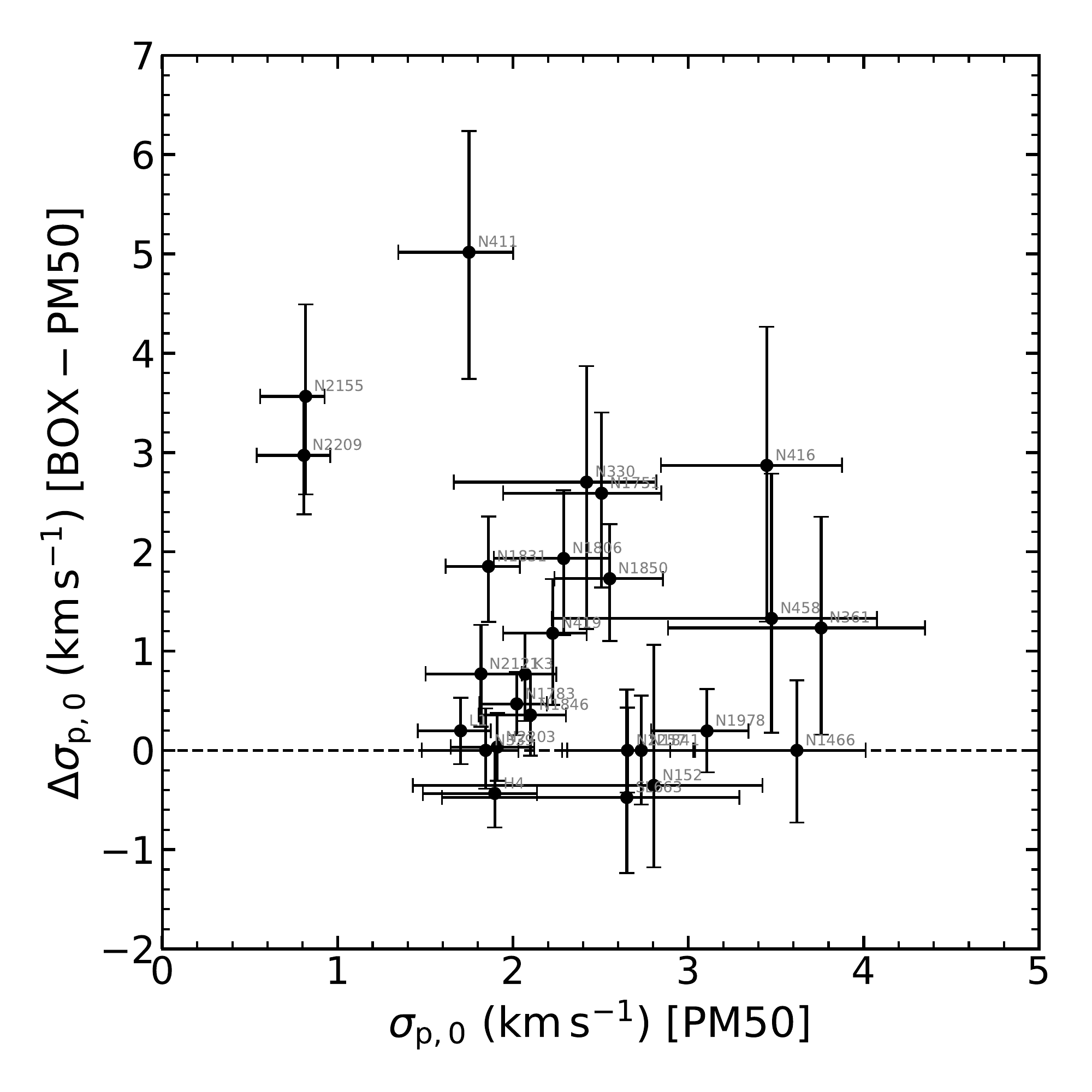}
   \includegraphics[width=0.47\textwidth]{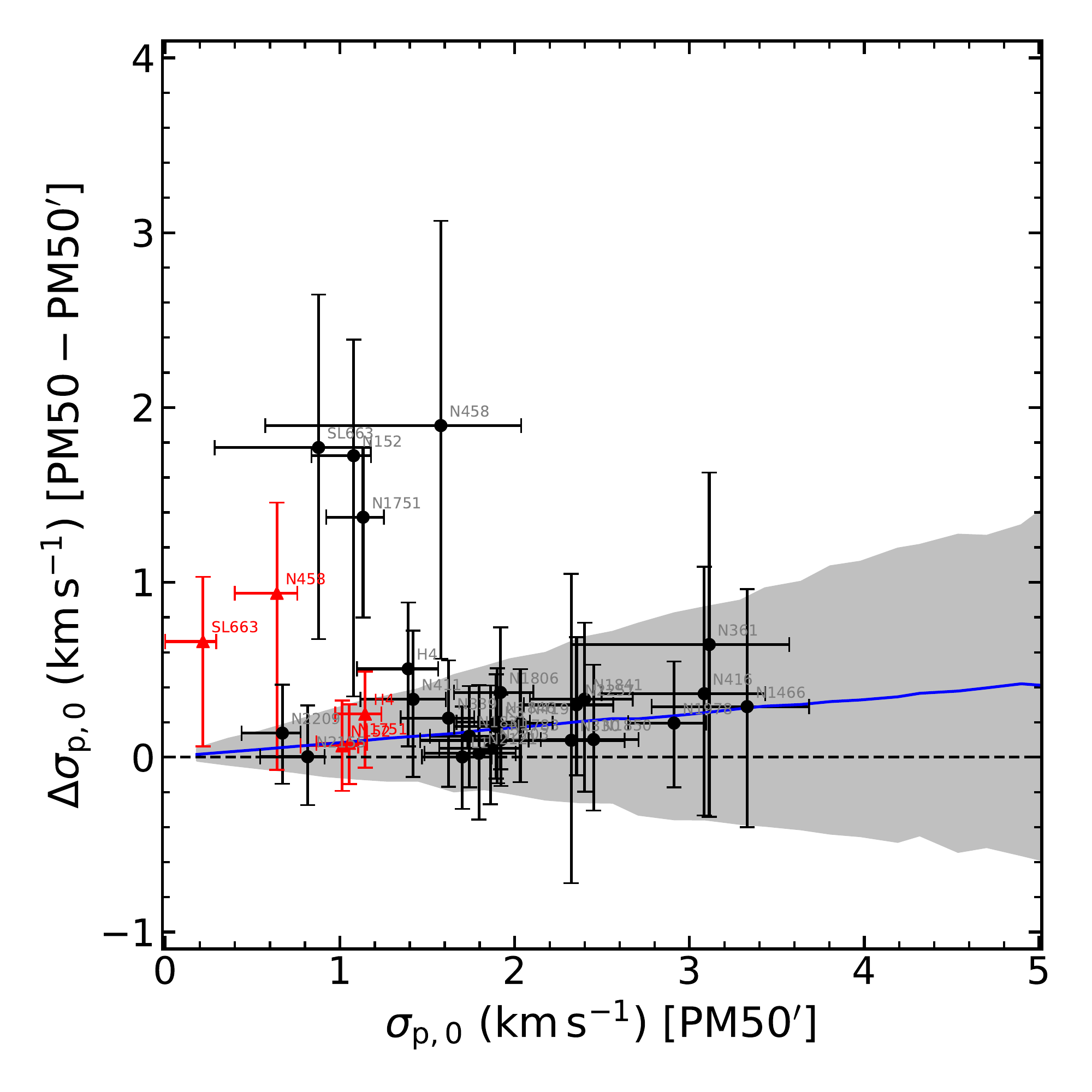}
   \caption{Comparisons of estimated $\sigma_{\rm p,0}$ between the BOX and PM50 methods (left panel), and between the PM50 and PM50$^\prime$ methods (right panel). The $\sigma_{\rm p,0}$ values used in the plots are listed in \autoref{tab:PM50}. In the right panel, the gray band denotes the 68\% ($\pm 1\sigma$) range of 1000 simulated results obtained by removing the most extreme star from the tail of Gaussian distributions consisting of 30 velocities with initial $\sigma_{\rm p,0}$ value ranging between 0.2--8.0~\kms\ in 0.2~\kms\ intervals. The blue line denotes the median results of these simulations. The red triangles denote the results of replicating the process of producing the PM50$^\prime$ sample to the five cases (NGC~152, NGC~458, NGC~1751, SL~663 and Hodge~4) whose black points lie outside the gray band.  These results imply that the PM50$^\prime$ samples for NGC~458 and SL~663 are not strictly consistent with Gaussian distributions while the PM50$^\prime$ samples for the other three are consistent. Full details are discussed in \autoref{sec:vd}.}
   \label{fig:comp_vd0_BOX_PM50}
\end{figure*}

The systemic velocity ($V_{\rm sys}$) and projected central velocity dispersion ($\sigma_{\rm p,0}$) listed in \autoref{tab:EM} are the two direct kinematic parameters determined by the EM analysis for each cluster. As discussed in \citetalias{Song:2019aa}, the `raw' EM estimates for these parameters rely critically on the membership probabilities, some of which may be skewed due to field contamination that may result from the relatively small samples being used.  Here we follow and expand upon our approach in \citetalias{Song:2019aa}, by calculating $V_{\rm sys}$ and $\sigma_{\rm p,0}$ using two different sampling methods, `BOX' and `PM50'.  Both use EM results---systemic velocity in the case of BOX, modified membership probabilities in the case of PM50---to refine the cluster kinematic parameters.  The value of these methods lies in their utility to better understand the effect that marginal outlier stars may have in the final kinematic results for a given cluster.

In the BOX method, $V_{\rm sys}$ and $\sigma_{\rm p,0}$ are computed from all stars in a rectangular area in the \vlos-$r$ plane (see the shaded area in Figure 9 of \citetalias{Song:2019aa}). This BOX area is bounded by the cluster center and the adopted K66 tidal radius (see \autoref{tab:structure}) in $x$-axis, while in $y$-axis the minimum and maximum systemic velocities are set by conservatively assuming the target cluster has a (large) total mass of $10^6$~$\rm M_{\odot}$. The value of the BOX method is that it is comparatively easy to specify and compute at the cost of effectively ignoring the EM analysis membership probabilities.

In the PM50 method, $V_{\rm sys}$ and $\sigma_{\rm p,0}$ are computed from probable member stars based on their EM membership probabilities (see \autoref{sec:final_sample}). A star is assumed to be a cluster member if its initial EM membership probability is greater than or equal to 50\% (i.e. $P_{M}\geq0.5$).
Unlike the full EM analysis in which membership probabilities are used to weight individual stars, the PM50 method implicitly assigns 100\% membership probabilities (therefore assumed to be certain members) to any stars with $P_{M}\geq0.5$, and assigns certain non-membership (0\% membership probability) to the rest stars in the sample. 

\autoref{tab:PM50} lists the BOX and PM50 results for all clusters in our sample.  We list the numbers of members assigned by each method, along with the corresponding values of $V_{\rm sys}$ and $\sigma_{\rm p,0}$ (columns 3--5 for BOX; columns 6--8 for PM50).
The left panel of \autoref{fig:comp_vd0_BOX_PM50} compares $\sigma_{\rm p,0}$ estimated from the BOX and PM50 methods. One obvious and unsurprising conclusion from these figures is that the BOX method systematically overestimates the dispersion compared to PM50.  Since the BOX method has no provision for flagging nonmembers, it is more likely to include random outlier stars whose inclusion depends on how the box areas are (arbitrarily) assigned.   

To explore how marginal outlier stars may affect the PM50 method, we have defined a third approach denoted PM50$^\prime$. For this method, we excluded one PM50 star exhibiting the most deviant velocity (in absolute value) from the PM50 systemic velocity, and re-applied the EM analysis to the remaining full kinematic sample of the cluster. Again, stars are taken to be cluster members if their re-determined EM membership probabilities are greater or equal to 50\%, and only such stars are used to compute $V_{\rm sys}$ and $\sigma_{\rm p,0}$ for the PM50$^\prime$ sample; these results are listed in columns 9--11 of \autoref{tab:PM50}. Note that the numbers of stars used in the PM50 and PM50$^\prime$ samples of a given cluster do not always differ by one, because when the most probable outlier is removed from the PM50 sample, the EM analysis reassigns membership probabilities that may reassign more than one star as members or nonmembers.

We compare the PM50 and PM50$^\prime$ results in the right panel of \autoref{fig:comp_vd0_BOX_PM50}.  For most clusters, the change in $\sigma_{\rm p,0}$ is within 1-$\sigma$ to the expectation of removing the most deviant star from a Gaussian distribution (gray bands in the right panel of \autoref{fig:comp_vd0_BOX_PM50}). We conclude that for most cases, the adopted PM50 distributions are consistent with being Gaussian and with the EM membership assignments for individual stars in these distributions.

However, for five clusters---NGC~152, NGC~458, NGC~1751, SL~663 and, marginally, Hodge~4---we find the PM50 results to be overly sensitive to the removal of a single (outlier) star in the EM analysis. These clusters also exhibit a large change in the number of likely members from their respective PM50 samples as a result of removing one extreme star (see \autoref{tab:PM50}). For these clusters, we have chosen to adopt their PM50$^\prime$ velocity dispersions from \autoref{tab:PM50}.

To test the internal consistency of adopting the PM50$^\prime$ results, we  conducted the same test as above for the five PM50$^\prime$ clusters.  The results are shown in the right panel of \autoref{fig:comp_vd0_BOX_PM50}.  We find that the PM50$^\prime$ samples for three of these clusters---NGC~152, NGC~1751 and Hodge~4---are now consistent with being Gaussian given their updated EM membership assignments.  Two clusters, SL~663 and NGC~458, remain anomalous.  Rather than iterate further on these clusters' distributions, we conclude that these two clusters' samples may have issues that affect our EM analyses of the systems.  We note that NGC~458 has the smallest stellar sample of all the clusters in our study, and that the structural parameters for SL~663 are particularly uncertain; both factors may lead the EM process astray.  This suggests that even the PM50$^{\prime}$ results for these two clusters may be systematically suspect.

The preferred velocity dispersion and number of cluster members for each cluster---based on either the PM50 or PM50$^\prime$ results---are highlighted with bold font in \autoref{tab:PM50}.

\begin{table*}
 \caption{Results of the EM analysis}
 \label{tab:EM}
 \begin{center}
 \begin{threeparttable}
 \begin{tabular}{clrrcccc}
\hline
    \multicolumn{1}{c}{Galaxy}  &  
    \multicolumn{1}{c}{Cluster}  &  
    \multicolumn{1}{c}{$N_{\rm total}$} &  
    \multicolumn{1}{c}{$N_{\rm tidal}$} & 
    \multicolumn{1}{c}{ $V_{\rm sys}$}  &  
    \multicolumn{1}{c}{$\sigma_{\rm p,\,0}$}  &  
    \multicolumn{1}{c}{$V_{\rm field}$}  &  
    \multicolumn{1}{c}{$\sigma_{\rm field}$}  
    \\
    \multicolumn{1}{c}{} &  
    \multicolumn{1}{c}{} &  
    \multicolumn{1}{c}{} &  
    \multicolumn{1}{c}{} &  
    \multicolumn{1}{c}{$\rm (km\,s^{-1})$}  &  
    \multicolumn{1}{c}{$\rm (km\,s^{-1})$}  &  
    \multicolumn{1}{c}{$\rm (km\,s^{-1})$}  &  
    \multicolumn{1}{c}{$\rm (km\,s^{-1})$}  \\ 
    \multicolumn{1}{c}{(1)} & 
    \multicolumn{1}{c}{(2)} & 
    \multicolumn{1}{c}{(3)} & 
    \multicolumn{1}{c}{(4)} & 
    \multicolumn{1}{c}{(5)} & 
    \multicolumn{1}{c}{(6)} & 
    \multicolumn{1}{c}{(7)} &
    \multicolumn{1}{c}{(8)} \\
\hline 
SMC & Kron~3 & 116 & 84 & $132.7^{+0.3}_{-0.4}$ & $2.1^{+0.2}_{-0.3}$ & $143.3^{+3.2}_{-2.4}$ & $23.1^{+1.3}_{-1.5}$ \\ 
SMC & Lindsay~1 & 99 & 92 & $140.5^{+0.2}_{-0.2}$ & $1.8^{+0.2}_{-0.3}$ & $137.7^{+3.3}_{-4.4}$ & $16.1^{+1.7}_{-2.3}$ \\ 
SMC & NGC~152 & 107 & 86 & $172.4^{+0.5}_{-0.9}$ & $2.8^{+1.2}_{-1.8}$ & $150.7^{+4.0}_{-2.5}$ & $25.2^{+1.3}_{-2.4}$ \\ 
SMC & NGC~330 & 47 & 28 & $153.0^{+0.7}_{-0.7}$ & $2.4^{+0.3}_{-0.7}$ & $138.5^{+3.0}_{-3.3}$ & $16.0^{+1.7}_{-2.6}$ \\ 
SMC & NGC~339 & 94 & 65 & $112.9^{+0.4}_{-0.3}$ & $1.8^{+0.2}_{-0.3}$ & $161.1^{+2.2}_{-2.5}$ & $18.0^{+1.6}_{-2.3}$ \\ 
SMC & NGC~361 & 119 & 70 & $170.3^{+0.9}_{-0.9}$ & $4.0^{+0.7}_{-1.4}$ & $138.6^{+1.3}_{-2.8}$ & $20.1^{+1.0}_{-1.2}$ \\ 
SMC & NGC~411 & 117 & 56 & $163.8^{+4.5}_{-0.3}$ & $1.7^{+16.8}_{-0.6}$ & $147.7^{+1.4}_{-3.5}$ & $23.3^{+0.8}_{-2.2}$ \\ 
SMC & NGC~416 & 114 & 40 & $155.0^{+1.0}_{-0.5}$ & $3.4^{+0.6}_{-0.7}$ & $149.2^{+2.3}_{-2.9}$ & $27.2^{+1.5}_{-1.9}$ \\ 
SMC & NGC~419 & 110 & 81 & $189.9^{+0.3}_{-0.2}$ & $2.2^{+0.3}_{-0.3}$ & $159.8^{+2.7}_{-2.6}$ & $21.8^{+1.4}_{-1.7}$ \\ 
SMC & NGC~458 & 79 & 24 & $149.0^{+0.8}_{-0.9}$ & $3.5^{+0.7}_{-2.2}$ & $146.2^{+1.8}_{-1.1}$ & $13.3^{+1.5}_{-1.3}$ \\ 
LMC & Hodge~4 & 112 & 112 & $312.7^{+0.6}_{-1.3}$ & $2.1^{+5.1}_{-0.6}$ & $299.8^{+2.0}_{-5.6}$ & $21.8^{+3.5}_{-3.9}$ \\ 
LMC & NGC~1466 & 27 & 25 & $202.5^{+0.5}_{-0.5}$ & $3.6^{+0.4}_{-0.6}$ & ... & ... \\ 
LMC & NGC~1751 & 113 & 113 & $240.4^{+0.7}_{-0.6}$ & $2.5^{+0.6}_{-0.7}$ & $247.1^{+3.4}_{-2.7}$ & $27.4^{+3.0}_{-3.1}$ \\ 
LMC & NGC~1783 & 111 & 111 & $279.6^{+0.2}_{-0.2}$ & $2.0^{+0.2}_{-0.2}$ & $281.9^{+3.9}_{-3.5}$ & $24.5^{+2.3}_{-2.4}$ \\ 
LMC & NGC~1806 & 120 & 120 & $229.6^{+0.4}_{-0.4}$ & $2.4^{+0.4}_{-0.6}$ & $265.7^{+3.9}_{-3.0}$ & $27.0^{+1.7}_{-2.3}$ \\ 
LMC & NGC~1831 & 102 & 95 & $276.8^{+0.2}_{-0.2}$ & $1.9^{+0.3}_{-0.3}$ & $284.8^{+2.8}_{-2.5}$ & $17.5^{+1.9}_{-3.0}$ \\ 
LMC & NGC~1841 & 69 & 64 & $210.8^{+0.3}_{-0.3}$ & $2.7^{+0.3}_{-0.5}$ & ... & ... \\ 
LMC & NGC~1846 & 196 & 81 & $239.2^{+0.2}_{-0.3}$ & $2.1^{+0.3}_{-0.4}$ & $269.4^{+1.5}_{-1.8}$ & $25.0^{+1.3}_{-1.4}$ \\ 
LMC & NGC~1850 & 155 & 87 & $248.9^{+0.4}_{-0.5}$ & $2.5^{+1.8}_{-0.4}$ & $257.4^{+2.8}_{-2.4}$ & $23.6^{+1.7}_{-2.6}$ \\ 
LMC & NGC~1978 & 145 & 86 & $293.1^{+0.3}_{-0.3}$ & $3.1^{+0.3}_{-0.4}$ & $283.7^{+3.0}_{-2.2}$ & $25.9^{+2.2}_{-2.1}$ \\ 
LMC & NGC~2121 & 109 & 72 & $237.0^{+0.3}_{-0.2}$ & $1.8^{+0.1}_{-0.4}$ & $262.8^{+2.7}_{-3.1}$ & $21.5^{+2.0}_{-2.3}$ \\ 
LMC & NGC~2155 & 110 & 104 & $315.0^{+0.1}_{-0.2}$ & $0.8^{+0.1}_{-0.3}$ & $310.4^{+2.1}_{-1.8}$ & $20.7^{+1.3}_{-1.6}$ \\ 
LMC & NGC~2203 & 96 & 77 & $252.8^{+0.3}_{-0.2}$ & $1.9^{+0.2}_{-0.3}$ & $244.2^{+4.3}_{-4.8}$ & $23.0^{+4.9}_{-5.2}$ \\ 
LMC & NGC~2209 & 113 & 112 & $251.2^{+0.1}_{-0.4}$ & $0.8^{+1.2}_{-0.2}$ & $251.4^{+2.1}_{-1.6}$ & $16.5^{+1.3}_{-1.3}$ \\ 
LMC & NGC~2257 & 94 & 65 & $301.8^{+0.3}_{-0.4}$ & $2.7^{+0.2}_{-0.4}$ & $320.4^{+1.3}_{-2.5}$ & $12.8^{+1.6}_{-2.0}$ \\ 
LMC & SL~663 & 113 & 113 & $301.1^{+1.4}_{-1.2}$ & $3.2^{+6.4}_{-2.2}$ & $300.1^{+2.2}_{-2.7}$ & $19.5^{+2.5}_{-1.6}$ \\ 
\hline
\end{tabular}
\begin{tablenotes}
\item 
\end{tablenotes}
\end{threeparttable}
\end{center}
\end{table*}

\begin{table*}
 \caption{Results of the BOX, PM50 and PM50$'$ methods.$^{\rm a}$}
 \label{tab:PM50}
 \begin{center}
 \begin{threeparttable}
 \begin{tabular}{llcrrcrrcrr}
\hline
    \multicolumn{1}{c}{Galaxy}  &  
    \multicolumn{1}{c}{Field}  &  
    \multicolumn{1}{c}{$N_{\rm BOX}$} & 
    \multicolumn{1}{c}{ $V_{\rm sys,BOX}$}  &  
    \multicolumn{1}{c}{$\sigma_{\rm p,0,BOX}$}  &  
    \multicolumn{1}{c}{$N_{\rm PM50}$} &  
    \multicolumn{1}{c}{ $V_{\rm sys, PM50}$}  &  
    \multicolumn{1}{c}{$\sigma_{\rm p,0,PM50}$} &
    \multicolumn{1}{c}{$N_{\rm PM50'}$} &  
    \multicolumn{1}{c}{ $V_{\rm sys,PM50'}$}  &  
    \multicolumn{1}{c}{$\sigma_{\rm p,0,PM50'}$} 
    \\
    \multicolumn{1}{c}{} &  
    \multicolumn{1}{c}{} &  
    \multicolumn{1}{c}{} &  
    \multicolumn{1}{c}{$\rm (km\,s^{-1})$}  &  
    \multicolumn{1}{c}{$\rm (km\,s^{-1})$}  &  
    \multicolumn{1}{c}{} &  
    \multicolumn{1}{c}{$\rm (km\,s^{-1})$}  &  
    \multicolumn{1}{c}{$\rm (km\,s^{-1})$}  &
    \multicolumn{1}{c}{} &  
    \multicolumn{1}{c}{$\rm (km\,s^{-1})$}  &  
    \multicolumn{1}{c}{$\rm (km\,s^{-1})$}
    \\ 
    \multicolumn{1}{c}{(1)} & 
    \multicolumn{1}{c}{(2)} & 
    \multicolumn{1}{c}{(3)} & 
    \multicolumn{1}{c}{(4)} & 
    \multicolumn{1}{c}{(5)} & 
    \multicolumn{1}{c}{(6)} & 
    \multicolumn{1}{c}{(7)} &
    \multicolumn{1}{c}{(8)} &
    \multicolumn{1}{c}{(9)} & 
    \multicolumn{1}{c}{(10)} &
    \multicolumn{1}{c}{(11)} \\
\hline 
SMC & Kron~3 & 51 & $132.5^{+0.3}_{-0.3}$ & $2.8^{+0.3}_{-0.4}$ & \bf 41 & $\bf 132.7^{+0.2}_{-0.3}$ & $\bf 2.1^{+0.2}_{-0.3}$ & 39 & $132.9^{+0.3}_{-0.2}$ & $1.9^{+0.2}_{-0.2}$\\ 
SMC & Lindsay~1 & 83 & $140.5^{+0.2}_{-0.2}$ & $1.9^{+0.2}_{-0.3}$ & \bf 80 & $\bf 140.5^{+0.2}_{-0.2}$ & $\bf 1.7^{+0.2}_{-0.2}$ & 79 & $140.5^{+0.2}_{-0.2}$ & $1.7^{+0.2}_{-0.2}$\\ 
SMC & NGC~152 & 27 & $172.8^{+0.4}_{-0.4}$ & $2.4^{+0.3}_{-0.5}$ & 21 & $172.6^{+0.5}_{-0.5}$ & $2.8^{+0.6}_{-1.4}$ & \bf 17 & $\bf 172.6^{+0.3}_{-0.3}$ & $\bf 1.1^{+0.1}_{-0.2}$\\ 
SMC & NGC~330 & 18 & $152.2^{+0.9}_{-0.9}$ & $5.1^{+0.9}_{-1.4}$ & \bf 14 & $\bf 153.0^{+0.6}_{-0.7}$ & $\bf 2.4^{+0.4}_{-0.8}$ & 13 & $152.8^{+0.6}_{-0.7}$ & $2.3^{+0.3}_{-0.9}$\\ 
SMC & NGC~339 & 35 & $112.9^{+0.3}_{-0.3}$ & $1.8^{+0.2}_{-0.3}$ & \bf 35 & $\bf 112.9^{+0.3}_{-0.3}$ & $\bf 1.8^{+0.2}_{-0.4}$ & 34 & $113.0^{+0.3}_{-0.3}$ & $1.6^{+0.1}_{-0.3}$\\ 
SMC & NGC~361 & 24 & $170.4^{+0.7}_{-0.7}$ & $5.0^{+0.7}_{-0.9}$ & \bf 20 & $\bf 170.5^{+0.6}_{-0.6}$ & $\bf 3.8^{+0.6}_{-0.9}$ & 20 & $170.9^{+0.5}_{-0.5}$ & $3.1^{+0.5}_{-0.8}$\\ 
SMC & NGC~411 & 35 & $162.8^{+0.7}_{-0.7}$ & $6.8^{+1.2}_{-1.3}$ & \bf 22 & $\bf 163.8^{+0.3}_{-0.3}$ & $\bf 1.7^{+0.3}_{-0.4}$ & 20 & $164.0^{+0.2}_{-0.2}$ & $1.4^{+0.2}_{-0.3}$\\ 
SMC & NGC~416 & 25 & $155.6^{+0.8}_{-0.8}$ & $6.3^{+1.3}_{-1.5}$ & \bf 19 & $\bf 155.0^{+0.6}_{-0.6}$ & $\bf 3.4^{+0.4}_{-0.6}$ & 18 & $154.7^{+0.5}_{-0.6}$ & $3.1^{+0.3}_{-0.6}$\\ 
SMC & NGC~419 & 49 & $189.9^{+0.3}_{-0.3}$ & $3.4^{+0.5}_{-0.7}$ & \bf 44 & $\bf 189.9^{+0.2}_{-0.2}$ & $\bf 2.2^{+0.2}_{-0.3}$ & 43 & $189.8^{+0.2}_{-0.2}$ & $2.0^{+0.2}_{-0.2}$\\ 
SMC & NGC~458 & 19 & $148.8^{+0.7}_{-0.7}$ & $4.8^{+0.7}_{-1.0}$ & 14 & $149.0^{+0.7}_{-0.8}$ & $3.5^{+0.6}_{-1.3}$ & \bf 10 & $\bf 148.8^{+0.4}_{-0.4}$ & $\bf 1.6^{+0.5}_{-1.0}$\\ 
LMC & Hodge~4 & 33 & $313.0^{+0.3}_{-0.3}$ & $1.5^{+0.1}_{-0.2}$ & 33 & $312.8^{+0.4}_{-0.3}$ & $1.9^{+0.2}_{-0.4}$ & \bf 29 & $\bf 313.1^{+0.3}_{-0.3}$ & $\bf 1.4^{+0.2}_{-0.3}$\\ 
LMC & NGC~1466 & 25 & $202.5^{+0.5}_{-0.5}$ & $3.6^{+0.4}_{-0.6}$ & \bf 25 & $\bf 202.5^{+0.5}_{-0.5}$ & $\bf 3.6^{+0.4}_{-0.6}$ & 24 & $202.7^{+0.5}_{-0.5}$ & $3.3^{+0.4}_{-0.5}$\\ 
LMC & NGC~1751 & 46 & $240.3^{+0.5}_{-0.6}$ & $5.1^{+0.6}_{-0.9}$ & 28 & $240.3^{+0.4}_{-0.4}$ & $2.5^{+0.3}_{-0.6}$ & \bf 21 & $\bf 241.3^{+0.2}_{-0.2}$ & $\bf 1.1^{+0.1}_{-0.2}$\\ 
LMC & NGC~1783 & 75 & $279.5^{+0.2}_{-0.2}$ & $2.5^{+0.2}_{-0.3}$ & \bf 66 & $\bf 279.6^{+0.2}_{-0.2}$ & $\bf 2.0^{+0.2}_{-0.2}$ & 65 & $279.5^{+0.2}_{-0.2}$ & $1.9^{+0.2}_{-0.2}$\\ 
LMC & NGC~1806 & 43 & $230.0^{+0.5}_{-0.5}$ & $4.2^{+0.6}_{-0.7}$ & \bf 35 & $\bf 229.7^{+0.3}_{-0.3}$ & $\bf 2.3^{+0.3}_{-0.4}$ & 33 & $230.0^{+0.3}_{-0.3}$ & $1.9^{+0.2}_{-0.3}$\\ 
LMC & NGC~1831 & 77 & $276.4^{+0.3}_{-0.3}$ & $3.7^{+0.4}_{-0.5}$ & \bf 64 & $\bf 276.8^{+0.2}_{-0.2}$ & $\bf 1.9^{+0.2}_{-0.2}$ & 62 & $276.9^{+0.2}_{-0.2}$ & $1.7^{+0.2}_{-0.2}$\\ 
LMC & NGC~1841 & 64 & $210.8^{+0.3}_{-0.3}$ & $2.7^{+0.3}_{-0.5}$ & \bf 64 & $\bf 210.8^{+0.3}_{-0.3}$ & $\bf 2.7^{+0.3}_{-0.5}$ & 63 & $210.6^{+0.3}_{-0.3}$ & $2.4^{+0.3}_{-0.3}$\\ 
LMC & NGC~1846 & 55 & $239.2^{+0.3}_{-0.3}$ & $2.5^{+0.3}_{-0.4}$ & \bf 53 & $\bf 239.2^{+0.2}_{-0.2}$ & $\bf 2.1^{+0.2}_{-0.3}$ & 51 & $239.1^{+0.2}_{-0.2}$ & $1.9^{+0.2}_{-0.2}$\\ 
LMC & NGC~1850 & 74 & $249.0^{+0.4}_{-0.4}$ & $4.3^{+0.4}_{-0.5}$ & \bf 63 & $\bf 248.8^{+0.3}_{-0.3}$ & $\bf 2.6^{+0.3}_{-0.3}$ & 62 & $248.9^{+0.3}_{-0.3}$ & $2.5^{+0.3}_{-0.3}$\\ 
LMC & NGC~1978 & 76 & $293.0^{+0.3}_{-0.3}$ & $3.3^{+0.3}_{-0.3}$ & \bf 75 & $\bf 293.1^{+0.3}_{-0.3}$ & $\bf 3.1^{+0.2}_{-0.3}$ & 74 & $293.2^{+0.3}_{-0.3}$ & $2.9^{+0.2}_{-0.3}$\\ 
LMC & NGC~2121 & 49 & $236.9^{+0.3}_{-0.3}$ & $2.6^{+0.4}_{-0.5}$ & \bf 43 & $\bf 237.0^{+0.2}_{-0.2}$ & $\bf 1.8^{+0.2}_{-0.3}$ & 42 & $236.9^{+0.3}_{-0.2}$ & $1.8^{+0.2}_{-0.3}$\\ 
LMC & NGC~2155 & 57 & $314.9^{+0.4}_{-0.4}$ & $4.4^{+0.9}_{-1.0}$ & \bf 35 & $\bf 315.0^{+0.2}_{-0.1}$ & $\bf 0.8^{+0.1}_{-0.3}$ & 34 & $315.0^{+0.2}_{-0.2}$ & $0.8^{+0.1}_{-0.3}$\\ 
LMC & NGC~2203 & 73 & $252.9^{+0.2}_{-0.2}$ & $1.9^{+0.2}_{-0.3}$ & \bf 72 & $\bf 252.8^{+0.2}_{-0.2}$ & $\bf 1.9^{+0.2}_{-0.3}$ & 71 & $252.9^{+0.2}_{-0.2}$ & $1.9^{+0.2}_{-0.3}$\\ 
LMC & NGC~2209 & 74 & $250.9^{+0.3}_{-0.3}$ & $3.8^{+0.5}_{-0.6}$ & \bf 52 & $\bf 251.2^{+0.2}_{-0.2}$ & $\bf 0.8^{+0.2}_{-0.3}$ & 51 & $251.3^{+0.1}_{-0.1}$ & $0.7^{+0.1}_{-0.2}$\\ 
LMC & NGC~2257 & 63 & $301.8^{+0.3}_{-0.3}$ & $2.7^{+0.3}_{-0.3}$ & \bf 63 & $\bf 301.8^{+0.3}_{-0.3}$ & $\bf 2.7^{+0.2}_{-0.3}$ & 61 & $301.6^{+0.3}_{-0.3}$ & $2.4^{+0.2}_{-0.3}$\\ 
LMC & SL~663 & 32 & $301.0^{+0.4}_{-0.4}$ & $2.2^{+0.3}_{-0.4}$ & 23 & $301.2^{+0.5}_{-0.5}$ & $2.6^{+0.6}_{-1.1}$ & \bf 20 & $\bf 301.0^{+0.3}_{-0.2}$ & $\bf 0.9^{+0.3}_{-0.6}$\\ 
\hline
\end{tabular}
\begin{tablenotes}
\item $^{\rm a}$ {As discussed in \autoref{sec:vd}, the PM50$^{\prime}$ results for NGC~458 and SL~663 should be used with caution.}
\end{tablenotes}
\end{threeparttable}
\end{center}
\end{table*}

\subsection{Cluster Mass and Mass-to-light Ratio}
\label{sec:MLv}

\begin{table*}
\caption{Mass, Luminosity and $M/L_V$.$^{\rm a}$}
\label{tab:MLv}
\begin{threeparttable}
\begin{tabular}{clllllllr}
\hline
    \multicolumn{1}{c}{Galaxy} & 
    \multicolumn{1}{c}{Cluster} & 
    \multicolumn{1}{c}{Method} & 
    \multicolumn{1}{c}{${M_{\rm tot}}$} & 
    \multicolumn{1}{c}{${L_{V,\,\rm tot}}$} & 
    \multicolumn{1}{c}{$M/L_{\rm V}$} & 
    \multicolumn{1}{c}{$\log{M_{\rm tot}}$} & 
    \multicolumn{1}{c}{$\log{L_{V,\,\rm tot}}$} & 
    \multicolumn{1}{c}{$\log{M/L_{\rm V}}$} \\
    \multicolumn{1}{c}{} & 
    \multicolumn{1}{c}{} & 
    \multicolumn{1}{c}{} & 
    \multicolumn{1}{c}{$(\times 10^5\ {\rm M}_{\sun})$} & 
    \multicolumn{1}{c}{$(\times 10^5\ {\rm L}_{\sun})$} & 
    \multicolumn{1}{c}{$({\rm M}_{\sun}\ {\rm L}_{\sun}^{-1})$} & 
    \multicolumn{1}{c}{$({\rm M}_{\sun})$} & 
    \multicolumn{1}{c}{$({\rm L}_{\sun})$} & 
    \multicolumn{1}{c}{$({\rm M}_{\sun}\ {\rm L}_{\sun}^{-1})$} \\
    \multicolumn{1}{c}{(1)} & 
    \multicolumn{1}{c}{(2)} & 
    \multicolumn{1}{c}{(3)} & 
    \multicolumn{1}{c}{(4)} & 
    \multicolumn{1}{c}{(5)} & 
    \multicolumn{1}{c}{(6)} &
    \multicolumn{1}{c}{(7)} &
    \multicolumn{1}{c}{(8)} &
    \multicolumn{1}{c}{(9)} \\
\hline
SMC & Kron~3 & PM50 & $0.77^{+0.15}_{-0.20}$ & $1.41^{+0.18}_{-0.14}$ & $0.55^{+0.11}_{-0.14}$ & $4.89^{+0.08}_{-0.13}$ & $5.15^{+0.05}_{-0.05}$ & $-0.26^{+0.08}_{-0.13}$ \\ 
SMC & Lindsay~1 & PM50 & $0.76^{+0.16}_{-0.20}$ & $0.87^{+0.09}_{-0.08}$ & $0.88^{+0.20}_{-0.24}$ & $4.88^{+0.08}_{-0.13}$ & $4.94^{+0.04}_{-0.04}$ & $-0.05^{+0.09}_{-0.14}$ \\ 
SMC & NGC~152 & PM50 & $1.67^{+0.83}_{-1.27}$ & $0.80^{+0.12}_{-0.10}$ & $2.09^{+0.94}_{-1.60}$ & $5.22^{+0.18}_{-0.62}$ & $4.90^{+0.06}_{-0.06}$ & $0.32^{+0.16}_{-0.63}$ \\ 
 &  & PM50$^{\prime}$ & $0.25^{+0.07}_{-0.10}$ & $0.80^{+0.13}_{-0.09}$ & $0.31^{+0.06}_{-0.13}$ & $4.39^{+0.10}_{-0.22}$ & $4.90^{+0.07}_{-0.06}$ & $-0.51^{+0.08}_{-0.23}$ \\ 
SMC & NGC~330 & PM50 & $0.54^{+0.22}_{-0.30}$ & $8.93^{+1.05}_{-0.86}$ & $0.06^{+0.02}_{-0.03}$ & $4.74^{+0.15}_{-0.34}$ & $5.95^{+0.05}_{-0.04}$ & $-1.22^{+0.13}_{-0.33}$ \\ 
SMC & NGC~339 & PM50 & $0.57^{+0.23}_{-0.23}$ & $0.72^{+0.28}_{-0.16}$ & $0.79^{+0.15}_{-0.29}$ & $4.76^{+0.15}_{-0.23}$ & $4.86^{+0.14}_{-0.10}$ & $-0.10^{+0.07}_{-0.20}$ \\ 
SMC & NGC~361 & PM50 & $2.15^{+1.36}_{-0.98}$ & $1.04^{+0.54}_{-0.31}$ & $2.07^{+0.75}_{-0.85}$ & $5.33^{+0.21}_{-0.26}$ & $5.02^{+0.18}_{-0.15}$ & $0.32^{+0.13}_{-0.23}$ \\ 
SMC & NGC~411 & PM50 & $0.30^{+0.10}_{-0.13}$ & $0.80^{+0.10}_{-0.08}$ & $0.38^{+0.11}_{-0.16}$ & $4.48^{+0.12}_{-0.24}$ & $4.90^{+0.05}_{-0.04}$ & $-0.42^{+0.11}_{-0.23}$ \\ 
SMC & NGC~416 & PM50 & $0.80^{+0.24}_{-0.30}$ & $1.12^{+0.13}_{-0.10}$ & $0.72^{+0.17}_{-0.25}$ & $4.90^{+0.11}_{-0.20}$ & $5.05^{+0.05}_{-0.04}$ & $-0.15^{+0.09}_{-0.19}$ \\ 
SMC & NGC~419 & PM50 & $0.64^{+0.14}_{-0.15}$ & $3.49^{+0.62}_{-0.51}$ & $0.18^{+0.05}_{-0.05}$ & $4.80^{+0.09}_{-0.12}$ & $5.54^{+0.07}_{-0.07}$ & $-0.74^{+0.10}_{-0.13}$ \\ 
SMC & NGC~458 & PM50 & $1.26^{+0.58}_{-0.73}$ & $1.24^{+0.36}_{-0.26}$ & $1.02^{+0.44}_{-0.62}$ & $5.10^{+0.16}_{-0.37}$ & $5.09^{+0.11}_{-0.10}$ & $0.01^{+0.16}_{-0.40}$ \\ 
 &  & PM50$^{\prime}$ & $0.26^{+0.21}_{-0.22}$ & $1.24^{+0.35}_{-0.24}$ & $0.21^{+0.15}_{-0.18}$ & $4.42^{+0.26}_{-0.80}$ & $5.09^{+0.11}_{-0.09}$ & $-0.68^{+0.23}_{-0.80}$ \\ 
LMC & Hodge~4 & PM50 & $3.39^{+229.01}_{-2.49}$ & $2.67^{+202.43}_{-1.82}$ & $1.27^{+0.28}_{-0.54}$ & $5.53^{+1.84}_{-0.58}$ & $5.43^{+1.89}_{-0.50}$ & $0.10^{+0.09}_{-0.24}$ \\ 
 &  & PM50$^{\prime}$ & $1.82^{+125.83}_{-1.33}$ & $2.67^{+228.50}_{-1.81}$ & $0.68^{+0.17}_{-0.27}$ & $5.26^{+1.84}_{-0.57}$ & $5.43^{+1.94}_{-0.49}$ & $-0.17^{+0.09}_{-0.22}$ \\ 
LMC & NGC~1466 & PM50 & $1.02^{+0.27}_{-0.31}$ & $0.94^{+0.09}_{-0.07}$ & $1.09^{+0.25}_{-0.33}$ & $5.01^{+0.10}_{-0.16}$ & $4.97^{+0.04}_{-0.03}$ & $0.04^{+0.09}_{-0.15}$ \\ 
LMC & NGC~1751 & PM50 & $0.76^{+0.23}_{-0.31}$ & $0.90^{+0.14}_{-0.12}$ & $0.84^{+0.27}_{-0.34}$ & $4.88^{+0.11}_{-0.23}$ & $4.96^{+0.06}_{-0.06}$ & $-0.08^{+0.12}_{-0.23}$ \\ 
 &  & PM50$^{\prime}$ & $0.15^{+0.03}_{-0.05}$ & $0.90^{+0.13}_{-0.11}$ & $0.17^{+0.04}_{-0.06}$ & $4.19^{+0.09}_{-0.19}$ & $4.96^{+0.06}_{-0.06}$ & $-0.77^{+0.10}_{-0.19}$ \\ 
LMC & NGC~1783 & PM50 & $0.98^{+0.17}_{-0.20}$ & $3.77^{+0.28}_{-0.27}$ & $0.26^{+0.04}_{-0.05}$ & $4.99^{+0.07}_{-0.10}$ & $5.58^{+0.03}_{-0.03}$ & $-0.58^{+0.07}_{-0.09}$ \\ 
LMC & NGC~1806 & PM50 & $0.76^{+0.18}_{-0.26}$ & $1.42^{+0.10}_{-0.10}$ & $0.54^{+0.13}_{-0.18}$ & $4.88^{+0.09}_{-0.18}$ & $5.15^{+0.03}_{-0.03}$ & $-0.27^{+0.09}_{-0.18}$ \\ 
LMC & NGC~1831 & PM50 & $0.41^{+0.09}_{-0.11}$ & $1.33^{+0.18}_{-0.17}$ & $0.31^{+0.07}_{-0.08}$ & $4.61^{+0.09}_{-0.14}$ & $5.12^{+0.06}_{-0.06}$ & $-0.51^{+0.09}_{-0.14}$ \\ 
LMC & NGC~1841 & PM50 & $1.04^{+2.53}_{-0.25}$ & $0.74^{+1.63}_{-0.02}$ & $1.40^{+0.49}_{-0.51}$ & $5.02^{+0.54}_{-0.12}$ & $4.87^{+0.50}_{-0.01}$ & $0.14^{+0.13}_{-0.20}$ \\ 
LMC & NGC~1846 & PM50 & $0.57^{+0.13}_{-0.14}$ & $1.68^{+0.35}_{-0.30}$ & $0.34^{+0.11}_{-0.10}$ & $4.75^{+0.09}_{-0.12}$ & $5.23^{+0.08}_{-0.09}$ & $-0.47^{+0.13}_{-0.15}$ \\ 
LMC & NGC~1850 & PM50 & $0.52^{+0.05}_{-0.17}$ & $6.56^{+1.30}_{-1.07}$ & $0.08^{+0.01}_{-0.03}$ & $4.71^{+0.04}_{-0.18}$ & $5.82^{+0.08}_{-0.08}$ & $-1.10^{+0.06}_{-0.20}$ \\ 
LMC & NGC~1978 & PM50 & $1.36^{+0.24}_{-0.29}$ & $3.41^{+0.30}_{-0.25}$ & $0.40^{+0.06}_{-0.08}$ & $5.13^{+0.07}_{-0.11}$ & $5.53^{+0.04}_{-0.03}$ & $-0.40^{+0.06}_{-0.10}$ \\ 
LMC & NGC~2121 & PM50 & $0.50^{+0.22}_{-0.19}$ & $0.79^{+0.42}_{-0.26}$ & $0.63^{+0.20}_{-0.21}$ & $4.69^{+0.16}_{-0.21}$ & $4.90^{+0.19}_{-0.17}$ & $-0.20^{+0.12}_{-0.18}$ \\ 
LMC & NGC~2155 & PM50 & $0.06^{+0.03}_{-0.04}$ & $0.22^{+0.14}_{-0.08}$ & $0.29^{+0.17}_{-0.18}$ & $3.81^{+0.16}_{-0.36}$ & $4.35^{+0.22}_{-0.18}$ & $-0.54^{+0.20}_{-0.42}$ \\ 
LMC & NGC~2203 & PM50 & $0.51^{+0.13}_{-0.13}$ & $0.81^{+0.12}_{-0.11}$ & $0.63^{+0.19}_{-0.18}$ & $4.70^{+0.10}_{-0.13}$ & $4.91^{+0.06}_{-0.06}$ & $-0.20^{+0.11}_{-0.14}$ \\ 
LMC & NGC~2209 & PM50 & $0.09^{+0.02}_{-0.06}$ & $0.26^{+0.02}_{-0.02}$ & $0.36^{+0.08}_{-0.21}$ & $3.97^{+0.09}_{-0.40}$ & $4.41^{+0.03}_{-0.03}$ & $-0.45^{+0.09}_{-0.39}$ \\ 
LMC & NGC~2257 & PM50 & $1.01^{+0.68}_{-0.38}$ & $0.51^{+0.39}_{-0.17}$ & $2.00^{+0.38}_{-0.53}$ & $5.01^{+0.22}_{-0.20}$ & $4.71^{+0.25}_{-0.17}$ & $0.30^{+0.08}_{-0.13}$ \\ 
LMC & SL~663 & PM50 & $3.32^{+778.97}_{-2.65}$ & $0.45^{+132.55}_{-0.33}$ & $7.31^{+5.34}_{-4.81}$ & $5.52^{+2.37}_{-0.70}$ & $4.66^{+2.47}_{-0.55}$ & $0.86^{+0.24}_{-0.47}$ \\ 
 &  & PM50$^{\prime}$ & $0.36^{+69.09}_{-0.33}$ & $0.45^{+164.23}_{-0.36}$ & $0.80^{+0.71}_{-0.70}$ & $4.56^{+2.28}_{-1.03}$ & $4.66^{+2.56}_{-0.67}$ & $-0.10^{+0.27}_{-0.87}$ \\ 
\hline
\end{tabular}
\begin{tablenotes}
\item $^{\rm a}$ {As discussed in \autoref{sec:vd}, the PM50$^{\prime}$ results for NGC~458 and SL~663 should be used with caution.}
\end{tablenotes}
\end{threeparttable}
\end{table*}

\begin{figure*}
   \centering    
   \includegraphics[width=0.525\textwidth]{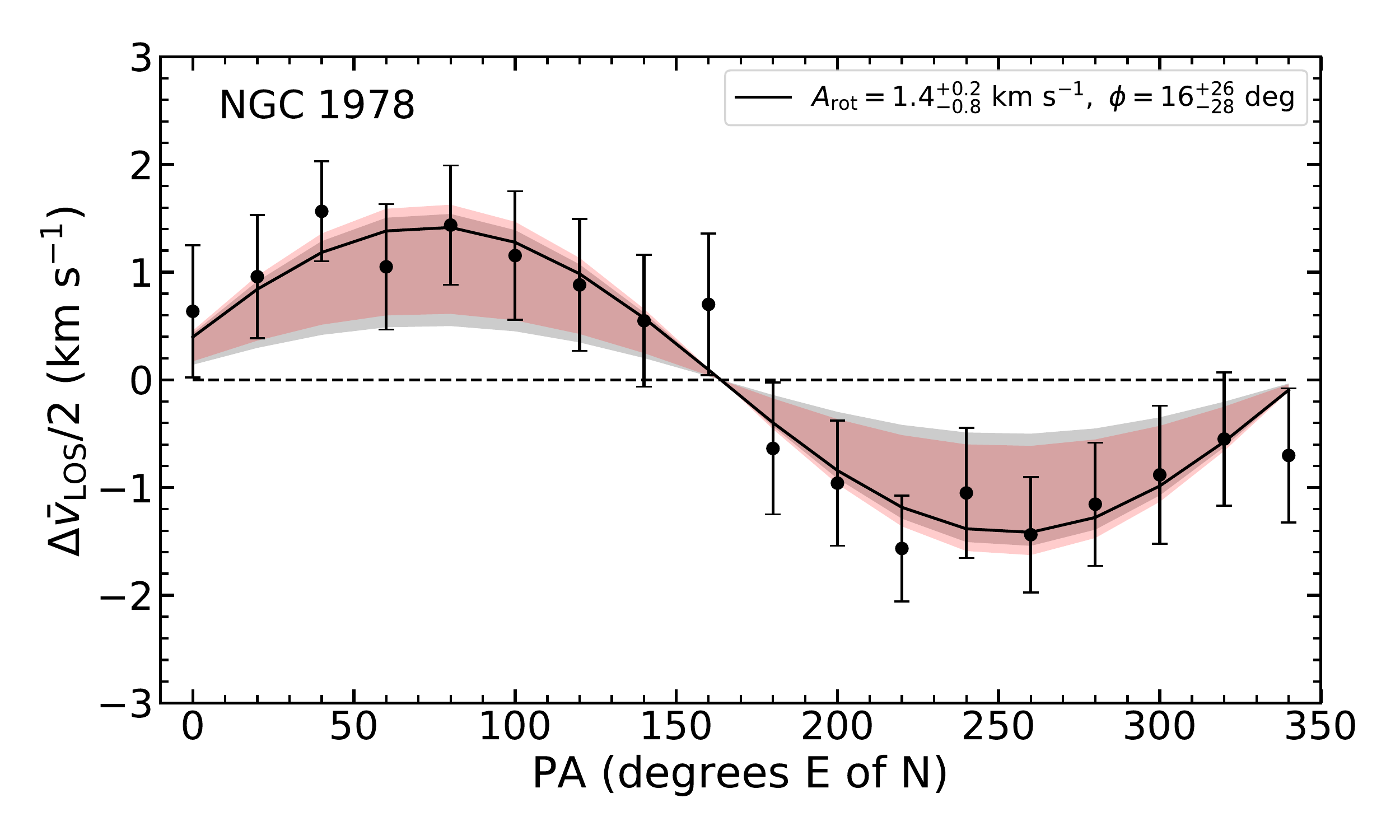}
   \includegraphics[width=0.415\textwidth]{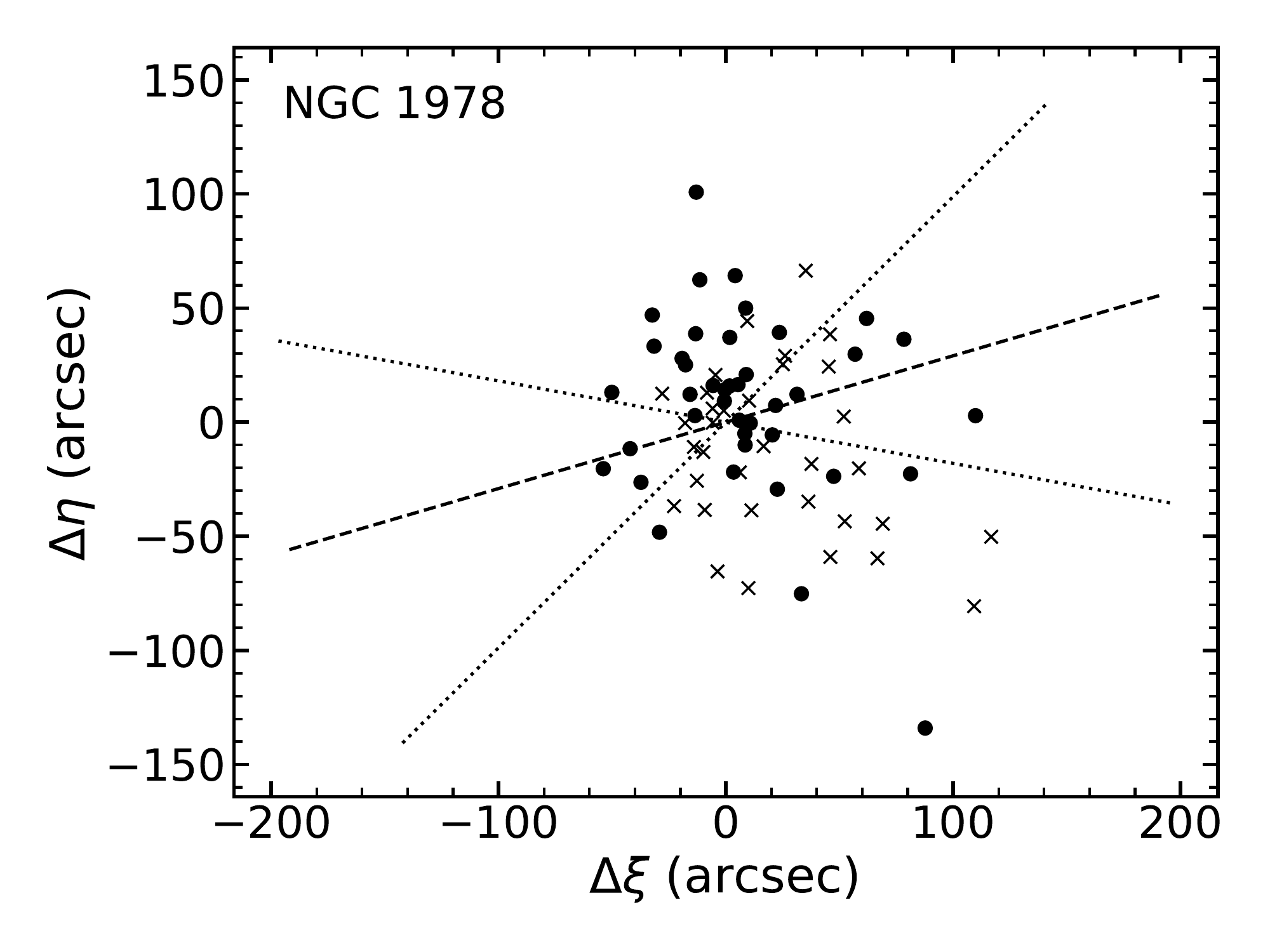}
   \caption{Simple rotation analysis for the stars with $P_{\rm M}\geq 0.5$ in NGC~1978 (see \autoref{sec:MLv}) using the cluster centers listed in \autoref{tab:structure}. In the left panel, we show $\Delta \bar{V}_{\rm los}/2$ as a function of the bisector position angle (PA), together with the best-fit sinusoidal model $\Delta \bar{V}_{\rm los}/2=A_{\rm rot}\sin{({\rm PA}+\phi)}$ (see \citetalias{Song:2019aa} for details). The best-fit parameters are listed in the figure legend. In the right panel, crosses (dots) indicate stars with velocities greater (less) than the systemic velocity. The best-fit rotation axis from the left panel is marked as a dash line in the right panel, and two dotted lines denote the 1$\sigma$ uncertainties.  Apart from NGC~1978 shown in this figure, the rotation signature of all other clusters in our sample is negligible ($\left|A_{\rm rot}\right|/\sigma_{\rm p,0}\lesssim0.3$).}
   \label{fig:rot}
\end{figure*}

\begin{figure*}
   \centering    
   \includegraphics[width=0.47\textwidth]{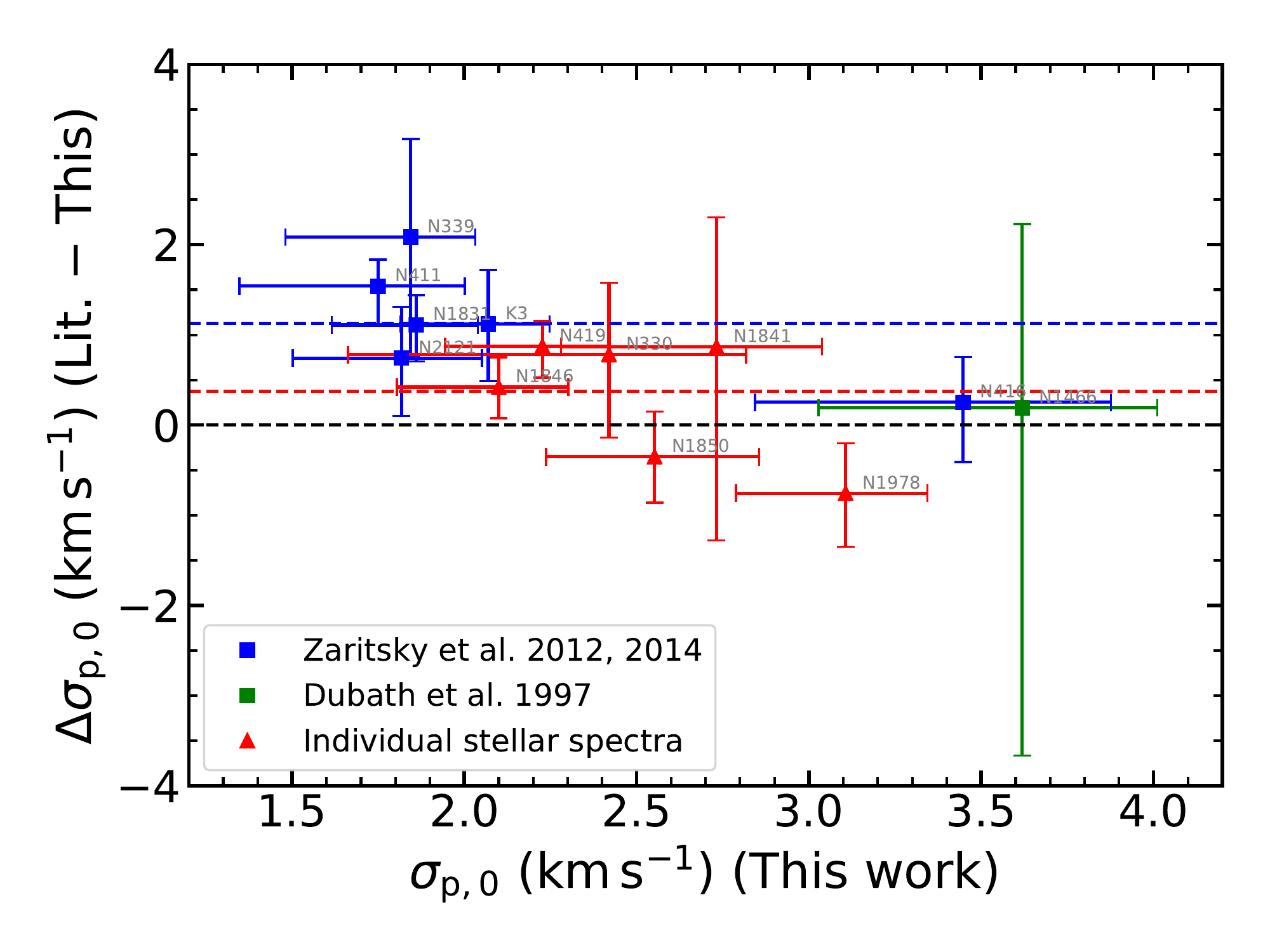}
   \includegraphics[width=0.47\textwidth]{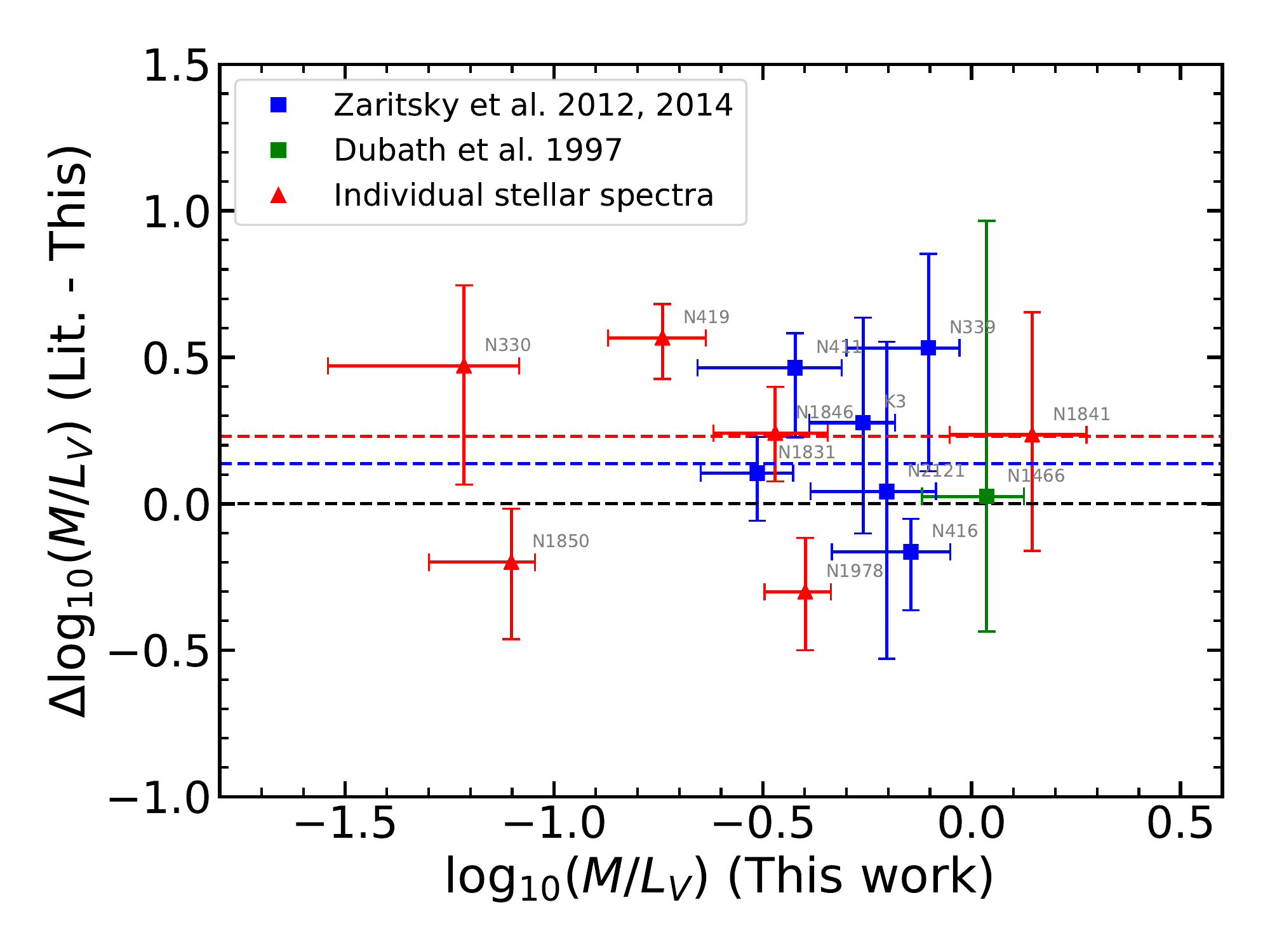}
   \caption{Differences between dynamical results---central velocity dispersions (left panel) and $M/L_V$ ratios (right panel)---found in previous studies compared to results of the present paper. In both panels, different symbols indicate different spectroscopic measurements used in those studies (see \autoref{sec:comp}). The blue and dashed horizontal lines denote the weighted-average offset values of the corresponding colored data points.}
   \label{fig:comp_vd0_MLR}
\end{figure*}

\begin{figure*}
   \centering
   \includegraphics[width=0.47\textwidth]{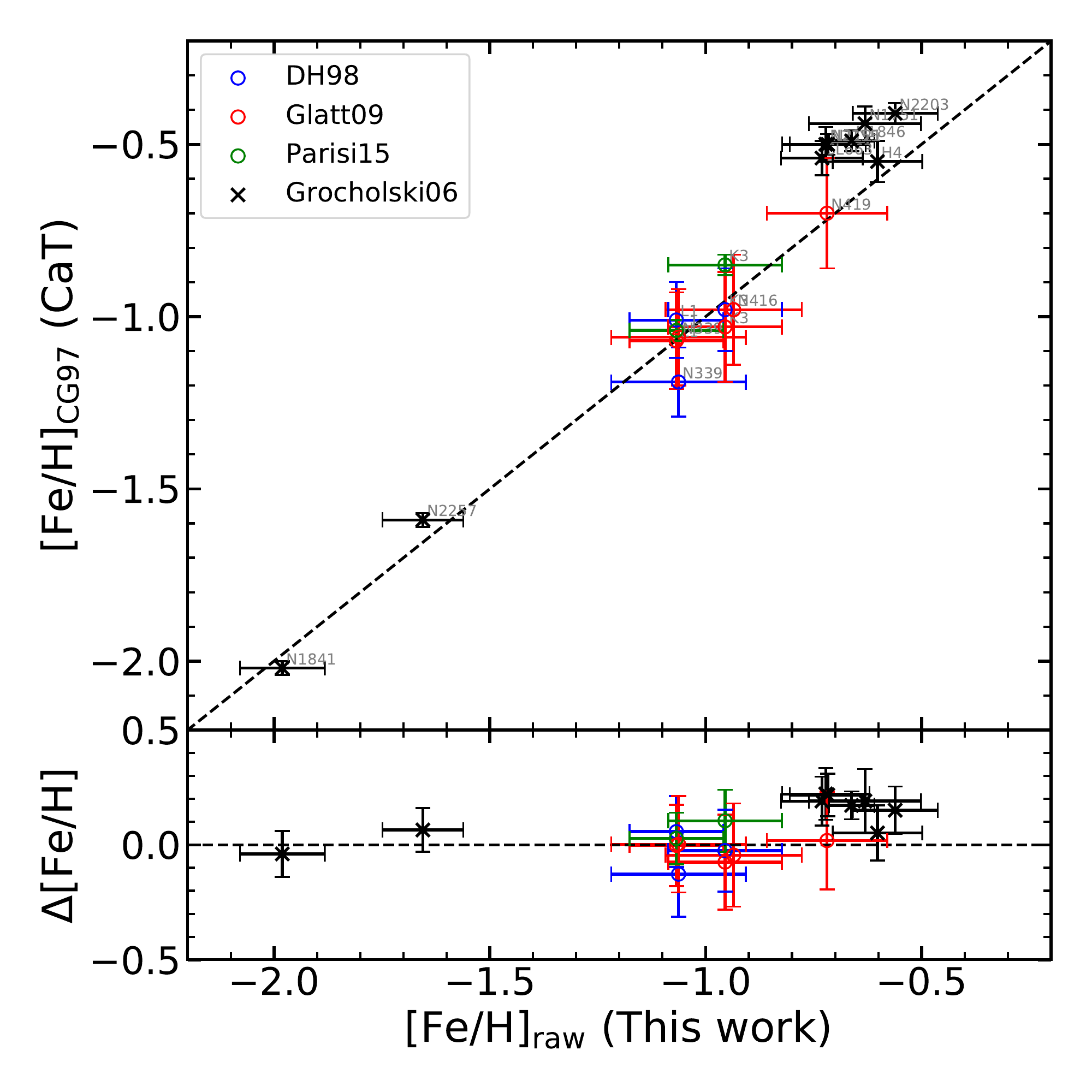}
   \includegraphics[width=0.47\textwidth]{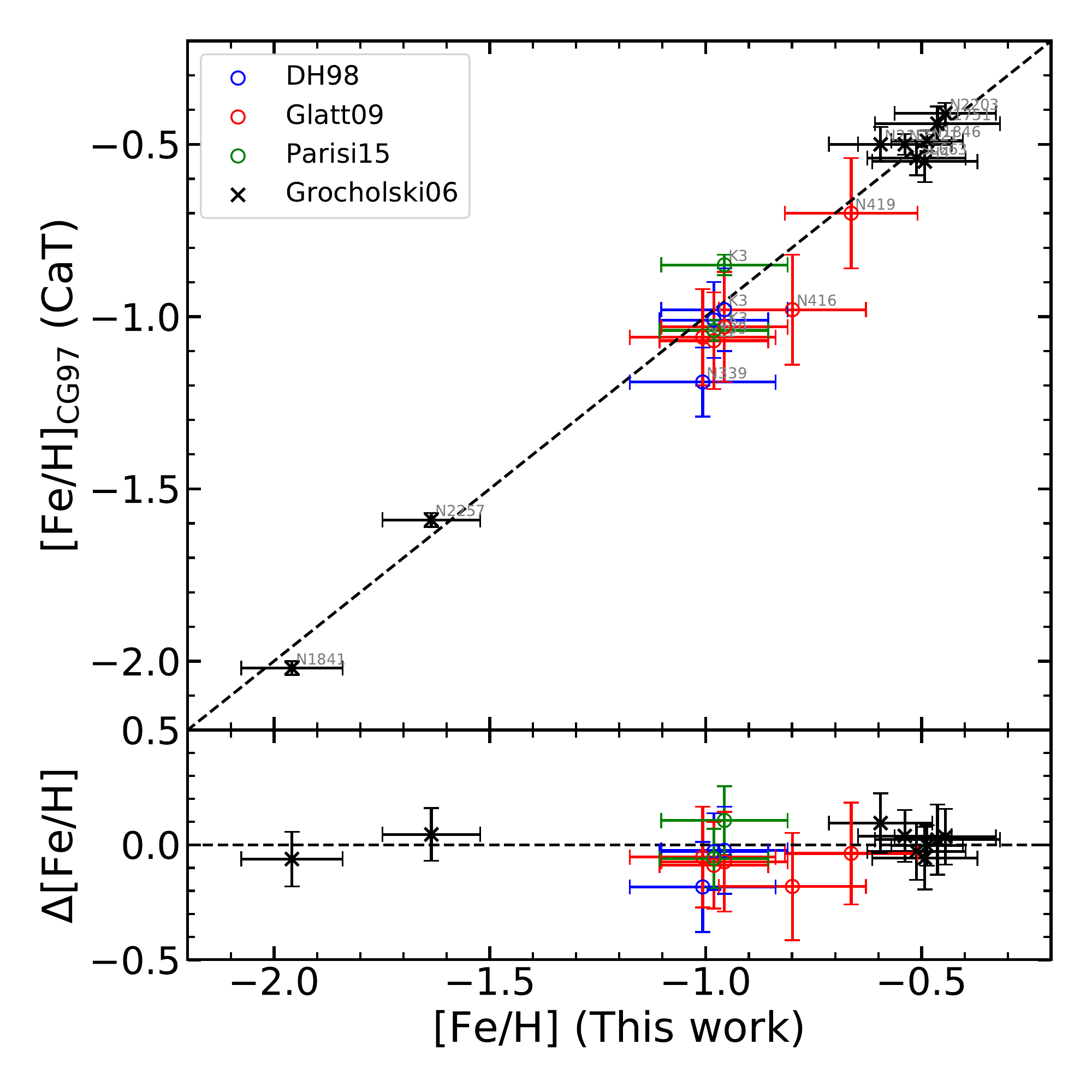}
   \caption{Comparisons of metallicities for clusters in common between our work and that of previous CaT studies \citep{Da-Costa:1998aa, Glatt:2009aa, Grocholski:2006aa, Parisi:2015aa}. In both panels, open circles denote the SMC clusters and crosses denote the LMC clusters of our sample (see \autoref{tab:FeH}). The raw cluster metallicities derived in \autoref{sec:rm_FeH} are used in the left panel, while right panel uses the
   cluster metallicities (as listed in \autoref{tab:FeH}) obtained after the calibration described in \autoref{sec:FeH}.}
   \label{fig:comp_FeH}
\end{figure*}

A cluster's total luminosity and dynamical mass can be derived by scaling the central surface brightness and the projected central velocity dispersion, respectively, with the dimensionless cluster structural profiles (i.e. the K66 models listed in \autoref{tab:structure}). The two required scaling parameters, central surface brightness and projected central velocity dispersion, were determined, respectively, from aperture photometry (as described in this section) and from the individual stellar velocities (see \autoref{sec:vd}).

To determine the central surface brightness ($\Sigma_{V,0}$), we integrated the dimensionless K66 model to a maximal reference radius within which an aperture luminosity was measured from the ground-based CCD photometry. The dimensionless K66 models were computed using the code \texttt{LIMEPY} \citep{Gieles:2015aa}, consistent with how we determine cluster masses (see below). The aperture magnitude and associated reference radius used for each cluster are listed in \autoref{tab:basic}, columns 3--4, with the sources identified in column 5. When transforming magnitude to luminosity, we adopted $M_{V,{\sun}}=4.85$ in addition to the distance modulus and extinction value listed in \autoref{tab:basic}. We could not find published $V$-band aperture photometry for SL~663, so instead we adopted the central surface surface brightness reported by \citet{McLaughlin:2005aa}. This value, and its uncertainty, is given in \autoref{tab:basic}. 

The total luminosity ($L_{V,\,\rm tot}$) was obtained by integrating the $\Sigma_{V,0}$-scaled K66 model to the K66 tidal radius listed in \autoref{tab:structure}. 
Similarly, the total cluster mass ($M_{\rm tot}$) was derived with the same dimensionless K66 model but scaled to $\sigma_{\rm p,0}$ fitted from the PM50 or PM50$^\prime$ sample (see \autoref{sec:vd} and \autoref{tab:PM50}).  The dynamical $M/L_V$ ratios of the clusters were then derived from the total masses and total $V$-band luminosity determined above.  

Uncertainties in cluster masses, luminosities and $M/L_V$ ratios were estimated using a bootstrapping technique that accounted for uncertainties in distance modulus and extinction  (\autoref{tab:basic}), the K66 structural parameters (see \autoref{tab:structure}), the $V$-band aperture magnitude (for $L_{V,\,\rm tot}$ and $M/L_V$; see \autoref{tab:basic}), and the central velocity dispersion (for $M_{\rm tot}$ and $M/L_V$; see \autoref{tab:PM50}). All relevant parameters were randomly sampled, assuming that the error distributions are joint-Gaussian to account for the asymmetric uncertainties (e.g. $c_{\rm K66}$ in \autoref{tab:structure}), or using simple Gaussian distributions for parameters with symmetric uncertainties (e.g. $V_{\rm ap}$ in \autoref{tab:basic}).  For each cluster, we created 1000 samples from the selected parameters to calculate $M_{\rm tot}$, $L_{V,\,\rm tot}$ and $M/L_V$. We took the 15.9-th and 84.1-th percentiles of the simulated values as the lower and upper 1$\sigma$ confidence boundaries of the respective parameter. \autoref{tab:MLv} lists the masses, luminosities and $M/L_V$ ratios for all 26 clusters in our sample.

Our dynamical analysis assumed that all clusters are exclusively pressure-supported systems.  However, for two cases---NGC~1978 and NGC~1846---the kinematic data suggest that the clusters may exhibit coherent rotation (see \citetalias{Song:2019aa} and \autoref{fig:rot}). For NGC~1846, we roughly addressed the dynamical effect of rotation in \citetalias{Song:2019aa}, concluding that rotation may be causing $\sim$9\% overestimation in the cluster's total mass and $M/L_V$. For NGC~1978, \citet{Fischer:1992ab} carried out a more sophisticated analysis in which they fitted single-mass rotating and non-rotating oblate spheroid models to the surface luminosity profiles and their radial velocity data. They found no significant differences in $M/L_V$s derived with these models and those derived with single-mass K66 models. However, the mass estimates for the cluster did differ systematically among the rotating and non-rotating models. Our data for NGC~1978, while clearly showing a rotation signature (see right panel of \autoref{fig:rot}), are limited by poor background determination as noted in \autoref{sec:sky_sub}.  For this reason, and because NGC~1978 requires a more involved dynamical modeling approach, we will defer a detailed analysis for this cluster to a later paper.

\subsection{Comparison with Previous Work}
\label{sec:comp}

Previously published studies have reported velocity dispersions and $\rm M/L$ ratios of 13 clusters in common with our work. 
Seven of these clusters were studied using integrated-light spectroscopy \citep[][]{Dubath:1997aa, Zaritsky:2012aa, Zaritsky:2014aa}, and another six clusters were studied from dynamical modelling using radial velocities of individual cluster member stars similar to this work \citep{Fischer:1992ab, Fischer:1993aa, Suntzeff:1992aa, Mackey:2013aa, Kamann:2018ab, Patrick:2020aa}. 

The left panel of \autoref{fig:comp_vd0_MLR} shows the difference in central velocity dispersions between these previous studies and our work (see \autoref{tab:PM50} and \autoref{tab:MLv}). Our dispersion estimates agree well with studies using individual stellar spectra (red triangles); the average difference (red dashed line) is $0.37\pm0.19$~\kms. As for the integrated-light studies (blue and green squares), only two clusters with relatively high central values (i.e. NGC~419 and NGC~1466) agree in the dispersions, while we measured lower values for the remaining five clusters. For those studied by \citet{Zaritsky:2012aa, Zaritsky:2014aa} (blue squares), we obtained an average difference of $1.13\pm0.20$~\kms\ in dispersion (blue dashed line).

In the right panel of \autoref{fig:comp_vd0_MLR}, we show the difference in $M/L_V$ ratios between these previous studies and our work (see \autoref{tab:PM50} and \autoref{tab:MLv}). In general, our measured $M/L_V$ ratios agree better with those studied by \citet{Zaritsky:2012aa, Zaritsky:2014aa}, compared to those using individual stellar spectra.
For the clusters studied by \citet{Zaritsky:2012aa, Zaritsky:2014aa} (blue squares), we obtain a weighted average of $0.14\pm0.08$ in $\Delta\log_{10}{(M/L_V)}$ ratio (blue dashed curves); while for the clusters studied using individual stellar spectra (red triangles), the weighted  average (red dashed curves) is $0.23\pm0.08$. These differences may be rooted from the various modeling methods used to estimate cluster $M/L_V$ ratio. Indeed, our results are in good agreement with those of the studies adopted the K66 models (i.e., NGC~1466, NGC~1841 and NGC~1850)\footnote{This discussion should not be construed to imply broad agreement between our work and that of \citet{Zaritsky:2012aa, Zaritsky:2014aa}.  As it happens, the only clusters in common between these studies have intermediate ages.  As we shall see in \autoref{sec:IMF}, our M/L estimates do not agree as well for younger and older clusters where our results are systematically lower and higher, respectively.}.

Overall, the variations in $M/L_V$ ratios between our work and previous studies seem reasonable given difference in  observational technique, data quality, sample size and analysis methods. This conclusion is consistent with the detailed case study for two clusters---NGC~419 and NGC~1846---in \citetalias{Song:2019aa}, where we compared our results with similar studies based on individual stellar velocity measurements \citep{Kamann:2018ab, Mackey:2013aa}.  We explored the effects of both poorly-estimated velocity uncertainties (for NGC~419) and different dynamical modeling (for NGC~1846) on the determination of $M/L_V$ ratios.  We presume that similar issues affect the comparisons of results described here, though for the present sample, and in particular for the integrated-light spectroscopy results \citep{Dubath:1997aa, Zaritsky:2012aa, Zaritsky:2014aa}, we are unable to carry out as detailed a comparison as in \citetalias{Song:2019aa}.

\begin{figure}
   \centering
   \includegraphics[width=0.47\textwidth]{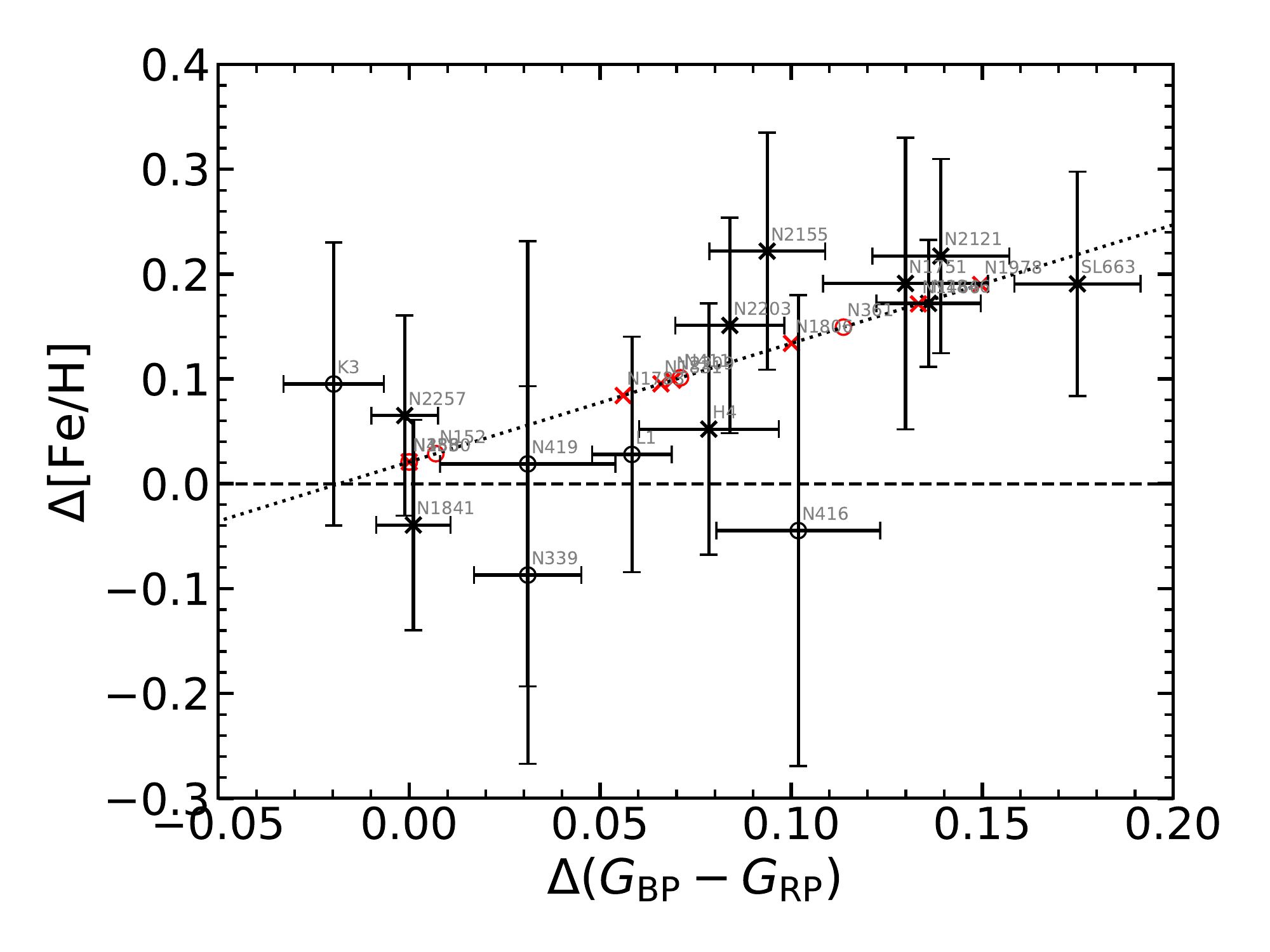}
   \caption{Metallicity differences for clusters in common with our work and previous CaT studies as a function of the color offsets raised in the  $T_{\rm eff}$ calculation (see \autoref{sec:Teff}). Open circles denote the SMC clusters and crosses denote the LMC clusters in our sample (see \autoref{tab:FeH}). The dotted line is the weighted linear fit to these data given by \autoref{eq:cor_FeH}.  This fitted line was used to place our cluster metallicities on the CaT system described in \autoref{sec:FeH} and \autoref{fig:comp_FeH}. For clusters in our sample lacking published CaT measurements, the adopted $\Delta{\rm [Fe/H]}$ values are denoted as red symbols along the dotted line. The calibrated cluster metallicities are listed in \autoref{tab:FeH}.}
   \label{fig:dFeH_dcolor}
\end{figure}

\subsection{Cluster Metallicity}
\label{sec:FeH}
As described in \autoref{sec:Teff}, we estimated effective temperature for each stellar target from their colors in order to break the strong temperature-metallicity degeneracy in our Bayesian fitting analysis (see \autoref{sec:spec_fit}). The stellar metallicities obtained in this analysis were used in \autoref{sec:rm_FeH} to flag chemically discrepant stars as non-members before estimating the `raw' mean metallicities of the clusters. 

An open question remains as to how well these `raw'  metallicities compare to previous chemical analyses.  One extensive and consistent source of cluster metallicities come from Ca II triplet (CaT) results that are typically calibrated to two [Fe/H] scales, i.e. \citet{Zinn:1984aa} (ZW84) and \citet{Carretta:1997aa} (CG97). For our target clusters, we have collected all the studies using either [Fe/H] scale, and then used the relation in \citet[][see their Equation 3]{Carretta:2001aa} to transfer any ZW84 [Fe/H] values onto the CG97 scale. In the left panel of \autoref{fig:comp_FeH}, we compared the [Fe/H] values of nine LMC clusters from \citet{Grocholski:2006aa} (crosses) and five SMC clusters from \citet{Da-Costa:1998aa, Glatt:2009aa, Parisi:2015aa} (open circles). For metal-poor clusters, our raw [Fe/H] values agree well with the CaT results; for metal-rich (and mostly LMC) clusters, our raw [Fe/H] values are systematically lower than those measured by CaT spectroscopy.

The systematic offset in metallicity is found to be related to the color offset between the colors used to calculate $T_{\rm eff}$ (see \autoref{sec:Teff}) and those published in the \textit{Gaia} DR2. \autoref{fig:dFeH_dcolor} shows the metallicity offsets between the CaT studies and our work as a function of the $G_{\rm BP}-G_{\rm RP}$ color offsets of the target clusters in common. The color offsets were calculated the same as that shown in \autoref{fig:dcolor} but in a cluster-by-cluster manner, and in the calculation we only considered the cluster members confirmed in \autoref{sec:rm_FeH}. For the three SMC clusters (Kron~3, Lindsay~1 and NGC~339) that have multiple CaT measurements available, we took the weighted average of their [Fe/H] values when calculating the metallicity offsets. We found a clear positive correlation between metallicity offsets and color offsets.  We fitted the following linear relation (see the dotted line in \autoref{fig:comp_FeH}) to calibrate our raw cluster metallicities onto the CaT CG97 scale:
\begin{equation}
\label{eq:cor_FeH}
{\rm [Fe/H]_{CG97}} = {\rm [Fe/H]_{raw}}+1.131\Delta(G_{\rm BP}-G_{\rm RP})+0.021.
\end{equation}
The final ${\rm [Fe/H]_{CG97}}$ values are listed in column 3 of \autoref{tab:FeH}. The weighted standard deviation in ${\rm [Fe/H]_{CG97}}$ about the fitted line is 0.064 dex; we have added this value to the final metallicity errors in quadrature (column 4 of \autoref{tab:FeH}). The comparison between our calibrated cluster metallicities and those in the CG97 scale is shown in the right panel of \autoref{fig:comp_FeH}. The plot shows that along the full metallicity range, our modified [Fe/H] values now agree well with those in the CG97 scale after the calibration using \autoref{eq:cor_FeH}.

\section{Discussion}
\label{sec:discuss}

\begin{table*}
 \caption{Age/Metallcity bins of the cluster sample}
 \label{tab:age_bin}
 \begin{center}
 \begin{threeparttable}
 \begin{tabular}{clcccl}
\hline
    \multicolumn{1}{c}{Bin}  &  
    \multicolumn{1}{c}{Color$^{\rm a}$}  &  
    \multicolumn{1}{c}{Age Range}  &  
    \multicolumn{1}{c}{Bin Age$^{\rm b}$}  &  
    \multicolumn{1}{c}{Bin ${\rm [Fe/H]}^{\rm b}$}  &
    \multicolumn{1}{c}{Clusters}  \\
    \multicolumn{1}{c}{} &  
    \multicolumn{1}{c}{} &  
    \multicolumn{1}{c}{$\rm (Gyr)$}  &  
    \multicolumn{1}{c}{$\rm (Gyr)$}  &  
    \multicolumn{1}{c}{$\rm (dex)$}  &
    \multicolumn{1}{c}{} \\ 
    \multicolumn{1}{c}{(1)} & 
    \multicolumn{1}{c}{(2)} & 
    \multicolumn{1}{c}{(3)} & 
    \multicolumn{1}{c}{(4)} & 
    \multicolumn{1}{c}{(5)} &
    \multicolumn{1}{c}{(6)} \\
\hline 
1 & Purple & 0.04--0.14 & 0.1 & --0.4 & NGC~330, 458, 1850 \\
2 & Blue & 0.7--1.5 & 1.0 & --0.4 & NGC~152, 411, 419, 1751, 1831, 2209\\
3 & Green & 1.5--2.2 & 1.8 & --0.4 & NGC~1783, 1806, 1846, 1978, 2203; Hodge~4\\
4 & Yellow & 2.9--3.2 & 3.0 & --0.4 & NGC~2121, 2155; SL~663\\
5 & Red & 6.3--8.1 & 7.0 & --0.7 & NGC~339, 361, 416; Kron~3, Lindsay~1\\
6 & Black & 12.7--13.8 & 12.5 & --1.7 & NGC~1466, 1841, 2257\\
\hline
\end{tabular}
\begin{tablenotes}
\item $^{\rm a}${These colors are used in Figures~\ref{fig:MLR_ssp}--\ref{fig:MLR_bottom_light} to represent clusters and model results associated with the age/metallicity bins listed here.}
\item $^{\rm b}${These bin age and metallicity values are used to produce the isochrone curves shown in \autoref{fig:MLR_ssp}, \autoref{fig:MLR_age_anders} and \autoref{fig:MLR_mass_anders}. They are set to equal grid point values in the \citet{Anders:2009aa} models. }
\end{tablenotes}
\end{threeparttable}
\end{center}
\end{table*}

Having produced $M/L_V$ values for 26 clusters spanning a range of ages (\autoref{tab:basic}), metallicities (\autoref{tab:FeH}) and masses (\autoref{tab:MLv}), we now turn to a discussion regarding the basic trends of $M/L_V$ with respect to these various parameters.  Part of that discussion involves a comparison of our results to $M/L_V$ values predicted by a reference set of Simple Stellar Population (SSP) models.  Dynamical effects can also lead to evolution of $M/L_V$ values in clusters, and we consider these effects as well.  Given that the SSP+dynamics models appear to explain our results reasonably well, we also explore how cluster $M/L_V$ values may be used to constrain other astrophysical parameters such as the stellar IMF and cluster disruption timescales in the Magellanic Clouds.

\subsection{$M/L_V$ Trends and comparing to SSP model}
\label{sec:MLR_trends}

\begin{figure*}
   \centering
   \includegraphics[width=0.47\textwidth]{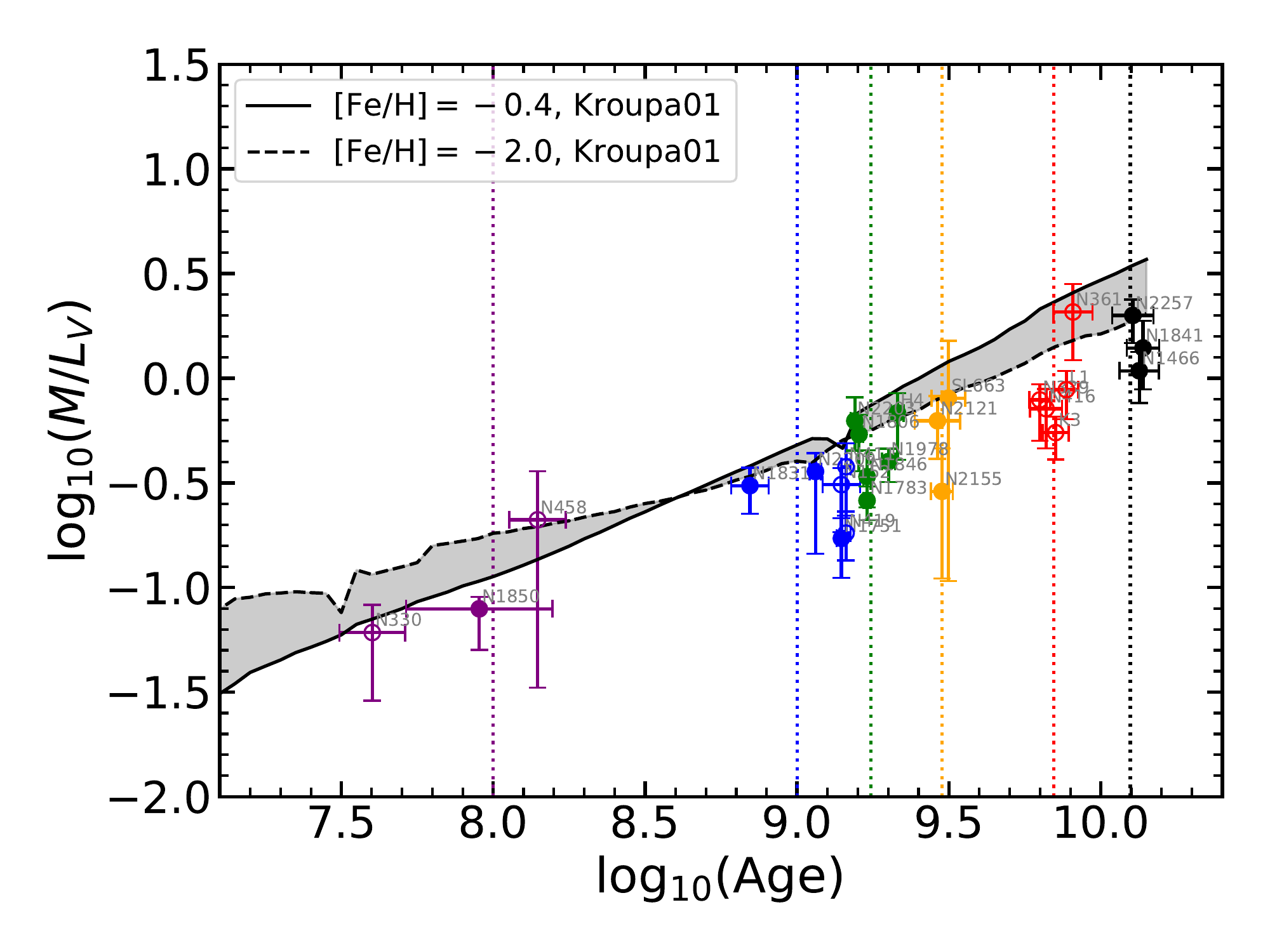}
   \includegraphics[width=0.47\textwidth]{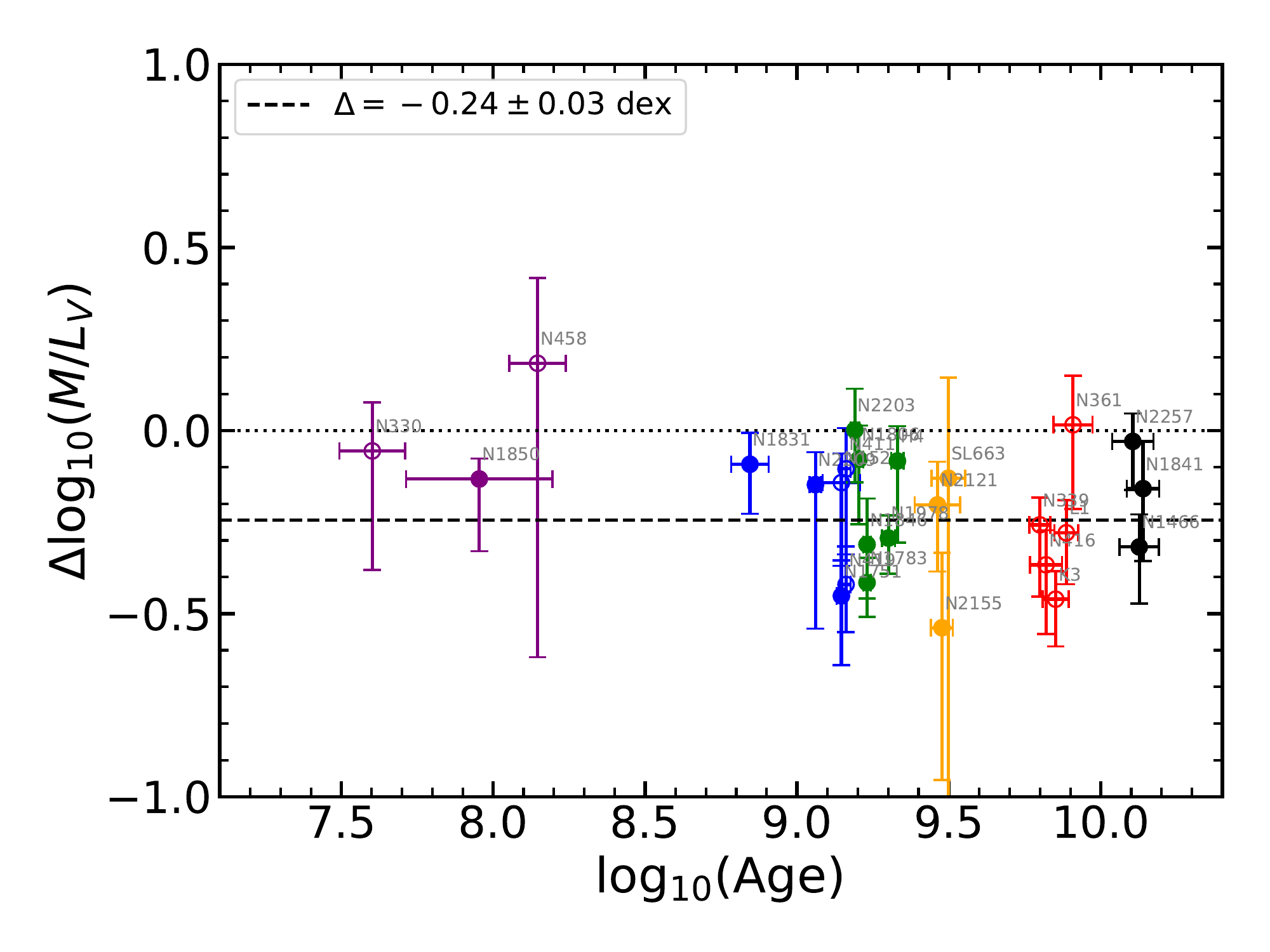}
   \includegraphics[width=0.47\textwidth]{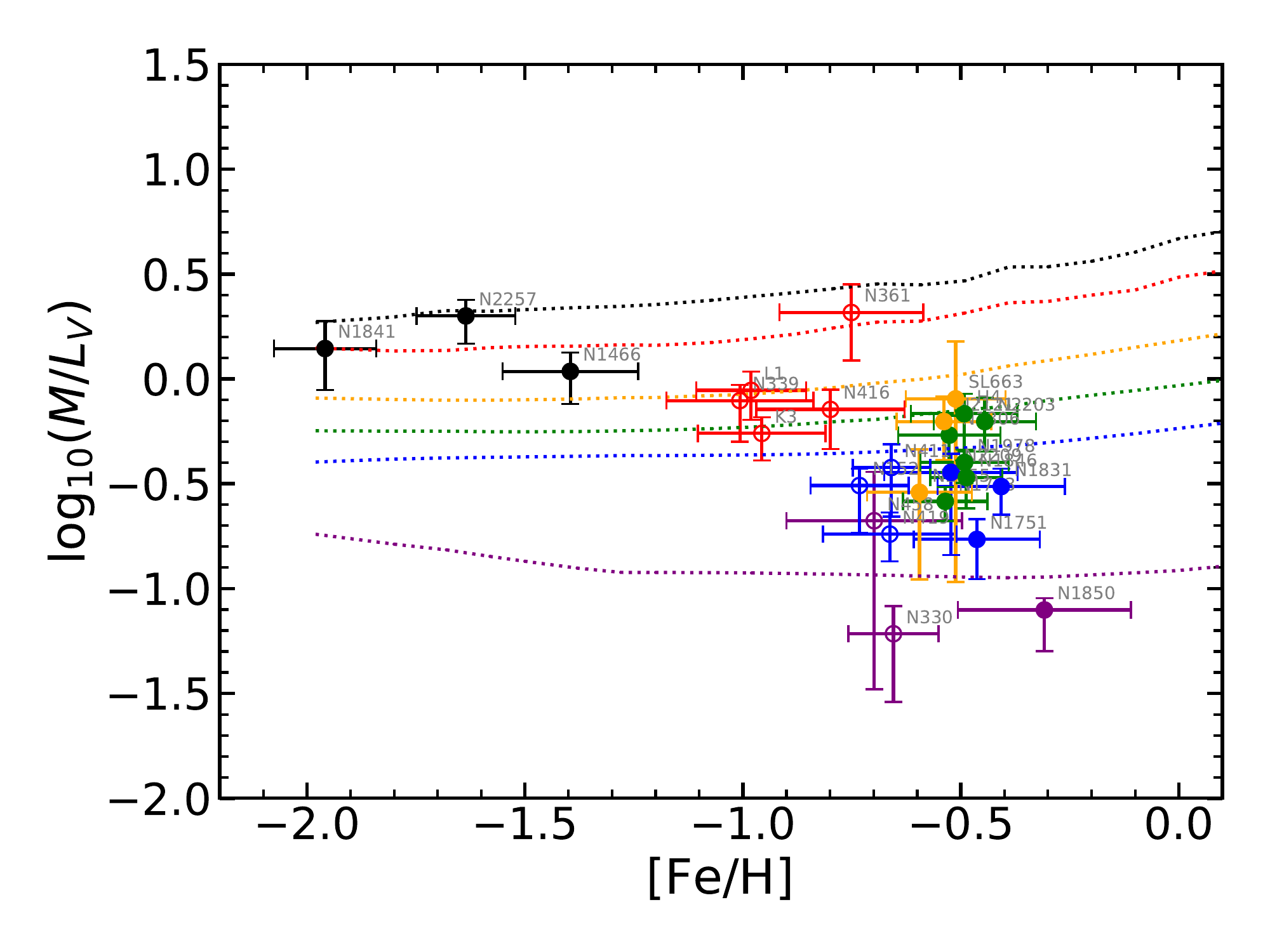}
   \includegraphics[width=0.47\textwidth]{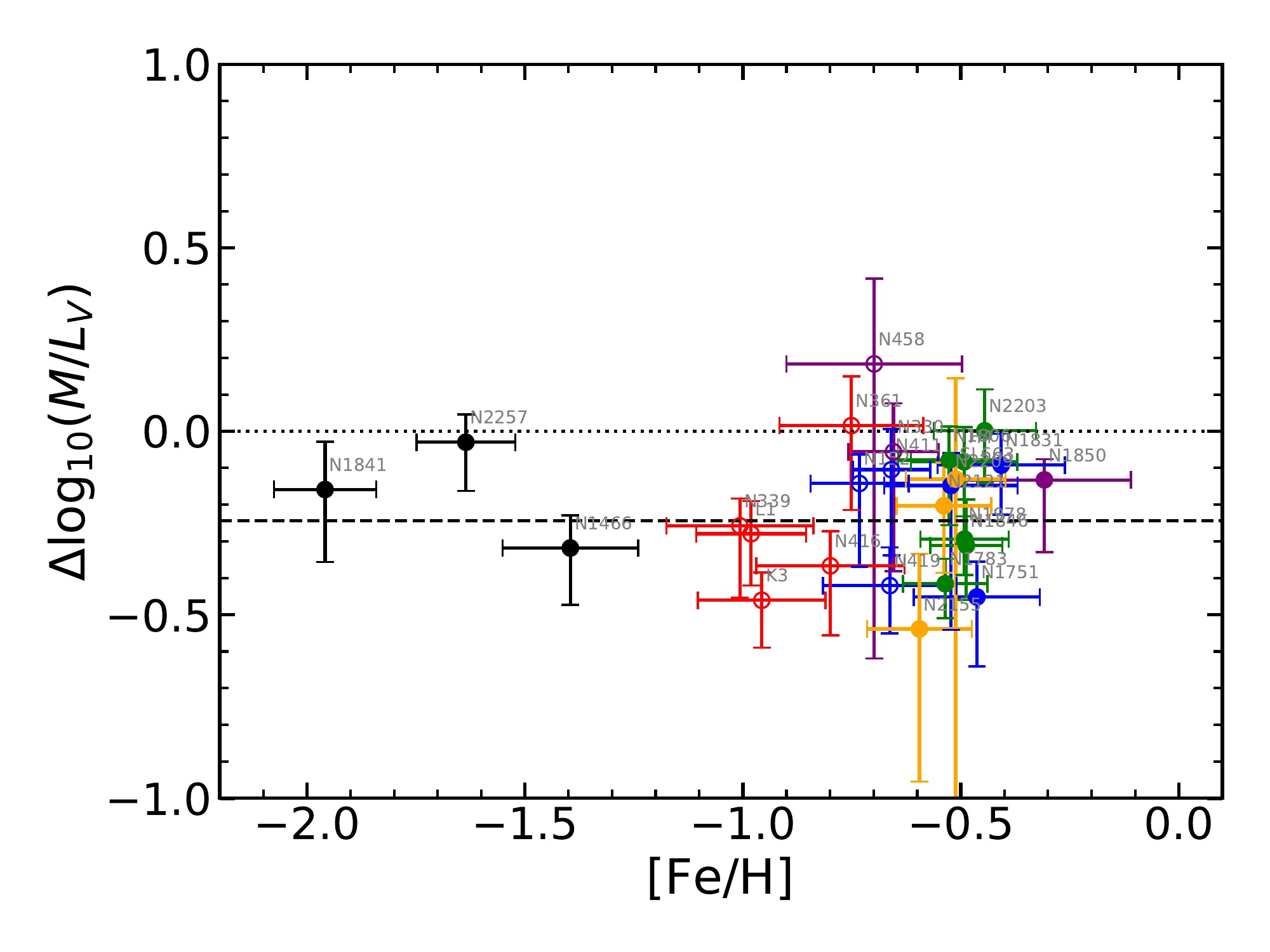}
   \includegraphics[width=0.47\textwidth]{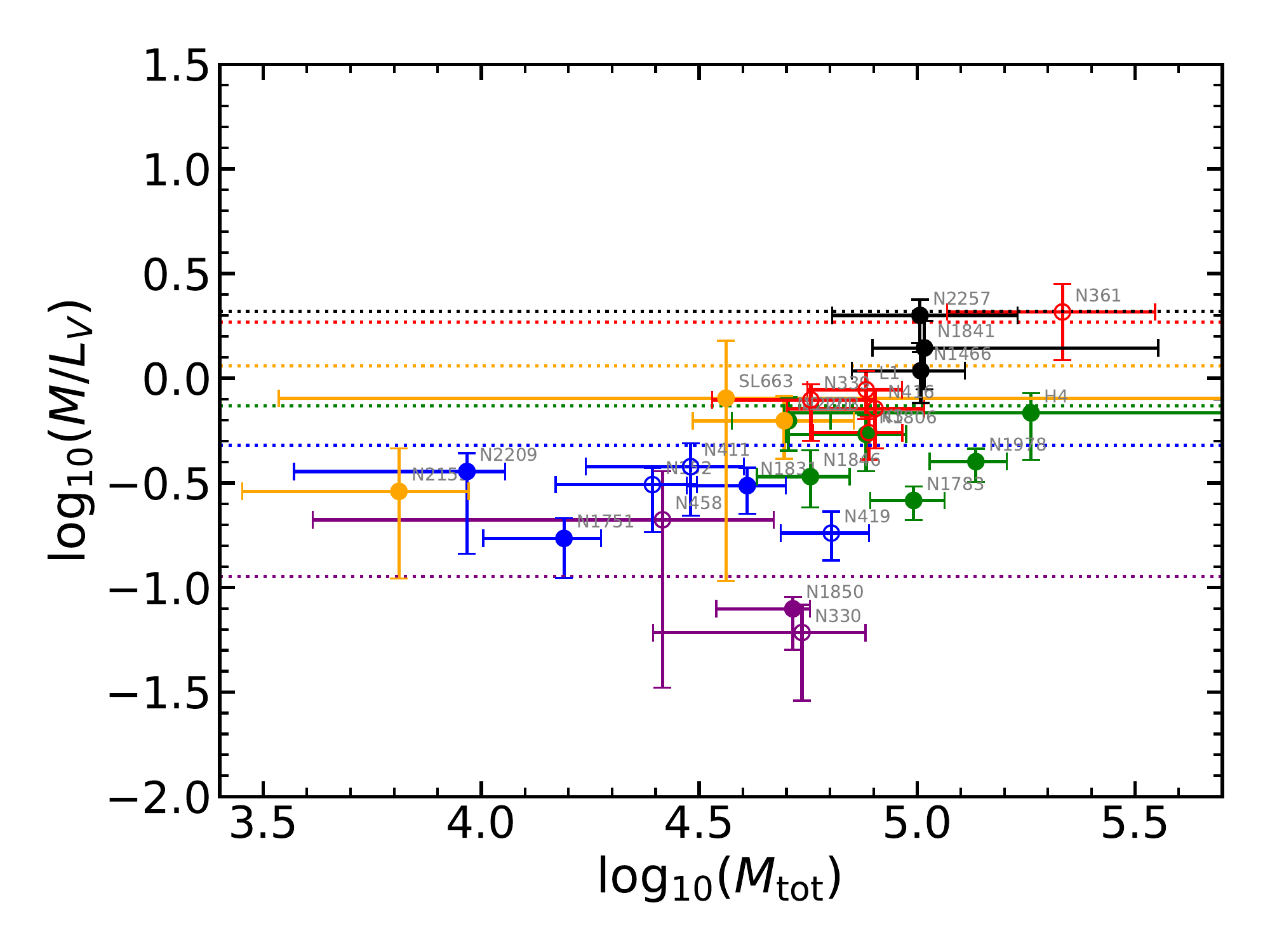}
   \includegraphics[width=0.47\textwidth]{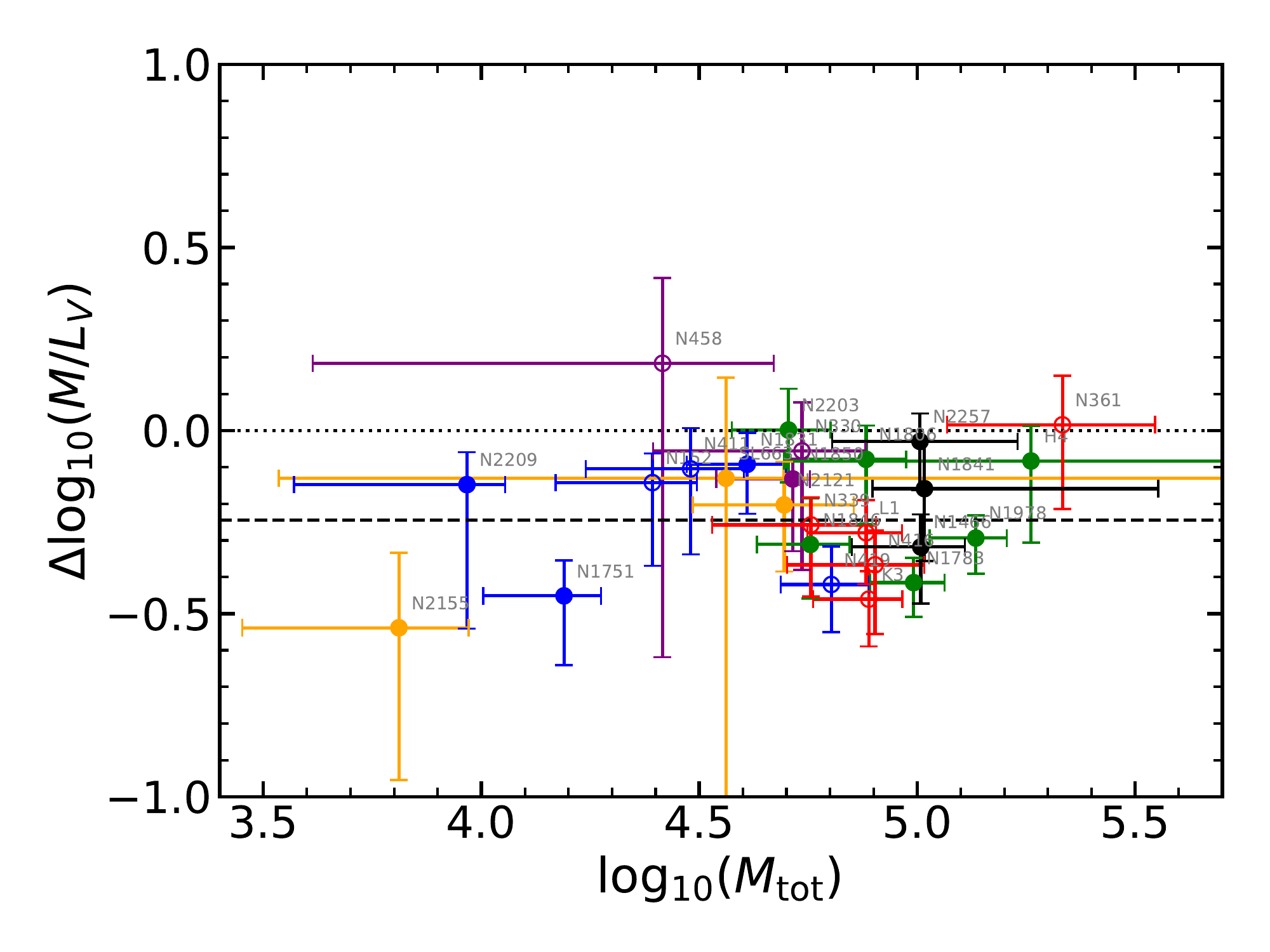}
   \caption{Left panels: Dynamical $M/L_V$ ratios of our target clusters in the LMC (filled circles) and SMC (open circles) as a function of age (top), metallicity (middle) and mass (bottom), over-plotted with evolutionary $M/L_V$ ratio isochrones from a set of FSPS SSP models (see \autoref{sec:MLR_trends} for details). The clusters are binned into distinct age groups denoted by different colors (see \autoref{tab:age_bin} for bin details). In the top left panel, the solid and dashed curves show $M/L_V$ evolutionary tracks for two  metallicities (see the legend). In the top and middle left panels, the colored dotted lines denote the ages used to generate the SSP isochrone curves (see column 4 of \autoref{tab:age_bin}). In the bottom left panel, the colored dotted lines are calculated for the same ages but in each case we adopted the bin metallicity of the set of clusters denoted by the same color (see column 5 of \autoref{tab:age_bin}).  Right panels: Differences of $M/L_V$ ratios in log space between our measurements and the SSP predictions. The $\Delta\log_{10}{(M/L_V)}$ values were calculated in a cluster-by-cluster manner using the age and metallicity of each cluster (we used the ages listed in \autoref{tab:basic} and metallicities listed in \autoref{tab:FeH}).  In each right-hand panel, the horizontal dashed line shows the weighted mean offset in $\Delta\log_{10}{(M/L_V)}$ for the entire sample.}
   \label{fig:MLR_ssp}
\end{figure*}

A fundamental prediction of stellar evolutionary models is that, barring any strong IMF variations or pathological dynamical effects, simple populations should become `darker' (i.e., higher $M/L_V$) with increasing age.  The top left panel of \autoref{fig:MLR_ssp} reveals such a trend as a clear positive correlation between $M/L_V$ ratio and age for our cluster sample. 

In the same panel, we have overplotted evolutionary tracks of a set of SSP models denoted by the solid and dashed black lines and the shaded gray area bounded by these two lines.  The tracks were produced using the Flexible Stellar Population Synthesis (FSPS) code \citep{Conroy:2009aa, Conroy:2010aa} employing Padova isochrones \citep{Girardi:2000aa, Marigo:2008aa}, a \citet{Kroupa:2001aa} IMF, and the BaSeL spectral library \citep{Lejeune:1997aa, Lejeune:1998aa, Westera:2002aa}. We will refer to the tracks produced by this combination of Kroupa/Padova/BaSeL models as the `reference' SSP models throughout this paper. The shaded area in the figure is meant to roughly represent the age-metallicity regime appropriate for MC clusters.  The colors of the data points denote age bins used to categorize the clusters. The colored dotted lines represent the approximate average ages of these bins (see \autoref{tab:age_bin} and the caption of \autoref{fig:MLR_ssp} for details regarding these age bins). 

We also tested other options in the FSPS code, such as using the MILES spectral library \citep{Sanchez-Blazquez:2006aa} with both MIST \citep{Dotter:2016aa, Choi:2016aa, Paxton:2011aa, Paxton:2013aa, Paxton:2015aa} and BaSTI theoretical isochrones \citep{Pietrinferni:2004aa}. In all cases, we found good agreement ($\lesssim 10\%$) with the adopted Padova/Kroupa/BaSeL models among predicted $M/L_V$ ratios using alternative isochrones and stellar libraries.

\autoref{fig:MLR_ssp} reveals that when compared to the adopted reference SSP models, our dynamical $M/L_V$ ratios tend to run lower than the predictions. This offset is highlighted in the top right panel of \autoref{fig:MLR_ssp}, where we show the difference between our measured $M/L_V$ ratios and the SSP predictions as a function of cluster age. The $\Delta\log_{10}{(M/L_V)}$ values were calculated uniquely for each cluster using its own age (see \autoref{tab:basic}) and metallicity (see \autoref{tab:FeH}). Across the entire sample, our $M/L_V$ ratios are $-0.24\pm0.03$ dex lower on average than the theoretical predictions in log space, with an error-weighted standard deviation of 0.16 dex \footnote{When excluding the two clusters with anomalous PM50$^{\prime}$ results---NGC~458 and SL~663 (see \autoref{sec:vd}), the offset in $\Delta\log_{10}{(M/L_V)}$ of the remaining cluster sample is $-0.25\pm0.03$ dex, with an error-weighted standard deviation of 0.16 dex. This shows that the two clusters do not influence the results of our entire cluster sample. The same conclusion applies to all the following statistical results presented in \autoref{sec:discuss}.}. There is a weak indication that the youngest clusters may more closely follow the SSP predictions.  Notably, the overall offset---about 70\%---cannot be accounted for by using different input models in the FSPS code.

As shown in the middle left panel of \autoref{fig:MLR_ssp}, we also find a trend of decreasing $M/L_V$ ratio with increasing cluster metallicity. This trend is not unexpected since the more metal-rich clusters in our sample are also younger and hence lower in $M/L_V$ ratio than those of the metal-poor counterparts; this is simply a manifestation of the well-known age-metallicity relation for MC clusters (see \autoref{fig:amr} and, \citealp[e.g.][]{Harris:2009aa, Parisi:2015aa}). In the middle right panel, the $\Delta\log_{10}{(M/L_V)}$ values are plotted against the metallicities for all clusters in our sample. 

The bottom left panel of \autoref{fig:MLR_ssp} shows a broad trend of increasing $M/L_V$ ratio with increasing cluster mass. This behavior is also seen among old (globular) clusters in both our Galaxy \citep{Mandushev:1991aa, Kimmig:2015aa} and M31 \citep{Strader:2011aa}, and were attributed by these authors to be due to dynamical evolutionary effects. 
The isochrones from the reference SSP models (colored dotted lines) remain constant with cluster mass and so do not predict a trend of $M/L_V$ with total mass.
We will return to this when we consider dynamical evolution effects in the MC clusters. The bottom right panel shows the offsets in $\Delta\log_{10}{(M/L_V)}$ with the SSP predictions calculated using the appropriate age and metallicity for each cluster.

\subsection{Dynamical Effects on Cluster $M/L_V$ Values}
\label{sec:MLR_dyn}

\begin{figure*}
   \centering
   \includegraphics[width=0.47\textwidth]{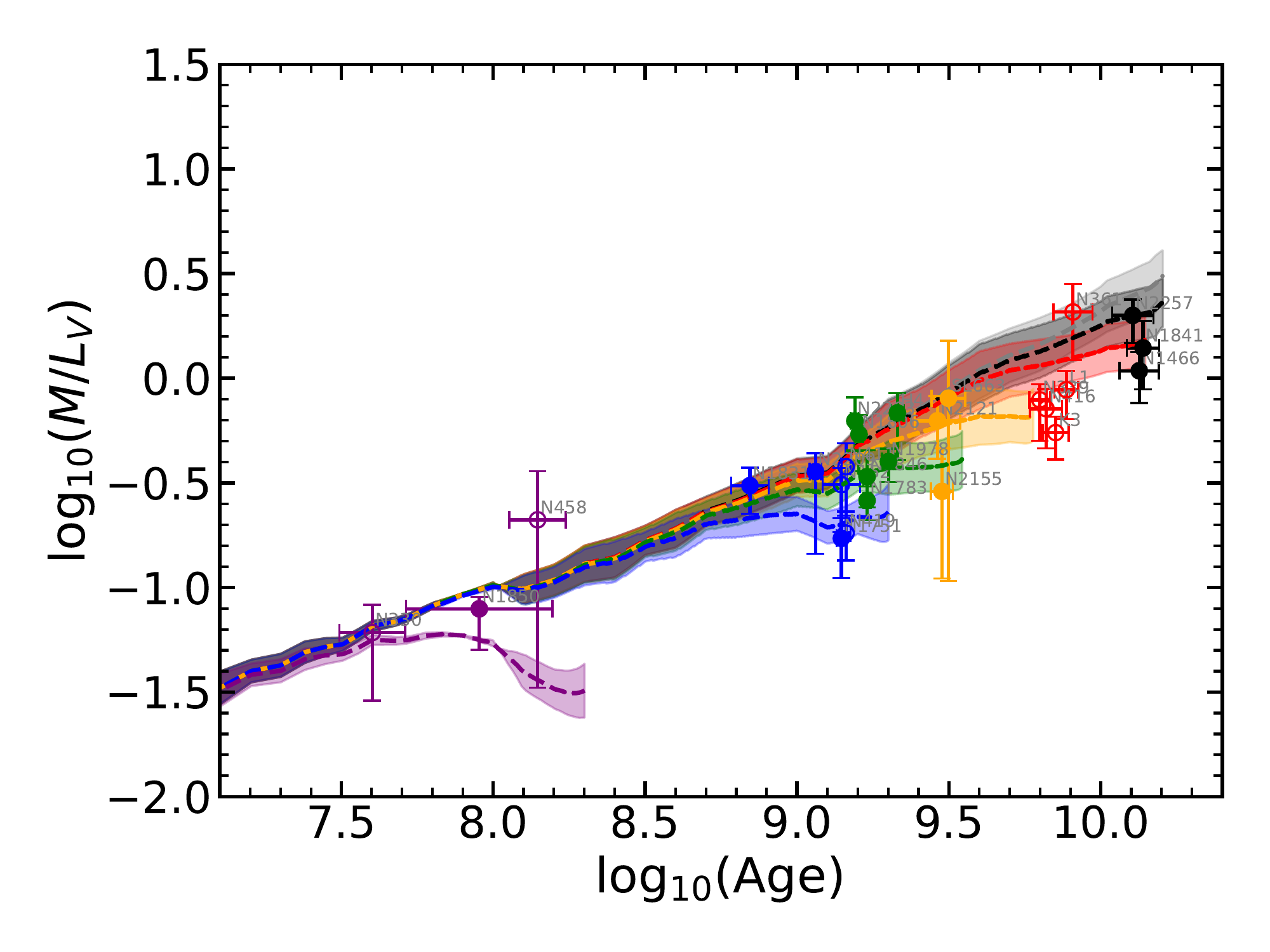}
   \includegraphics[width=0.47\textwidth]{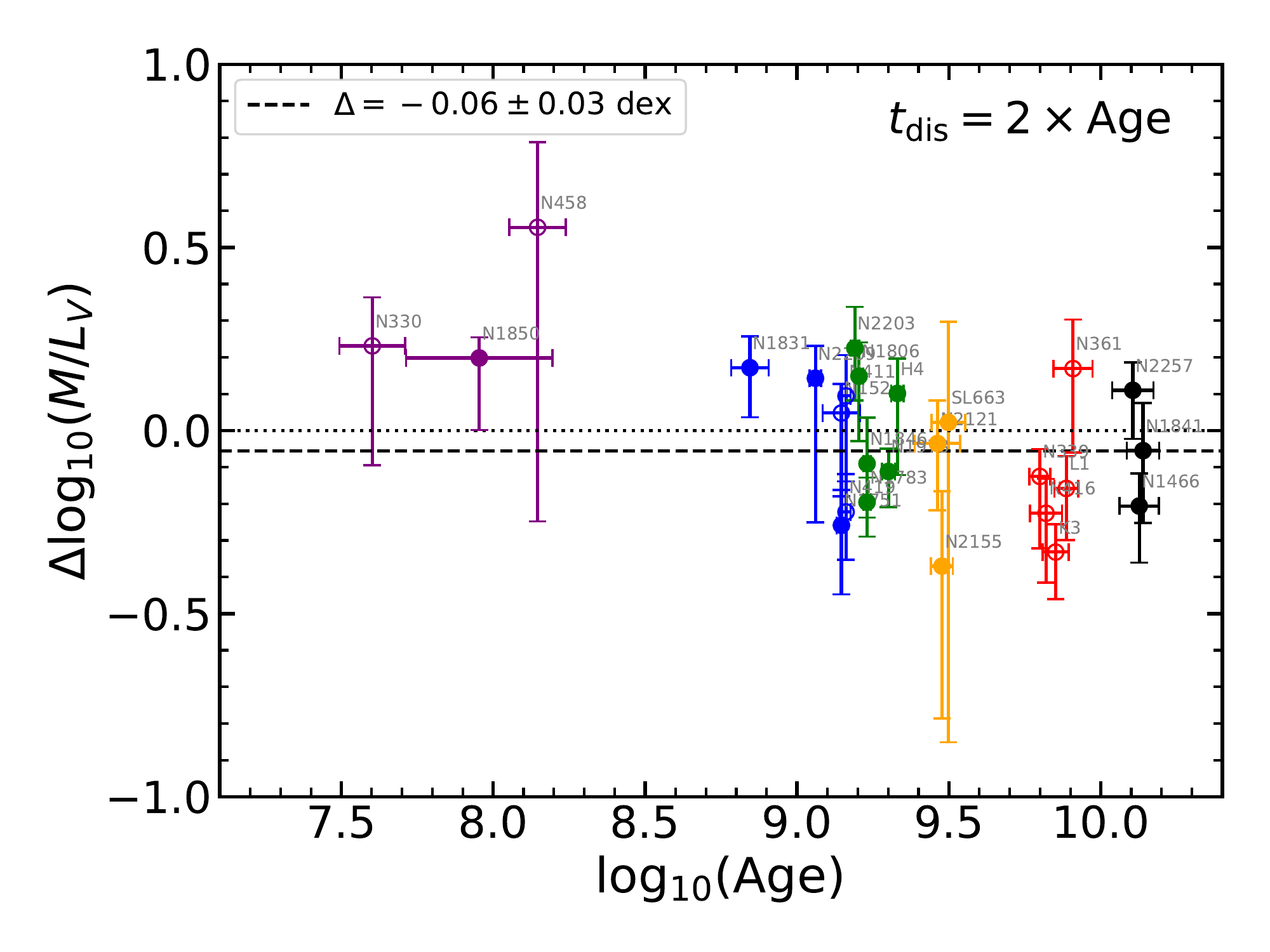}
   \includegraphics[width=0.47\textwidth]{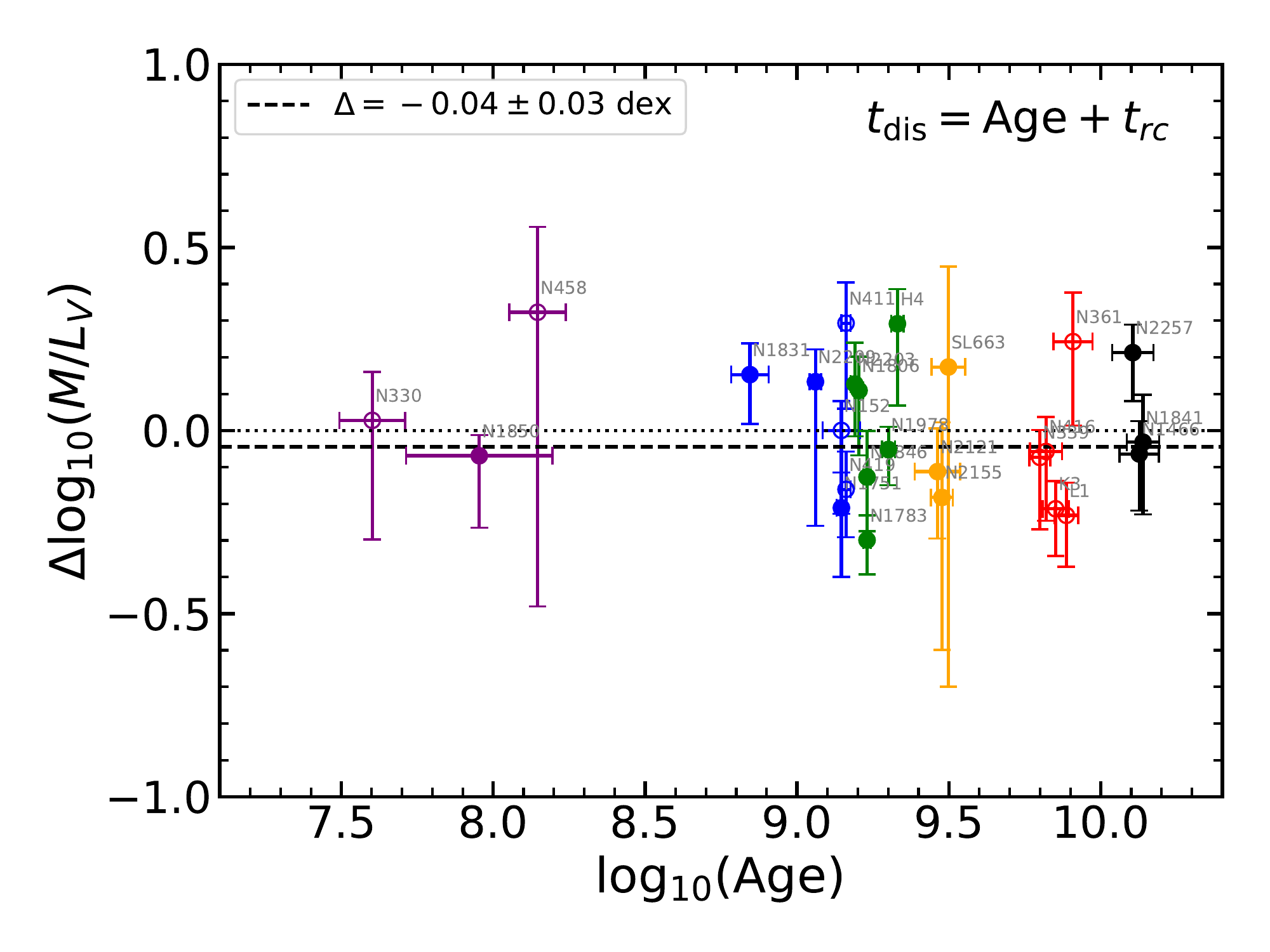}
   \includegraphics[width=0.47\textwidth]{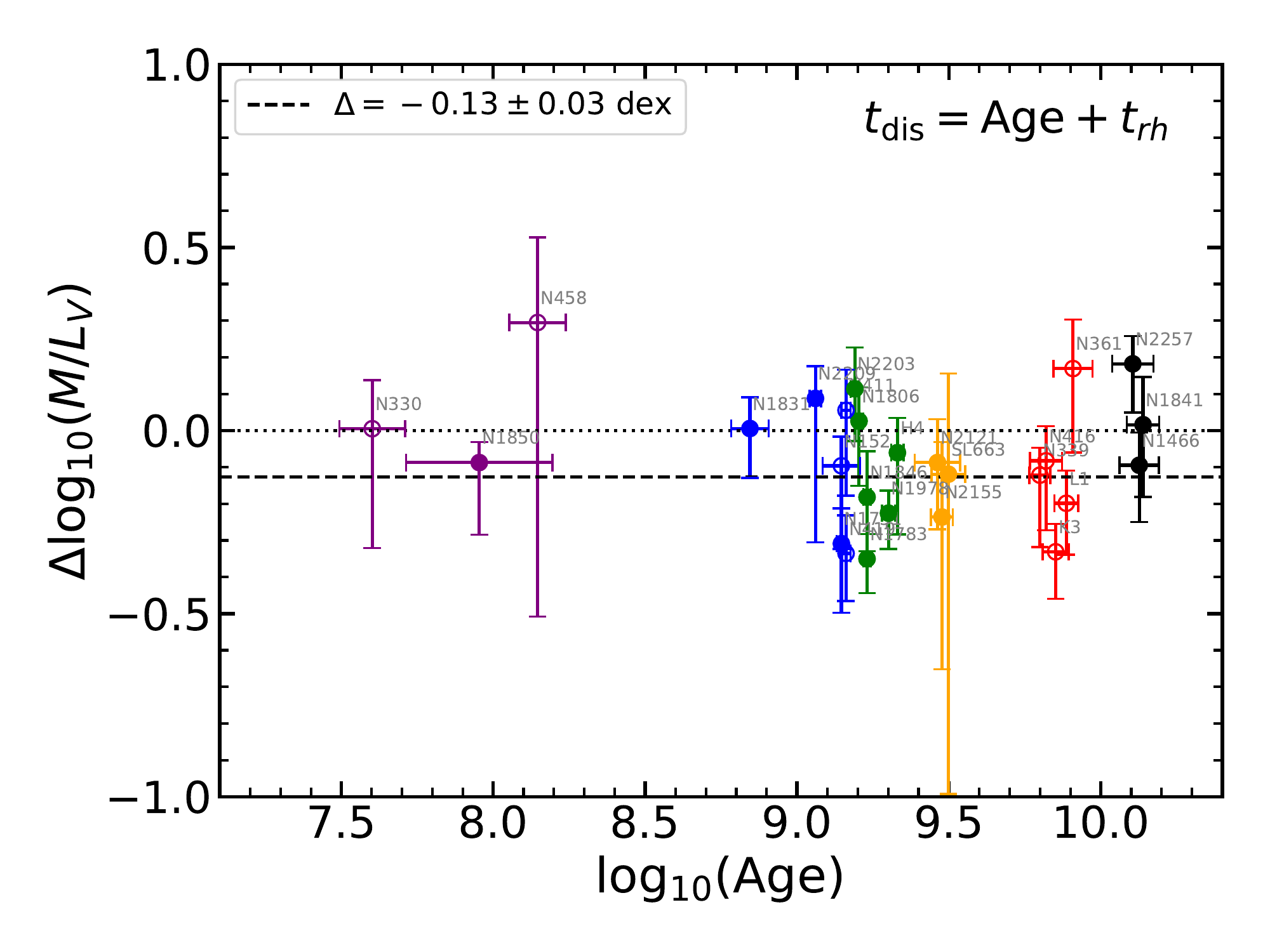}
   \caption{The same as the upper panels of \autoref{fig:MLR_ssp}, but now the colored bands denote results from the synthesis models \citep{Anders:2009aa} that account for cluster dynamical evolution (see \autoref{sec:MLR_anders}). In the top left panel, the color bands denote a range of evolutionary tracks over metallicity from -2.0 (lower band boundaries) to -0.7 (upper band boundaries) dex, and the colored dashed curves denote the mean evolutionary tracks in this metallicity range. Different colors correspond to disruption times of twice the bin ages listed in \autoref{tab:age_bin}. In the remaining panels, the $\Delta\log_{10}{(M/L_V)}$ values were calculated in a cluster-by-cluster manner using the age and metallicity of each cluster and compared with the \citet{Anders:2009aa} models under the assumption that the total disruption time is double the age of the cluster (top right), the sum of the age of the cluster and its current core relaxation time (bottom left), and the sum of the age of the cluster and its current half-mass relaxation time (bottom right), respectively. }
   \label{fig:MLR_age_anders}
\end{figure*}

\subsubsection{Mass underestimates from single-mass models}
\label{sec:MLR_K66}

Our dynamical analysis employs single-mass models---specifically K66 models---that assume equal-mass particles and a constant cluster $M/L_V$ ratio at all radii. Such models do not account for dynamical effects associated with energy equipartition that lead to observable features such as mass segregation and, hence, spatial evolution of $M/L_V$ over time. Due to these effects, we can expect that any single-mass models, such as the K66 models we used, will tend to underestimate the total mass, especially when the kinematic tracers (such as our observed RGs) are more massive than the mean mass of cluster members and therefore kinematically colder and more concentrated in the inner regions of the cluster.

\citet{Sollima:2015aa} explored this effect quantitatively by comparing different analytic models (including the K66 model) used to simulated observations obtained from a suite of $N$-body simulations of star clusters in different stages of their evolution. 
For clusters with high degree of relaxation (i.e., half-mass relaxation time scale $t_{rh}=0.12$ Gyr), they found that the cluster mass can be underestimated up to 50\% of the true value. This accounts for a correction of 0.30 dex in $\log_{10}{(M/L_V)}$ and could, in principle, fully explain the offset seen in \autoref{sec:MLR_trends}. However, it is unlikely that most of our clusters are highly relaxed since some are relatively young and most are comparatively low-density systems.

A more applicable estimation using a longer relaxation timescale (i.e., $t_{rh}=4.97$ Gyr) was also studied by \citet{Sollima:2015aa}. For this case, they found that K66 models can underestimate the true mass of about 10--20\%, depending on the initial cluster mass, the strength of the tidal field and the radial extent of the kinematics tracers (the RGs in our case) used in the dynamical analysis. This is consistent with the work by \citet{Henault-Brunet:2019aa}, who used mock data from a star-by-star $N$-body simulation of M4 to compare mass modelling techniques, including ones using K66 models. They found that the K66 model underestimates the true cluster mass by about 17\%. This corresponds to an offset in $\log_{10}{(M/L_V)}$ of 0.08 dex, insufficient to fully account for the offset between our observed $M/L_V$ and the reference SSP models (though it can reduce the offset to about half of what is observed).  We conclude that our adoption of K66 models tends to underestimate the true cluster mass, but this does not by itself account for  the offset we observe between our measured $M/L_V$ ratios and the reference SSP models (top right panel of \autoref{fig:MLR_ssp}).

\subsubsection{External dynamical effects}
\label{sec:MLR_anders}

After a bound star cluster forms from a dense gas cloud and survives the so-called `infant mortality' stage (the timescale of about 10 Myr for unbound clusters to totally dissolve, \citealp[e.g,][]{Lada:2003aa,Whitmore:2004aa}), its evolution will be driven by both internal and external dynamical effects.  The internal effects include those described in \autoref{sec:MLR_K66} as well as changes in the masses of stars due to mass loss or binary mergers \citep[e.g,][]{Portegies-Zwart:2010aa,Renaud:2018aa}.   External effects can include tidal perturbations from the host galaxy due to either impulsive effects---e.g. encounters with giant molecular clouds, spiral arms or other clusters---or secular evolution arising from a changing tidal field as a cluster orbits within a galaxy \citep[see e.g.][]{Krumholz:2019aa}.  In the reference SSP models used here, generally only the internal effects related to stellar evolution are considered. Ignoring cluster dynamical evolution---both the internal dynamics and the external tidal effects---can lead to overestimated $M/L_V$ values over time (we have seen this to be the case for internal dynamical evolution in \autoref{sec:MLR_K66}).

To test how external dynamical effects may influence cluster $M/L_V$ ratios, we have adapted a suite of evolutionary synthesis models developed by \citet{Anders:2009aa}. Their models are built from the GALEV code \citep[see, e.g.][]{Kotulla:2009aa} using the Padova isochrones, a \citet{Kroupa:2001aa} IMF and BaSeL spectral library (the same as the SSP models we adopted in \autoref{sec:MLR_trends}). These models account for dynamical evolution of star cluster by introducing a mass-dependent parameter---the total cluster disruption time, $t_{95\%}$, defined as the time when 95\% of the initial cluster mass is unbound. This timescale attempts to parameterize the mass-function evolution found in $N$-body simulations of stars clusters dissolving in tidal fields \citep{Baumgardt:2003aa}.

In the top left panel of \autoref{fig:MLR_age_anders}, we compare our empirical $M/L_V$ ratios as a function of age with the \citet{Anders:2009aa} models. For purposes of comparison, we present the \citet{Anders:2009aa} models as colored bands denoting disruption times set to twice the mean ages of each age group (see the figure caption and \autoref{tab:age_bin} for details). The models reveal that external dynamical evolution causes the $M/L_V$ ratio to increase slowly then decrease over time as a cluster ages.  This is in contrast to the SSP predictions where cluster $M/L_V$ values steadily increase with time (the light gray band in the top left panel of \autoref{fig:MLR_age_anders} and, more clearly, in the top left panel of \autoref{fig:MLR_ssp}). 

The remaining panels of \autoref{fig:MLR_age_anders} plot $\Delta\log_{10}{(M/L_V)}$ as a function of age in a cluster-by-cluster manner. The $\Delta\log_{10}{(M/L_V)}$ in each panel are determined against the \citet{Anders:2009aa} models assuming different disruption times in an increasing trend, i.e. the disruption time is assumed to be twice the age of a given cluster (top right), the sum of the cluster age and its current core relaxation time (bottom left), and the sum of the cluster age and its current half-mass relaxation time (bottom right), respectively. The core and half-mass relaxation times of each cluster were derived by using the equations listed in \citet{Mackey:2013aa}, except that we adopted the Coulomb logarithm as ${\Lambda}\sim {0.11N}$ with $N$ being the total number of stars in the cluster \citep{Giersz:1994aa}. All three panels show that our measured $M/L_V$ values agree better with models accounting for cluster evolution than the reference SSP models (as shown in the top right panel of \autoref{fig:MLR_ssp}), as $\Delta\log_{10}{(M/L_V)}$ shrinking from -0.24 to a range between -0.13 and -0.04 according to different assumptions. The differences among these three disruption-timescale assumptions indicate that the clusters in our sample are in quite different evolutionary stages. For instance, the four youngest clusters are more likely dynamically unevolved as given their comparatively long relaxation times relative to their current ages.

Clearly, if we combine mass underestimation discussed in \autoref{sec:MLR_K66}, the offset in $\Delta\log_{10}{(M/L_V)}$ can become negligible, implicitly assuming the two dynamical effects are at least partly independent. The results shown in \autoref{fig:MLR_age_anders} allow us to conclude that internal/external dynamical effects plausibly account for the offset between our measured $M/L_V$ and the SSP models shown in the upper panels of \autoref{fig:MLR_ssp}.

\subsection{Cluster dissolution in the LMC and SMC}
\label{sec:MLR_anders_fit}

\begin{figure*}
\centering
 \includegraphics[width=0.47\textwidth]{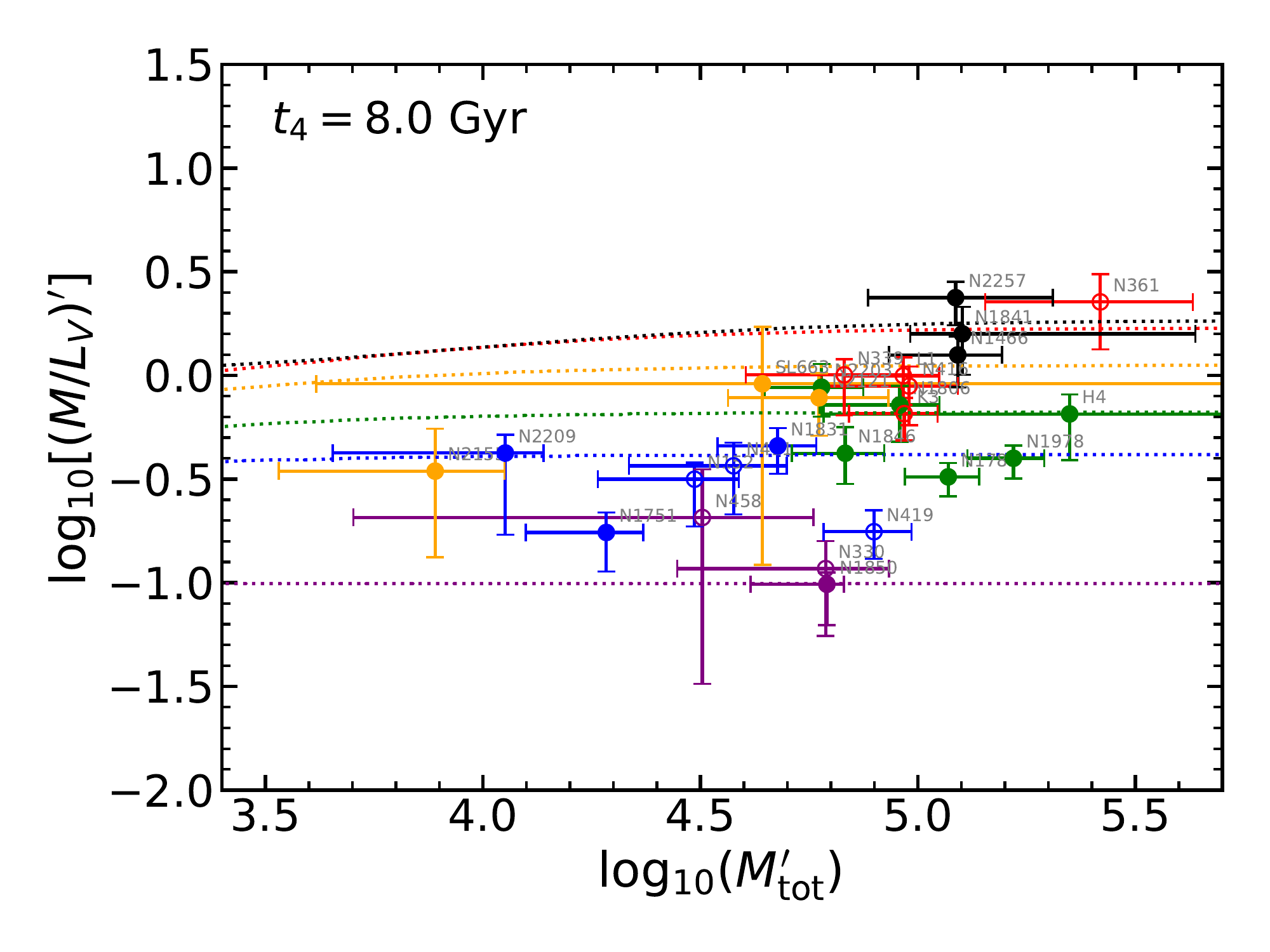}
 \includegraphics[width=0.47\textwidth]{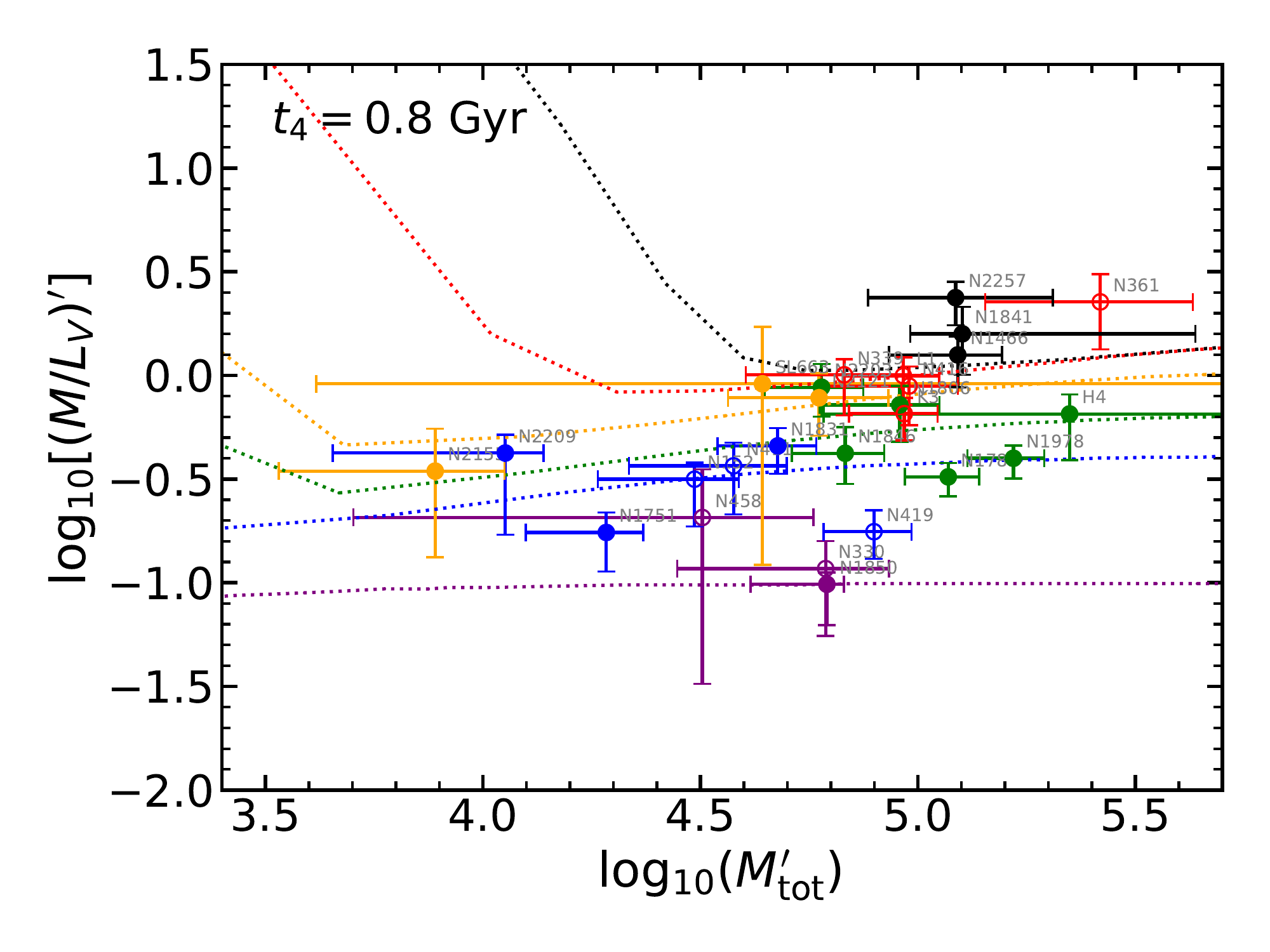}
 \caption{Time evolution of $M/L_V$ ratios as a function of (present-day) cluster mass for two fixed local gravitational field strengths (characterized by the characterized disruption time $t_4$; see \autoref{sec:MLR_anders_fit}). Both panels contain the same data as the bottom left panel in \autoref{fig:MLR_ssp}, but here the colored dotted lines are isochrones calculated from the synthesis models \citep{Anders:2009aa} accounting for dynamical evolution of star clusters. The bin ages and metallicities used to produce the isochrone curves are listed in \autoref{tab:age_bin}.  Small shifts have been applied to the data using the reference SSP models as described in \autoref{sec:MLR_anders_fit} so as to match the ages/metallicities of the associated isochrones (see \autoref{tab:age_bin}; compare to the lower-left panel of \autoref{fig:MLR_ssp}).  }
\label{fig:MLR_mass_anders}
\end{figure*}

For star clusters in a tidal field, the total disruption time depends on the a cluster's initial mass as $t_{\rm {95\%}}\propto M_{i}^{\gamma}$, based on both $N$-body simulations \citep{Baumgardt:2003aa} and observations \citep[e.g.][]{Boutloukos:2003aa, Lamers:2005ab}. The index $\gamma$ has been found to be 0.62 and $0.60 \pm 0.02$, respectively, from $N$-body simulations \citep{Baumgardt:2003aa} and observations of solar-neighborhood open clusters \citep{Lamers:2005aa}. A scaling factor, $t_4$, specifies the disruption time of a $10^4\ M_{\odot}$ star cluster within its host galaxy. 

Adopting a model in which both $t_4$ and the cluster formation rate (CFR) are constant, \citet{Boutloukos:2003aa} derived $\log{t_4}=9.90\pm0.20$ and $\gamma=0.61\pm0.08$ from the analysis of 314 SMC clusters located within 4 kpc from that galaxy's center. For the LMC, \citet{Parmentier:2008aa} used the same approach to constrain $t_4$ but with $\gamma$ set to a fixed value of 0.62. They concluded only that $t_4 \geq 1$ Gyr, principally due to an apparent steady increase of the CFR in the LMC over the past 5 Gyr, negating one of the assumptions of the analysis.

We show here that we can use our $M/L_V$ results to constrain $t_4$ in both galaxies.  We start with the \citet{Anders:2009aa} models as shown in the left panel of \autoref{fig:MLR_age_anders}.  For a given age and metallicity, we can read off a $M/L_V$ corresponding to a given disruption time.  Using Equation 2 from \citet{Anders:2009aa}, we can write for a given cluster
\begin{equation}
\label{eq:tdis_anders}
M_{\rm tot}^{\prime}(t) = f(t)\cdot M_{\rm i} = f(t)\cdot 10^4\,M_\odot\cdot \left(\frac{t_{95\%}}{t_4}\right)^{1/\gamma}\cdot \left[\frac{\mu_{\rm ev}(t_{95\%})}{\mu_{\rm ev}(t_4)}\right]^{-1},
\end{equation} 
where $t$ is the cluster's age.
The parameter $\mu_{\rm ev}$ specifies the mass loss of a cluster due to standard stellar evolutionary effects (e.g. mass loss).  The function $f(t)$ specifies the remaining bound mass fraction of a clusters when both stellar evolution and dynamical effects are considered.  Given a value for $t_4$ and $\gamma$, we can use this procedure to generate isochrones in the $(M/L_V)^{\prime}$-$M_{\rm tot}^{\prime}$ plane.

The left panel of \autoref{fig:MLR_mass_anders} shows a set of such isochrones for $t_4=8.0$ Gyr \citep[the measured value for the SMC by][]{Boutloukos:2003aa}, while the right panel shows results for $t_4 = 0.8$ Gyr \citep[consistent with the lower limit value for the LMC clusters by][]{Parmentier:2008aa}. In both cases, we have adopted $\gamma=0.62$ \citep{Boutloukos:2003aa}. The isochrone colors correspond to the adopted mean ages and metallicities for the bins in which the clusters have been assigned (see the caption).  Since the clusters of a given age bin vary in age and metallicity, we have shifted their positions in \autoref{fig:MLR_mass_anders} (relative to the lower left panel of \autoref{fig:MLR_ssp}) by using the reference SSP model \citep[i.e., the model with a disruption time of 200 Gyr in][]{Anders:2009aa} to determine the small shifts in $M$ and $L_V$ associated with the shift in age and metallicity of each cluster to the corresponding bin values.   The `prime' notation ($M_{\rm tot}^{\prime}$ and $(M/L_V)^{\prime}$) is meant to emphasize that the plotted values have been adjusted from the results shown in \autoref{fig:MLR_ssp}.

In both panels of \autoref{fig:MLR_mass_anders}, we can see that $(M/L_V)^{\prime}$ is constant for the highest-mass and youngest clusters as these systems have not yet attained internal energy equipartition; thus their $(M/L_V)^{\prime}$ values are nearly the same as expected for the reference SSP models (see \autoref{fig:MLR_ssp}). For the $t_4 = 0.8$ Gyr models (right panel), it can also be seen that as age increases, $(M/L_V)^{\prime}$ can increase with decreasing mass.
The reflects the fact that, at any given age, low-mass clusters will have lost more low-mass stars due to energy equipartition \citep{Kruijssen:2008aa}. 
For the lowest mass clusters, 
(e.g. the black and red dotted lines in the right panel of \autoref{fig:MLR_mass_anders}) the increasing fraction of bound stellar remnants near the end stages of cluster dissolution \citep{Anders:2009aa} causes a rapid increase in $(M/L_V)^{\prime}$.  This also implies that clusters found near the minima of the isochrones are very close to complete dissolution.

Note that for the larger value of $t_4$ (8.0 Gyr; left panel in \autoref{fig:MLR_mass_anders}), the isochrones tend to run above the data for clusters of corresponding age.  For the smaller $t_4$ value (0.8 Gyr; right panel), the isochrones systematically match the cluster data better in both the SMC and LMC. The reduced $\chi^2$ values are 5.15 and 2.16 for the cases of $t_4=8.0$ Gyr and 0.8 Gyr, respectively, with a degree of freedom of 25. This indicates that most clusters in our sample have evolved in a relatively strong tidal field with a small $t_4$ value. However, there are two possible exceptions to this conclusion. First, the oldest clusters (in black) appear to agree better with the larger $t_4$ value. These clusters, all associated with the LMC, are located furthest from the galaxy center. This suggests that they may be subjected to a milder tidal field and hence take considerably longer to disrupt.  The second exception is NGC~361 that appears to deviate from the lower-$t_4$ models. This particular cluster has a small kinematic sample that is notably sensitive to contamination by outliers (see \autoref{sec:vd}) and, hence, a potentially large systematic overestimate of its $(M/L_V)^{\prime}$ value. 

Our estimate of $t_4$ based on \autoref{fig:MLR_mass_anders} does not rely on sample completeness, nor any sorts of assumptions regarding the cluster mass functions and cluster formation rates in the MCs.  This suggests that we can use our results to test assumptions used in other studies that have used these clusters to estimate $t_4$.  For the LMC, our estimated value of $t_4 \sim 1$ Gyr is similar to the lower limit found by \citet{Parmentier:2008aa}.  That limit resulted from the assumption of a non-constant cluster formation that had a minimum around 5 Gyr ago.   For the SMC, our estimated value of $t_4$ is considerably shorter than the value of 8.0 Gyr found by \citet{Boutloukos:2003aa} and who assumed a constant CFR.  Our results bring this assumption into question but do not allow us to specify an alternate form of the CFR in the SMC.

\begin{figure*}
   \centering
   \includegraphics[width=0.47\textwidth]{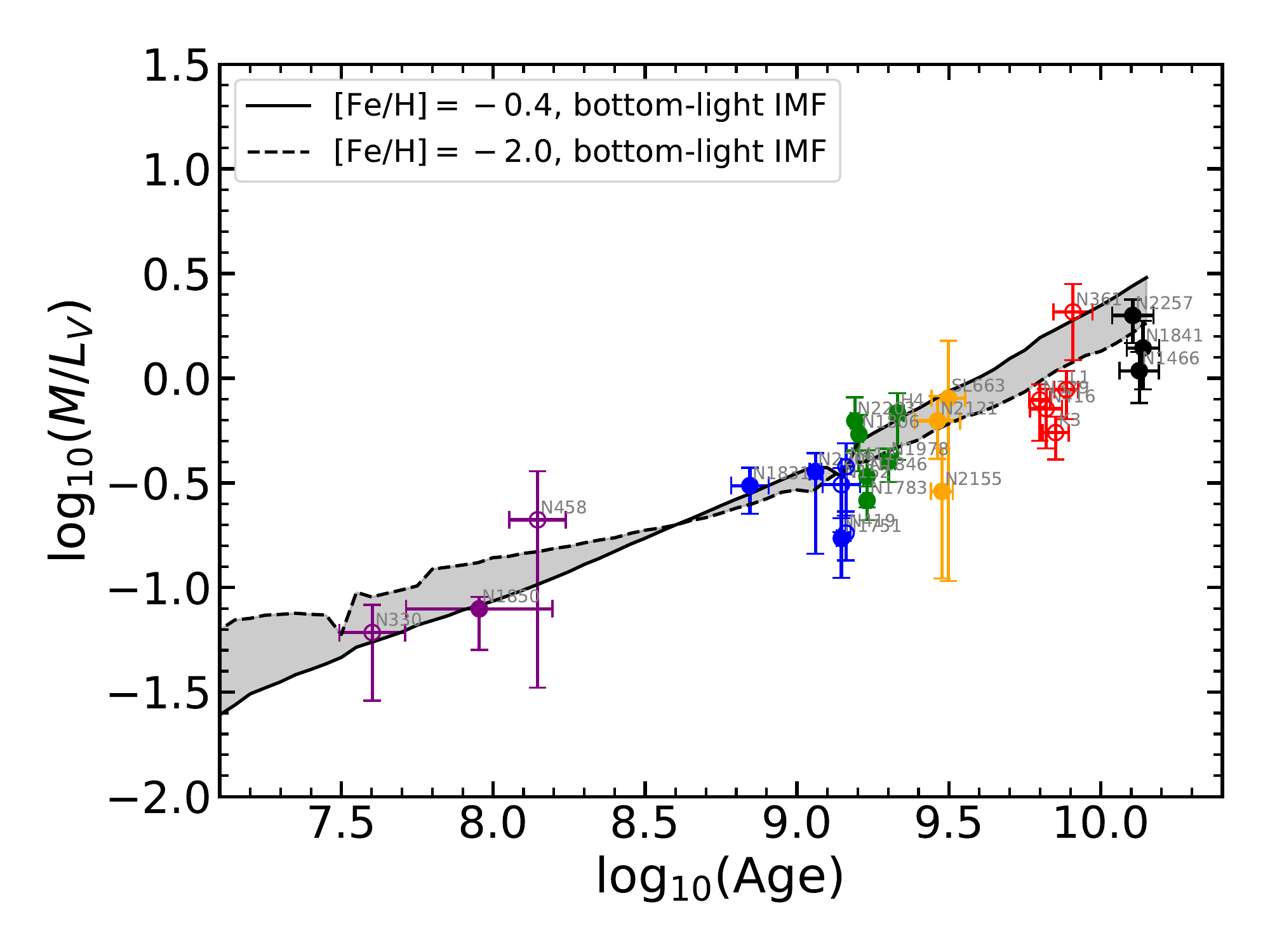}
   \includegraphics[width=0.47\textwidth]{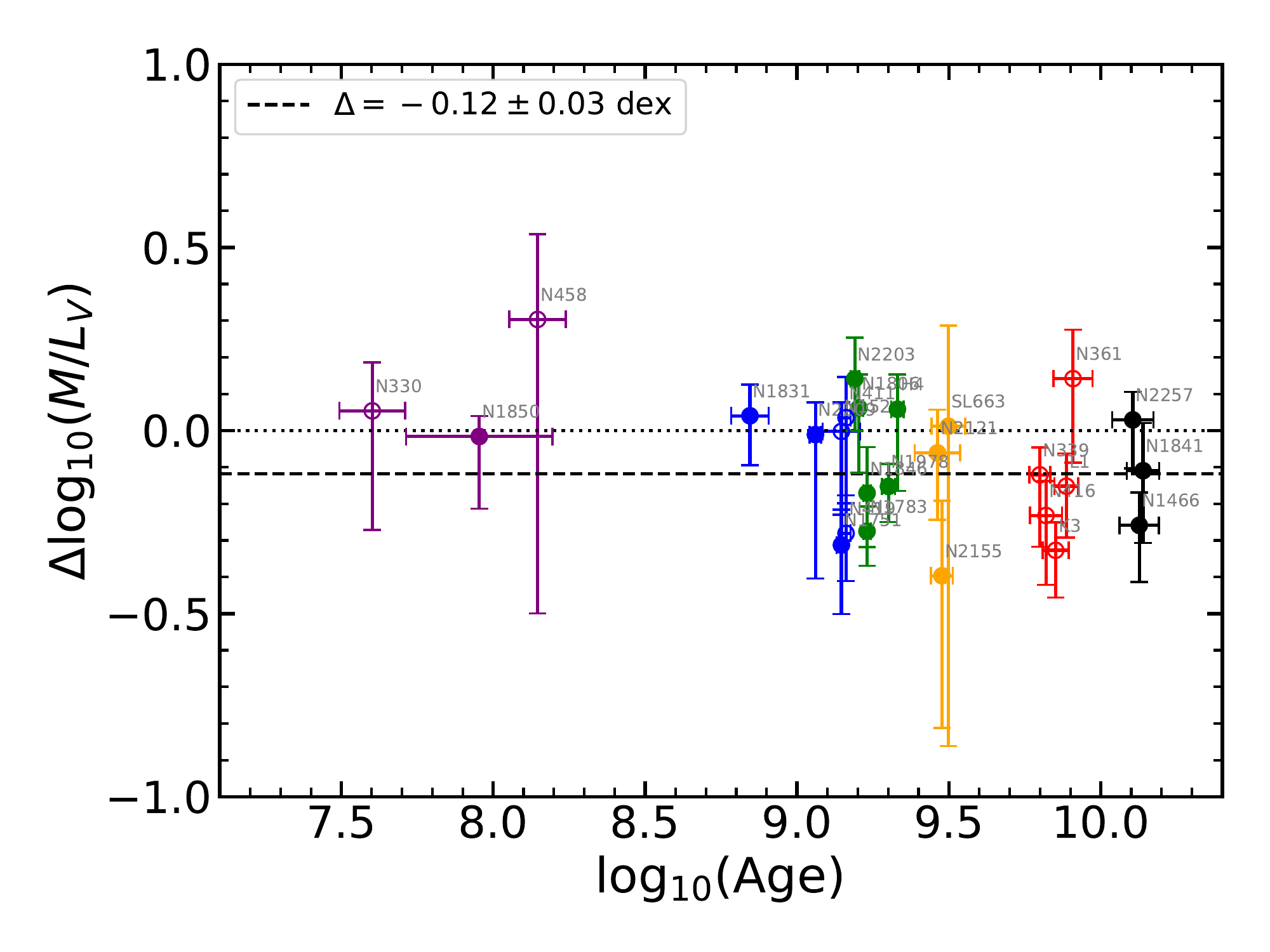}
   \includegraphics[width=0.47\textwidth]{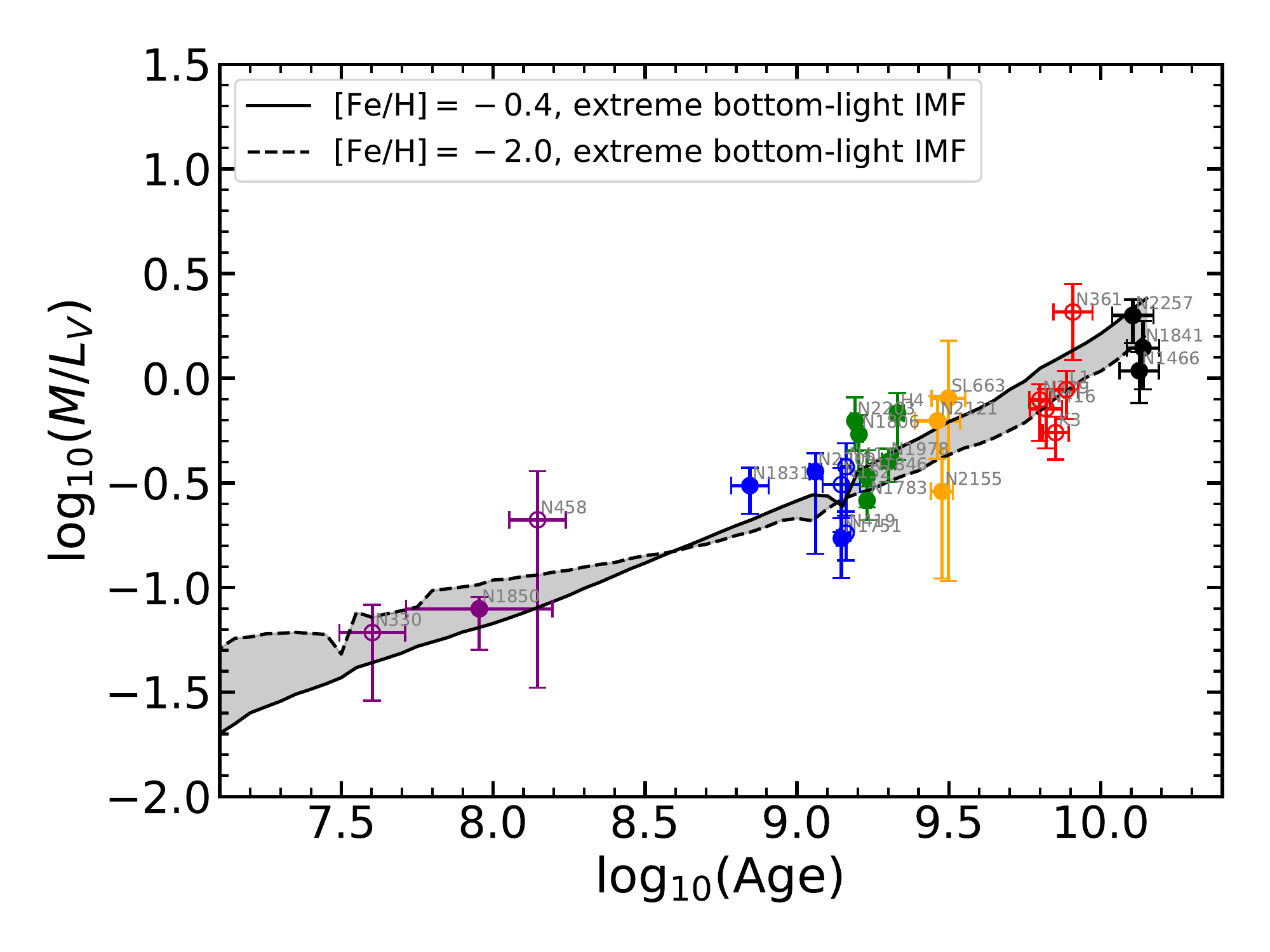}
   \includegraphics[width=0.47\textwidth]{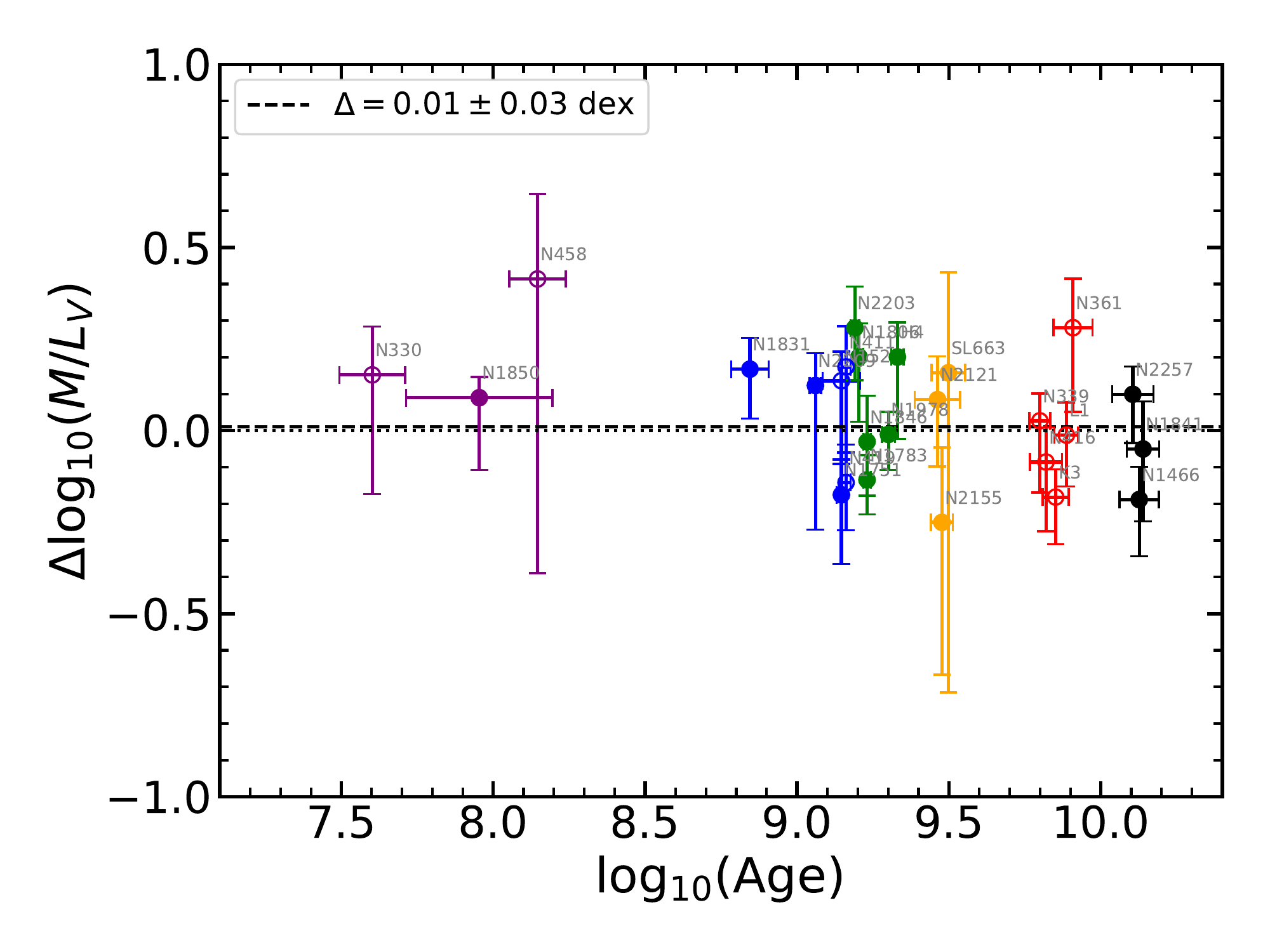}
   \caption{The same as the upper panels of \autoref{fig:MLR_ssp}, but now the solid and dashed curves and their enclosed gray band denote the evolutionary tracks of SSP models with a bottom-light (top panels) and an extreme bottom-light IMFs (bottom panels), respectively. These two IMFs are defined in \autoref{sec:IMF}.  The $\Delta\log_{10}{(M/L_V)}$ values in the right panels were again calculated in a cluster-by-cluster manner using the age and metallicity of each cluster similar to those shown in \autoref{fig:MLR_ssp}.   The colors of the points correspond to the age bins described in \autoref{tab:age_bin}.
   }
   \label{fig:MLR_bottom_light}
\end{figure*}

The LMC cluster NGC~2155 lies closest to the minimum of its corresponding isochrone in the right panel of \autoref{fig:MLR_mass_anders}, indicating, as noted above, that this cluster may be close to complete dissolution \citep{Anders:2009aa}. This is particularly interesting because NGC~2155 is one of the oldest clusters (at 3.0 Gyr; see \autoref{tab:basic}) in the LMC found at the young edge of the well-known cluster age gap of that galaxy \citep{Bertelli:1992aa, Girardi:1995aa, Olszewski:1996aa}. Given this location in \autoref{fig:MLR_mass_anders}, the \citet{Anders:2009aa} models allow us to estimate that the cluster has lost 80--95\%\ of its initial mass. If we assume for NGC~2155 a factor of 10 mass loss over its lifetime, its initial mass would have been around $10^5 M_\odot$, similar to the globular-like LMC clusters in our sample (NGC~1466, NGC~1841 and NGC~2157)\footnote{A similar conclusion may hold for SL~663, the other cluster in our sample of similar age to NGC~2155, but its $M/L_V$ and mass uncertainties (see \autoref{fig:MLR_mass_anders}) make any estimate of its dynamical state or its total mass loss quite uncertain.}. This suggests that many present-day intermediate-age LMC clusters may have started out similar in mass to objects we now consider to be globular clusters, but they have succumbed in a comparatively short time to the disruptive tidal field of the LMC disk due to a considerably shorter value of $t_4$ in that part of the galaxy.

\subsection{Variations in the Stellar Initial Mass Function?}
\label{sec:IMF}

\begin{figure}
   \centering
   \includegraphics[width=0.47\textwidth]{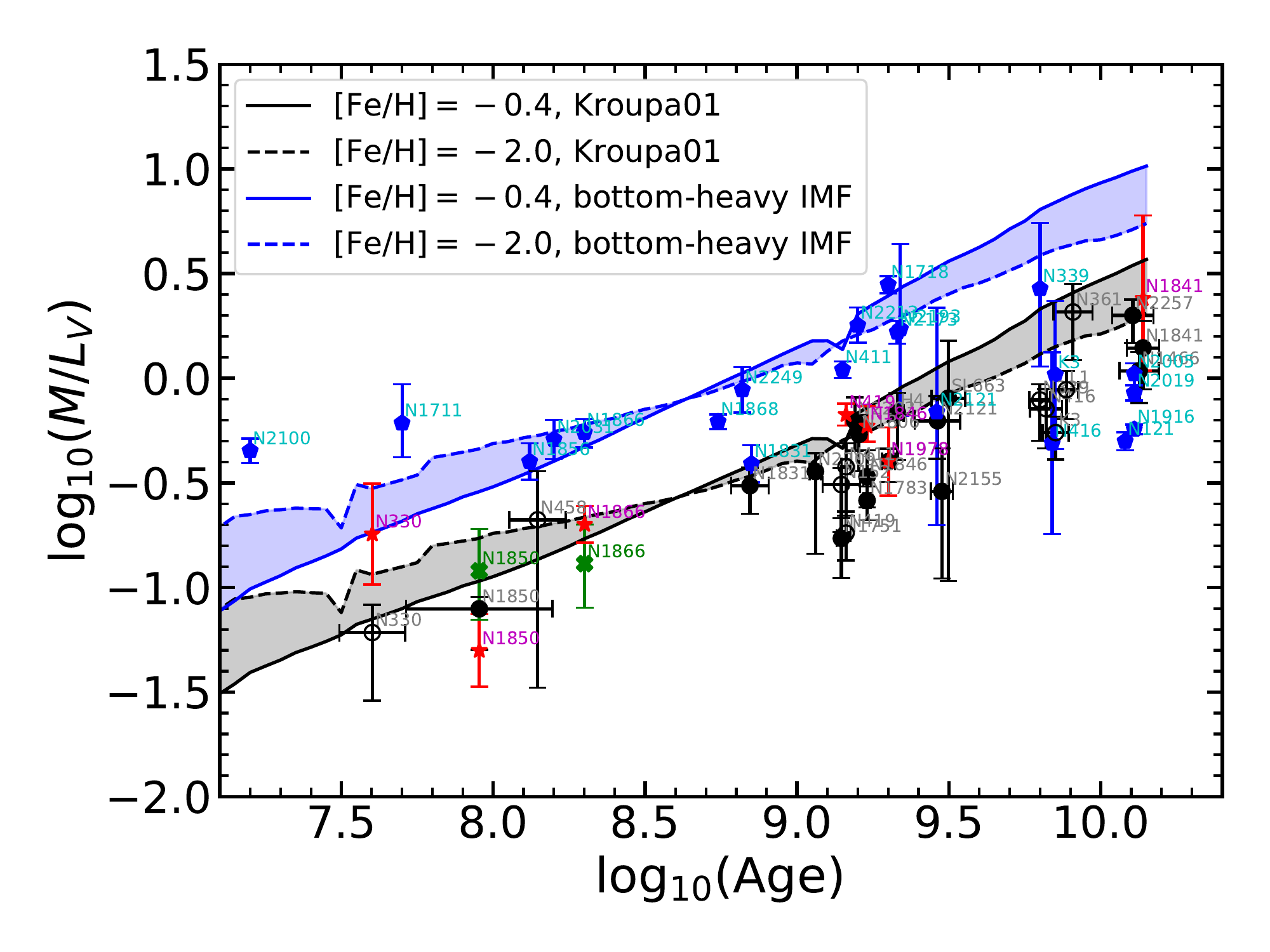}
   \caption{The same as the top left panel of \autoref{fig:MLR_ssp}, but with all $M/L_V$ values from this work shown in black. Over-plotted are the $M/L_V$ results from previously published studies using either integrated-light spectroscopy (in blue; \citealt{Zaritsky:2012aa, Zaritsky:2014aa}) or individual stellar spectra (in red and green). The green data points denote the re-calculated results of \citet{McLaughlin:2005aa} by fitting K66 models to previously published velocity dispersion results (i.e., NGC~1866 from \citealt{Fischer:1992aa} and NGC~1850 from \citealt{Fischer:1993aa}), and hence are labelled differently from other individual-star studies shown in red \citep{Suntzeff:1992aa, Fischer:1992ab, Mackey:2013aa, Kamann:2018ab, Patrick:2020aa}. We also plot the evolutionary tracks of SSP models with a bottom-heavy IMF (blue curves and band) to illustrate the trend of integrated-light results for clusters young than about 3~Gyr (blue pentagons). 
   }
   \label{fig:MLR_comp}
\end{figure}

Up to this point we have only considered the \citet{Kroupa:2001aa} IMF in the reference SSP models we have employed so far in our analysis.  
For an IMF of the form $dN/dM \propto M^{-\alpha}$, the \citet{Kroupa:2001aa} IMF has three mass ranges:
less than $0.5~{\rm M_{\odot}}$, between 0.5 and 1.0~${\rm M_{\odot}}$, and greater than $1.0~{\rm M_{\odot}}$, respectively characterized by $\alpha_1$, $\alpha_2$, and $\alpha_3$ \citep[see also][]{Strader:2011aa}. In the FSPS code \citep{Conroy:2009aa, Conroy:2010aa} with an initial mass range between 0.08 and 100~$\rm M_{\odot}$, the default values for the \citet{Kroupa:2001aa} IMF  are $\alpha_1=1.3$ and $\alpha_2=\alpha_3=2.3$.

Of course, owing to the immense range in $M/L_V$ among stars of different mass, the integrated $M/L_V$ ratio of a stellar population depends sensitively on the form of the initial mass function (IMF). Broadly speaking, `bottom heavy' IMFs---ones with relatively more low-mass stars than the Kroupa IMF---will produce larger $M/L_V$ values, while `bottom light' IMFs will result in lower $M/L_V$ values.  We explore here the extent to which IMF variations alone may account for the observed offset of $M/L_V$ relative to the reference SSP models (upper panels of \autoref{fig:MLR_ssp}). 

Since the observed $M/L_V$ of the clusters in our sample are smaller than SSP model predictions, the argument above implies that a bottom light IMF is needed to reconcile the data and models in the absence of any other effects---such as dynamical evolution as discussed above---which may alter $M/L_V$. We follow the discussion of  \citet{Dalgleish:2020aa} who defined two simple power-law mass functions (MF).  In both cases, $\alpha_3=2.3$ (the same as the Kroupa IMF), but for lower stellar masses between 0.08 and 1.0~$\rm M_{\odot}$, $\alpha_1=\alpha_2=1.3$ for their so-called `bottom-light' MF, and $\alpha_1=\alpha_2=0.3$ for their `extreme bottom-light' MF. In \autoref{fig:MLR_bottom_light} we illustrate how the SSP models are altered by adopting these as IMFs.  Both improve the agreement of the models and cluster data; in the case of the `extreme bottom-light' IMF, the net offset of the data and models is nearly completely accounted for.  

This conclusion contrasts with that of
\citet{Zaritsky:2012aa, Zaritsky:2013aa, Zaritsky:2014aa} who used integrated-light spectroscopy to estimate $M/L_V$ ratios for a sample of MC clusters.  In \autoref{fig:MLR_comp} we plot along with our results these integrated-light measurements as well as $M/L_V$ estimates of MC clusters based on individual-star spectroscopy of intrinsic precision similar to that of our study (\citep{Fischer:1992aa, Fischer:1992ab, Fischer:1993aa, Suntzeff:1992aa, McLaughlin:2005aa, Mackey:2013aa, Kamann:2018ab, Patrick:2020aa}. This comparison highlights some key points.  First, all individual-star results seem to agree systematically over the full range of ages explored by the data.  Moreover, these data run roughly parallel the model expectations. Second, the integrated-light measurements appear to define a relation that is considerably shallower than the SSP models or the individual-star measurements.  Third, modified SSP models based on a bottom-heavy IMF ($\alpha_1=\alpha_2=2.8$ and $\alpha_3=2.3$, plotted in \autoref{fig:MLR_comp}) agree well with the \citet{Zaritsky:2012aa, Zaritsky:2014aa} integrated-light results for clusters younger than about 3 Gyr.  For older clusters, the integrated-light and individual-star $M/L_V$ results broadly agree (see \autoref{sec:comp}). 

Integrated-light studies tend to favor high-concentration clusters with moderate to bright central surface brightnesses, while individual-star studies tend to employ more diffuse and larger clusters where obtaining spectra of distinct stars is more feasible.  Moreover, the integrated-light spectra tend to only consist of contributions from the innermost regions of the clusters.  These may point to a possible physical distinction between the clusters that reveals real IMF variations. 
However, the systematic tendency of the integrated-light results to run high compared to individual-star results, particularly for clusters younger than about 3 Gyr, suggests a more prosaic explanation. For instance, systematic effects in the integrated-light studies have the potential to inflate the line widths (such as blends and focus errors), leading to overestimates of the velocity dispersion \citep{Zaritsky:2012aa}.
Clearly, more studies of clusters observed using both techniques would help explore the nature of this apparent discrepancy.

\section{Summary and Conclusions}
\label{sec:summary}
In this paper, we have presented \textit{Magellan}/M2FS observations of (mostly) red giants in and around 26 Magellanic Cloud star clusters (10 in the SMC and 16 in the LMC) chosen to span the range from $\sim$100 Myr to $\sim$13 Gyr in age, and from $-2.0$ to $-0.4$ in [Fe/H].
We employed an improved version of the data reduction process (described in \citetalias{Song:2019aa}) to extract from the raw data 3137 stellar spectra of 2901 distinct targets. Using stellar effective temperatures estimated with \textit{Gaia} DR2 $G$-band magnitudes (see \autoref{sec:Teff}), we applied a Bayesian methodology to obtain radial velocities, metallicities and surface gravities from these spectra (see \autoref{sec:spec_fit} and \citealt{Song:2017aa}). These parameters were used to identify peculiar targets, such as C stars, binary/blended stars, extreme M supergiants, etc. (see \autoref{sec:rm_all}).  Combined with previously published velocities of individual stars in the clusters of our sample, we have produced a kinematic dataset of 2787 stars suitable for dynamical and chemical analyses.

Using this sample, we have determined membership probabilities of individual stars in each cluster using an Expectation-Maximization (EM) algorithm \citep{Walker:2015aa, Walker:2015ab, Song:2019aa} with the assumption that cluster members are spatially and kinematically distributed as expected for a single-mass K66 model \citep{King:1966aa}. The EM algorithm we used assumes that superimposed on the cluster is a spatially uniform field population that follows a kinematically much broader Gaussian distribution than the cluster population. In order to properly account for the influence of likely non-members, we followed the same approaches developed in \citetalias{Song:2019aa} to assign cluster membership probabilities for the stars in each cluster sample  (\autoref{sec:mem}). We found that for five clusters, the so-called PM50 samples (comprised of stars with membership probabilities greater than 50\%) still have potential field interlopers.  We developed a related approach, PM50$'$, to identify cases where significant contamination by a single unflagged non-member appears to be present (see \autoref{sec:vd}). Using the cluster members in the resulting PM50 or, for a few clusters, PM50$'$ samples, we obtained projected central velocity dispersion of each cluster.  From these, we have derived total masses, $M/L_V$ ratios and mean metallicities for all 26 clusters in our sample. 

Our results exhibit readily understandable trends of $M/L_V$ with cluster age, metallicity and mass (see \autoref{sec:MLR_trends}). When compared with the reference SSP models, we found that our empirical $M/L_V$ values are about 40\% ($-0.24$ dex in $\Delta\log_{10}{(M/L_V)}$) lower than model predictions over the full range of ages exhibited by the clusters in our sample. We explored the origin of this offset by considering two specific dynamical effects (\autoref{sec:MLR_dyn}). First, the single-mass K66 model we adopted do not account for energy equipartition and, hence, mass segregation within clusters.  Consequently, these models will tend to underestimate a cluster's total mass compared to more realistic multi-mass models. $N$-body simulations of star clusters \citep{Sollima:2015aa, Henault-Brunet:2019aa} suggest that this only partially accounts for the offset as we found (about 0.08 dex). 
Second, using modified SSP models that account for cluster evolution in a tidal field \citep{Anders:2009aa}, the $M/L_V$ offset is significantly reduced for reasonable cluster disruption timescale assumptions. 

The observed masses and $M/L_V$ ratios of the clusters in our sample were used to try to constrain timescales of cluster dissolution in the MCs using a simple tidal disruption model (see \autoref{sec:MLR_anders_fit}; also \citealt{Baumgardt:2003aa, Boutloukos:2003aa, Parmentier:2008aa}). Our results suggest that external tidal dynamical effects lead to relatively rapid dissolution in both galaxies ($t_4 \sim 0.8$ Gyr, where $t_4$ is time to disrupt half of a sample of clusters with initial mass $10^4 M_\odot$).  One exception is, perhaps, clusters located well outside the main body of the LMC where a longer $t_4$ is indicated.  For the LMC, this model is consistent with the assumption of a non-constant cluster formation rate (CFR) \citep{Parmentier:2008aa}. Our analysis suggests that the LMC cluster NGC~2155---with an age of 3.0 Gyr---may be close to total dissolution, having already lost 65--90\% of its initial mass. More detailed kinematic studies of this cluster and other LMC clusters near the temporally more recent edge of the `age gap' \citep{Bertelli:1992aa, Girardi:1995aa, Olszewski:1996aa} may provide important constraints on the CFR of this galaxy. 
For the SMC, our analysis suggests a short mean cluster disruption timescale (possibly non-constant CFR in the past), somewhat in contrast to the conclusions of \citep[e.g.][]{Boutloukos:2003aa} who assumed a constant CFR and comparatively long cluster disruption timescale. 

We also considered the effect of varying stellar IMFs among star clusters to explain the offset between observed and model $M/L_V$ values (\autoref{sec:IMF}) and found that an extreme bottom-light IMF could, by itself, almost fully account for the offset (\autoref{fig:MLR_bottom_light}). In contrast, the fact that our observed $M/L_V$ ratios run lower than SSP models with a \citet{Kroupa:2001aa} IMF strongly disfavors a bottom-heavy IMF for these clusters.

A key conclusion of this study is that both dynamical effects and IMF variations can account for the $M/L_V$ values we have measured for our cluster sample (see \autoref{fig:MLR_age_anders}, \autoref{fig:MLR_mass_anders} and \autoref{fig:MLR_bottom_light}).  It is worth noting, however, that while dynamical effects {\it must} be affecting the evolution of MC star clusters, it is not as clear that IMF variations can or must be present.  Thus, we favor the dynamical modifications to SSP models as described in \autoref{sec:MLR_anders} and \autoref{sec:MLR_anders_fit} as the more likely reason to account that our $M/L_V$ results run systematically below the predictions from the reference SSP models.  To the extent that IMF variations may be present, our results require these to be in the form of a bottom-light mass function since bottom-heavy IMFs would cause $M/L_V$ values to run higher than the reference models.  As long as dynamical or minor IMF variations are allowed, we find that present-day SSP models such as those used for our analysis do a remarkably good job of explaining the integrated $M/L_V$ values we observe for MC clusters. 

We end by noting that our conclusion that SSP models perform well is tempered by a few factors.  First, the dynamical models in our analysis are fairly simple.  More realistic models could take into account a spread of stellar masses, anisotropy and possible rotation within clusters. This increased sophistication would come at the price of additional parameters that are likely only weakly constrained by our cluster datasets.  Second, we have not attempted to account for binary stars in our analysis.  We have noted the presence of a some binaries of moderate velocity amplitude in two clusters in our sample (see \autoref{sec:combine_samples}).  Such systems are `easy' cases since they are most likely to be flagged as non-members.  More problematic are binaries with small velocity amplitudes comparable to the clusters' internal dispersions of a few km/s \citep[see][]{Spencer:2018aa}.  Producing empirical constraints on the frequency of such binaries would be a formidable task.  One can imagine that missed binaries would systematically inflate the observed cluster velocity dispersions---and hence their $M/L$ ratios---making the offset between models and data larger than 40\% shown in \autoref{fig:MLR_ssp}.  But the magnitude of the effect remains poorly constrained given the large uncertainties in the relevant binary-population parameters \citep[see][]{Spencer:2018aa}.

\section*{Acknowledgements}
We thank the anonymous referee for helpful comments.  
Y.-Y.S. and M.M. were supported by U.S.\ National Science Foundation (NSF) grants AST-1312997, AST-1726457 and AST-1815403.  
M.G.W. acknowledges support from NSF grant AST-1813881.
I.U.R.\ acknowledges support from NSF grants AST-1613536, AST-1815403, and
PHY-1430152 (Physics Frontier Center/JINA-CEE).
E.O. is partially supported by NSF grant AST-1815767.
M.R. received funding from the European Union’s Horizon 2020 research and innovation programme under the Marie Sklodowska-Curie grant agreement No.~665593 awarded to the Science and Technology Facilities Council.
This paper includes data gathered with the 6.5-meter \textit{Magellan} Telescopes located at Las Campanas Observatory, Chile.
We thank Jeff Crane, Steve Shectman and Ian Thompson for invaluable contributions to the design, construction and support of M2FS.  
We thank the M2FS Team members and telescope operators, especially Meghin Spencer, Daniela Barrientos, Valentino Gonzalez and Terese Hansen, for obtaining the spectroscopic data at the {\it Magellan}/Clay telescope.

\section*{Data availability}
The full version of \autoref{tab:sample_all} in this article is in its online supplementary material.


\bibliographystyle{mnras}
\bibliography{StarCluster.bib} 

\begin{thebibliography}{}
\makeatletter
\relax
\def\mn@urlcharsother{\let\do\@makeother \do\$\do\&\do\#\do\^\do\_\do\%\do\~}
\def\mn@doi{\begingroup\mn@urlcharsother \@ifnextchar [ {\mn@doi@}
  {\mn@doi@[]}}
\def\mn@doi@[#1]#2{\def\@tempa{#1}\ifx\@tempa\@empty \href
  {http://dx.doi.org/#2} {doi:#2}\else \href {http://dx.doi.org/#2} {#1}\fi
  \endgroup}
\def\mn@eprint#1#2{\mn@eprint@#1:#2::\@nil}
\def\mn@eprint@arXiv#1{\href {http://arxiv.org/abs/#1} {{\tt arXiv:#1}}}
\def\mn@eprint@dblp#1{\href {http://dblp.uni-trier.de/rec/bibtex/#1.xml}
  {dblp:#1}}
\def\mn@eprint@#1:#2:#3:#4\@nil{\def\@tempa {#1}\def\@tempb {#2}\def\@tempc
  {#3}\ifx \@tempc \@empty \let \@tempc \@tempb \let \@tempb \@tempa \fi \ifx
  \@tempb \@empty \def\@tempb {arXiv}\fi \@ifundefined
  {mn@eprint@\@tempb}{\@tempb:\@tempc}{\expandafter \expandafter \csname
  mn@eprint@\@tempb\endcsname \expandafter{\@tempc}}}

\bibitem[\protect\citeauthoryear{{Alcaino}}{{Alcaino}}{1978}]{Alcaino:1978aa}
{Alcaino} G.,  1978, \aaps, \href
  {https://ui.adsabs.harvard.edu/abs/1978A&AS...34..431A} {34, 431}

\bibitem[\protect\citeauthoryear{{Alcaino}, {Alvarado}, {Borissova}  \&
  {Kurtev}}{{Alcaino} et~al.}{2003}]{Alcaino:2003aa}
{Alcaino} G.,  {Alvarado} F.,  {Borissova} J.,   {Kurtev} R.,  2003, \mn@doi
  [\aap] {10.1051/0004-6361:20030069}, \href
  {https://ui.adsabs.harvard.edu/abs/2003A&A...400..917A} {400, 917}

\bibitem[\protect\citeauthoryear{{Anders}, {Lamers}  \& {Baumgardt}}{{Anders}
  et~al.}{2009}]{Anders:2009aa}
{Anders} P.,  {Lamers} H.~J.~G.~L.~M.,   {Baumgardt} H.,  2009, \mn@doi [\aap]
  {10.1051/0004-6361/200810615}, \href
  {https://ui.adsabs.harvard.edu/abs/2009A&A...502..817A} {502, 817}

\bibitem[\protect\citeauthoryear{{Anders} et~al.,}{{Anders}
  et~al.}{2019}]{Anders:2019aa}
{Anders} F.,  et~al., 2019, \mn@doi [\aap] {10.1051/0004-6361/201935765}, \href
  {https://ui.adsabs.harvard.edu/abs/2019A&A...628A..94A} {628, A94}

\bibitem[\protect\citeauthoryear{{Baumgardt}}{{Baumgardt}}{2017}]{Baumgardt:2017aa}
{Baumgardt} H.,  2017, \mn@doi [\mnras] {10.1093/mnras/stw2488}, \href
  {https://ui.adsabs.harvard.edu/abs/2017MNRAS.464.2174B} {464, 2174}

\bibitem[\protect\citeauthoryear{{Baumgardt} \& {Hilker}}{{Baumgardt} \&
  {Hilker}}{2018}]{Baumgardt:2018aa}
{Baumgardt} H.,  {Hilker} M.,  2018, \mn@doi [\mnras] {10.1093/mnras/sty1057},
  \href {https://ui.adsabs.harvard.edu/abs/2018MNRAS.478.1520B} {478, 1520}

\bibitem[\protect\citeauthoryear{{Baumgardt} \& {Makino}}{{Baumgardt} \&
  {Makino}}{2003}]{Baumgardt:2003aa}
{Baumgardt} H.,  {Makino} J.,  2003, \mn@doi [\mnras]
  {10.1046/j.1365-8711.2003.06286.x}, \href
  {https://ui.adsabs.harvard.edu/abs/2003MNRAS.340..227B} {340, 227}

\bibitem[\protect\citeauthoryear{{Baumgardt}, {Hilker}, {Sollima}  \&
  {Bellini}}{{Baumgardt} et~al.}{2019}]{Baumgardt:2019aa}
{Baumgardt} H.,  {Hilker} M.,  {Sollima} A.,   {Bellini} A.,  2019, \mn@doi
  [\mnras] {10.1093/mnras/sty2997}, \href
  {https://ui.adsabs.harvard.edu/abs/2019MNRAS.482.5138B} {482, 5138}

\bibitem[\protect\citeauthoryear{{Beers}, {Flynn}  \& {Gebhardt}}{{Beers}
  et~al.}{1990}]{Beers:1990aa}
{Beers} T.~C.,  {Flynn} K.,   {Gebhardt} K.,  1990, \mn@doi [\aj]
  {10.1086/115487}, \href
  {https://ui.adsabs.harvard.edu/abs/1990AJ....100...32B} {100, 32}

\bibitem[\protect\citeauthoryear{{Bell}, {McIntosh}, {Katz}  \&
  {Weinberg}}{{Bell} et~al.}{2003}]{Bell:2003aa}
{Bell} E.~F.,  {McIntosh} D.~H.,  {Katz} N.,   {Weinberg} M.~D.,  2003, \mn@doi
  [\apjs] {10.1086/378847}, \href
  {https://ui.adsabs.harvard.edu/abs/2003ApJS..149..289B} {149, 289}

\bibitem[\protect\citeauthoryear{{Bellini} et~al.,}{{Bellini}
  et~al.}{2014}]{Bellini:2014aa}
{Bellini} A.,  et~al., 2014, \mn@doi [\apj] {10.1088/0004-637X/797/2/115},
  \href {https://ui.adsabs.harvard.edu/abs/2014ApJ...797..115B} {797, 115}

\bibitem[\protect\citeauthoryear{{Bernard}}{{Bernard}}{1975}]{Bernard:1975aa}
{Bernard} A.,  1975, \aap, \href
  {https://ui.adsabs.harvard.edu/abs/1975A&A....40..199B} {40, 199}

\bibitem[\protect\citeauthoryear{{Bertelli}, {Mateo}, {Chiosi}  \&
  {Bressan}}{{Bertelli} et~al.}{1992}]{Bertelli:1992aa}
{Bertelli} G.,  {Mateo} M.,  {Chiosi} C.,   {Bressan} A.,  1992, \mn@doi [\apj]
  {10.1086/171163}, \href
  {https://ui.adsabs.harvard.edu/abs/1992ApJ...388..400B} {388, 400}

\bibitem[\protect\citeauthoryear{{Bica}, {Claria}, {Dottori}, {Santos}  \&
  {Piatti}}{{Bica} et~al.}{1996}]{Bica:1996aa}
{Bica} E.,  {Claria} J.~J.,  {Dottori} H.,  {Santos} J.~F.~C. J.,   {Piatti}
  A.~E.,  1996, \mn@doi [\apjs] {10.1086/192251}, \href
  {https://ui.adsabs.harvard.edu/abs/1996ApJS..102...57B} {102, 57}

\bibitem[\protect\citeauthoryear{{Blanton} \& {Roweis}}{{Blanton} \&
  {Roweis}}{2007}]{Blanton:2007aa}
{Blanton} M.~R.,  {Roweis} S.,  2007, \mn@doi [\aj] {10.1086/510127}, \href
  {https://ui.adsabs.harvard.edu/abs/2007AJ....133..734B} {133, 734}

\bibitem[\protect\citeauthoryear{{Boutloukos} \& {Lamers}}{{Boutloukos} \&
  {Lamers}}{2003}]{Boutloukos:2003aa}
{Boutloukos} S.~G.,  {Lamers} H.~J.~G.~L.~M.,  2003, \mn@doi [\mnras]
  {10.1046/j.1365-8711.2003.06083.x}, \href
  {https://ui.adsabs.harvard.edu/abs/2003MNRAS.338..717B} {338, 717}

\bibitem[\protect\citeauthoryear{{Bressan}, {Marigo}, {Girardi}, {Salasnich},
  {Dal Cero}, {Rubele}  \& {Nanni}}{{Bressan} et~al.}{2012}]{Bressan:2012aa}
{Bressan} A.,  {Marigo} P.,  {Girardi} L.,  {Salasnich} B.,  {Dal Cero} C.,
  {Rubele} S.,   {Nanni} A.,  2012, \mn@doi [\mnras]
  {10.1111/j.1365-2966.2012.21948.x}, \href
  {https://ui.adsabs.harvard.edu/abs/2012MNRAS.427..127B} {427, 127}

\bibitem[\protect\citeauthoryear{{Bruzual} \& {Charlot}}{{Bruzual} \&
  {Charlot}}{2003}]{Bruzual:2003aa}
{Bruzual} G.,  {Charlot} S.,  2003, \mn@doi [\mnras]
  {10.1046/j.1365-8711.2003.06897.x}, \href
  {https://ui.adsabs.harvard.edu/abs/2003MNRAS.344.1000B} {344, 1000}

\bibitem[\protect\citeauthoryear{{Cardelli}, {Clayton}  \& {Mathis}}{{Cardelli}
  et~al.}{1989}]{Cardelli:1989aa}
{Cardelli} J.~A.,  {Clayton} G.~C.,   {Mathis} J.~S.,  1989, \mn@doi [\apj]
  {10.1086/167900}, \href
  {https://ui.adsabs.harvard.edu/abs/1989ApJ...345..245C} {345, 245}

\bibitem[\protect\citeauthoryear{{Carretta} \& {Gratton}}{{Carretta} \&
  {Gratton}}{1997}]{Carretta:1997aa}
{Carretta} E.,  {Gratton} R.~G.,  1997, \mn@doi [\aaps] {10.1051/aas:1997116},
  \href {https://ui.adsabs.harvard.edu/abs/1997A&AS..121...95C} {121, 95}

\bibitem[\protect\citeauthoryear{{Carretta}, {Cohen}, {Gratton}  \&
  {Behr}}{{Carretta} et~al.}{2001}]{Carretta:2001aa}
{Carretta} E.,  {Cohen} J.~G.,  {Gratton} R.~G.,   {Behr} B.~B.,  2001, \mn@doi
  [\aj] {10.1086/322116}, \href
  {https://ui.adsabs.harvard.edu/abs/2001AJ....122.1469C} {122, 1469}

\bibitem[\protect\citeauthoryear{{Carvalho}, {Saurin}, {Bica}, {Bonatto}  \&
  {Schmidt}}{{Carvalho} et~al.}{2008}]{Carvalho:2008aa}
{Carvalho} L.,  {Saurin} T.~A.,  {Bica} E.,  {Bonatto} C.,   {Schmidt} A.~A.,
  2008, \mn@doi [\aap] {10.1051/0004-6361:20079298}, \href
  {https://ui.adsabs.harvard.edu/abs/2008A&A...485...71C} {485, 71}

\bibitem[\protect\citeauthoryear{{Chen} et~al.,}{{Chen}
  et~al.}{2012}]{Chen:2012aa}
{Chen} Y.-M.,  et~al., 2012, \mn@doi [\mnras]
  {10.1111/j.1365-2966.2011.20306.x}, \href
  {https://ui.adsabs.harvard.edu/abs/2012MNRAS.421..314C} {421, 314}

\bibitem[\protect\citeauthoryear{{Choi}, {Dotter}, {Conroy}, {Cantiello},
  {Paxton}  \& {Johnson}}{{Choi} et~al.}{2016}]{Choi:2016aa}
{Choi} J.,  {Dotter} A.,  {Conroy} C.,  {Cantiello} M.,  {Paxton} B.,
  {Johnson} B.~D.,  2016, \mn@doi [\apj] {10.3847/0004-637X/823/2/102}, \href
  {https://ui.adsabs.harvard.edu/abs/2016ApJ...823..102C} {823, 102}

\bibitem[\protect\citeauthoryear{{Conroy} \& {Gunn}}{{Conroy} \&
  {Gunn}}{2010}]{Conroy:2010aa}
{Conroy} C.,  {Gunn} J.~E.,  2010, \mn@doi [\apj]
  {10.1088/0004-637X/712/2/833}, \href
  {https://ui.adsabs.harvard.edu/abs/2010ApJ...712..833C} {712, 833}

\bibitem[\protect\citeauthoryear{{Conroy}, {Gunn}  \& {White}}{{Conroy}
  et~al.}{2009}]{Conroy:2009aa}
{Conroy} C.,  {Gunn} J.~E.,   {White} M.,  2009, \mn@doi [\apj]
  {10.1088/0004-637X/699/1/486}, \href
  {https://ui.adsabs.harvard.edu/abs/2009ApJ...699..486C} {699, 486}

\bibitem[\protect\citeauthoryear{{Correnti}, {Goudfrooij}, {Kalirai},
  {Girardi}, {Puzia}  \& {Kerber}}{{Correnti} et~al.}{2014}]{Correnti:2014aa}
{Correnti} M.,  {Goudfrooij} P.,  {Kalirai} J.~S.,  {Girardi} L.,  {Puzia}
  T.~H.,   {Kerber} L.,  2014, \mn@doi [\apj] {10.1088/0004-637X/793/2/121},
  \href {https://ui.adsabs.harvard.edu/abs/2014ApJ...793..121C} {793, 121}

\bibitem[\protect\citeauthoryear{{Correnti}, {Goudfrooij}, {Bellini}, {Kalirai}
   \& {Puzia}}{{Correnti} et~al.}{2017}]{Correnti:2017aa}
{Correnti} M.,  {Goudfrooij} P.,  {Bellini} A.,  {Kalirai} J.~S.,   {Puzia}
  T.~H.,  2017, \mn@doi [\mnras] {10.1093/mnras/stx010}, \href
  {https://ui.adsabs.harvard.edu/abs/2017MNRAS.467.3628C} {467, 3628}

\bibitem[\protect\citeauthoryear{{Crowl}, {Sarajedini}, {Piatti}, {Geisler},
  {Bica}, {Clari{\'a}}  \& {Santos}}{{Crowl} et~al.}{2001}]{Crowl:2001aa}
{Crowl} H.~H.,  {Sarajedini} A.,  {Piatti} A.~E.,  {Geisler} D.,  {Bica} E.,
  {Clari{\'a}} J.~J.,   {Santos} Jo{\~a}o F.~C. J.,  2001, \mn@doi [\aj]
  {10.1086/321128}, \href
  {https://ui.adsabs.harvard.edu/abs/2001AJ....122..220C} {122, 220}

\bibitem[\protect\citeauthoryear{{Da Costa} \& {Hatzidimitriou}}{{Da Costa} \&
  {Hatzidimitriou}}{1998}]{Da-Costa:1998aa}
{Da Costa} G.~S.,  {Hatzidimitriou} D.,  1998, \mn@doi [\aj] {10.1086/300340},
  \href {https://ui.adsabs.harvard.edu/abs/1998AJ....115.1934D} {115, 1934}

\bibitem[\protect\citeauthoryear{{Dalgleish} et~al.,}{{Dalgleish}
  et~al.}{2020}]{Dalgleish:2020aa}
{Dalgleish} H.,  et~al., 2020, \mn@doi [\mnras] {10.1093/mnras/staa091}, \href
  {https://ui.adsabs.harvard.edu/abs/2020MNRAS.492.3859D} {492, 3859}

\bibitem[\protect\citeauthoryear{{Dias}, {Alessi}, {Moitinho}  \&
  {L{\'e}pine}}{{Dias} et~al.}{2002}]{Dias:2002aa}
{Dias} W.~S.,  {Alessi} B.~S.,  {Moitinho} A.,   {L{\'e}pine} J.~R.~D.,  2002,
  \mn@doi [\aap] {10.1051/0004-6361:20020668}, \href
  {https://ui.adsabs.harvard.edu/abs/2002A&A...389..871D} {389, 871}

\bibitem[\protect\citeauthoryear{{Dolphin}}{{Dolphin}}{2000}]{Dolphin:2000aa}
{Dolphin} A.~E.,  2000, \mn@doi [\pasp] {10.1086/316630}, \href
  {https://ui.adsabs.harvard.edu/abs/2000PASP..112.1383D} {112, 1383}

\bibitem[\protect\citeauthoryear{{Dotter}}{{Dotter}}{2016}]{Dotter:2016aa}
{Dotter} A.,  2016, \mn@doi [\apjs] {10.3847/0067-0049/222/1/8}, \href
  {https://ui.adsabs.harvard.edu/abs/2016ApJS..222....8D} {222, 8}

\bibitem[\protect\citeauthoryear{{Dubath}, {Meylan}  \& {Mayor}}{{Dubath}
  et~al.}{1997}]{Dubath:1997aa}
{Dubath} P.,  {Meylan} G.,   {Mayor} M.,  1997, \aap, \href
  {https://ui.adsabs.harvard.edu/abs/1997A&A...324..505D} {324, 505}

\bibitem[\protect\citeauthoryear{{Evans} et~al.,}{{Evans}
  et~al.}{2018}]{Evans:2018aa}
{Evans} D.~W.,  et~al., 2018, \mn@doi [\aap] {10.1051/0004-6361/201832756},
  \href {https://ui.adsabs.harvard.edu/abs/2018A&A...616A...4E} {616, A4}

\bibitem[\protect\citeauthoryear{{Ferraro}, {Mucciarelli}, {Carretta}  \&
  {Origlia}}{{Ferraro} et~al.}{2006}]{Ferraro:2006aa}
{Ferraro} F.~R.,  {Mucciarelli} A.,  {Carretta} E.,   {Origlia} L.,  2006,
  \mn@doi [\apjl] {10.1086/506178}, \href
  {https://ui.adsabs.harvard.edu/abs/2006ApJ...645L..33F} {645, L33}

\bibitem[\protect\citeauthoryear{{Fischer}, {Welch}, {Cote}, {Mateo}  \&
  {Madore}}{{Fischer} et~al.}{1992a}]{Fischer:1992aa}
{Fischer} P.,  {Welch} D.~L.,  {Cote} P.,  {Mateo} M.,   {Madore} B.~F.,
  1992a, \mn@doi [\aj] {10.1086/116107}, \href
  {https://ui.adsabs.harvard.edu/abs/1992AJ....103..857F} {103, 857}

\bibitem[\protect\citeauthoryear{{Fischer}, {Welch}  \& {Mateo}}{{Fischer}
  et~al.}{1992b}]{Fischer:1992ab}
{Fischer} P.,  {Welch} D.~L.,   {Mateo} M.,  1992b, \mn@doi [\aj]
  {10.1086/116299}, \href
  {https://ui.adsabs.harvard.edu/abs/1992AJ....104.1086F} {104, 1086}

\bibitem[\protect\citeauthoryear{{Fischer}, {Welch}  \& {Mateo}}{{Fischer}
  et~al.}{1993}]{Fischer:1993aa}
{Fischer} P.,  {Welch} D.~L.,   {Mateo} M.,  1993, \mn@doi [\aj]
  {10.1086/116483}, \href
  {https://ui.adsabs.harvard.edu/abs/1993AJ....105..938F} {105, 938}

\bibitem[\protect\citeauthoryear{{Gaia Collaboration} et~al.,}{{Gaia
  Collaboration} et~al.}{2018}]{Gaia-Collaboration:2018aa}
{Gaia Collaboration} et~al., 2018, \mn@doi [\aap]
  {10.1051/0004-6361/201833051}, \href
  {https://ui.adsabs.harvard.edu/abs/2018A&A...616A...1G} {616, A1}

\bibitem[\protect\citeauthoryear{{Gieles} \& {Zocchi}}{{Gieles} \&
  {Zocchi}}{2015}]{Gieles:2015aa}
{Gieles} M.,  {Zocchi} A.,  2015, \mn@doi [\mnras] {10.1093/mnras/stv1848},
  \href {https://ui.adsabs.harvard.edu/abs/2015MNRAS.454..576G} {454, 576}

\bibitem[\protect\citeauthoryear{{Giersz} \& {Heggie}}{{Giersz} \&
  {Heggie}}{1994}]{Giersz:1994aa}
{Giersz} M.,  {Heggie} D.~C.,  1994, \mn@doi [\mnras]
  {10.1093/mnras/268.1.257}, \href
  {https://ui.adsabs.harvard.edu/abs/1994MNRAS.268..257G} {268, 257}

\bibitem[\protect\citeauthoryear{{Girardi}, {Chiosi}, {Bertelli}  \&
  {Bressan}}{{Girardi} et~al.}{1995}]{Girardi:1995aa}
{Girardi} L.,  {Chiosi} C.,  {Bertelli} G.,   {Bressan} A.,  1995, \aap, \href
  {https://ui.adsabs.harvard.edu/abs/1995A&A...298...87G} {298, 87}

\bibitem[\protect\citeauthoryear{{Girardi}, {Bressan}, {Bertelli}  \&
  {Chiosi}}{{Girardi} et~al.}{2000}]{Girardi:2000aa}
{Girardi} L.,  {Bressan} A.,  {Bertelli} G.,   {Chiosi} C.,  2000, \mn@doi
  [\aaps] {10.1051/aas:2000126}, \href
  {https://ui.adsabs.harvard.edu/abs/2000A&AS..141..371G} {141, 371}

\bibitem[\protect\citeauthoryear{{Girardi}, {Bertelli}, {Bressan}, {Chiosi},
  {Groenewegen}, {Marigo}, {Salasnich}  \& {Weiss}}{{Girardi}
  et~al.}{2002}]{Girardi:2002aa}
{Girardi} L.,  {Bertelli} G.,  {Bressan} A.,  {Chiosi} C.,  {Groenewegen}
  M.~A.~T.,  {Marigo} P.,  {Salasnich} B.,   {Weiss} A.,  2002, \mn@doi [\aap]
  {10.1051/0004-6361:20020612}, \href
  {https://ui.adsabs.harvard.edu/abs/2002A&A...391..195G} {391, 195}

\bibitem[\protect\citeauthoryear{{Glatt} et~al.,}{{Glatt}
  et~al.}{2008}]{Glatt:2008ab}
{Glatt} K.,  et~al., 2008, \mn@doi [\aj] {10.1088/0004-6256/136/4/1703}, \href
  {https://ui.adsabs.harvard.edu/abs/2008AJ....136.1703G} {136, 1703}

\bibitem[\protect\citeauthoryear{{Glatt} et~al.,}{{Glatt}
  et~al.}{2009}]{Glatt:2009aa}
{Glatt} K.,  et~al., 2009, \mn@doi [\aj] {10.1088/0004-6256/138/5/1403}, \href
  {https://ui.adsabs.harvard.edu/abs/2009AJ....138.1403G} {138, 1403}

\bibitem[\protect\citeauthoryear{{Goudfrooij}, {Gilmore}, {Kissler-Patig}  \&
  {Maraston}}{{Goudfrooij} et~al.}{2006}]{Goudfrooij:2006aa}
{Goudfrooij} P.,  {Gilmore} D.,  {Kissler-Patig} M.,   {Maraston} C.,  2006,
  \mn@doi [\mnras] {10.1111/j.1365-2966.2006.10314.x}, \href
  {https://ui.adsabs.harvard.edu/abs/2006MNRAS.369..697G} {369, 697}

\bibitem[\protect\citeauthoryear{{Goudfrooij}, {Puzia}, {Kozhurina-Platais}  \&
  {Chandar}}{{Goudfrooij} et~al.}{2009}]{Goudfrooij:2009aa}
{Goudfrooij} P.,  {Puzia} T.~H.,  {Kozhurina-Platais} V.,   {Chandar} R.,
  2009, \mn@doi [\aj] {10.1088/0004-6256/137/6/4988}, \href
  {https://ui.adsabs.harvard.edu/abs/2009AJ....137.4988G} {137, 4988}

\bibitem[\protect\citeauthoryear{{Goudfrooij}, {Puzia}, {Kozhurina-Platais}  \&
  {Chandar}}{{Goudfrooij} et~al.}{2011}]{Goudfrooij:2011aa}
{Goudfrooij} P.,  {Puzia} T.~H.,  {Kozhurina-Platais} V.,   {Chandar} R.,
  2011, \mn@doi [\apj] {10.1088/0004-637X/737/1/3}, \href
  {https://ui.adsabs.harvard.edu/abs/2011ApJ...737....3G} {737, 3}

\bibitem[\protect\citeauthoryear{{Goudfrooij} et~al.,}{{Goudfrooij}
  et~al.}{2014}]{Goudfrooij:2014aa}
{Goudfrooij} P.,  et~al., 2014, \mn@doi [\apj] {10.1088/0004-637X/797/1/35},
  \href {https://ui.adsabs.harvard.edu/abs/2014ApJ...797...35G} {797, 35}

\bibitem[\protect\citeauthoryear{{Grocholski}, {Cole}, {Sarajedini}, {Geisler}
  \& {Smith}}{{Grocholski} et~al.}{2006}]{Grocholski:2006aa}
{Grocholski} A.~J.,  {Cole} A.~A.,  {Sarajedini} A.,  {Geisler} D.,   {Smith}
  V.~V.,  2006, \mn@doi [\aj] {10.1086/507303}, \href
  {https://ui.adsabs.harvard.edu/abs/2006AJ....132.1630G} {132, 1630}

\bibitem[\protect\citeauthoryear{{Grocholski}, {Sarajedini}, {Olsen}, {Tiede}
  \& {Mancone}}{{Grocholski} et~al.}{2007}]{Grocholski:2007aa}
{Grocholski} A.~J.,  {Sarajedini} A.,  {Olsen} K. A.~G.,  {Tiede} G.~P.,
  {Mancone} C.~L.,  2007, \mn@doi [\aj] {10.1086/519735}, \href
  {https://ui.adsabs.harvard.edu/abs/2007AJ....134..680G} {134, 680}

\bibitem[\protect\citeauthoryear{{Gunn} \& {Griffin}}{{Gunn} \&
  {Griffin}}{1979}]{Gunn:1979aa}
{Gunn} J.~E.,  {Griffin} R.~F.,  1979, \mn@doi [\aj] {10.1086/112477}, \href
  {https://ui.adsabs.harvard.edu/abs/1979AJ.....84..752G} {84, 752}

\bibitem[\protect\citeauthoryear{{Harris} \& {Zaritsky}}{{Harris} \&
  {Zaritsky}}{2009}]{Harris:2009aa}
{Harris} J.,  {Zaritsky} D.,  2009, \mn@doi [\aj]
  {10.1088/0004-6256/138/5/1243}, \href
  {https://ui.adsabs.harvard.edu/abs/2009AJ....138.1243H} {138, 1243}

\bibitem[\protect\citeauthoryear{{H{\'e}nault-Brunet}, {Gieles}, {Sollima},
  {Watkins}, {Zocchi}, {Claydon}, {Pancino}  \&
  {Baumgardt}}{{H{\'e}nault-Brunet} et~al.}{2019}]{Henault-Brunet:2019aa}
{H{\'e}nault-Brunet} V.,  {Gieles} M.,  {Sollima} A.,  {Watkins} L.~L.,
  {Zocchi} A.,  {Claydon} I.,  {Pancino} E.,   {Baumgardt} H.,  2019, \mn@doi
  [\mnras] {10.1093/mnras/sty3187}, \href
  {https://ui.adsabs.harvard.edu/abs/2019MNRAS.483.1400H} {483, 1400}

\bibitem[\protect\citeauthoryear{{Illingworth}}{{Illingworth}}{1976}]{Illingworth:1976ab}
{Illingworth} G.,  1976, \mn@doi [\apj] {10.1086/154152}, \href
  {https://ui.adsabs.harvard.edu/abs/1976ApJ...204...73I} {204, 73}

\bibitem[\protect\citeauthoryear{{Jeon}, {Nemec}, {Walker}  \& {Kunder}}{{Jeon}
  et~al.}{2014}]{Jeon:2014aa}
{Jeon} Y.-B.,  {Nemec} J.~M.,  {Walker} A.~R.,   {Kunder} A.~M.,  2014, \mn@doi
  [\aj] {10.1088/0004-6256/147/6/155}, \href
  {https://ui.adsabs.harvard.edu/abs/2014AJ....147..155J} {147, 155}

\bibitem[\protect\citeauthoryear{{Johnson}}{{Johnson}}{1927}]{Johnson:1927aa}
{Johnson} R.~C.,  1927, \mn@doi [Philosophical Transactions of the Royal
  Society of London Series A] {10.1098/rsta.1927.0005}, \href
  {https://ui.adsabs.harvard.edu/abs/1927RSPTA.226..157J} {226, 157}

\bibitem[\protect\citeauthoryear{{Kamann} et~al.,}{{Kamann}
  et~al.}{2016}]{Kamann:2016aa}
{Kamann} S.,  et~al., 2016, \mn@doi [\aap] {10.1051/0004-6361/201527065}, \href
  {https://ui.adsabs.harvard.edu/abs/2016A&A...588A.149K} {588, A149}

\bibitem[\protect\citeauthoryear{{Kamann} et~al.,}{{Kamann}
  et~al.}{2018a}]{Kamann:2018aa}
{Kamann} S.,  et~al., 2018a, \mn@doi [\mnras] {10.1093/mnras/stx2719}, \href
  {https://ui.adsabs.harvard.edu/abs/2018MNRAS.473.5591K} {473, 5591}

\bibitem[\protect\citeauthoryear{{Kamann} et~al.,}{{Kamann}
  et~al.}{2018b}]{Kamann:2018ab}
{Kamann} S.,  et~al., 2018b, \mn@doi [\mnras] {10.1093/mnras/sty1958}, \href
  {https://ui.adsabs.harvard.edu/abs/2018MNRAS.480.1689K} {480, 1689}

\bibitem[\protect\citeauthoryear{{Kauffmann} et~al.,}{{Kauffmann}
  et~al.}{2003}]{Kauffmann:2003aa}
{Kauffmann} G.,  et~al., 2003, \mn@doi [\mnras]
  {10.1046/j.1365-8711.2003.06291.x}, \href
  {https://ui.adsabs.harvard.edu/abs/2003MNRAS.341...33K} {341, 33}

\bibitem[\protect\citeauthoryear{{Kerber}, {Santiago}  \& {Brocato}}{{Kerber}
  et~al.}{2007}]{Kerber:2007aa}
{Kerber} L.~O.,  {Santiago} B.~X.,   {Brocato} E.,  2007, \mn@doi [\aap]
  {10.1051/0004-6361:20066128}, \href
  {https://ui.adsabs.harvard.edu/abs/2007A&A...462..139K} {462, 139}

\bibitem[\protect\citeauthoryear{{Kimmig}, {Seth}, {Ivans}, {Strader},
  {Caldwell}, {Anderton}  \& {Gregersen}}{{Kimmig}
  et~al.}{2015}]{Kimmig:2015aa}
{Kimmig} B.,  {Seth} A.,  {Ivans} I.~I.,  {Strader} J.,  {Caldwell} N.,
  {Anderton} T.,   {Gregersen} D.,  2015, \mn@doi [\aj]
  {10.1088/0004-6256/149/2/53}, \href
  {https://ui.adsabs.harvard.edu/abs/2015AJ....149...53K} {149, 53}

\bibitem[\protect\citeauthoryear{{King}}{{King}}{1948}]{King:1948aa}
{King} R.~B.,  1948, \mn@doi [\apj] {10.1086/145078}, \href
  {https://ui.adsabs.harvard.edu/abs/1948ApJ...108..429K} {108, 429}

\bibitem[\protect\citeauthoryear{{King}}{{King}}{1962}]{King:1962aa}
{King} I.,  1962, \mn@doi [\aj] {10.1086/108756}, \href
  {https://ui.adsabs.harvard.edu/abs/1962AJ.....67..471K} {67, 471}

\bibitem[\protect\citeauthoryear{{King}}{{King}}{1966}]{King:1966aa}
{King} I.~R.,  1966, \mn@doi [\aj] {10.1086/109857}, \href
  {https://ui.adsabs.harvard.edu/abs/1966AJ.....71...64K} {71, 64}

\bibitem[\protect\citeauthoryear{{Kotulla}, {Fritze}, {Weilbacher}  \&
  {Anders}}{{Kotulla} et~al.}{2009}]{Kotulla:2009aa}
{Kotulla} R.,  {Fritze} U.,  {Weilbacher} P.,   {Anders} P.,  2009, \mn@doi
  [\mnras] {10.1111/j.1365-2966.2009.14717.x}, \href
  {https://ui.adsabs.harvard.edu/abs/2009MNRAS.396..462K} {396, 462}

\bibitem[\protect\citeauthoryear{{Kroupa}}{{Kroupa}}{2001}]{Kroupa:2001aa}
{Kroupa} P.,  2001, \mn@doi [\mnras] {10.1046/j.1365-8711.2001.04022.x}, \href
  {https://ui.adsabs.harvard.edu/abs/2001MNRAS.322..231K} {322, 231}

\bibitem[\protect\citeauthoryear{{Kruijssen}}{{Kruijssen}}{2008}]{Kruijssen:2008aa}
{Kruijssen} J.~M.~D.,  2008, \mn@doi [\aap] {10.1051/0004-6361:200810237},
  \href {https://ui.adsabs.harvard.edu/abs/2008A&A...486L..21K} {486, L21}

\bibitem[\protect\citeauthoryear{{Krumholz}, {McKee}  \& {Bland
  -Hawthorn}}{{Krumholz} et~al.}{2019}]{Krumholz:2019aa}
{Krumholz} M.~R.,  {McKee} C.~F.,   {Bland -Hawthorn} J.,  2019, \mn@doi
  [\araa] {10.1146/annurev-astro-091918-104430}, \href
  {https://ui.adsabs.harvard.edu/abs/2019ARA&A..57..227K} {57, 227}

\bibitem[\protect\citeauthoryear{{Lada} \& {Lada}}{{Lada} \&
  {Lada}}{2003}]{Lada:2003aa}
{Lada} C.~J.,  {Lada} E.~A.,  2003, \mn@doi [\araa]
  {10.1146/annurev.astro.41.011802.094844}, \href
  {https://ui.adsabs.harvard.edu/abs/2003ARA&A..41...57L} {41, 57}

\bibitem[\protect\citeauthoryear{{Lamers}, {Gieles}  \& {Portegies
  Zwart}}{{Lamers} et~al.}{2005a}]{Lamers:2005aa}
{Lamers} H.~J.~G.~L.~M.,  {Gieles} M.,   {Portegies Zwart} S.~F.,  2005a,
  \mn@doi [\aap] {10.1051/0004-6361:20041476}, \href
  {https://ui.adsabs.harvard.edu/abs/2005A&A...429..173L} {429, 173}

\bibitem[\protect\citeauthoryear{{Lamers}, {Gieles}, {Bastian}, {Baumgardt},
  {Kharchenko}  \& {Portegies Zwart}}{{Lamers} et~al.}{2005b}]{Lamers:2005ab}
{Lamers} H.~J.~G.~L.~M.,  {Gieles} M.,  {Bastian} N.,  {Baumgardt} H.,
  {Kharchenko} N.~V.,   {Portegies Zwart} S.,  2005b, \mn@doi [\aap]
  {10.1051/0004-6361:20042241}, \href
  {https://ui.adsabs.harvard.edu/abs/2005A&A...441..117L} {441, 117}

\bibitem[\protect\citeauthoryear{{Lane} et~al.,}{{Lane}
  et~al.}{2010}]{Lane:2010ab}
{Lane} R.~R.,  et~al., 2010, \mn@doi [\mnras]
  {10.1111/j.1365-2966.2010.16874.x}, \href
  {https://ui.adsabs.harvard.edu/abs/2010MNRAS.406.2732L} {406, 2732}

\bibitem[\protect\citeauthoryear{{Larsen}, {Brodie}, {Sarajedini}  \&
  {Huchra}}{{Larsen} et~al.}{2002}]{Larsen:2002aa}
{Larsen} S.~S.,  {Brodie} J.~P.,  {Sarajedini} A.,   {Huchra} J.~P.,  2002,
  \mn@doi [\aj] {10.1086/344110}, \href
  {https://ui.adsabs.harvard.edu/abs/2002AJ....124.2615L} {124, 2615}

\bibitem[\protect\citeauthoryear{{Lee} et~al.,}{{Lee}
  et~al.}{2008a}]{Lee:2008aa}
{Lee} Y.~S.,  et~al., 2008a, \mn@doi [\aj] {10.1088/0004-6256/136/5/2022},
  \href {https://ui.adsabs.harvard.edu/abs/2008AJ....136.2022L} {136, 2022}

\bibitem[\protect\citeauthoryear{{Lee} et~al.,}{{Lee}
  et~al.}{2008b}]{Lee:2008ab}
{Lee} Y.~S.,  et~al., 2008b, \mn@doi [\aj] {10.1088/0004-6256/136/5/2050},
  \href {https://ui.adsabs.harvard.edu/abs/2008AJ....136.2050L} {136, 2050}

\bibitem[\protect\citeauthoryear{{Lejeune}, {Cuisinier}  \& {Buser}}{{Lejeune}
  et~al.}{1997}]{Lejeune:1997aa}
{Lejeune} T.,  {Cuisinier} F.,   {Buser} R.,  1997, \mn@doi [\aaps]
  {10.1051/aas:1997373}, \href
  {https://ui.adsabs.harvard.edu/abs/1997A&AS..125..229L} {125, 229}

\bibitem[\protect\citeauthoryear{{Lejeune}, {Cuisinier}  \& {Buser}}{{Lejeune}
  et~al.}{1998}]{Lejeune:1998aa}
{Lejeune} T.,  {Cuisinier} F.,   {Buser} R.,  1998, \mn@doi [\aaps]
  {10.1051/aas:1998405}, \href
  {https://ui.adsabs.harvard.edu/abs/1998A&AS..130...65L} {130, 65}

\bibitem[\protect\citeauthoryear{{Lupton}, {Gunn}  \& {Griffin}}{{Lupton}
  et~al.}{1987}]{Lupton:1987aa}
{Lupton} R.~H.,  {Gunn} J.~E.,   {Griffin} R.~F.,  1987, \mn@doi [\aj]
  {10.1086/114395}, \href
  {https://ui.adsabs.harvard.edu/abs/1987AJ.....93.1114L} {93, 1114}

\bibitem[\protect\citeauthoryear{{Lupton}, {Fall}, {Freeman}  \&
  {Elson}}{{Lupton} et~al.}{1989}]{Lupton:1989aa}
{Lupton} R.~H.,  {Fall} S.~M.,  {Freeman} K.~C.,   {Elson} R. A.~W.,  1989,
  \mn@doi [\apj] {10.1086/168110}, \href
  {https://ui.adsabs.harvard.edu/abs/1989ApJ...347..201L} {347, 201}

\bibitem[\protect\citeauthoryear{{Mackey} \& {Gilmore}}{{Mackey} \&
  {Gilmore}}{2003a}]{Mackey:2003aa}
{Mackey} A.~D.,  {Gilmore} G.~F.,  2003a, \mn@doi [\mnras]
  {10.1046/j.1365-8711.2003.06021.x}, \href
  {https://ui.adsabs.harvard.edu/abs/2003MNRAS.338...85M} {338, 85}

\bibitem[\protect\citeauthoryear{{Mackey} \& {Gilmore}}{{Mackey} \&
  {Gilmore}}{2003b}]{Mackey:2003ab}
{Mackey} A.~D.,  {Gilmore} G.~F.,  2003b, \mn@doi [\mnras]
  {10.1046/j.1365-8711.2003.06022.x}, \href
  {https://ui.adsabs.harvard.edu/abs/2003MNRAS.338..120M} {338, 120}

\bibitem[\protect\citeauthoryear{{Mackey}, {Da Costa}, {Ferguson}  \&
  {Yong}}{{Mackey} et~al.}{2013}]{Mackey:2013aa}
{Mackey} A.~D.,  {Da Costa} G.~S.,  {Ferguson} A.~M.~N.,   {Yong} D.,  2013,
  \mn@doi [\apj] {10.1088/0004-637X/762/1/65}, \href
  {https://ui.adsabs.harvard.edu/abs/2013ApJ...762...65M} {762, 65}

\bibitem[\protect\citeauthoryear{{Ma{\'\i}z Apell{\'a}niz} \&
  {Weiler}}{{Ma{\'\i}z Apell{\'a}niz} \&
  {Weiler}}{2018}]{Maiz-Apellaniz:2018aa}
{Ma{\'\i}z Apell{\'a}niz} J.,  {Weiler} M.,  2018, \mn@doi [\aap]
  {10.1051/0004-6361/201834051}, \href
  {https://ui.adsabs.harvard.edu/abs/2018A&A...619A.180M} {619, A180}

\bibitem[\protect\citeauthoryear{{Mandushev}, {Staneva}  \&
  {Spasova}}{{Mandushev} et~al.}{1991}]{Mandushev:1991aa}
{Mandushev} G.,  {Staneva} A.,   {Spasova} N.,  1991, \aap, \href
  {https://ui.adsabs.harvard.edu/abs/1991A&A...252...94M} {252, 94}

\bibitem[\protect\citeauthoryear{{Maraston}}{{Maraston}}{2005}]{Maraston:2005aa}
{Maraston} C.,  2005, \mn@doi [\mnras] {10.1111/j.1365-2966.2005.09270.x},
  \href {https://ui.adsabs.harvard.edu/abs/2005MNRAS.362..799M} {362, 799}

\bibitem[\protect\citeauthoryear{{Maraston} et~al.,}{{Maraston}
  et~al.}{2013}]{Maraston:2013aa}
{Maraston} C.,  et~al., 2013, \mn@doi [\mnras] {10.1093/mnras/stt1424}, \href
  {https://ui.adsabs.harvard.edu/abs/2013MNRAS.435.2764M} {435, 2764}

\bibitem[\protect\citeauthoryear{{Marigo}, {Girardi}, {Bressan}, {Groenewegen},
  {Silva}  \& {Granato}}{{Marigo} et~al.}{2008}]{Marigo:2008aa}
{Marigo} P.,  {Girardi} L.,  {Bressan} A.,  {Groenewegen} M.~A.~T.,  {Silva}
  L.,   {Granato} G.~L.,  2008, \mn@doi [\aap] {10.1051/0004-6361:20078467},
  \href {https://ui.adsabs.harvard.edu/abs/2008A&A...482..883M} {482, 883}

\bibitem[\protect\citeauthoryear{{Martocchia} et~al.,}{{Martocchia}
  et~al.}{2018}]{Martocchia:2018aa}
{Martocchia} S.,  et~al., 2018, \mn@doi [\mnras] {10.1093/mnras/sty916}, \href
  {https://ui.adsabs.harvard.edu/abs/2018MNRAS.477.4696M} {477, 4696}

\bibitem[\protect\citeauthoryear{{Mateo}, {Welch}  \& {Fischer}}{{Mateo}
  et~al.}{1991}]{Mateo:1991aa}
{Mateo} M.,  {Welch} D.,   {Fischer} P.,  1991, in {Haynes} R.,  {Milne} D.,
  eds,  IAU Symposium Vol. 148, The Magellanic Clouds. p.~191

\bibitem[\protect\citeauthoryear{{Mateo}, {Bailey}, {Crane}, {Shectman},
  {Thompson}, {Roederer}, {Bigelow}  \& {Gunnels}}{{Mateo}
  et~al.}{2012}]{Mateo:2012aa}
{Mateo} M.,  {Bailey} J.~I.,  {Crane} J.,  {Shectman} S.,  {Thompson} I.,
  {Roederer} I.,  {Bigelow} B.,   {Gunnels} S.,  2012, in \procspie. p. 84464Y,
  \mn@doi{10.1117/12.926448}

\bibitem[\protect\citeauthoryear{{McLaughlin} \& {van der Marel}}{{McLaughlin}
  \& {van der Marel}}{2005}]{McLaughlin:2005aa}
{McLaughlin} D.~E.,  {van der Marel} R.~P.,  2005, \mn@doi [\apjs]
  {10.1086/497429}, \href
  {https://ui.adsabs.harvard.edu/abs/2005ApJS..161..304M} {161, 304}

\bibitem[\protect\citeauthoryear{{Mermilliod}}{{Mermilliod}}{1981}]{Mermilliod:1981aa}
{Mermilliod} J.~C.,  1981, \aap, \href
  {https://ui.adsabs.harvard.edu/abs/1981A&A....97..235M} {97, 235}

\bibitem[\protect\citeauthoryear{{Meylan} \& {Mayor}}{{Meylan} \&
  {Mayor}}{1986}]{Meylan:1986aa}
{Meylan} G.,  {Mayor} M.,  1986, \aap, \href
  {https://ui.adsabs.harvard.edu/abs/1986A&A...166..122M} {166, 122}

\bibitem[\protect\citeauthoryear{{Milone}, {Bedin}, {Piotto}  \&
  {Anderson}}{{Milone} et~al.}{2009}]{Milone:2009aa}
{Milone} A.~P.,  {Bedin} L.~R.,  {Piotto} G.,   {Anderson} J.,  2009, \mn@doi
  [\aap] {10.1051/0004-6361/200810870}, \href
  {https://ui.adsabs.harvard.edu/abs/2009A&A...497..755M} {497, 755}

\bibitem[\protect\citeauthoryear{{Milone} et~al.,}{{Milone}
  et~al.}{2018}]{Milone:2018aa}
{Milone} A.~P.,  et~al., 2018, \mn@doi [\mnras] {10.1093/mnras/sty661}, \href
  {https://ui.adsabs.harvard.edu/abs/2018MNRAS.477.2640M} {477, 2640}

\bibitem[\protect\citeauthoryear{{Mucciarelli}, {Origlia}, {Ferraro},
  {Maraston}  \& {Testa}}{{Mucciarelli} et~al.}{2006}]{Mucciarelli:2006aa}
{Mucciarelli} A.,  {Origlia} L.,  {Ferraro} F.~R.,  {Maraston} C.,   {Testa}
  V.,  2006, \mn@doi [\apj] {10.1086/504969}, \href
  {https://ui.adsabs.harvard.edu/abs/2006ApJ...646..939M} {646, 939}

\bibitem[\protect\citeauthoryear{{Mucciarelli}, {Carretta}, {Origlia}  \&
  {Ferraro}}{{Mucciarelli} et~al.}{2008}]{Mucciarelli:2008aa}
{Mucciarelli} A.,  {Carretta} E.,  {Origlia} L.,   {Ferraro} F.~R.,  2008,
  \mn@doi [\aj] {10.1088/0004-6256/136/1/375}, \href
  {https://ui.adsabs.harvard.edu/abs/2008AJ....136..375M} {136, 375}

\bibitem[\protect\citeauthoryear{{Mucciarelli}, {Origlia}  \&
  {Ferraro}}{{Mucciarelli} et~al.}{2010}]{Mucciarelli:2010aa}
{Mucciarelli} A.,  {Origlia} L.,   {Ferraro} F.~R.,  2010, \mn@doi [\apj]
  {10.1088/0004-637X/717/1/277}, \href
  {https://ui.adsabs.harvard.edu/abs/2010ApJ...717..277M} {717, 277}

\bibitem[\protect\citeauthoryear{{Mucciarelli}, {Dalessandro}, {Ferraro},
  {Origlia}  \& {Lanzoni}}{{Mucciarelli} et~al.}{2014}]{Mucciarelli:2014aa}
{Mucciarelli} A.,  {Dalessandro} E.,  {Ferraro} F.~R.,  {Origlia} L.,
  {Lanzoni} B.,  2014, \mn@doi [\apjl] {10.1088/2041-8205/793/1/L6}, \href
  {https://ui.adsabs.harvard.edu/abs/2014ApJ...793L...6M} {793, L6}

\bibitem[\protect\citeauthoryear{{O'Donnell}}{{O'Donnell}}{1994}]{ODonnell:1994aa}
{O'Donnell} J.~E.,  1994, \mn@doi [\apj] {10.1086/173713}, \href
  {https://ui.adsabs.harvard.edu/abs/1994ApJ...422..158O} {422, 158}

\bibitem[\protect\citeauthoryear{{Olszewski}, {Suntzeff}  \&
  {Mateo}}{{Olszewski} et~al.}{1996}]{Olszewski:1996aa}
{Olszewski} E.~W.,  {Suntzeff} N.~B.,   {Mateo} M.,  1996, \mn@doi [\araa]
  {10.1146/annurev.astro.34.1.511}, \href
  {https://ui.adsabs.harvard.edu/abs/1996ARA&A..34..511O} {34, 511}

\bibitem[\protect\citeauthoryear{{Parisi}, {Geisler}, {Clari{\'a}},
  {Villanova}, {Marcionni}, {Sarajedini}  \& {Grocholski}}{{Parisi}
  et~al.}{2015}]{Parisi:2015aa}
{Parisi} M.~C.,  {Geisler} D.,  {Clari{\'a}} J.~J.,  {Villanova} S.,
  {Marcionni} N.,  {Sarajedini} A.,   {Grocholski} A.~J.,  2015, \mn@doi [\aj]
  {10.1088/0004-6256/149/5/154}, \href
  {https://ui.adsabs.harvard.edu/abs/2015AJ....149..154P} {149, 154}

\bibitem[\protect\citeauthoryear{{Parmentier} \& {de Grijs}}{{Parmentier} \&
  {de Grijs}}{2008}]{Parmentier:2008aa}
{Parmentier} G.,  {de Grijs} R.,  2008, \mn@doi [\mnras]
  {10.1111/j.1365-2966.2007.12602.x}, \href
  {https://ui.adsabs.harvard.edu/abs/2008MNRAS.383.1103P} {383, 1103}

\bibitem[\protect\citeauthoryear{{Patrick} et~al.,}{{Patrick}
  et~al.}{2020}]{Patrick:2020aa}
{Patrick} L.~R.,  et~al., 2020, \mn@doi [\aap] {10.1051/0004-6361/201936741},
  \href {https://ui.adsabs.harvard.edu/abs/2020A&A...635A..29P} {635, A29}

\bibitem[\protect\citeauthoryear{{Paxton}, {Bildsten}, {Dotter}, {Herwig},
  {Lesaffre}  \& {Timmes}}{{Paxton} et~al.}{2011}]{Paxton:2011aa}
{Paxton} B.,  {Bildsten} L.,  {Dotter} A.,  {Herwig} F.,  {Lesaffre} P.,
  {Timmes} F.,  2011, \mn@doi [\apjs] {10.1088/0067-0049/192/1/3}, \href
  {https://ui.adsabs.harvard.edu/abs/2011ApJS..192....3P} {192, 3}

\bibitem[\protect\citeauthoryear{{Paxton} et~al.,}{{Paxton}
  et~al.}{2013}]{Paxton:2013aa}
{Paxton} B.,  et~al., 2013, \mn@doi [\apjs] {10.1088/0067-0049/208/1/4}, \href
  {https://ui.adsabs.harvard.edu/abs/2013ApJS..208....4P} {208, 4}

\bibitem[\protect\citeauthoryear{{Paxton} et~al.,}{{Paxton}
  et~al.}{2015}]{Paxton:2015aa}
{Paxton} B.,  et~al., 2015, \mn@doi [\apjs] {10.1088/0067-0049/220/1/15}, \href
  {https://ui.adsabs.harvard.edu/abs/2015ApJS..220...15P} {220, 15}

\bibitem[\protect\citeauthoryear{{Pessev}, {Goudfrooij}, {Puzia}  \& {Chand
  ar}}{{Pessev} et~al.}{2006}]{Pessev:2006aa}
{Pessev} P.~M.,  {Goudfrooij} P.,  {Puzia} T.~H.,   {Chand ar} R.,  2006,
  \mn@doi [\aj] {10.1086/505625}, \href
  {https://ui.adsabs.harvard.edu/abs/2006AJ....132..781P} {132, 781}

\bibitem[\protect\citeauthoryear{{Pessev}, {Goudfrooij}, {Puzia}  \& {Chand
  ar}}{{Pessev} et~al.}{2008}]{Pessev:2008aa}
{Pessev} P.~M.,  {Goudfrooij} P.,  {Puzia} T.~H.,   {Chand ar} R.,  2008,
  \mn@doi [\mnras] {10.1111/j.1365-2966.2008.12935.x}, \href
  {https://ui.adsabs.harvard.edu/abs/2008MNRAS.385.1535P} {385, 1535}

\bibitem[\protect\citeauthoryear{{Pietrinferni}, {Cassisi}, {Salaris}  \&
  {Castelli}}{{Pietrinferni} et~al.}{2004}]{Pietrinferni:2004aa}
{Pietrinferni} A.,  {Cassisi} S.,  {Salaris} M.,   {Castelli} F.,  2004,
  \mn@doi [\apj] {10.1086/422498}, \href
  {https://ui.adsabs.harvard.edu/abs/2004ApJ...612..168P} {612, 168}

\bibitem[\protect\citeauthoryear{{Portegies Zwart}, {McMillan}  \&
  {Gieles}}{{Portegies Zwart} et~al.}{2010}]{Portegies-Zwart:2010aa}
{Portegies Zwart} S.~F.,  {McMillan} S. L.~W.,   {Gieles} M.,  2010, \mn@doi
  [\araa] {10.1146/annurev-astro-081309-130834}, \href
  {https://ui.adsabs.harvard.edu/abs/2010ARA&A..48..431P} {48, 431}

\bibitem[\protect\citeauthoryear{{Pryor} \& {Meylan}}{{Pryor} \&
  {Meylan}}{1993}]{Pryor:1993aa}
{Pryor} C.,  {Meylan} G.,  1993, in {Djorgovski} S.~G.,  {Meylan} G.,  eds,
  Astronomical Society of the Pacific Conference Series Vol. 50, Structure and
  Dynamics of Globular Clusters. p.~357

\bibitem[\protect\citeauthoryear{{Renaud}}{{Renaud}}{2018}]{Renaud:2018aa}
{Renaud} F.,  2018, \mn@doi [\nar] {10.1016/j.newar.2018.03.001}, \href
  {https://ui.adsabs.harvard.edu/abs/2018NewAR..81....1R} {81, 1}

\bibitem[\protect\citeauthoryear{{S{\'a}nchez-Bl{\'a}zquez}
  et~al.,}{{S{\'a}nchez-Bl{\'a}zquez} et~al.}{2006}]{Sanchez-Blazquez:2006aa}
{S{\'a}nchez-Bl{\'a}zquez} P.,  et~al., 2006, \mn@doi [\mnras]
  {10.1111/j.1365-2966.2006.10699.x}, \href
  {https://ui.adsabs.harvard.edu/abs/2006MNRAS.371..703S} {371, 703}

\bibitem[\protect\citeauthoryear{{Sollima}, {Baumgardt}, {Zocchi}, {Balbinot},
  {Gieles}, {H{\'e}nault-Brunet}  \& {Varri}}{{Sollima}
  et~al.}{2015}]{Sollima:2015aa}
{Sollima} A.,  {Baumgardt} H.,  {Zocchi} A.,  {Balbinot} E.,  {Gieles} M.,
  {H{\'e}nault-Brunet} V.,   {Varri} A.~L.,  2015, \mn@doi [\mnras]
  {10.1093/mnras/stv1079}, \href
  {https://ui.adsabs.harvard.edu/abs/2015MNRAS.451.2185S} {451, 2185}

\bibitem[\protect\citeauthoryear{{Song}, {Mateo}, {Walker}  \&
  {Roederer}}{{Song} et~al.}{2017}]{Song:2017aa}
{Song} Y.-Y.,  {Mateo} M.,  {Walker} M.~G.,   {Roederer} I.~U.,  2017, \mn@doi
  [\aj] {10.3847/1538-3881/aa6eaa}, \href
  {https://ui.adsabs.harvard.edu/abs/2017AJ....153..261S} {153, 261}

\bibitem[\protect\citeauthoryear{{Song}, {Mateo}, {Mackey}, {Olszewski},
  {Roederer}, {Walker}  \& {Bailey}}{{Song} et~al.}{2019}]{Song:2019aa}
{Song} Y.-Y.,  {Mateo} M.,  {Mackey} A.~D.,  {Olszewski} E.~W.,  {Roederer}
  I.~U.,  {Walker} M.~G.,   {Bailey} J.~I.,  2019, \mn@doi [\mnras]
  {10.1093/mnras/stz2502}, \href
  {https://ui.adsabs.harvard.edu/abs/2019MNRAS.490..385S} {490, 385}

\bibitem[\protect\citeauthoryear{{Spencer}, {Mateo}, {Olszewski}, {Walker},
  {McConnachie}  \& {Kirby}}{{Spencer} et~al.}{2018}]{Spencer:2018aa}
{Spencer} M.~E.,  {Mateo} M.,  {Olszewski} E.~W.,  {Walker} M.~G.,
  {McConnachie} A.~W.,   {Kirby} E.~N.,  2018, \mn@doi [\aj]
  {10.3847/1538-3881/aae3e4}, \href
  {https://ui.adsabs.harvard.edu/abs/2018AJ....156..257S} {156, 257}

\bibitem[\protect\citeauthoryear{{Strader}, {Smith}, {Larsen}, {Brodie}  \&
  {Huchra}}{{Strader} et~al.}{2009}]{Strader:2009aa}
{Strader} J.,  {Smith} G.~H.,  {Larsen} S.,  {Brodie} J.~P.,   {Huchra} J.~P.,
  2009, \mn@doi [\aj] {10.1088/0004-6256/138/2/547}, \href
  {https://ui.adsabs.harvard.edu/abs/2009AJ....138..547S} {138, 547}

\bibitem[\protect\citeauthoryear{{Strader}, {Caldwell}  \& {Seth}}{{Strader}
  et~al.}{2011}]{Strader:2011aa}
{Strader} J.,  {Caldwell} N.,   {Seth} A.~C.,  2011, \mn@doi [\aj]
  {10.1088/0004-6256/142/1/8}, \href
  {https://ui.adsabs.harvard.edu/abs/2011AJ....142....8S} {142, 8}

\bibitem[\protect\citeauthoryear{{Suntzeff}, {Schommer}, {Olszewski}  \&
  {Walker}}{{Suntzeff} et~al.}{1992}]{Suntzeff:1992aa}
{Suntzeff} N.~B.,  {Schommer} R.~A.,  {Olszewski} E.~W.,   {Walker} A.~R.,
  1992, \mn@doi [\aj] {10.1086/116356}, \href
  {https://ui.adsabs.harvard.edu/abs/1992AJ....104.1743S} {104, 1743}

\bibitem[\protect\citeauthoryear{{Tojeiro}, {Wilkins}, {Heavens}, {Panter}  \&
  {Jimenez}}{{Tojeiro} et~al.}{2009}]{Tojeiro:2009aa}
{Tojeiro} R.,  {Wilkins} S.,  {Heavens} A.~F.,  {Panter} B.,   {Jimenez} R.,
  2009, \mn@doi [\apjs] {10.1088/0067-0049/185/1/1}, \href
  {https://ui.adsabs.harvard.edu/abs/2009ApJS..185....1T} {185, 1}

\bibitem[\protect\citeauthoryear{{VandenBerg}, {Brogaard}, {Leaman}  \&
  {Casagrand e}}{{VandenBerg} et~al.}{2013}]{VandenBerg:2013aa}
{VandenBerg} D.~A.,  {Brogaard} K.,  {Leaman} R.,   {Casagrand e} L.,  2013,
  \mn@doi [\apj] {10.1088/0004-637X/775/2/134}, \href
  {https://ui.adsabs.harvard.edu/abs/2013ApJ...775..134V} {775, 134}

\bibitem[\protect\citeauthoryear{{Vazdekis}, {S{\'a}nchez-Bl{\'a}zquez},
  {Falc{\'o}n-Barroso}, {Cenarro}, {Beasley}, {Cardiel}, {Gorgas}  \&
  {Peletier}}{{Vazdekis} et~al.}{2010}]{Vazdekis:2010aa}
{Vazdekis} A.,  {S{\'a}nchez-Bl{\'a}zquez} P.,  {Falc{\'o}n-Barroso} J.,
  {Cenarro} A.~J.,  {Beasley} M.~A.,  {Cardiel} N.,  {Gorgas} J.,   {Peletier}
  R.~F.,  2010, \mn@doi [\mnras] {10.1111/j.1365-2966.2010.16407.x}, \href
  {https://ui.adsabs.harvard.edu/abs/2010MNRAS.404.1639V} {404, 1639}

\bibitem[\protect\citeauthoryear{{Wagner-Kaiser} et~al.,}{{Wagner-Kaiser}
  et~al.}{2017}]{Wagner-Kaiser:2017aa}
{Wagner-Kaiser} R.,  et~al., 2017, \mn@doi [\mnras] {10.1093/mnras/stx1702},
  \href {https://ui.adsabs.harvard.edu/abs/2017MNRAS.471.3347W} {471, 3347}

\bibitem[\protect\citeauthoryear{{Walker}, {Olszewski}  \& {Mateo}}{{Walker}
  et~al.}{2015a}]{Walker:2015aa}
{Walker} M.~G.,  {Olszewski} E.~W.,   {Mateo} M.,  2015a, \mn@doi [\mnras]
  {10.1093/mnras/stv099}, \href
  {https://ui.adsabs.harvard.edu/abs/2015MNRAS.448.2717W} {448, 2717}

\bibitem[\protect\citeauthoryear{{Walker}, {Mateo}, {Olszewski}, {Bailey},
  {Koposov}, {Belokurov}  \& {Evans}}{{Walker} et~al.}{2015b}]{Walker:2015ab}
{Walker} M.~G.,  {Mateo} M.,  {Olszewski} E.~W.,  {Bailey} John~I. I.,
  {Koposov} S.~E.,  {Belokurov} V.,   {Evans} N.~W.,  2015b, \mn@doi [\apj]
  {10.1088/0004-637X/808/2/108}, \href
  {https://ui.adsabs.harvard.edu/abs/2015ApJ...808..108W} {808, 108}

\bibitem[\protect\citeauthoryear{{Watkins}, {van der Marel}, {Bellini}  \&
  {Anderson}}{{Watkins} et~al.}{2015}]{Watkins:2015aa}
{Watkins} L.~L.,  {van der Marel} R.~P.,  {Bellini} A.,   {Anderson} J.,  2015,
  \mn@doi [\apj] {10.1088/0004-637X/803/1/29}, \href
  {https://ui.adsabs.harvard.edu/abs/2015ApJ...803...29W} {803, 29}

\bibitem[\protect\citeauthoryear{{Weiler}}{{Weiler}}{2018}]{Weiler:2018aa}
{Weiler} M.,  2018, \mn@doi [\aap] {10.1051/0004-6361/201833462}, \href
  {https://ui.adsabs.harvard.edu/abs/2018A&A...617A.138W} {617, A138}

\bibitem[\protect\citeauthoryear{{Wenger} et~al.,}{{Wenger}
  et~al.}{2000}]{Wenger:2000aa}
{Wenger} M.,  et~al., 2000, \mn@doi [\aaps] {10.1051/aas:2000332}, \href
  {https://ui.adsabs.harvard.edu/abs/2000A&AS..143....9W} {143, 9}

\bibitem[\protect\citeauthoryear{{Westera}, {Lejeune}, {Buser}, {Cuisinier}  \&
  {Bruzual}}{{Westera} et~al.}{2002}]{Westera:2002aa}
{Westera} P.,  {Lejeune} T.,  {Buser} R.,  {Cuisinier} F.,   {Bruzual} G.,
  2002, \mn@doi [\aap] {10.1051/0004-6361:20011493}, \href
  {https://ui.adsabs.harvard.edu/abs/2002A&A...381..524W} {381, 524}

\bibitem[\protect\citeauthoryear{{Whitmore}}{{Whitmore}}{2004}]{Whitmore:2004aa}
{Whitmore} B.~C.,  2004, in {Lamers} H. J.~G.~L.~M.,  {Smith} L.~J.,   {Nota}
  A.,  eds,  Astronomical Society of the Pacific Conference Series Vol. 322,
  The Formation and Evolution of Massive Young Star Clusters. p.~419
  (\mn@eprint {arXiv} {astro-ph/0403709})

\bibitem[\protect\citeauthoryear{{Zacharias}, {Monet}, {Levine}, {Urban},
  {Gaume}  \& {Wycoff}}{{Zacharias} et~al.}{2004}]{Zacharias:2004aa}
{Zacharias} N.,  {Monet} D.~G.,  {Levine} S.~E.,  {Urban} S.~E.,  {Gaume} R.,
  {Wycoff} G.~L.,  2004, in American Astronomical Society Meeting Abstracts. p.
  48.15

\bibitem[\protect\citeauthoryear{{Zaritsky}, {Harris}, {Thompson}, {Grebel}  \&
  {Massey}}{{Zaritsky} et~al.}{2002}]{Zaritsky:2002aa}
{Zaritsky} D.,  {Harris} J.,  {Thompson} I.~B.,  {Grebel} E.~K.,   {Massey} P.,
   2002, \mn@doi [\aj] {10.1086/338437}, \href
  {https://ui.adsabs.harvard.edu/abs/2002AJ....123..855Z} {123, 855}

\bibitem[\protect\citeauthoryear{{Zaritsky}, {Harris}, {Thompson}  \&
  {Grebel}}{{Zaritsky} et~al.}{2004}]{Zaritsky:2004aa}
{Zaritsky} D.,  {Harris} J.,  {Thompson} I.~B.,   {Grebel} E.~K.,  2004,
  \mn@doi [\aj] {10.1086/423910}, \href
  {https://ui.adsabs.harvard.edu/abs/2004AJ....128.1606Z} {128, 1606}

\bibitem[\protect\citeauthoryear{{Zaritsky}, {Colucci}, {Pessev}, {Bernstein}
  \& {Chandar}}{{Zaritsky} et~al.}{2012}]{Zaritsky:2012aa}
{Zaritsky} D.,  {Colucci} J.~E.,  {Pessev} P.~M.,  {Bernstein} R.~A.,
  {Chandar} R.,  2012, \mn@doi [\apj] {10.1088/0004-637X/761/2/93}, \href
  {https://ui.adsabs.harvard.edu/abs/2012ApJ...761...93Z} {761, 93}

\bibitem[\protect\citeauthoryear{{Zaritsky}, {Colucci}, {Pessev}, {Bernstein}
  \& {Chandar}}{{Zaritsky} et~al.}{2013}]{Zaritsky:2013aa}
{Zaritsky} D.,  {Colucci} J.~E.,  {Pessev} P.~M.,  {Bernstein} R.~A.,
  {Chandar} R.,  2013, \mn@doi [\apj] {10.1088/0004-637X/770/2/121}, \href
  {https://ui.adsabs.harvard.edu/abs/2013ApJ...770..121Z} {770, 121}

\bibitem[\protect\citeauthoryear{{Zaritsky}, {Colucci}, {Pessev}, {Bernstein}
  \& {Chandar}}{{Zaritsky} et~al.}{2014}]{Zaritsky:2014aa}
{Zaritsky} D.,  {Colucci} J.~E.,  {Pessev} P.~M.,  {Bernstein} R.~A.,
  {Chandar} R.,  2014, \mn@doi [\apj] {10.1088/0004-637X/796/2/71}, \href
  {https://ui.adsabs.harvard.edu/abs/2014ApJ...796...71Z} {796, 71}

\bibitem[\protect\citeauthoryear{{Zinn} \& {West}}{{Zinn} \&
  {West}}{1984}]{Zinn:1984aa}
{Zinn} R.,  {West} M.~J.,  1984, \mn@doi [\apjs] {10.1086/190947}, \href
  {https://ui.adsabs.harvard.edu/abs/1984ApJS...55...45Z} {55, 45}

\bibitem[\protect\citeauthoryear{{van den Bergh}}{{van den
  Bergh}}{1981}]{van-den-Bergh:1981aa}
{van den Bergh} S.,  1981, \aaps, \href
  {https://ui.adsabs.harvard.edu/abs/1981A&AS...46...79V} {46, 79}

\makeatother
\end{thebibliography}



\appendix

\section{Determining Cluster Centers}
\label{sec:cluster_center}

\begin{figure*}
   \centering
    \includegraphics[width=0.47\textwidth]{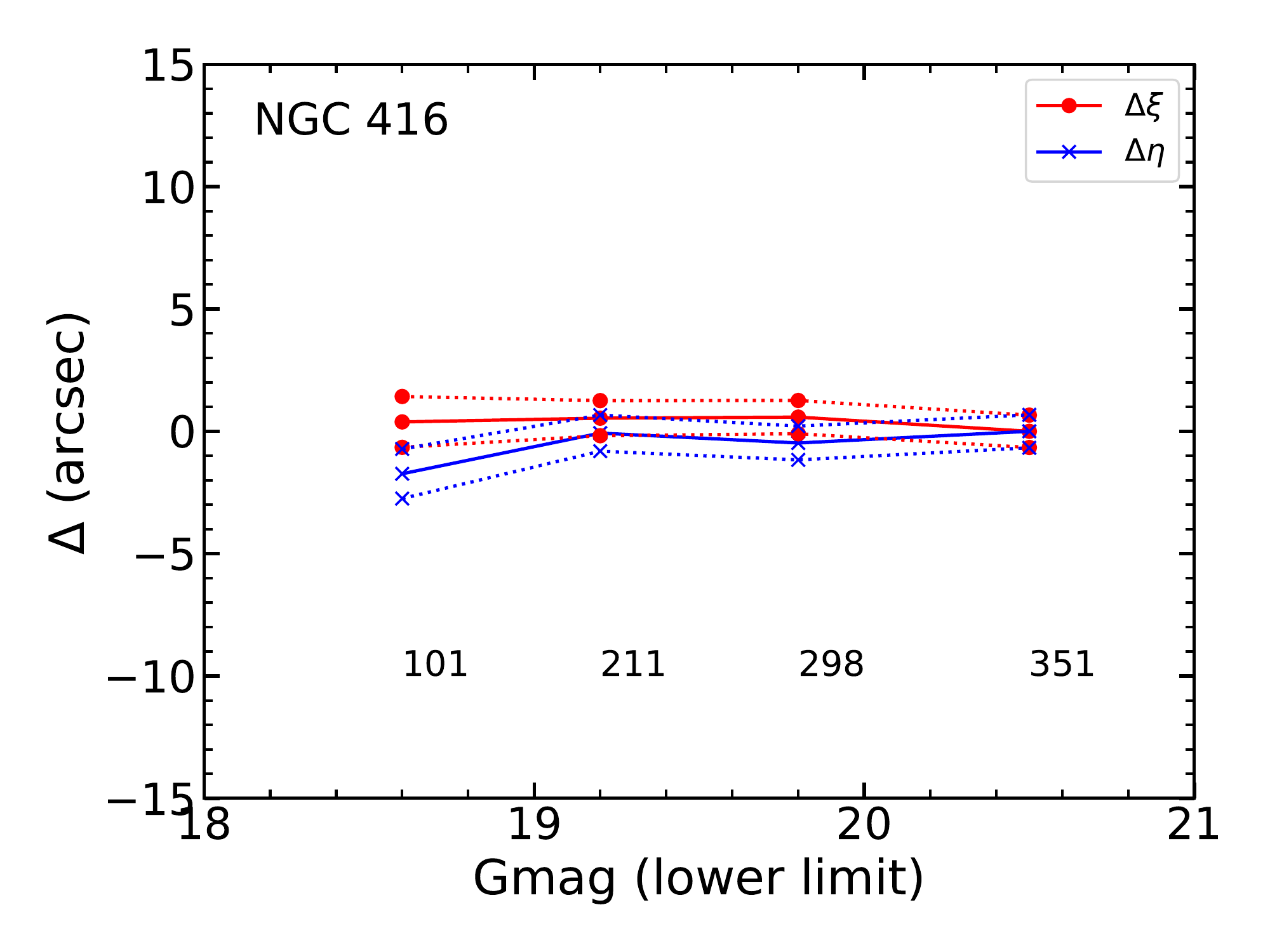}
    \includegraphics[width=0.47\textwidth]{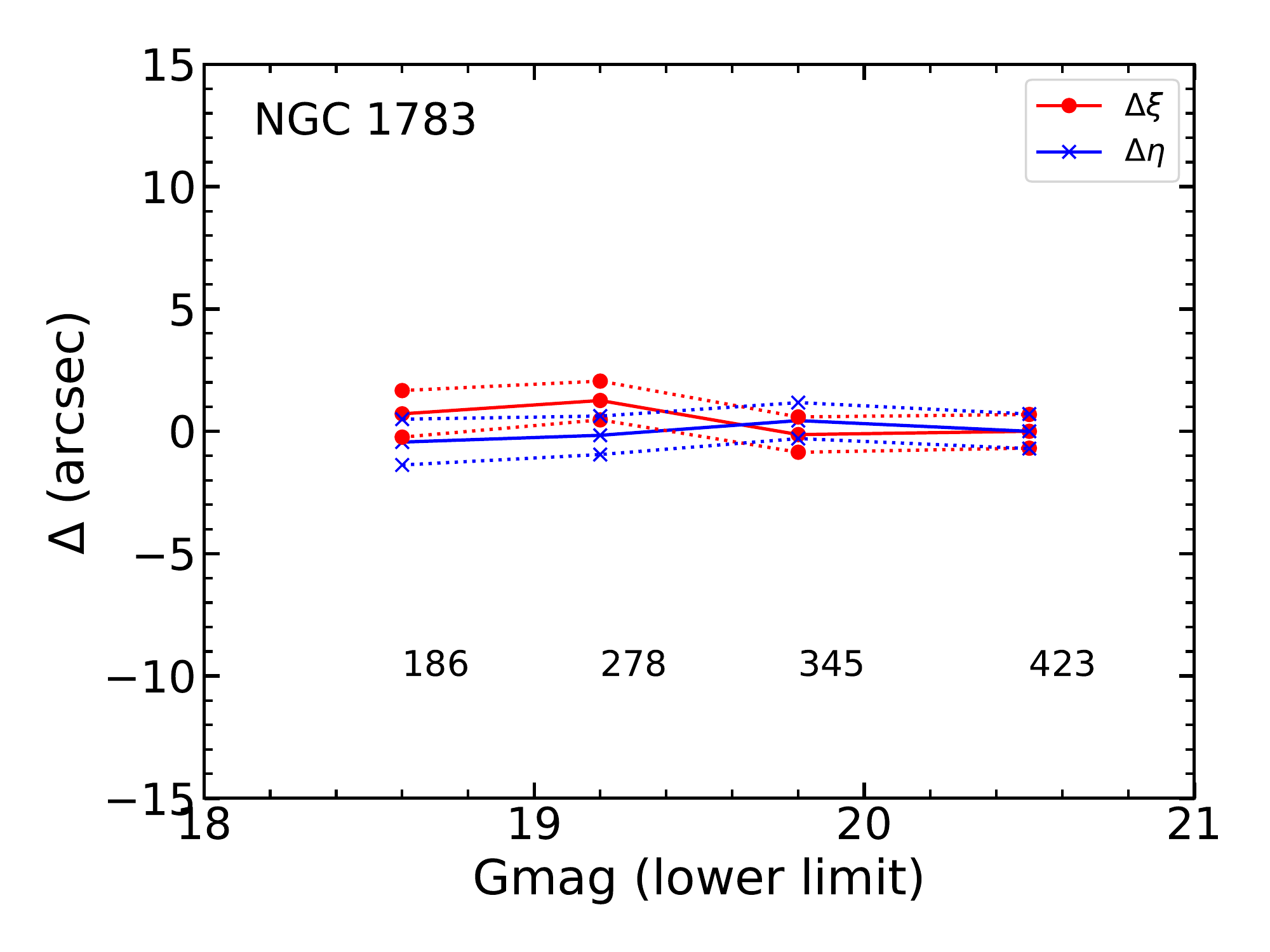}
    \includegraphics[width=0.47\textwidth]{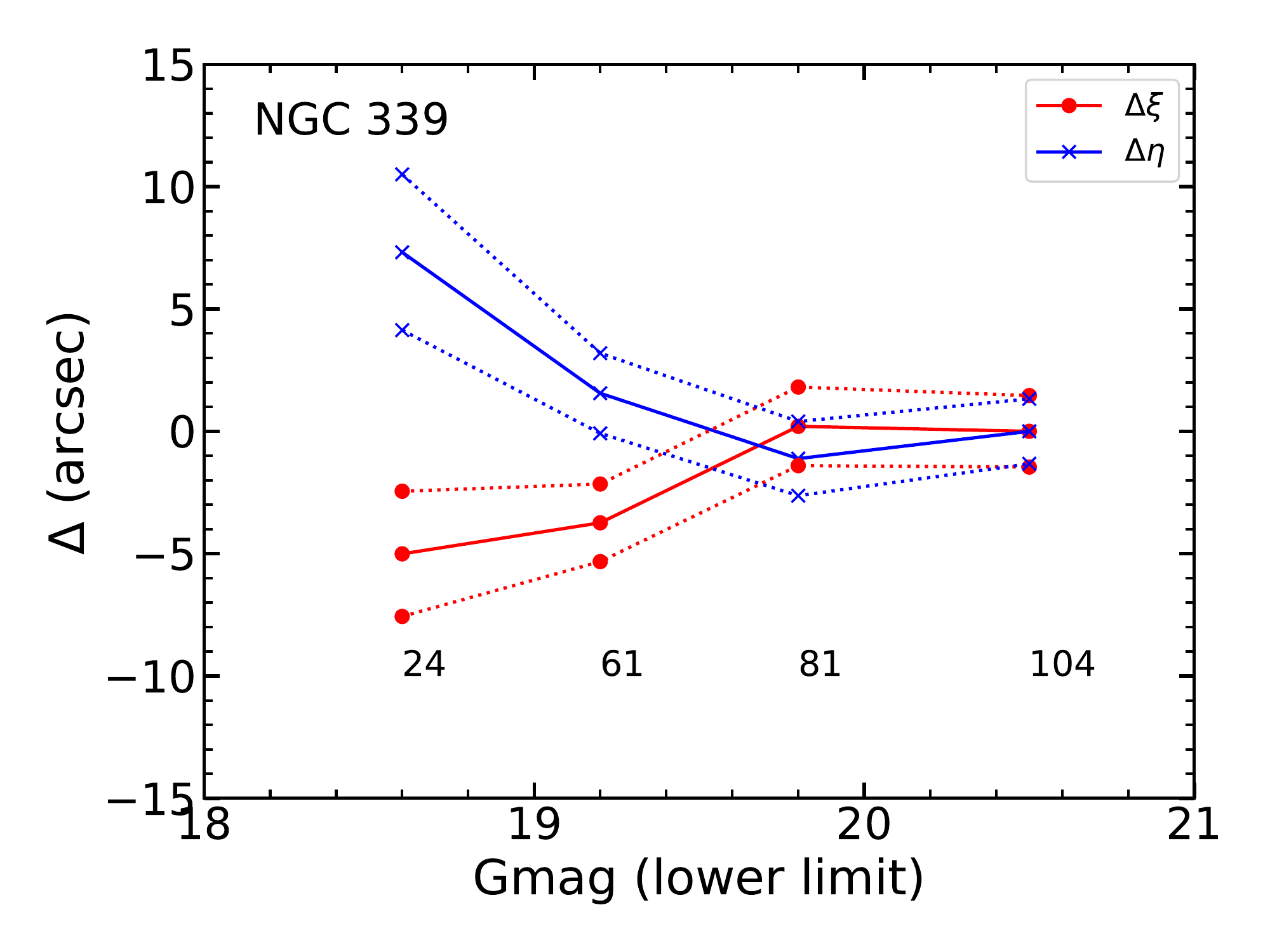}
    \includegraphics[width=0.47\textwidth]{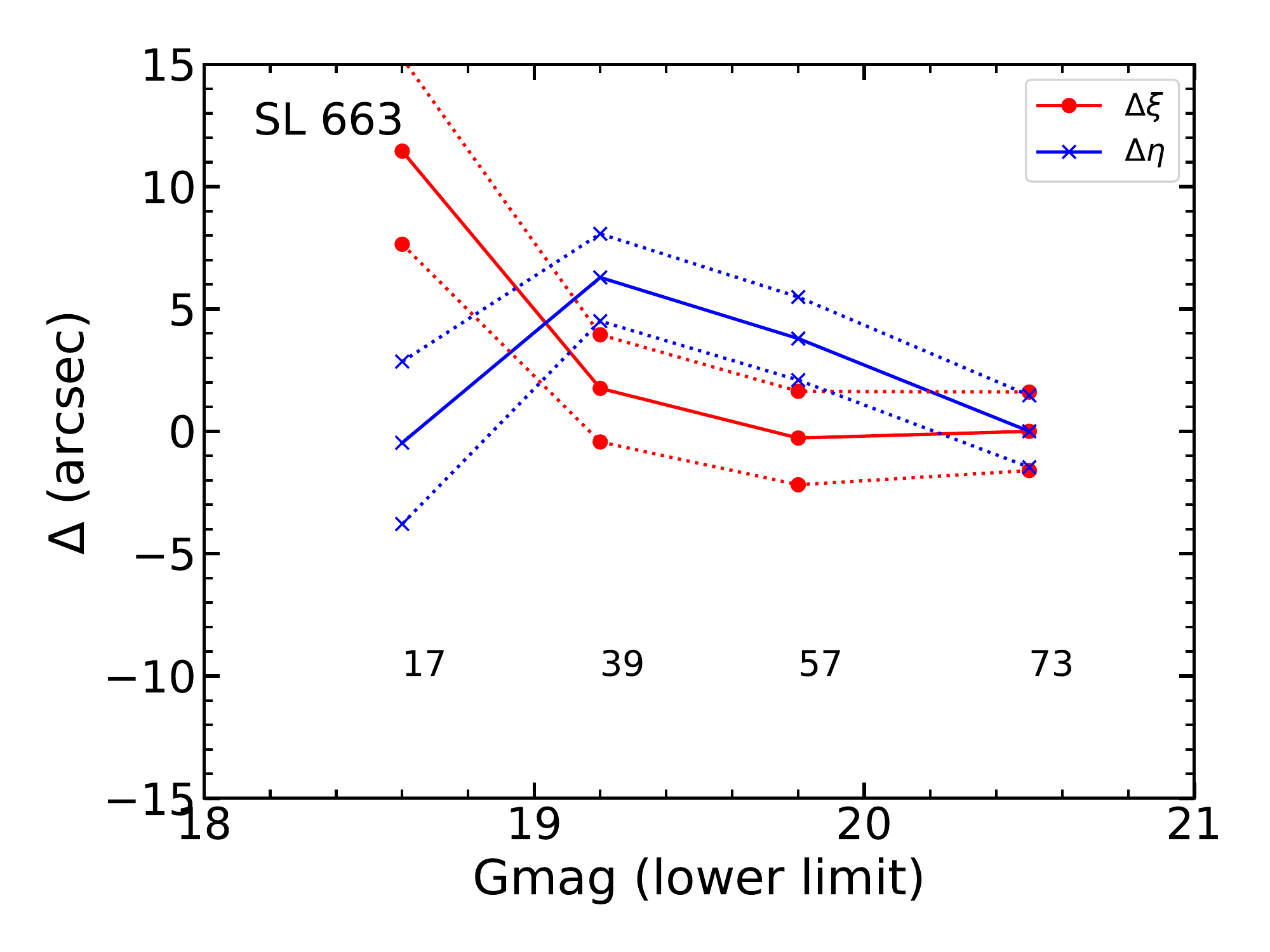}
   \caption{Residuals in RA ($\Delta\xi$, red dots) and DEC ($\Delta\eta$, blue dots) as a function of \textit{Gaia} DR2 limiting magnitude for targets in the MC clusters of this study. Four representative clusters---NGC~416, NGC~1783, NGC~339 and SL~663---are presented to illustrate the stability and precision of the cluster centers determined as described in Appendix~\ref{sec:cluster_center}. In all panels, the solid red and blue lines represents the best-fit value of RA ($\xi$) or DEC ($\eta$), respectively, while the dotted lines show the corresponding 1-$\sigma$ error ranges. The numbers below each set of symbols in each panel give the number stars within 30 arcsec in radius from the cluster center to that magnitude limit.  The upper two clusters have comparatively large samples of stars in the \textit{Gaia} DR2 and their center positions remain fairly stable as fainter targets are introduced.  The lower panels are for sparser clusters for which the center positions vary more strongly with magnitude. These plots suggest that cluster centers determined in this manner become stable when samples of 70 or more stars are used.}
   \label{fig:std_centers}
\end{figure*}

\begin{figure*}
   \centering
    \includegraphics[width=0.47\textwidth]{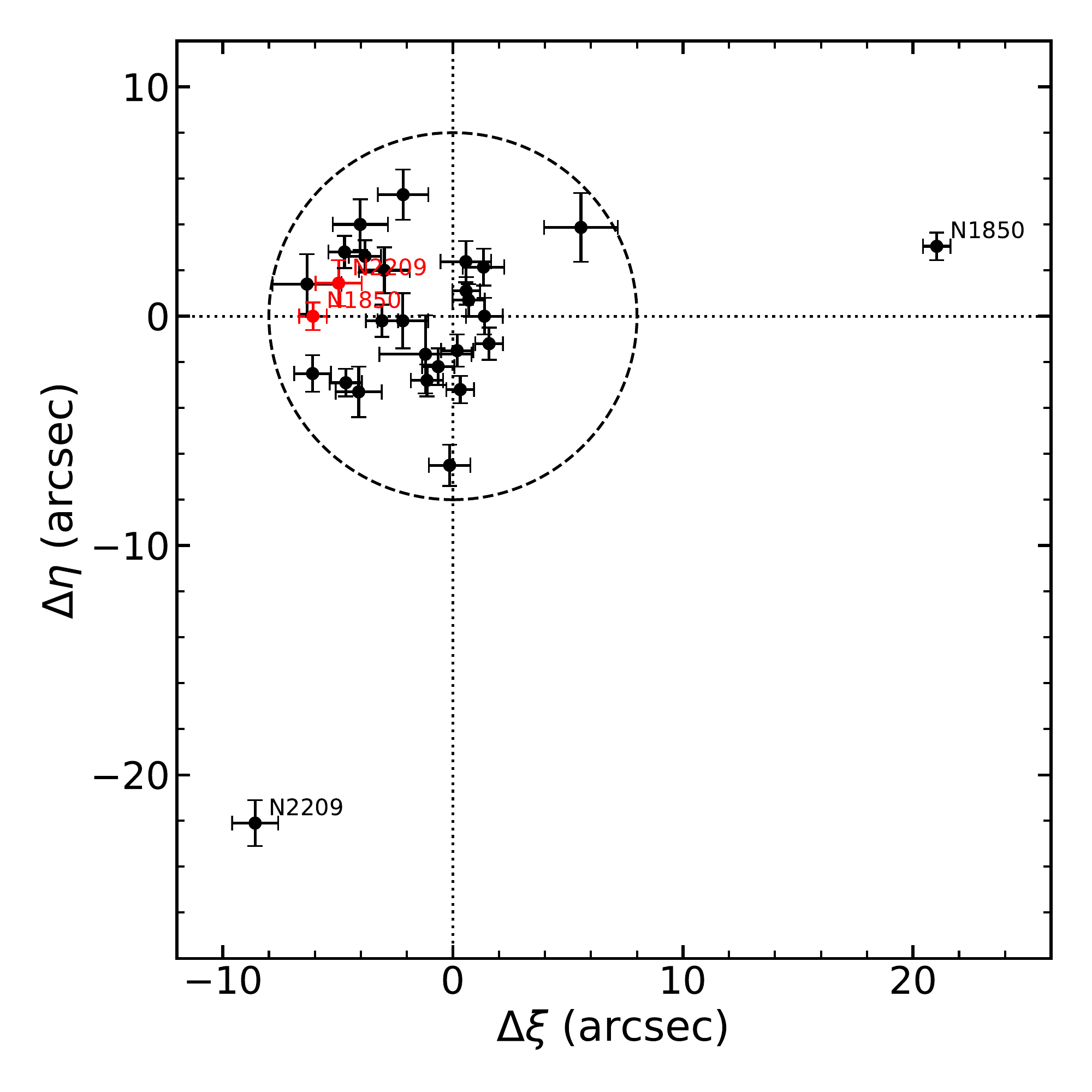}
    \includegraphics[width=0.47\textwidth]{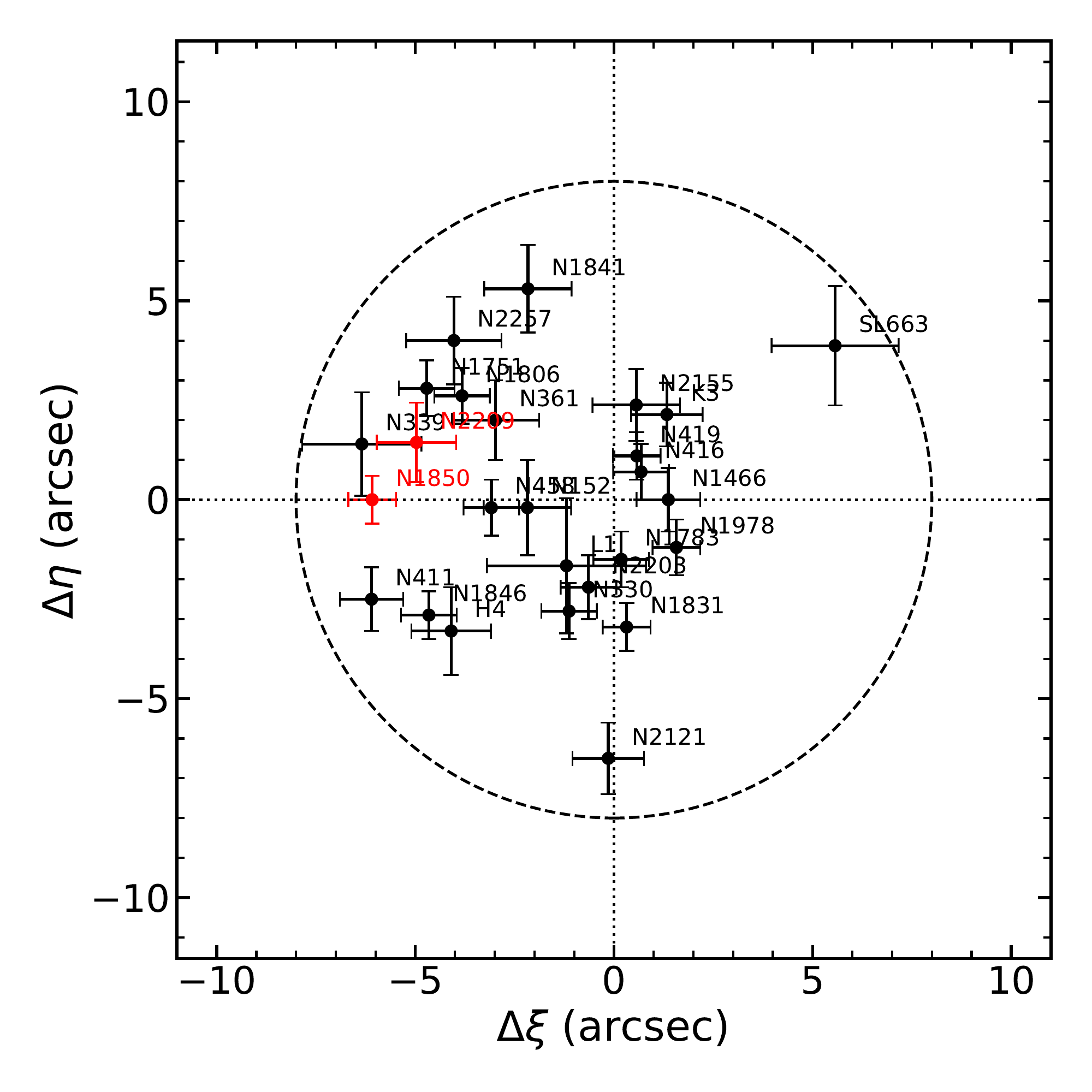}
   \caption{A comparison in the ($\Delta\xi$, $\Delta\eta$) plane of the cluster centers determined using the \textit{Gaia} DR2 catalog (Appendix~\ref{sec:cluster_center}) with centers provided by SIMBAD \citep[][in black]{Wenger:2000aa}, where $\Delta$ is defined as the SIMBAD minus the \textit{Gaia} DR2 center positions. The red dots correspond to our centers compared with center coordinates from other studies (\citealt{Milone:2018aa} for NGC~1850 and \citealt{Mucciarelli:2006aa} for NGC~2209). The circles in both panels are 8 arcsec in radius. The left panel shows all the clusters studied in this work, including the two most deviant cases, NGC~1850 and NGC~2209.  The right panel highlights the distribution of central position residuals for all clusters in our sample. Further details are provided in Appendix~\ref{sec:cluster_center}.}
   \label{fig:comp_centers}
\end{figure*}

As noted in \autoref{sec:cluster_select}, accurate cluster centers are crucial for both photometric and dynamical analysis.  \citet{Carvalho:2008aa} showed that errors in the cluster center can cause significant offsets in the resulting structural parameters, especially the central surface brightness, measured from the surface brightness profile. We explore this further at the end of this section to determine specifically how sensitive our final masses are to the centering errors in the cluster centers. As we shall show, the net impact of centering errors described below appears to be generally small, mostly negligible given other sources of errors for the derived dynamical properties of the clusters in our sample.

Many previous studies have determined cluster centers from photometric data as part of their analysis of the surface brightness profiles \citep[e.g.][]{Carvalho:2008aa, Glatt:2009aa, Mackey:2013aa}. However, data on cluster centers is not available in the literature for all clusters in our sample.  To obtain this information a more convenient approach is to take cluster centers from online astronomical databases \citep[e.g.][]{Mackey:2003aa, Mackey:2003ab}, such as SIMBAD \citep{Wenger:2000aa}. A major limitation of this approach, however, is that the methodology, accuracy and uncertainties of the centers cited in this approach are generally not provided nor are they assured of being internally consistent.   Some center positions listed for NGC~1850 and NGC~2209, for example, appear to be significantly offset from their locations on DSS images.  For this reason, we chose to re-determine the centers of all the clusters in our sample using the \textit{Gaia} DR2, as described next.

For a given cluster, we first selected all stars within 7 arcmin from the center coordinate listed on SIMBAD, brighter than 20.5 mag in $G$-band and with parallax less than 0.2 mas. This selection radius is considerably larger than the core (and often tidal) radii of our clusters.  The selected stars were then used to compute an initial estimate of the position of the cluster center based on the mean positions in RA and Dec.   
The position was improved over four iterations (in a process similar to the description in \citealt{Glatt:2009aa}) as the selection radius was decreased.  For example, a second position estimate was calculated using all stars within a radius of 4 arcmin from the initial guess position.  In subsequent iterations, the selection radii were halved from the previous iteration; the last iteration used a selection radius of 0.5 arcmin (for reference, most core radii of the clusters in our sample are smaller than this value; see \autoref{tab:structure}). The primary reason for simply calculating the mean positions is that the \textit{Gaia} DR2 remains impressively complete near the centers of the target clusters and hence there is a strong positional weighting inherent in the stellar samples used to determine the cluster centers.

We adopted the coordinates of the last iteration for each cluster.  The resulting uncertainties of the cluster centers using this procedure were estimated using the standard deviation of RA or Dec of all stars in the last iteration divided by the square root of the sample size.  These uncertainties are typically between 0.6--2.0 arcsec in both RA and Dec. The adopted cluster centers are listed in columns 3 and 4 of \autoref{tab:structure}.  

To address any bias or funneling effect in our center determination, we re-derived three extra sets of cluster centers with the different lower magnitude limit of stars---instead of 20.5 mag, we also used 18.6, 19.2 and 19.8 mag. \autoref{fig:std_centers} provides some examples of the residuals in RA (red dots), and DEC (blue crosses) from the centers determined using 20.5 mag as the lower magnitude limit. For most clusters (e.g. the top two panels of \autoref{fig:std_centers}), the standard deviations of the four best-fit values in either RA or DEC are below 3 arcsec, indicating good stability. For a few clusters, i.e. Kron~3, Lindsay~1, NGC~339, NGC~361, NGC~411 and NGC~1850, at least one of their standard deviations in RA and Dec falls between 3 and 5 arcsec, which is mostly due to the small number of stars brighter than 18.6 mag (see, e.g. the bottom left panel of \autoref{fig:std_centers}). The worst case is SL~663, for which the values are 6.7 and 4.3 arcsec in RA and DEC, respectively. As shown in the bottom right panel of \autoref{fig:std_centers}, the results can also be attributed to the small number of stars brighter than 18.6 mag. 

\autoref{fig:comp_centers} compares the derived cluster centers with those provided by SIMBAD in the ($\Delta\xi$, $\Delta\eta$) plane. The $\Delta\xi$ and $\Delta\eta$ values represent the positional offsets in RA and DEC, respectively, from the SIMBAD centers to the \textit{Gaia} DR2 centers. For NGC~1850 and NGC~2209, the new centers are offset more than 20 arcsec from the old ones (see the left panel of \autoref{fig:comp_centers}), but they are consistent with the central locations on the DSS images. Except for these two outliers, the remaining cluster center offsets are all within 8 arcsec (see the right panel of \autoref{fig:comp_centers}), which is typically multiple times smaller than the King radii listed in \autoref{tab:structure}, column 8. We have also compared the new centers of NGC~1850 and NGC~2209 with those from other sources not cited by SIMBAD \citep{Milone:2018aa,Mucciarelli:2006aa}, and found that they agreed to the same level of precision as the other clusters in our sample (see the red dots in \autoref{fig:comp_centers}).

To further test the influence of cluster center offsets on the central velocity dispersion, we applied the dynamical analysis described in \autoref{sec:results} to the samples of NGC~419 and NGC~1846 published in \citetalias{Song:2019aa}. Using 100 random positions that are 10 arcsec from the original cluster center, the derived central velocity dispersions are negligibly different from the reported dispersions reported in \citetalias{Song:2019aa}. For this study, we adopt the cluster centers derived from the \textit{Gaia} DR2 data.

\section{Background subtraction}
\label{sec:sky_sub_app}

\begin{figure*}
   \centering
    \includegraphics[width=0.47\textwidth]{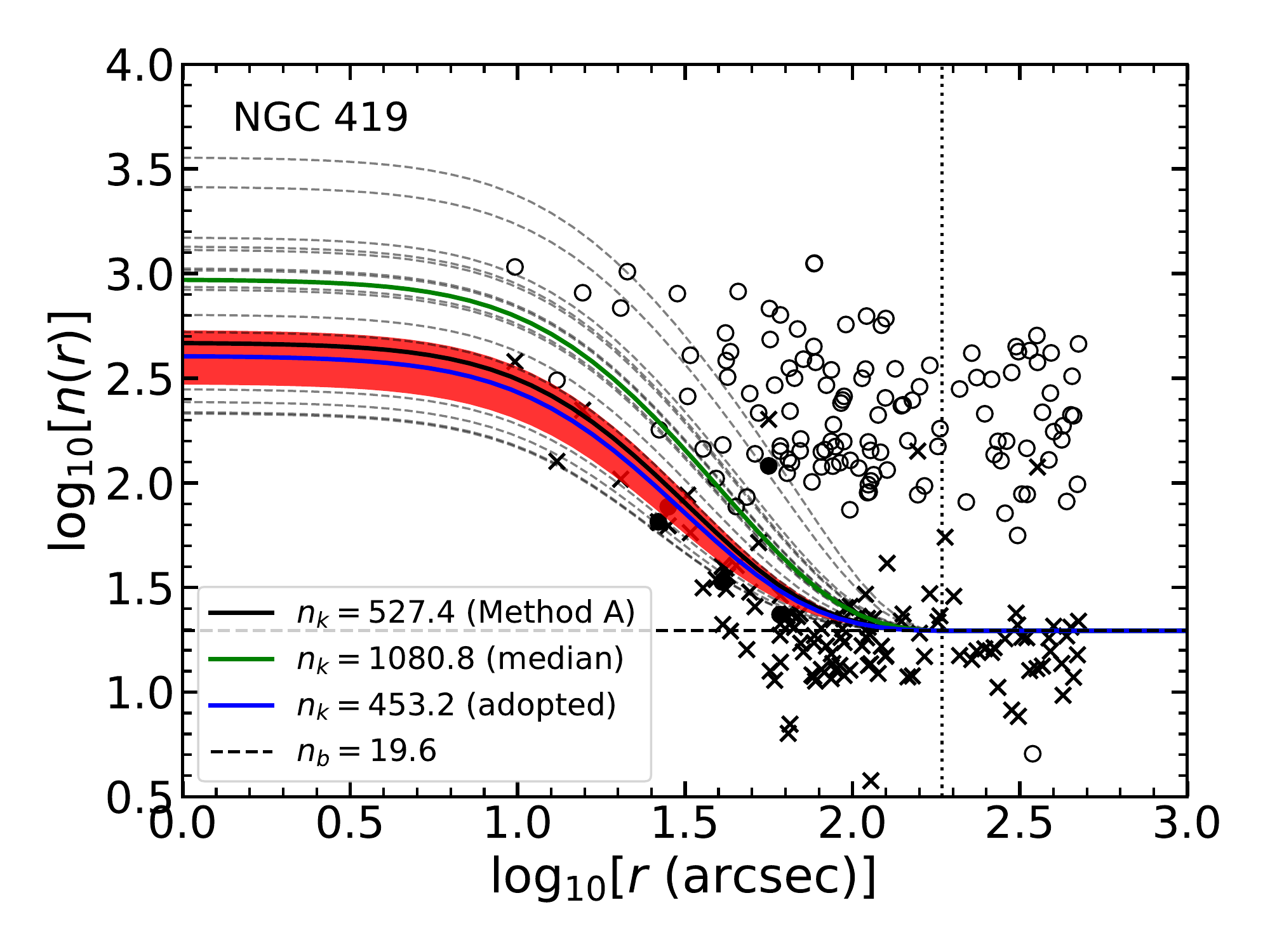}
    \includegraphics[width=0.47\textwidth]{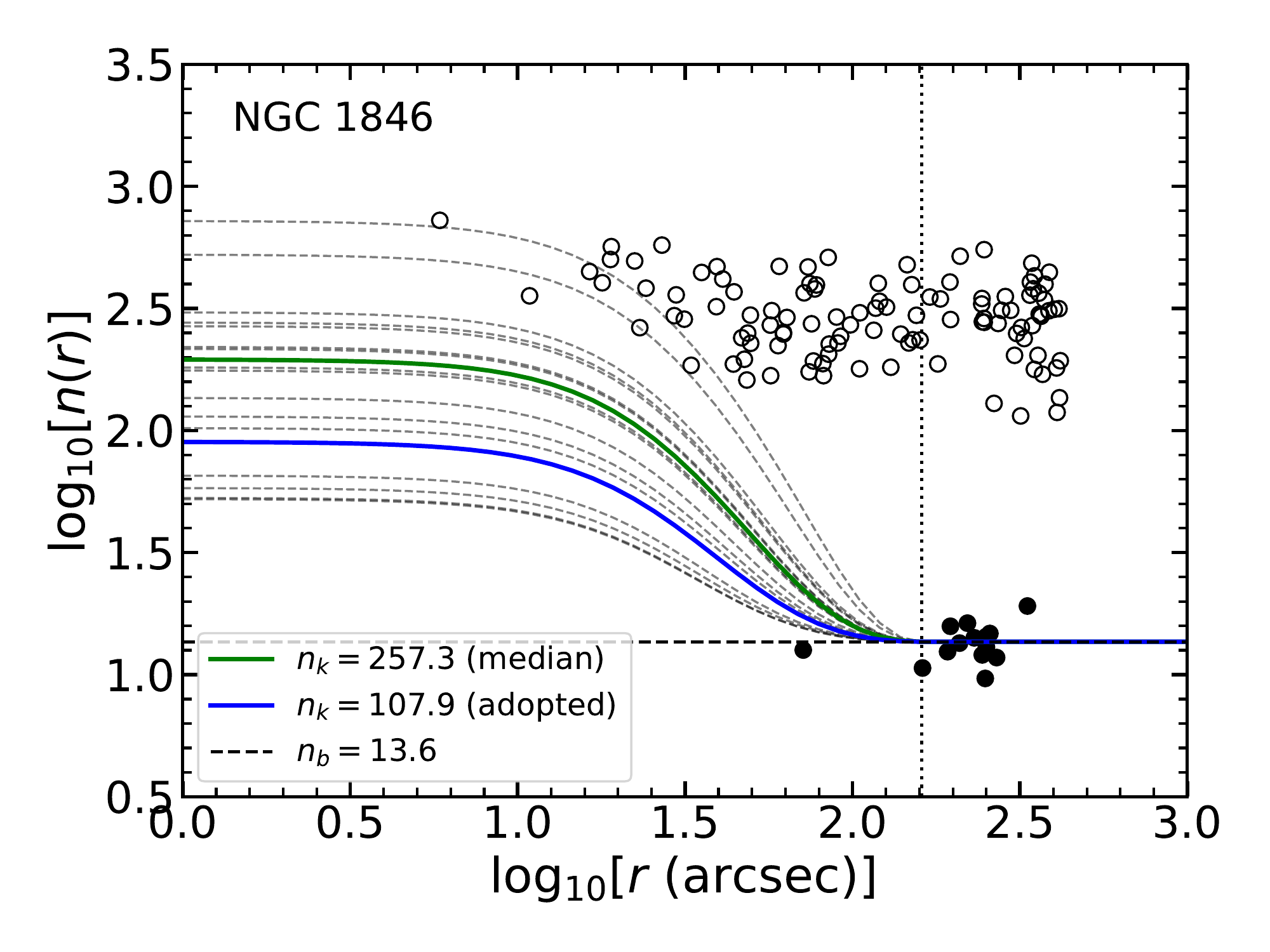}
   \caption{Two examples of background subtraction. In both panels, symbols represent the median total counts of target spectra (open circles) and dedicated sky fibers (filled circles). The crosses in the left panel represent target fibers when the telescope was offset as described in \autoref{sec:sky_obs}.  For NGC~419 ({\it left panel}), the black curve shows the best-fit profile determined using \autoref{eq:K62} (defined as `Method A' in \autoref{sec:sky_sub}) which employs both the dedicated sky fibers and offset-sky observations for that cluster. The y-intercepts of the horizontal dashed line and the solid black curve denote the values of the background and central mean counts ($n_b$ and $n_k$, respectively; see \autoref{eq:K62}) of the fitted profile. The red band shows the central 68-percentile (taken as $\pm 1\sigma$) determined from a bootstrap error estimator that takes into account observational scatter on the estimated value of $n_k$.  The {\it right panel} illustrates the application of \autoref{eq:nk_methodB} (Method B, \autoref{sec:sky_sub}) to our data for NGC~1846. Here, the background mean counts, $n_b$, are determined from the dedicated sky fibers (filled circles) located outside the cluster tidal radius, $r_t$ (denoted as a vertical dotted line for both clusters). The central background profile counts, $n_k$,  are obtained from pairwise applications of \autoref{eq:nk_methodB} with the 17 clusters in the sample for which Method A was applicable (see \autoref{tab:obs}).  The dashed gray profiles each represent one application of \autoref{eq:nk_methodB} to a specific cluster.  The green solid line follows the median (central) curve.   The left panel also shows Method B results applied to NGC~419 where we use the outermost offset-sky fibers to estimate $n_b$ and then apply \autoref{eq:nk_methodB} as for NGC~1846.   Note that the median profile (green line) runs above the black curve obtained using Method A.  When this same procedure is applied to all Method-A clusters, we find the best average agreement between Method A and Method B profiles to correspond to the `minus 1-$\sigma$' profile (the fifth-lowest profile of the 17 plotted, shown as a blue curve in both profiles).
   }
   \label{fig:K62}
\end{figure*}

For Method A, we model the background in the vicinity of a cluster as 
\begin{equation}
    n(r)=n_k\left[\frac{1}{\sqrt{1+(r/r_0)^2}}-\frac{1}{\sqrt{1+(r_t/r_0)^2}}\right]^2+n_b,
\label{eq:K62}
\end{equation}
where $n(r)$ is the median counts obtained from all background sources that contribute to our spectra, $n_k$ is the peak counts that come from a K62 profile with structural parameters equal to those of the cluster being analyzed (see \autoref{tab:structure}), and $n_b$ is the counts the extended background from non-cluster sources (for instance, unresolved light from LMC/SMC field stars, telluric emission,  moonlight).  If spectra from dedicated sky fibers and offset observations are both available, we can solve for $n_k$ and $n_b$ given the cluster structural parameters. The key for this method is to have enough sky positions near the cluster center---typically from the offset observations---to effectively constrain $n_k$. 

Strictly speaking, $n_k$ and $n_b$ in \autoref{eq:K62} should be determined for each wavelength to account for spectral variations of the background with distance from a cluster's center.  In practice, our background sampling is too sparse in the inner regions of nearly all of our target clusters to attempt this.  We therefore took the spectral shape of the background spectrum to be equal to the mean of the spectra from all the dedicated sky fibers (that is, $(n_k/n_b)_\lambda = {\rm Constant}$ for all wavelengths). Clusters for which Method A was applied are flagged with an `A' in \autoref{tab:obs}.

When no offset observations, are available, determining $n_k$ and $n_b$ is less precise due to poor or non-existent sampling of the background near the a cluster's core.  For these cases we used `Method B' (\citetalias{Song:2019aa}) in which $n_k$ is estimated for a given `Target' cluster by assuming that the ratio of this parameter for the cluster divided by the value of the parameter in a `Reference' cluster is the same as the ratio of the photometric central surface brightnesses, $\Sigma_{V,0}$, of the clusters adjusted by exposure time.  That is, 
\begin{equation}
    \frac{n_{k,\, \rm Target}}{n_{k,\,\rm Reference}}=\frac{\Sigma_{V,\,0,\,\rm Target}}{\Sigma_{V,\,0,\,\rm Reference}} \cdot \frac{t_{\rm exp,\,Target}}{t_{\rm exp,\,Reference}}.
 \label{eq:nk_methodB}
\end{equation}
In practice, a Reference cluster is one for which $n_k$ was determined using Method A; if we have $N$ Reference clusters, we can produce a distribution of estimates for $n_{k,\,\rm Target}$.

Method B is somewhat crude in that it ignores possible transparency variations between observations of different clusters and it relies on the precision of the central surface brightnesses of the clusters in the sample, typically about 10-50\%.  But the method has the advantage of being applicable to all the clusters in the sample.  A corollary benefit is that we can use any clusters that are suitable for background determination using \autoref{eq:K62} (Method A) and apply \autoref{eq:nk_methodB} (Method B) as a check on how well the latter method works.

\autoref{fig:K62} shows the application of \autoref{eq:K62} (Method A) to NGC~419, one of the clusters we analyzed in \citetalias{Song:2019aa}.  In this case, $n_b$ and $n_k$ can be determined reliably from dedicated sky fibers and offset observations; the resulting background profile is shown as a black solid curve in the left panel of \autoref{fig:K62}. The right panel of this figure illustrates how we apply Method B (\autoref{eq:nk_methodB}) for the case of NGC~1846, the other cluster in \citetalias{Song:2019aa}.  Here, $n_b$ is well-determined from the dedicated sky fibers (filled circles in \autoref{fig:K62}) in the field surrounding the cluster, while $n_k$ is poorly constrained due to the lack of offset observations to sample the background in the inner parts of the cluster.  In this instance, $n_k$ has been estimated by applying Method B (\autoref{eq:nk_methodB}) in a pairwise manner to all 17 clusters for which Method A could be applied.  These background profiles are shown as dashed gray solid lines.  These profiles all have the same value of $n_b$ as derived from the NGC~1846 dedicated sky fibers located far from the cluster center, but all have distinct values of $n_k$ values derived by applying \autoref{eq:nk_methodB}. 

As previously noted, we can apply Method B to clusters for which Method A is also applicable.  In the case of NGC~419 (left panel, \autoref{fig:K62}), the median background profile of the set of profiles obtained using Method B (shown as a green solid curve) runs significantly above the one profile obtained using Method A (black line).  Applying this test to all the clusters in our sample for which Method A was applicable, we found that the background profile (shown as the blue solid curve in \autoref{fig:K62}) was consistently located below the median profile. 

\section{Rejected and/or Unusual M2FS spectra}

\subsection{Examples of M2FS spectra rejected by the SK cut}
\label{sec:spec_PSK}

\begin{figure}
   \centering
   \includegraphics[width=0.47\textwidth]{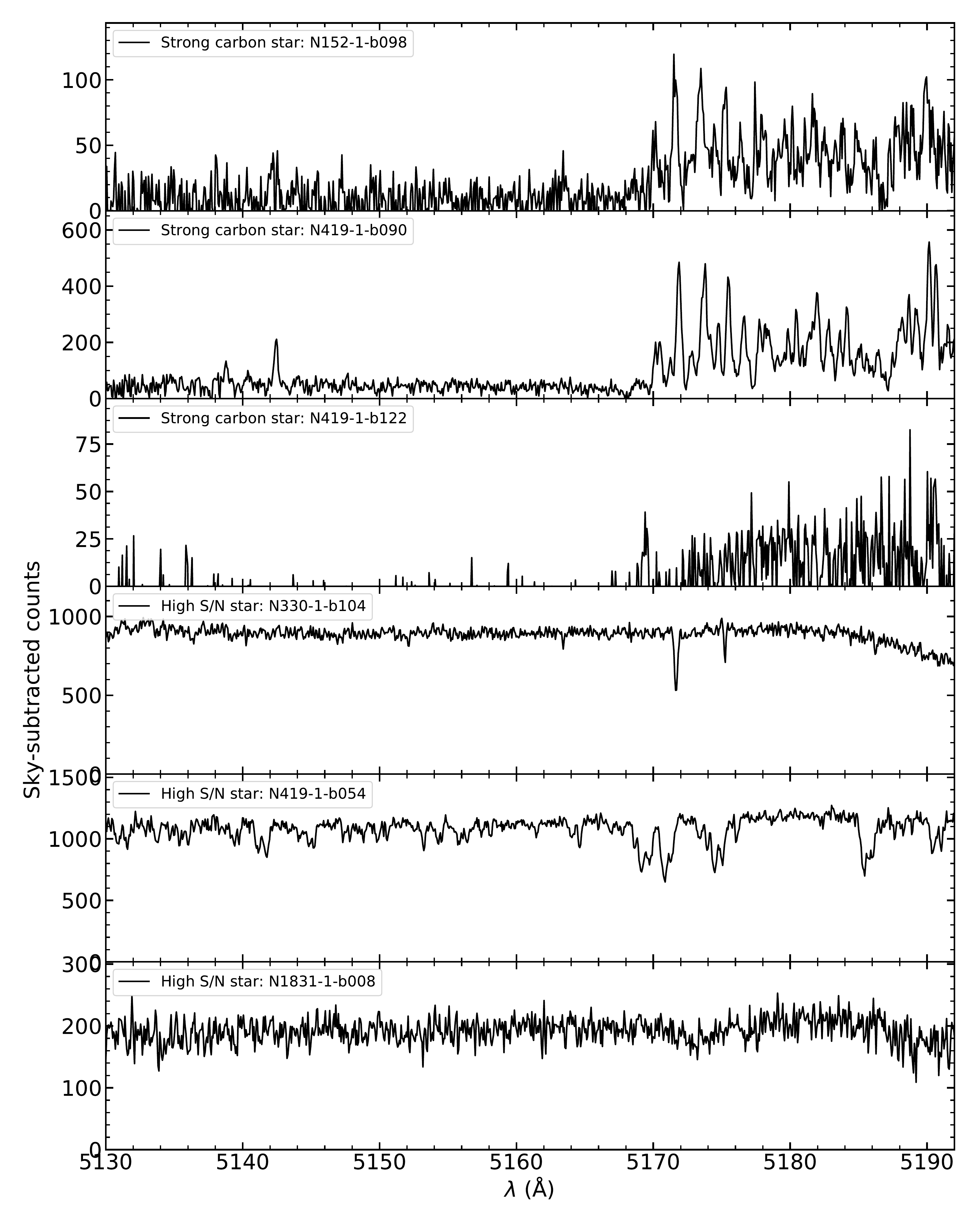}
   \caption{Examples of M2FS spectra of stars rejected by the SK cut described in \autoref{sec:rm_PSK}. The three top panels appear to be strong C stars at low S/N (more on these in \autoref{sec:rm_odd}).  The bottom three panels show stars that have high mean S/N values, but were still rejected by the SK cut due to having too few or blended lines.}
   \label{fig:spec_PSK}
\end{figure}

As described in \autoref{sec:rm_PSK}, all but three of the stars rejected for skew/kurtosis have low average S/N of $1.2 \pm 0.1$. Among these remaining 68 stars, three appear to have spectra consistent with those of carbon (C) stars (more on these in Appendix~\ref{sec:spec_odd}) at low average S/N ($\sim$1.2 for these three; see \autoref{fig:spec_PSK}). The additional three spectra flagged in the SK cut but that have fairly high mean S/N ($25\pm 8$) either contain very few identifiable spectral features or possibly have broadened or blended features.  Not surprisingly, the Bayesian analysis was unable to settle on single well-defined LOS velocities in these cases, resulting in large skew/kurtosis indices. For reference, the spectra of these three higher S/N objects that failed the `SK cut' are also shown in \autoref{fig:spec_PSK}.  

\begin{figure}
   \centering
    \includegraphics[width=0.47\textwidth]{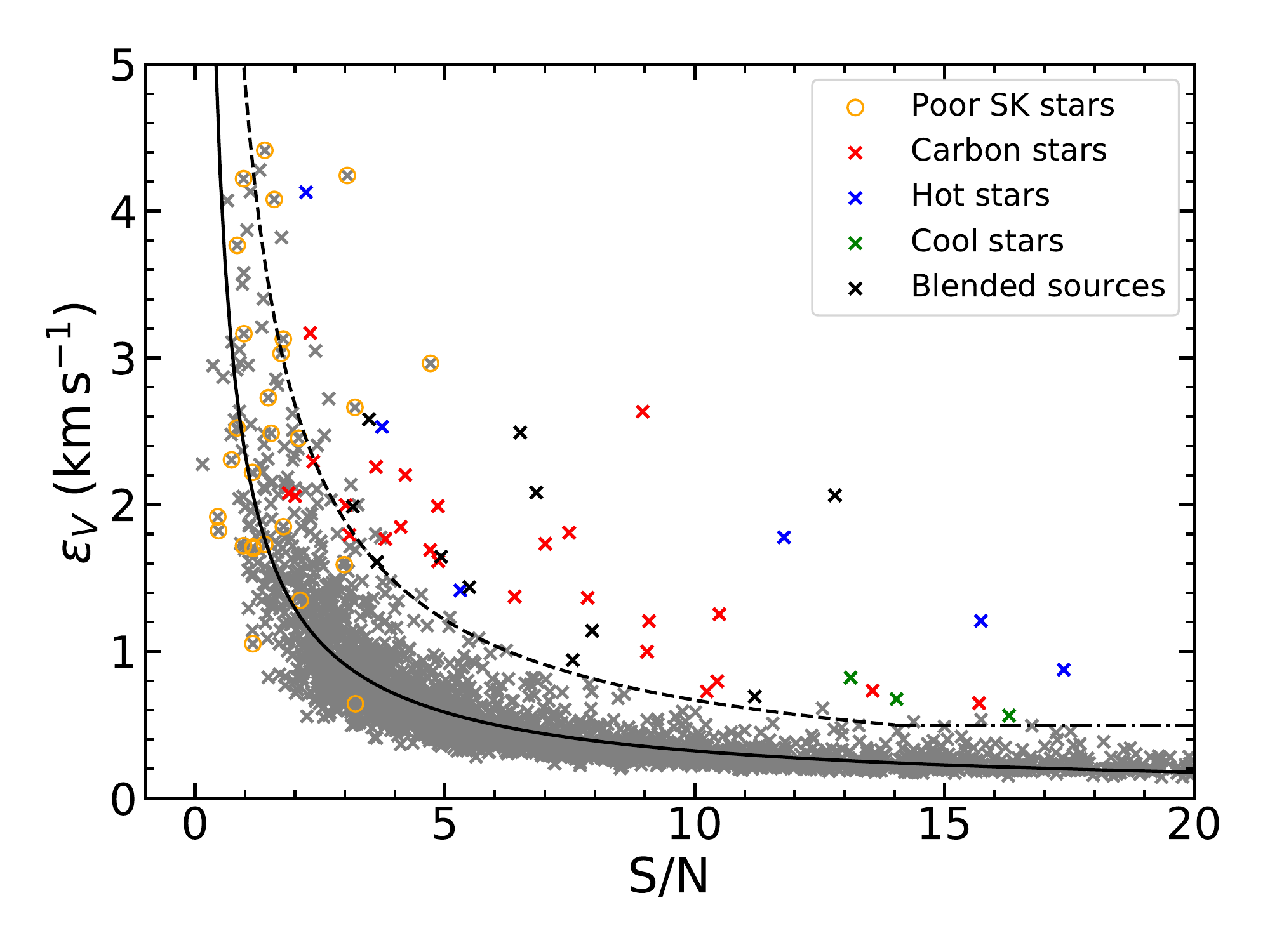}     
   \caption{A similar plot as \autoref{fig:e_vlos}, but here in linear space and with anomalous stars highlighted by colored symbols. The meanings of these symbols are given in the legend. Different symbols correspond to different types of anomalous stars identified in \autoref{sec:rm_odd}.  The cases denoted as `poor SK stars' are discussed in \autoref{sec:rm_PSK}.}
   \label{fig:e_vlos_odd}
\end{figure}

\begin{figure}
   \centering
   \includegraphics[width=0.47\textwidth]{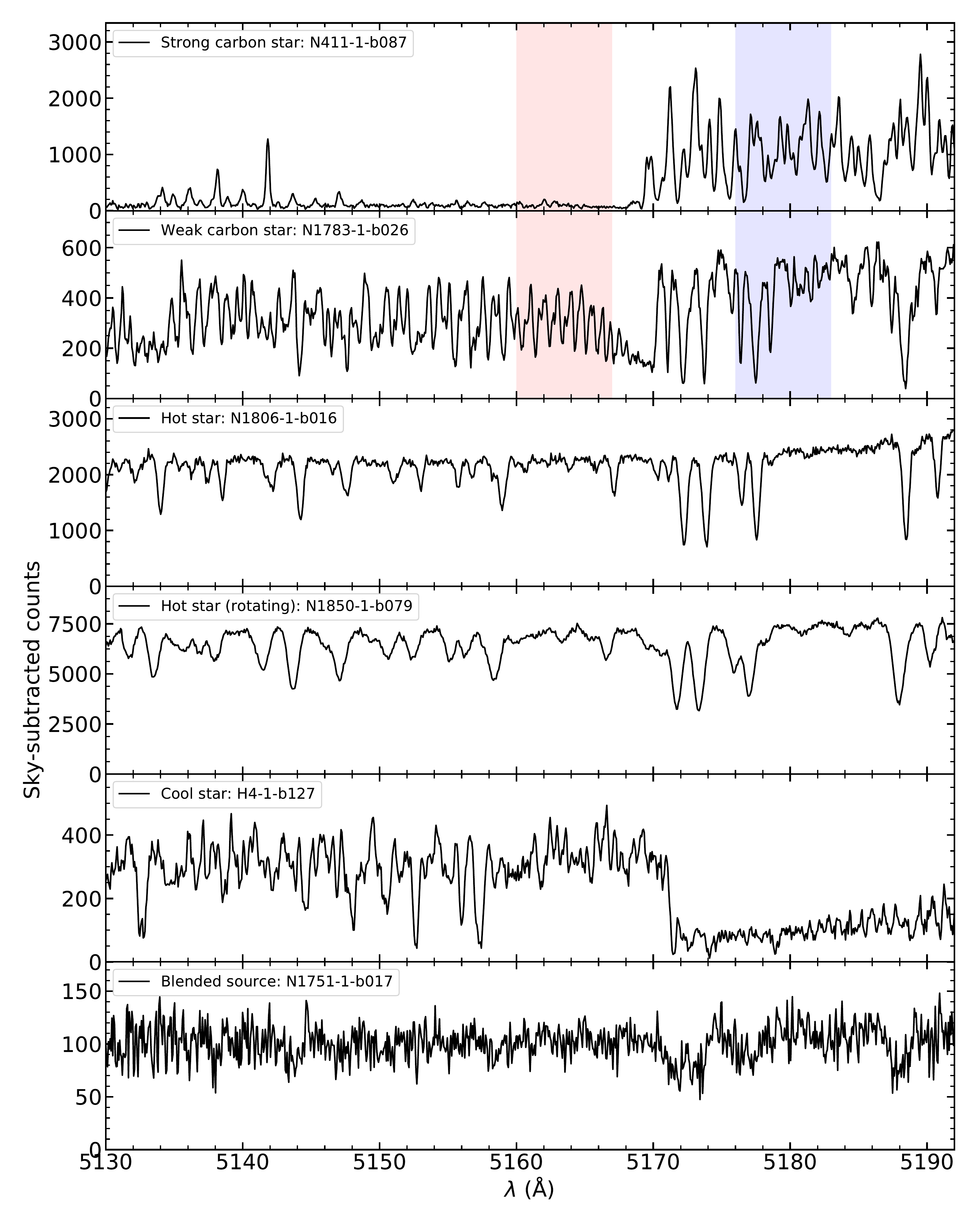}
   \caption{Examples of M2FS spectra of stars with anomalous velocity uncertainties as identified in \autoref{fig:e_vlos_odd}. From top to bottom, we show the spectra of a strong C star, a weak C star, a star with high surface temperature (hot star), a similarly hot star but with moderate rotation ($v \sin{i} \sim 30$~\kms), an M-type star with low surface temperature (cool star), and a blended source. For the two C stars, we indicate the two bands used to identify such stars as outlined in Appendix~\ref{sec:spec_odd}. A total of 62 targets with anomalous velocity uncertainties (as described in \autoref{sec:rm_PSK} and identified in \autoref{fig:e_vlos_odd}) are excluded from our cluster dynamical analysis. }
   \label{fig:spec_odd}
\end{figure}

\begin{figure}
   \centering
   \includegraphics[width=0.47\textwidth]{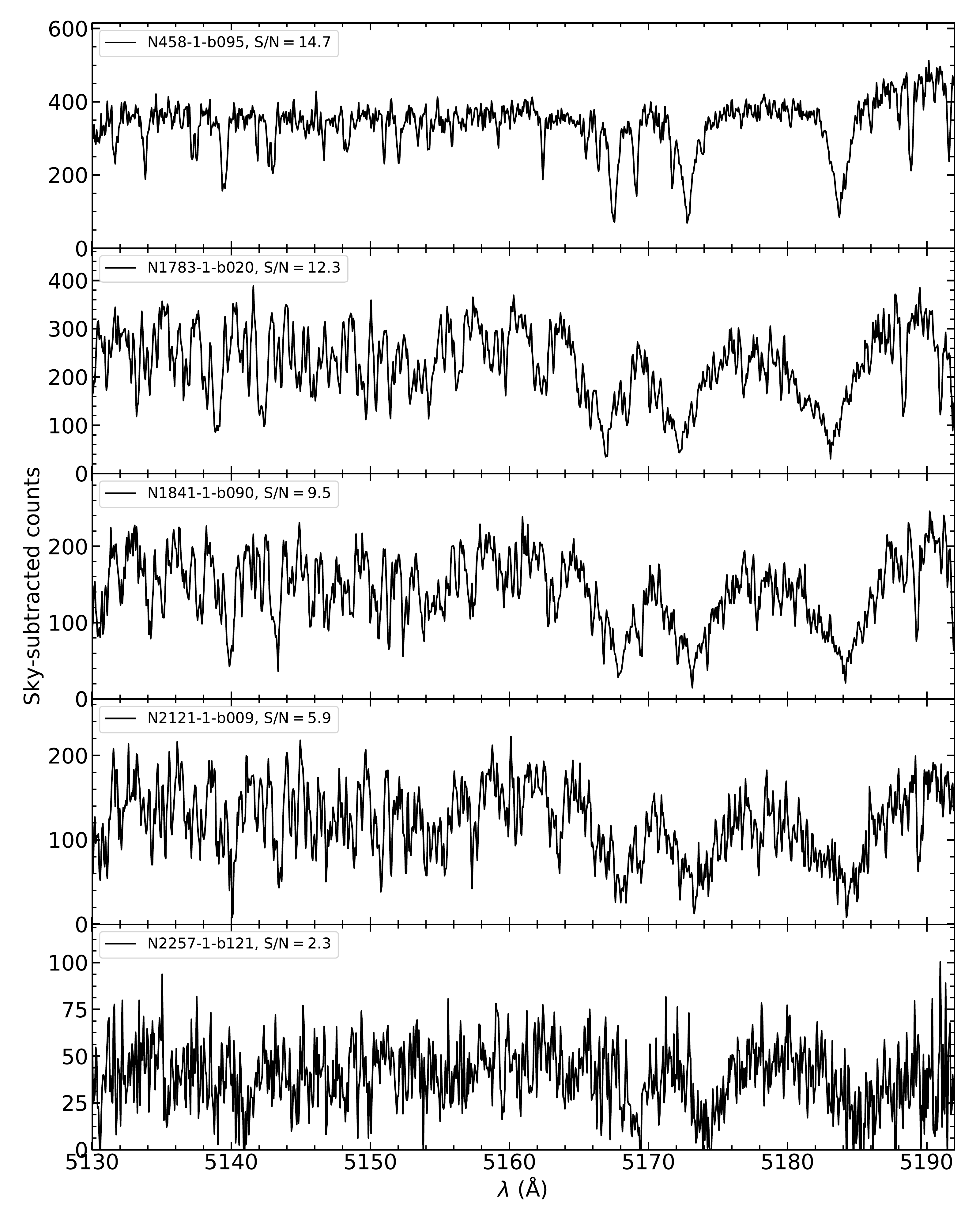}
   \caption{Example M2FS spectra of dwarf stars confirmed in our sample. From top to bottom, we show dwarf stars spanning a similar S/N range as those giants shown in \autoref{fig:spec_sample}; these dwarf stars and giants also have comparable colors. As discussed in \autoref{sec:rm_dwarf} and \autoref{fig:hist_logg}, 84 dwarf stars with spectra similar to these were removed from the sample used to for the dynamical analyses the clusters in this study.}
   \label{fig:spec_dwarf}
\end{figure}

\subsection{Examples of M2FS spectra with anomalous velocity uncertainties}
\label{sec:spec_odd}

As discussed in \autoref{sec:rm_odd}, we inspected each of the 62 stars with anomalous velocity uncertainties and found that we could classify them into the following distinct categories:

\vskip1em

\noindent {\it Carbon Stars}. A total of 21 C stars were identified based on the presence of a clear C$_2$ Swan band feature at $\lambda$5165 \AA\ \citep{Johnson:1927aa,King:1948aa}.  Among these stars, 17 have quite strong absorption features (the first spectrum in \autoref{fig:spec_odd} shows a typical example), while the remaining four C stars have relatively weak Swan-band absorption (the second spectrum in \autoref{fig:spec_odd} shows a typical case).
Since the spectral library we are using \citep{Lee:2008aa, Lee:2008ab} does not model C stars, it is unsurprising that our Bayesian analysis returned LOS velocities with large uncertainties relative to other stars at similar S/N.

\vskip0.5em

\noindent {\it Hot Stars}.  A total of eight stars appear to be hot stars that were forced to an incorrect low temperature using the method to fix $T_{\rm eff}$ described in \autoref{sec:Teff}.  We chose not to single out these stars and allow their $T_{\rm eff}$ to float as this would be inconsistent with the strategy described in \autoref{sec:Teff}. Two examples, one also exhibiting moderate rotation, are shown in \autoref{fig:spec_odd}. 

\vskip0.5em

\noindent {\it Cool Stars}. A total of three stars appear to be cool giant stars with the feature of TiO bands above $\sim$5180 \AA\ (the fifth spectrum in \autoref{fig:spec_odd} shows a typical example). Similar to hot stars, cool stars were forced to an incorrect temperature---in this case too high---as described in \autoref{sec:Teff}. Moreover, the spectral library we used does not contain stars sufficiently cool that exhibit TiO bands. For these reasons, it is likely unsurprising that the Bayesian analysis returned poor LOS velocity uncertainties for these cases.

\vskip0.5em

\noindent {\it Blended Sources}. A total of 12 stars are confirmed as sources with the features of more than one star (see the sixth spectrum in \autoref{fig:spec_odd}). The spectra in this category provide poor LOS velocities for a given S/N.  We did not carry out dual-star fits as in \citetalias{Song:2019aa} because these stars represent a fairly small sample that may have a different error distribution from the single-star fitted spectra.

\vskip0.5em

\noindent {\it Statistical Tail}.  A total of 18 stars appear to represent the tail of the distribution of the velocity errors relative to the fit line/power law evident in \autoref{fig:e_vlos_odd}.   These stars represent about 0.6\% of the total sample of spectra, roughly consistent with what a $3\sigma$ cut would achieve for a normal distribution.

\vskip1em

Due to the relatively high frequency of C stars in our sample of rejected spectra, we developed a more objective means of identifying candidate C stars regardless of where they lie in \autoref{fig:e_vlos_odd}. Specifically, we measured the ratio of the fluxes in all the spectra of our sample on the red and blue sides of the C$_2$ band head (the first two spectra in \autoref{fig:spec_odd} illustrate where these bands are located).  This ratio cleanly identified all C stars in the velocity-error rejection region of \autoref{fig:e_vlos_odd}, as well as seven apparent C stars below the rejection boundary.  Four of these are are strong C stars similar to the top spectrum in \autoref{fig:spec_odd}, and three are weaker C stars with spectra similar to the second panel of this figure. For consistency, we have removed these four strong C-stars from our sample but have retained the three weak C-stars since they show clean atomic features similar to non-C stars of similar color.  

This method of identifying C stars found numerous even weaker C stars.  We have chosen to accept these in the present sample as they are mildly affected by the C$_2$ features and in all cases their spectra appear to provide good LOS velocity estimates. We plan to discuss this expanded C-star sample in more detail in a future paper.  

\subsection{Examples of M2FS spectra from dwarf stars confirmed in our sample.}
\label{sec:spec_dwarf}
In \autoref{sec:rm_dwarf}, we have removed 84 stars with surface gravity parameters $\log{g}\geq3.2$.
The distinctive spectral signatures of these stars are evident by comparing the dwarf-star spectra in \autoref{fig:spec_dwarf} with spectra of giants of similar colors and S/N ratios shown in \autoref{fig:spec_sample}. As expected, the proper motions of these stars, as provided in \textit{Gaia} DR2, exhibit a large spread that is offset from the mean proper motions of the MCs.



\bsp	
\label{lastpage}
\end{document}